\documentclass[openright,twoside,11pt]{book}

\pdfoutput=1

\usepackage[USletter]{vmargin} 

\usepackage[utf8]{inputenc}           
\usepackage{amsmath}
\usepackage{amsfonts}
\usepackage{amssymb}                    
\usepackage{amsthm}
\usepackage{bbm}
\usepackage{graphicx}                   
\usepackage{nonfloat}
\usepackage{fancyhdr}                   
\usepackage{makeidx}                    
\usepackage{tocbibind}                  
\usepackage{hyperref}

\hypersetup{%
pdftitle  = {Symmetric extension of bipartite quantum states and its use in quantum key distribution with two-way postprocessing},
pdfauthor = {Geir Ove Myhr},
pdfkeywords = {quantum state, symmetric extension, quantum key distribution, QKD, extendible, extendable, extensible, extendibility, extendability, extensibility, Bell-diagonal, semidefinite programming, advantage distillation},
colorlinks,
linkcolor=linkblue,
filecolor=linkblue,
urlcolor=linkblue,
citecolor=linkblue,
bookmarksopen=true,
bookmarksopenlevel=0,
bookmarksnumbered=true
}
\makeindex

\setmargins{40mm}{22mm}{139mm}{204mm}{14pt}{11mm}{0pt}{11mm}

\newcommand{\captionfonts}{\small}
\makeatletter  
\long\def\@makecaption#1#2{%
  \vskip\abovecaptionskip
  \sbox\@tempboxa{{\captionfonts #1: #2}}%
  \ifdim \wd\@tempboxa >\hsize
    {\captionfonts #1: #2\par}
  \else
    \hbox to\hsize{\hfil\box\@tempboxa\hfil}%
  \fi
  \vskip\belowcaptionskip}
\makeatother   

\newcommand{\im}{\mathrm{i}}
\newcommand{\e}{\mathrm{e}}

\newcommand{\diff}[2]{\frac{\mathrm{d}#1}{\mathrm{d}#2}}

\newcommand{\etal}{{\it et al.~}}

\input{Qcircuit}

\renewcommand{\ket}[1]{|#1\rangle}
\renewcommand{\bra}[1]{\langle #1 |}
\newcommand{\proj}[1]{\ket{#1}\bra{#1}}
\newcommand{\ketbra}[2]{\ket{#1}\bra{#2}}
\newcommand{\braket}[2]{\langle #1 | #2 \rangle}

\newcommand{\calone}{\mathbbm{1}}
\newcommand{\oa}{\ensuremath{\oplus}}
\newcommand{\ob}{\ensuremath{\ominus}}
\newcommand{\oc}{\ensuremath{\odot}}
\newcommand{\od}{\ensuremath{\oslash}}
\newcommand{\of}{\ensuremath{\otimes}}
\newcommand{\og}{\ensuremath{\circledcirc}}
\newcommand{\oh}{\ensuremath{\circledast}}
\newcommand{\oi}{\ensuremath{\circledS}}
\newcommand{\oj}{\ensuremath{\copyright}}

\newcommand{\emphindex}[1]{\emph{#1}\index{#1}}
\newcommand{\textindex}[1]{#1\index{#1}}
\newcommand{\swap}{P_{BB^\prime}}
\newcommand{\swapBE}{P_{BE}}
\newcommand{\spec}[1]{\vec{\lambda}(#1)}

\DeclareMathOperator{\tr}{tr}
\DeclareMathOperator{\rank}{rank}

\DeclareMathOperator{\mathspan}{span}
\DeclareMathOperator{\diag}{diag}

\newtheorem{theorem}{Theorem}
\newtheorem{lemma}[theorem]{Lemma}
\newtheorem{corollary}[theorem]{Corollary}
\newtheorem{proposition}[theorem]{Proposition}
\newtheorem{conjecture}[theorem]{Conjecture}

\theoremstyle{definition}
\newtheorem{defn}[theorem]{Definition}
\newtheorem{example}[theorem]{Example}
\newtheorem{remark}[theorem]{Remark}

\begin{document}

\definecolor{linkblue}{rgb}{0,0,0.7}
\definecolor{linkred}{rgb}{0.7,0,0}
\definecolor{linkgreen}{rgb}{0,0.5,0}

\index{density matrix|see{density operator}}
\index{EPR pair|see{Bell state}}
\index{Local Operations and Classical Communication|see{LOCC}}

\index{RCAD|see{repetition code advantage \linebreak distillation}}
\index{LAD|see{linear advantage distillation}}
\index{SDP|see{semidefinite program}}
\index{echelon form|see{row-reduced echelon \linebreak form}}
\index{QKD|see{quantum key distribution}}
\index{positive operator-valued measure|see{\linebreak POVM}}
\index{i.i.d.|see{independent and identically \linebreak distributed}}



\frontmatter


\thispagestyle{empty}
\pdfbookmark[0]{Front}{frontpage}

\noindent
\begin{center}
\vspace*{3cm}
\noindent
{\huge
    \begin{center}
      \mbox{Symmetric extension of bipartite quantum states}\\
      and its use in quantum key distribution\\
      with two-way postprocessing
    \end{center}
}

\vspace{1cm}

\noindent
{\LARGE
    \begin{center}
      Der Naturwissenschaftlichen Fakult{\"a}t\\
      der Friedrich-Alexander-Universit{\"a}t Erlangen-N{\"u}rnberg\\
      zur\\
      Erlangung des Doktorgrades Dr.~rer.~nat.
    \end{center}
}

\vspace{3cm}

\noindent
{\LARGE
    \begin{center}
      vorgelegt von\\
      Geir Ove Myhr\\
      aus Oslo
    \end{center}
}

\end{center}

\pagebreak
\thispagestyle{empty}

\noindent

{\large
\begin{flushleft}
Als Dissertation genehmigt von der \\
Naturwissenschaftlichen Fakult{\"a}t der \\
Universit{\"a}t Erlangen-N{\"u}rnberg

\vspace{\stretch{1}}

\begin{tabular}{ll}
Tag der m{\"u}ndlichen Pr{\"u}fung: &       08/11/2010\\
Vorsitzender der Promotionskommission: &    Prof. Dr. Rainer Fink\\
Erstberichterstatter: &                     Prof. Dr. Norbert L{\"u}tkenhaus\\
Zweitberichterstatter: &                    Prof. Dr. Andreas Knauf
\end{tabular}
\end{flushleft}
} 

\cleardoublepage
\pagestyle{fancy}
\chapter{Summary}
\markboth{Summary}{} 

Just like we can divide the set of bipartite quantum states into separable states and entangled states, we can divide it into states with and without a symmetric extension.
The states with a symmetric extension---which includes all the separable states---behave classically in many ways, while the states without a symmetric extension---which are all entangled---have the potential to exhibit quantum effects.
The set of states with a symmetric extension is closed under local quantum operations assisted by one-way classical communication (1-LOCC) just like the set of separable states is closed under local operations assisted by two-way classical communication (LOCC).
Because of this, states with a symmetric extension often play the same role in a one-way communication setting as the separable states play in a two-way communication setting.

We show that any state with a symmetric extension can be decomposed into a convex combination of states that have a \emph{pure} symmetric extension.
A necessary condition for a state to have a pure symmetric extension is that the spectra of the local and global density matrices are equal.
This condition is also sufficient for two qubits, but not for any larger systems.

We present a conjectured necessary and sufficient condition for two-qubit states with a symmetric extension. 
Proofs are provided for some classes of states: rank-two states, states on the symmetric subspace, Bell-diagonal states and states that are invariant under $S^\dagger \otimes S$, where $S$ is a phase gate.
We also show how the symmetric extension problem for multi-qubit Bell-diagonal states can be simplified and the simplified problem implemented as a semidefinite program.

Quantum key distribution protocols such as the six-state protocol and the BB84 protocol effectively gives Alice and Bob Bell-diagonal states that they measure in the standard basis to obtain a raw key which they may then process further to obtain a secret error-free key.
When the raw key has a high error rate, the underlying Bell-diagonal state has a symmetric extension that must be broken with a two-way advantage distillation step before one-way processing can finish the job and generate a secret key.
We analyze the currently used advantage distillation step and some generalizations of it and show that all these steps fail to break the symmetric extension for states at the current key distillation threshold.
This shows that these generalizations cannot improve the currently best known threshold of the six-state protocol from 27.6 \%.

\chapter{Zusammenfassung}
\markboth{Zusammenfassung}{} 

Analog zur Spaltung der Menge von zweiteiligen Quantenzuständen in separable und verschränkte Zustände können wir diese Menge auch in Zustände mit und ohne eine symmetrische Erweiterung spalten.
Die Zustände mit einer symmetrischen Erweiterung --- die alle separablen Zustände umfassen --- 
verhalten sich in vieler Hinsicht klassisch, 
während die Zustände ohne eine symmetrische Erweiterung 
--- die alle verschränkt sind ---
das Potenzial haben, Quanteneffekte zu zeigen.
Die Menge von Zuständen mit symmetrischer Erweiterung ist unter lokalen Quantenoperationen und klassischer Einwege-Kommunikation (1-LOCC) abgeschlossen, 
genau wie die Menge der separablen Zustände unter lokalen Quantenoperationen und klassischer Zweiwege-Kommunikation (LOCC) abgeschlossen ist.
Aus diesem Grund spielen Zustände mit einer symmetrischen Erweiterung häufig die gleiche Rolle in einem Aufbau mit Einwege-Kommunikation 
wie separable Zustände in einem Aufbau mit Zweiwege-Kommunikation.

Wir zeigen, dass jeder Zustand mit einer symmetrischen Erweiterung in eine konvexe Kombination von Zuständen, die eine \emph{reine} symmetrische Erweiterung haben, zerlegt werden kann.
Eine notwendige Bedingung für einen Zustand, eine reine symmetrische Erweiterung zu haben, ist, dass die Spektren der lokalen und globalen Dichtematrizen gleich sind.
Für zwei Qubits, aber nicht f\"ur größere Systeme, ist diese Bedingung auch hinreichend.

Wir stellen eine vermutete notwendige und hinreichende Bedingung f\"ur zwei-Qubit-Zustände mit einer symmetrischen Erweiterung auf.
Beweise werden für einige Klassen von Zuständen angeführt: Zustände von Rang zwei, Zustände auf dem symmetrischen Unterraum, Bell-diagonale Zustände und Zustände, die invariant unter $S^\dagger \otimes S$ sind, wobei $S$ ein Phasengatter ist.
Wir zeigen auch, wie das Problem der symmetrischen Erweiterungen für multi-Qubit Bell-diagonale Zustände vereinfacht und wie das vereinfachte Problem als semidefinites Programm implementiert werden kann.

Protokolle für Quantenschlüsselverteilung, wie das \emph{six-state}-Protokoll und das BB84-Protokoll,
geben Alice und Bob effektiv Zugriff auf Bell-diagonale Zustände.
Diese messen sie dann in der Standardbasis, um einen Rohschlüssel zu erhalten,
welchen sie weiterverarbeiten können, um einen geheimen, fehlerfreien Schlüssel zu erhalten.
Wenn der Rohschlüssel eine hohe Fehlerrate aufweist, hat der zugrunde liegende Bell-diagonale Zustand eine symmetrische Erweiterung, 
die mit Zweiwege-Vorteilsdestillation gebrochen werden muss, bevor durch weiterführende Einwege-Kommunikation ein geheimer Schl\"ussel generiert werden kann.

Wir analysieren die derzeit verwendeten Vorteilsdestillations-Protokolle und ei\-ni\-ge Ver\-all\-ge\-mei\-ne\-rungen 
und zeigen, dass keins dieser Protokolle für Zustände an den aktuellen Schlüsseldestillationschwellen die symmetrische Erweiterung brechen kann.
Dies zeigt, dass diese Verallgemeinerungen die derzeit höchste Schwelle des \emph{ six-state}-Protokolls von 27.6\% nicht verbessern können.

\cleardoublepage
\chapter*{Preface}
\addcontentsline{toc}{chapter}{Preface} 
\markboth{Preface}{} 
\label{ch:preface}

Most of the material included in this thesis can be found in two papers published in Phys. Rev. A and one upcoming paper. 
In the first paper \cite{myhr09a}, written together with Andrew Doherty, Joe Renes and Norbert L{\"u}tkenhaus, the symmetric extension problem is solved analytically for (two-qubit) Bell-diagonal states. 
These results can be found in this thesis in sections \ref{sec:simplifying} and \ref{sec:sdpimplementation} where we have generalized the setting to multi-qubit Bell-diagonal states and sections \ref{sec:belldiagsymext-symmetry} and \ref{sec:belldiagsymext-sdp}.
This paper also covers the results form Chapter \ref{ch:rcad} which says that at or above the threshold introduced by Chau \cite{chau02a}, repetition code advantage distillation will produce states with a symmetric extension.
In the second paper \cite{myhr09b}, together with Norbert L{\"u}tkenhaus, two-qubit states with a symmetric extension are analyzed. 
Section \ref{sec:symext-properties} and chapters \ref{ch:twoqubits}--\ref{ch:degradable} come from this paper.
The upcoming paper covers sections \ref{sec:simplifying} and \ref{sec:sdpimplementation} in its current form, chapters \ref{ch:lad}, \ref{ch:k2analytical}, and \ref{ch:numerical}, and appendix \ref{app:adcorrespondence}.
In addition, some of this work has been reported in conference proceedings \cite{myhr09c}.

The major part of the funding for my PhD has been through a personal PhD scholarship from the Research Council of Norway. 
This project was called ``Quantum correlations for secret key distribution'' 
and had project No.~166842/V30.
Additional funding sources have been 
the European Union through the IST Integrated Project SECOQC and the IST-FET Integrated Project QAP, Ontario Centres of Excellence, the NSERC Innovation Platform Quantum Works the NSERC Discovery Grant.

Throughout the work with my PhD I have interacted with many people and I am thankful for all conversations and support that I have received.
There are many more than those I could possibly name here.

First, I wish to thank my supervisor, Norbert L{\"u}tkenhaus for taking me on as a student and guiding me through the PhD.
Also, many thanks to everyone at the Institute for Quantum Computing at the University of Waterloo and everyone else who made my years in Waterloo enjoyable.

Some of the results in this thesis has come about with the help of others. 
The biggest part is probably from Andrew Doherty and Joe Renes who solved the symmetric extension problem for Bell-diagonal states before I became a part of the collaboration, and it was through this that I learned the details of using semidefinite programming for solving symmetric extension problems.
The proof of theorem \ref{thm:extremal} was missing an essential piece until Marco Piani showed me how to fill in the hole.
For the numerical solution of semidefinite programs, Tobias Moroder's advice and initial Matlab scripts using Yalmip \cite{yalmip} and the SDPT3 solver \cite{sdpt3} has been an invaluable starting point.
Toby Cubitt's Matlab scripts \cite{cubittscripts} has saved me the trouble of having to implement some basic quantum information functions myself.
The German version of the summary would have been full of mistakes if it had not been for Hauke H{\"a}seler's corrections.

Throughout the work with my PhD I have enjoyed the hospitality of and discussions with Johannes Skaar, Jon Magne Leinaas, Jan Myrheim, Michael M. Wolf and Matthias Christandl during visits.
I have also enjoyed discussions with Barbara Kraus, Antonio Ac{\'i}n, and all the members of the Optical Quantum Communication Theory group at the IQC.

Finally, a big thank you to my wife Kjersti for her love and patience and to my son Marius for always greeting me with a big smile when I come home.

\cleardoublepage
\tableofcontents

\listoffigures

\cleardoublepage

\pagestyle{fancy}


\mainmatter

\chapter{Introduction}
\label{ch:intro}

In \textindex{quantum key distribution} (QKD), we use the physical principles of quantum mechanics to distribute identical random bits to two (or sometimes more) cooperating parties while ensuring that no other party gains any knowledge about those random bits.
The string of random bits that is produced is called a \emphindex{secret key}.
The creation of a secret key is a \textindex{cryptographic primitive} that can then be used to enable secret communication, message authentication, or other cryptographic tasks. 

A fundamental question in quantum key distribution is what the maximum rate of secret key is that can be generated, given an observed error rate\footnote{
Other observable parameters such as loss can also be taken into account. 
In this thesis we do not consider loss explicitly, but as long as there are no imperfections  
in the signal states or measurements we can consider only the signals that actually arrive in the BB84 and six-state protocol.
With for example a multi-photon component in the signal states \cite{brassard00a} or a basis-dependent detector efficiency\cite{makarov06a,qi07a}, or for the B92 protocol \cite{bennett92b}, it is crucial to take the loss into account.
} in a given QKD setup.
A security proof will give a lower bound, since it typically will give an eavesdropper some extra power in order to simplify the problem and prove that a key can be generated at a certain rate. 
There are also various upper bounds on the key rate from \textindex{entanglement} theory and explicit attacks.

An even more fundamental question is when a key can be generated at all and when it cannot.
In protocols like BB84 \cite{bennett84a} and the six-state protocol \cite{bruss98a,bechmann99a} the error can be summarized using one parameter, the \textindex{quantum bit error rate} (QBER).
Using \textindex{preprocessing} followed by \textindex{error correction} to correct errors in the raw key and \textindex{privacy amplification} to remove any knowledge an eavesdropper may have about the key, the limit for provable secure key has been pushed to 12.9\% \cite{smith08a} for the BB84 protocol. 
These techniques only need to use one-way communication from a sender to a receiver. 
At 14.6 \% no secret key may be generated with one-way communication, since at this QBER an eavesdropper and the receiver may be in symmetric situation. 
This comes from a property of the underlying quantum state called a \textindex{symmetric extension}.

It was discovered by Gottesman and Lo \cite{gottesman03a} that by using a two-way procedure derived from classical \textindex{advantage distillation} \cite{maurer93a} before error correction and privacy amplification, a key could be distilled even when it was provably impossible with only one-way communication.
Chau \cite{chau02a} then showed that this technique could be used to distill a key for any QBER below 20.0 \% in the BB84 protocol.
The upper bound when allowing two-way communication is 25.0 \%, which is when the underlying quantum state may be separable\index{separable state}.
For QBERs between 20\% and 25\% it is not known whether a secret key can be produced at a finite rate or not.
The work presented in this thesis is motivated by this \textindex{gap}.

\begin{figure}
  \resizebox{0.8\textwidth}{!}{
  \includegraphics{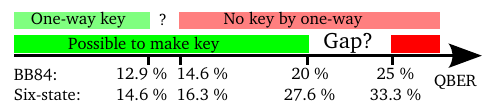}
  }
 \caption{Different QBER regimes for the BB84 and six-state protocols}
 \label{fig:thresholds}
\end{figure}

The ultimate goal would be to close the gap (see Fig. \ref{fig:thresholds}).
This may be done either by proving that a secret key can be distilled all the way up to 25 \%, 
or proving that no key can be distilled above 20 \%, 
or establishing a new threshold which is both a lower and upper bound. 

For the six-state protocol, the picture is similar and the relevant QBERs for both protocols are illustrated in figure \ref{fig:thresholds}.
The one-way lower bound of 14.6~\% is taken from \cite{kern08a}.
The one-way upper bound of 16.3~\% is taken from \cite{renner05b} and is slightly below the symmetric extension bound of 16.7~\%.

There have been previous attempts at shrinking the gap.
Ac{\'i}n et al.~showed in \cite{acin06a} that the 20 \% limit that Chau had found was indeed as high as possible with the prescribed key distillation procedure, since there is an explicit attack that puts an eavesdropper in a similar situation as the receiver after the two-way \textindex{advantage distillation} part of the procedure.
Portmann tried to change the advantage distillation to detect errors using random parities on larger blocks, but could only prove a BB84 secret key rate up to 16.9 \% using that technique \cite{portmann05a}.
Finally, Bae and Ac{\'i}n attempted to improve the threshold by adding noise in the beginning of the procedure, 
allowing coherent quantum operation on one side or measuring in a different basis, 
but without success \cite{bae07a}.

In this thesis we seek to learn from how a key can be distilled by two-way processing when it is impossible to distill by one-way processing.
In this regime, the underlying quantum state starts out with a symmetric extension, but through a sequence of two-way steps the symmetric extension is broken and the rest of the protocol can proceed using the same one-way techniques that on its own could not possibly distill key above 14.6~\%.
This has led us to investigate which states have a symmetric extension, a question which may also be interesting for other areas of quantum information.
Furthermore, we have looked at different types of advantage distillation and find evidence that they fail to break a symmetric extension and can therefore not be used to shrink the gap from below.

Some of the results apply both to the BB84 protocol and the six-state protocol, but some only apply to the six-state protocol.
The six-state protocol is easier to analyze since the measurements give full tomographic information about the quantum states. 
In particular, the analytical computation in chapter \ref{ch:k2analytical} and the numerical computations in chapter \ref{ch:numerical} are done for the six-state protocol only.

The rest of the thesis is organized as follows. 
Chapter \ref{ch:prelim} contains preliminaries on quantum mechanics, quantum key distribution, and semidefinite programming. 
Part \ref{part:symext}, consisting of chapters \ref{ch:symext}--\ref{ch:degradable}, contains results related to states with a symmetric extension.
Part \ref{part:qkd} , consisting of chapters \ref{ch:lad}--\ref{ch:numerical}, analyzes QKD with different kinds of advantage distillation and find that all the resulting states have a symmetric extension.

\chapter{Preliminaries}
\label{ch:prelim}

\section{Quantum mechanics}

The quantum mechanics of a closed (isolated) system is described mathematically by assigning to it a separable complex \textindex{Hilbert space} $\mathcal{H}$. 
We will only deal with systems with a finite Hilbert space dimension and in this context any complex Hilbert space reduces to a complex vector space with an inner product.
We denote vectors in the Hilbert space using the bra-ket notation.
A vector with the label $\psi$ is denoted $\ket{\psi}$ and its dual vector in the dual Hilbert space is denoted $\bra{\psi}$.
The inner product between two vectors $\ket{\psi_1}$ and $\ket{\psi_2}$ is written $\braket{\psi_1}{\psi_2}$.

\subsection{Quantum states}
\index{quantum state}

The physical state of a quantum system where we have complete information is described by a ray (i.e. a one-dimensional subspace) in the Hilbert space.
Such a state is called a \emphindex{pure state}.
Usually, we choose a unit vector in a ray to represent the state. 
With this representation, if the unit vector $\ket{\psi}$ represents a state, also the vector $\e^{\im \phi} \ket{\psi}$, represents the same state.
To avoid this ambiguity, we may use the one-dimensional projector $\proj{\psi}$ to represent the state.
Then, unlike the unit vector representation, different projectors correspond to different states.

In many cases we do not have complete information, either because we have a statistical mixture of pure states or because the system is part of a bigger system which is in a pure state.
An \emphindex{ensemble} of pure states $\{(p_i,\ket{\psi_i})\}$ is a set of pure states $\{\ket{\psi_i}\}$ where each has an associated probability $p_i$.
To any ensemble, we have a corresponding \emphindex{mixed state}, a quantum state with incomplete knowledge.
The mixed state corresponding to the ensemble 
$\{(p_i,\ket{\psi_i})\}$
is 
\begin{equation}
 \label{eq:mixedstatefromensemble}
 \rho = \sum_i p_i \ket{\psi_i}.
\end{equation}
There is a continuum of different ensembles that realizes any given mixed state, but there is no physical way to distinguish them, so they all correspond to the same mixed state.

The operator $\rho$ is called a \emphindex{density operator} or a \emph{density matrix}.
In general, we define a density operator or density matrix as a positive semidefinite operator with trace 1.
Any density operator represent a pure or mixed quantum state.
The density operator representing the pure state $\ket{\psi}$ is simply $\rho = \proj{\psi}$. 
Since any positive semidefinite operator is also a Hermitian, it can be decomposed using the spectral decomposition
\begin{equation}
 \rho = \sum_i \lambda_i \proj{\phi_i}.
\end{equation}
The eigenvalues and eigenvectors form an ensemble $\{ \lambda_i, \phi_i \}$, but the only property that makes this somewhat special compared to other ensembles realizing the same density matrix is that it consists of orthogonal states.

In addition to distinguishing between pure and mixed states, we can also quantify their purity. 
There are several functions including entropies that can do this, and they are usually a function of the eigenvalues only.
One quantity that will be useful later is the \emphindex{purity}, defined as $\tr[\rho^2]$.
A purity of one means that the state is pure and otherwise it is mixed.

The density operator is our preferred representation of a quantum state, but since the unit vector representation of pure states makes calculations both easier and clearer we will often use that for pure states.

\subsection{Unitary evolution}

A quantum system starting out in a prepared state will change over time.
For a pure state in a closed system the exact time evolution is given by a Hamiltonian operator $H$ through the \textindex{Sch{\"o}dinger equation},
\begin{equation}
 \im \hbar \diff{\ket{\psi}}{t} = H \ket{\psi}.
\end{equation}
For a mixed state, the evolution can be computed by decomposing it into an ensemble of pure states.
Alternatively, it can be described by the \textindex{von Neumann equation},
\begin{equation}
 \im \hbar \diff{\rho}{t} = H \rho - \rho H.
\end{equation}
Since these equations essentially give the time derivative of the state ($\ket{\psi}$ or $\rho$) as a system-dependent function of itself, 
this gives the evolution of any starting state for any later time.

For our purposes, we will only be interested in the initial state and the state at some later time. 
The evolution is then best described by a unitary operator,
\begin{equation}
 \ket{\psi} = U \ket{\psi_0},
\end{equation}
and
\begin{equation}
 \rho = U \rho_0 U^\dagger,
\end{equation}
where
\begin{equation}
 U = \exp(-\im H t/\hbar)
\end{equation}
when $H$ is time independent.
For a given time interval it is possible to achieve any unitary operator $U$ by choosing an appropriate Hamiltonian $H$. 



\subsection{Measurements}
\label{sec:measurements}

In classical systems we can have measurements that do not change the classical state of the system.
In practice some systems may be more delicate than others, but there is no fundamental principle that excludes such perfect measurements.
A mathematical description of a classical measurement therefore only need to specify what result it would get for any possible state of the system.
Except in the trivial case where the measurement result is independent of the system state, 
a quantum measurement cannot avoid disturbing the system to some degree.
For a full description of a quantum measurement we therefore need to specify the measurement results for all possible states, but also how the state changes after a measurement.
While we can also allow a classical measurement to give probabilistic outcomes for a given system state, even an ideal quantum measurement gives probabilistic outcomes on many states and we can therefore only describe their probability.

We first describe ideal measurements, i.e. measurements that contain no other imperfections than those imposed by quantum mechanics.
Such measurements are sometimes called \emphindex{generalized measurements} \cite{nielsen00a} to distinguish them from the special case of \textindex{projective measurement}s.
A $p$-outcome measurement is described by a set of $p$ \emphindex{Kraus operators} \cite{kraus71a,kraus83a} (called \emphindex{measurement operators} in \cite{nielsen00a}), $\{K_m\}_{m=1}^p$ that satisfy
\begin{equation}
 \label{eq:krausoperatorcompleteness}
 \sum_{m=1}^p K_m^\dagger K_m  = \calone.
\end{equation}
When a density operator $\rho$ is measured, the probability for outcome $m$ is
\begin{equation}
  p_m = \tr[K_m\rho K_m^\dagger]
\end{equation}
and condition \eqref{eq:krausoperatorcompleteness} is simply that the probabilities sum up to 1.
The state after the measurement is then
\begin{equation}
 \rho_m = \frac{1}{p_m} K_m\rho K_m^\dagger.
\end{equation}
Because of the form of this, we will sometimes compute the unnormalized state after measurement
\begin{equation}
 \tilde{\rho}_m = K_m\rho K_m^\dagger
\end{equation}
and remember that the trace of the unnormalized state is the probability of this outcome.

If we add a result-dependent unitary operator to the left of each Kraus operator, that does not change the probability for any of the outcomes, but it changes the resulting state.
That is, the set $\{U_m K_m\}$ gives the same measurement statistics as the set $\{K_m\}$.

Measurements may include random choices made by the measurement device. 
Say that a measurement device makes a random choice between $q$ different measurements.
Measurement $k$ is described by the Kraus operators $\{ K_m^{(k)} \}_{m=1}^p$ and is chosen with probability $q_k$.
The measurement including the random choice can be described as a measurement with $pq$ outcomes described by the measurement operators $\{ \sqrt{q_k} K_m^{(k)} \}$.
This kind of measurement is used in the BB84 and six-state QKD protocols, where there is a random choice between different measurement bases.
A trivial example of this kind of a measurement is one that does nothing to the state and gives a random outcome $k$ with probability $q_k$. 
This measurement is described by the Kraus operators $\{ \sqrt{q_k} \calone  \}$.

The measurements described by Kraus operators are ideal in the sense that the state after the measurement is not disturbed more than necessary. 
If a pure state is measured, the post-measurement state is also pure. 
A real imperfect measurement device may of course add some noise and turn a pure state into a mixed state. 

A special class of measurements are \emph{projective measurements}\index{projective measurement}. 
Each Kraus operator is then an orthogonal projector---$K_m^2 = K_m$ and $K_m^\dagger = K_m$--- 
and in addition they project onto orthogonal subspaces, $K_m K_n = \delta_{mn} K_m$.
A projective measurement is usually represented by an Hermitian \emphindex{observable} operator,
\begin{equation}
 \label{eq:observable}
 O = \sum_m \lambda_m K_m.
\end{equation}
The distinct eigenvalues $\lambda_m$ label the outcomes and the state gets projected onto the eigenspace by the projector $K_m$.
We say that we measure the operator $O$ and get outcome $\lambda_m$.
One example that we will use later is the Pauli operator $\sigma_z := \proj{0} - \proj{1}$, which can give outcomes $\lambda_1 = +1$ or $\lambda_2 = -1$ and will project the state to $\ket{0}$ or $\ket{1}$, respectively.
We can also be interested in the expectation value of an operator $O$ on a given state $\rho$, i.e. the average value of the eigenvalue $\lambda_m$ when measured many times. 
This has the straightforward expression,
\begin{equation}
 \langle O \rangle_\rho := \tr[O \rho].
\end{equation}

In many cases we are only interested in the measurement outcomes and not in the state after after the measurement. 
In that case it is more convenient to use a \emph{positive operator-valued measure (POVM)\index{POVM}} to describe the measurement. 
Any set of positive semidefinite operators $\{F_m\}$ such that 
\begin{equation}
 \sum_m F_m = \calone
\end{equation}
is a POVM, and its individual operators are called \emph{POVM elements}\index{POVM element}.
The probability of obtaining outcome $m$ is 
\begin{equation}
 p_m = \tr[F_m \rho].
\end{equation}
With a given set of Kraus operators $\{K_m\}$, 
a POVM is constructed by taking the operators $F_m = K_m^\dagger K_m$. 
With a given POVM $\{F_m\}$, one set of Kraus operator that gives the same measurement statistics is $\{K_m = \sqrt{F_m}\}$

One advantage of using POVMs instead of Kraus operators is that it is easy to coarse grain results.
Say that a physical measurement can be described by a $p$-element POVM $\{F_m\}_{m=1}^p$.
The $p$ outcomes can be divided into $q < p$ subsets $\{\mathcal{S}_n\}_{n=1}^q$ and we are only interested in knowing in which subset the result was.
We can think about this as a $q$-outcome measurement. 
The $q$ POVM elements for this measurement are
\begin{equation}
 G_n = \sum_{m \in \mathcal{S}_n} F_m.
\end{equation}


\subsection{Composite systems}
\label{sec:entanglement}

So far we have looked at a quantum system as a whole. 
In many cases they consist of distinct subsystems and in the context of quantum information theory, these subsystems may even be in remote locations.
When we have $n$ subsystems, each with its Hilbert space $\mathcal{H}_i$, the Hilbert space of the joint system is the tensor product of the individual Hilbert spaces,
\begin{equation}
 \mathcal{H}_\text{total} = \mathcal{H}_1 \otimes \mathcal{H}_2 \otimes \cdots \otimes \mathcal{H}_n.
\end{equation}
If each of the $n$ subsystems is in a pure state $\ket{\psi_i}$, the state vector of the joint system is
\begin{equation}
 \label{eq:productvector}
 \ket{\psi} = \ket{\psi_1} \otimes \ket{\psi_2} \otimes \cdots \otimes \ket{\psi_n},
\end{equation}
which we will sometimes write without the tensor product symbol as $\ket{\psi_1}\ket{\psi_2} \cdots \ket{\psi_n}$ or even just as one single ket $\ket{\psi_1 \psi_2 \cdots \psi_n}$.
Note that most vectors in $\mathcal{H}_\text{total}$ are not of the product form \eqref{eq:productvector} but rather linear combinations of those.

We will now restrict the discussion to two subsystems $A$ and $B$ with Hilbert space $\mathcal{H}_{AB} = \mathcal{H}_A \otimes \mathcal{H}_B$.
Most properties can be generalized to more subsystems by considering $A$ and $B$ as being subdivided into smaller subsystems.

We can ignore one part of a composite quantum system by taking the \emphindex{partial trace} over it (we also say that we \emph{trace out} the system).
Let us say that we have a density operator $\rho^{AB}$ on $\mathcal{H}_{AB}$ and we want a description of the state of the $B$ system.
We then take the partial trace over system $A$ to obtain the \emphindex{reduced density operator}, \emphindex{reduced state}, or simply \emphindex{reduction} to $B$.
We write this as
\begin{equation}
 \rho^B = \tr_A[\rho^{AB}].
\end{equation}
The partial trace is defined on product operators by,
\begin{equation}
 \tr_A [M \otimes N] = \tr[M] N
\end{equation}
and is a linear operator. 
By decomposing a general density operator $\rho^{AB}$ using orthogonal bases 
$\{ \ket{a_i} \}$ and $\{ \ket{b_i} \}$,
\begin{equation}
 \rho^{AB} =  \sum_{ijkl} \rho_{ij,kl} \ketbra{a_i b_j}{a_k b_l} 
   = \sum_{ijkl} \rho_{ij,kl} \ketbra{a_i}{a_k} \otimes \ketbra{b_j}{b_l}
\end{equation}
the definition yields
\begin{align*}
  \tr_A[\rho^{AB}] 
  &= \sum_{ijkl} \rho_{ij,kl} \tr_A [ \ketbra{a_i}{a_k} \otimes \ketbra{b_j}{b_l} ]\\
  &= \sum_{ijkl} \rho_{ij,kl} \tr [\ketbra{a_i}{a_k}] \ketbra{b_j}{b_l}\\
  &= \sum_{ijkl} \rho_{ij,il} \ketbra{b_j}{b_l}.
\end{align*}
Note that unless a pure state is a product state $\ket{\psi} = \ket{\psi_1} \otimes \ket{\psi_2}$, its reduced states are mixed states.

A \emphindex{product state} is a state of the form
\begin{equation}
 \rho^{AB}_\textrm{product} = \rho^A \otimes \rho^B.
\end{equation}
In this case, the states of the $A$ and $B$ systems are independent.
The class of states that are convex combinations of product states,
\begin{equation}
 \label{eq:separable}
 \rho^{AB}_\textrm{sep} = \sum_i p_i \rho_i^A \otimes \rho_i^B,
\end{equation}
are called \emph{separable states}\index{separable state}. 
Here, the states of the two subsystems are not independent, but all correlations between the two systems are classical.
There are many states that are not separable, and they are called \emphindex{entangled}.
Entangled states can be used to perform tasks that are impossible to recreate classically, and quantum key distribution is one of them.

\subsection{Qubits and Pauli operators}
\label{sec:qubitsandpaulis}

Many of the quantum systems that we use in this thesis are built up from \emph{qubits}\index{qubit}.
A qubit is the simplest nontrivial quantum system in that the Hilbert space is two-dimensional.
The two orthogonal unit vectors $\ket{0}$ and $\ket{1}$ are called the \emphindex{computational basis} of the Hilbert space.

When dealing with qubits, it is convenient to use the 
\emph{Pauli operators}\index{Pauli operator},
\begin{subequations}
\begin{align}
\sigma_x &:= \phantom{-\im} \ketbra{0}{1} + \ketbra{1}{0} = \begin{bmatrix} 0 & 1\\ 1 & 0 \end{bmatrix},\\
\sigma_y &:= -\im \ketbra{0}{1} + \im \ketbra{1}{0} = \begin{bmatrix} 0 & -\im \\ \im & 0 \end{bmatrix},\\
\sigma_z &:= \phantom{-\im} \ketbra{0}{0} - \ketbra{1}{1} = \begin{bmatrix} 1 & 0\\ 0 & 1 \end{bmatrix}.
\end{align}
\end{subequations}
They all have eigenvalues $+1$ and $-1$. 
Their spectral decomposition is 
\begin{subequations}
\begin{align}
\sigma_x &:= \proj{+} - \proj{-},\\
\sigma_y &:= \proj{+\im} - \proj{-\im},\\
\sigma_z &:= \proj{0} - \proj{1},
\end{align}
\end{subequations}
where 
$\ket{\pm} := \frac{1}{\sqrt{2}}(\ket{0} \pm \ket{1})$ 
and 
$\ket{\pm \im} := \frac{1}{\sqrt{2}}(\ket{0} \pm \im \ket{1})$.

The three Pauli operators are often indexed as 
$\sigma_1 := \sigma_x$, $\sigma_2 := \sigma_y$, $\sigma_3 := \sigma_z$
The set of Pauli operators are often extended to include the identity operator $\calone$, 
and it is then sometimes written as $\sigma_0$.
Together, these four operators form a Hermitian operator basis for the qubit space.

Computations including Pauli operators are often simplified using their commutation and anticommutation relations and the fact that they are square to the identity operator, 
$\sigma_i^2 = \calone$.
Of course, all operators commute with itself and $\sigma_0 = \calone$, 
$\sigma_i \calone = \calone \sigma_i$.
Otherwise, they all anticommute,
$\sigma_i \sigma_j = - \sigma_j \sigma_i$ for $i \ne j \in {1,2,3}$.
For example, 
$\sigma_x \sigma_y \sigma_x = -\sigma_x \sigma_x \sigma_y = -\sigma_y$,
where we first used the anticommutation relation between $\sigma_x$ and $\sigma_y$ and then that $\sigma_x \sigma_x = \calone$.

The final useful property is that products of Pauli operators are proportional to Pauli operators,
\begin{subequations}
\begin{align}
 \sigma_x \sigma_y = \im \sigma_z,\\
 \sigma_y \sigma_z = \im \sigma_x,\\
 \sigma_z \sigma_x = \im \sigma_y.
\end{align}
\end{subequations}

Any density operator on a single qubit can be written as
\begin{equation}
 \frac{1}{2} ( \calone + r_x \sigma_x + r_y \sigma_y + r_z \sigma_z ),
\end{equation}
with three real parameters $r_i$.
An operator of this form is positive semidefinite whenever 
$r_x^2 + r_y^2 + r_z^2 \leq 1$.
This means that the possible density operators form a sphere in the parameters space, 
and this sphere is called the \emphindex{Bloch sphere}.
The 3-component vector $(r_x, r_y, r_z)$ is called the \emphindex{Bloch vector} of the state.
The components of the Bloch vector can be found from $r_i = \tr [\sigma_i \rho]$.

Similar to the Bloch vector for one qubit is the $4 \times 4$ $R$-matrix\index{R-matrix@$R$-matrix} for two qubits. 
Its components are 
\begin{equation}
 r_{ij} = \tr[(\sigma_i \otimes \sigma_j) \rho], \quad i,j \in \{0,1,2,3\}.
\end{equation}
Any two-qubit state can be written in terms of these parameters,
\begin{equation}
 \rho = \frac{1}{4} \sum_{i=0}^3 \sum_{j=0}^3 r_{ij} \sigma_i \otimes \sigma_j.
\end{equation}
A special case is when the $R$-matrix\index{R-matrix@$R$-matrix} is diagonal, i.e. the state is of the form 
\begin{equation}
 \rho = \frac{1}{4} \sum_{i=0}^3 r_{ii} \sigma_i \otimes \sigma_i.
\end{equation}
This is called a \emphindex{Bell-diagonal state}, since its spectral decomposition is of the form
\begin{equation}
 \rho = p_I \proj{\Phi^+} + p_x \proj{\Psi^+} + p_y \proj{\Psi^-} + p_z \proj{\Phi^-},
\end{equation}
where 
$\ket{\Phi^\pm} = \frac{1}{\sqrt{2}} (\ket{00} \pm \ket{11})$
and
$\ket{\Psi^\pm} = \frac{1}{\sqrt{2}} (\ket{01} \pm \ket{10})$
are the \textindex{maximally entangled} \emph{Bell states}\index{Bell state}.
The eigenvalues of a \textindex{Bell-diagonal state} are related to the diagonal elements of the $R$-matrix\index{R-matrix@$R$-matrix} as
\begin{subequations}
\begin{align}
\label{eq:belleigenvaluesfromRmatrixfirst}
p_I &= \frac{1}{4}(1+ r_{11} - r_{22} + r_{33}), \\
p_x &= \frac{1}{4}(1+ r_{11} + r_{22} - r_{33}), \\
p_y &= \frac{1}{4}(1- r_{11} - r_{22} - r_{33}), \\
\label{eq:belleigenvaluesfromRmatrixlast}
p_z &= \frac{1}{4}(1- r_{11} + r_{22} + r_{33}). 
\end{align}
\end{subequations}

For $N$ qubits, we can write any state as 
\begin{equation}
 \rho = \frac{1}{2^N} \sum_{i_1=0}^3 \cdots \sum_{i_N=0}^3 
  \alpha_{i_1 \cdots i_N} \sigma_{i_1} \otimes \cdots \otimes \sigma_{i_N}.
\end{equation}
The parameters $\alpha_{i_1 \cdots i_N}$ are then
\begin{equation}
 \alpha_{i_1 \cdots i_N} = \tr[(\sigma_{i_1} \otimes \cdots \otimes \sigma_{i_N}) \rho].
\end{equation}

The Pauli operators are also unitary, so they also describe evolution of qubit states. 
Each Pauli operator corresponds to a rotation of an angle $\pi$, $\sigma_x$ around the $r_x$ axis, $\sigma_y$ around the $r_y$ axis, and $\sigma_z$ around the $r_z$ axis. 
Since $\sigma_x$ interchanges $\ket{0}$ and $\ket{1}$ it is called a \textindex{bit-flip operator}.
Similarly $\sigma_z$ is called a \textindex{phase-flip operator} since it interchanges
$\alpha \ket{0} + \beta \ket{1}$ with $\alpha \ket{0} - \beta \ket{1}$.
Since a global phase of a state vector is irrelevant, the Pauli operator $\sigma_y = \im \sigma_x \sigma_z$ is both phase and bit flip.


Finally, the Pauli operators are used to describe measurements on qubits.
Since they are Hermitian operators they are observables and describe a projective two-outcome measurement, each outcome corresponding to one of their eigenvalue.
For example, a measurement of $\sigma_x$ can get the two outcomes $+1$ or $-1$ and the state after measurement is then $\proj{+}$ in the first case and $\proj{-}$ in the second.
The expectation value $\langle \sigma_i \rangle = \tr[\sigma_i \rho] = r_i$ is the a component of the Bloch vector.
Tensor products of Pauli operators only have the two eigenvalues $+1$ and $-1$ and therefore describe measurements with only two outcomes.


\section{Quantum key distribution (QKD)}

In \emphindex{quantum key distribution} the quantum mechanical nature of fundamental information carriers is exploited to generate a secret key shared by a number of cooperating parties. 
In this thesis we will only consider two cooperating parties and by convention they are called Alice and Bob.
By the same convention, any eavesdropper is called Eve.
In general, Eve is assumed to control anything and anyone who is not under Alice and Bob's control.
Alice and Bob can communicate classically on a public, but authenticated channel. 
That is, Eve can listen to all classical communication between Alice and Bob, but cannot change it.
The authentication can be implemented by a small pre-shared secret key, so we do not really need to make any unrealistic assumption about the physical channel.
In addition to the classical channel, Alice and Bob has some way of exchanging quantum states.
This \textindex{quantum channel} has no built-in authentication and Eve is allowed to do anything on it that quantum mechanics allows.
We assume no computational or technological limitations on Eve.
In practice the quantum channels are usually optical fibers or free space and the quantum signals that are sent through them are pulses of weak light.

In this situation, Alice and Bob can use the quantum channel to distribute quantum signals which are the measured. 
When this is done right, they can use some of their measurement results and the laws of quantum mechanics to estimate Eve's knowledge about the other measurement results. 
If they find that Eve has sufficiently little information about their results, they can coordinate their effort using the classical channel and produce a secret key from the measurement data.

\subsection{Prepare and measure QKD}

The first QKD protocol to be described \cite{bennett84a} was a prepare and measure (P\&M) protocol.
In such a protocol, Alice prepares an ensemble of $n$ quantum states 
$\{(p_i,\ket{\psi_i})\}_{i=1}^n$,
where at least some of the $\ket{\psi_i}$ are nonorthogonal. 
She chooses a state $\ket{\psi_i}$ according to the probability distribution $p_i$ and sends it to Bob over the insecure quantum channel.
Bob measures the incoming state using a POVM with $m$  POVM elements\footnote{
In many protocols, Bob chooses randomly between a set of measurements. 
In this context, where we assume that we know the details about how our devices are working, we include that random choice as part of the measurement device (see section \ref{sec:measurements}).
In other contexts, such as post-quantum \cite{barrett05a} and device-independent \cite{acin07a} QKD, it is important that this choice is made outside the apparatus and we would have to describe a measurement for each choice.
},
$\{F_j\}_{j=1}^m$.
This is repeated many times.

After Alice has sent the signals and Bob has acknowledged receiving and measuring them, 
they choose a random subset of the signals where Alice announces which signal she sent and Bob announces which measurement result he got.
With this information, Alice and Bob can estimate the conditional probability distributions $p_{j|i}$---the probabilities that Bob gets measurement outcome $j$ when Alice sends state $i$---and the joint probability distribution $p_{ij}$.
This phase is called \emphindex{parameter estimation}.
In this thesis we assume that Alice and Bob have enough data available that they can estimate $p_{ij}$ arbitrarily well. 
In practice, the finite size of the datasets may make the parameter estimation more involved \cite{scarani08a,cai09a}.

This joint probability distribution tells them whether the data is well enough correlated and with sufficiently low information leakage to Eve.
If it is, they can proceed with postprocessing in order to generate a secret key.
If this is not the case, they will abort the protocol.
If the data passes the test, the joint probability distribution $p_{ij}$ is used to decide various parameters in the postprocessing which will ultimately decide the length of the secret key.
Alice and Bob exchange messages about their data and use this information to distill a secret key.
The details of this \textindex{postprocessing} will depend on the protocol

\subsection{Entanglement based QKD}
\label{sec:entbasedQKD}
\index{entanglement}

A seemingly different approach to quantum key distribution is based on bipartite entangled states \cite{ekert91a}.
An untrusted source produces bipartite entangled states and sends one subsystem to Alice and the other to Bob.
Alice measures her part of the quantum state with a POVM $\{A_i\}_{i=1}^n$ and Bob measures his with a POVM $\{ B_j\}_{j=1}^m$.

The \textindex{parameter estimation} is similar to the prepare and measure case.
After both parties have measured the quantum systems and acknowledged that to the other party, they randomly choose a subset where they announce their measurement results. 
Using this data, they can estimate the joint probability distribution $q_{ij}$, the probability that Alice gets outcome $i$ and Bob gets outcome $j$.
They can use the joint probability distribution $q_{ij}$ to estimate the joint quantum state that they received from the untrusted source over the untrusted channel.
The quantum state can be any $\rho^{AB}$ that satisfies
\begin{equation}
 \label{eq:entbasedstateestimation}
 q_{ij} = \tr[(A_i \otimes B_j)\rho^{AB}] 
\end{equation}
In some protocols, Alice and Bob's measurements are tomographically complete, so the joint quantum state is fully determined by the joint probability distribution.
In other protocols they can only estimate some parameters of the quantum states while others are left open, giving an equivalence class of possible quantum states.

If all the quantum states compatible with the measurement data are sufficiently entangled (i.e is well correlated with low information leakage), Alice and Bob proceed with \textindex{postprocessing} to generate a secret key.

\subsection{Equivalence between prepare and measure and entanglement based}
\label{sec:pmentequivalence}

In a prepare and measure scheme, there is never any entanglement or even bipartite systems involved.
In the entanglement based scheme, there is full symmetry between Alice and Bob who both passively receive quantum signals from a third-party source.
Nevertheless, a prepare and measure scheme is equivalent to an entanglement based scheme where the source is placed next to Alice's measurement device.
The two setups are indistinguishable from Eve's point of view \cite{bennett92c}.

When the entanglement source is placed next to the measurement device, 
they form a state preparation device.
Let the source produce some pure bipartite entangled state 
$\ket{\psi}_{AA'} = \sum_{rs} \alpha_{rs} \ket{r} \otimes \ket{s}$. 
Alice can measure using a POVM where each POVM element is proportional to a one-dimensional projector\footnote{
She could also measure using any POVM, with the only consequence that in general, POVM elements that are not proportional to a one-dimensional projector will be equivalent to Alice preparing a mixed signal state. 
}, 
$A_i = k_i \proj{\phi_i}$. 
She then gets outcome $i$ with probability 
$p_i = k_i\tr[(\proj{\phi_i}_A \otimes \calone_{A'})\proj{\psi}_{AA'}]$ 
and effectively prepares the state 
$\ket{\psi_i}_{A'} = (\bra{\phi_i}_A \otimes \calone_{A'})\ket{\psi}_{AA'} = \sum_{rs} \braket{\phi_i}{r} \alpha_{rs} \ket{s}_{A'}$ 
in system $A'$ which is subsequently sent to Bob over the channel.
The time ordering of Alice measuring system $A$ and sending system $A'$ to Bob is not important. 
If Bob has received the system (and even if he measured it), Alice's measurement effectively prepares Bob's signal state at a distance.

When the signal ensemble 
$\{(p_i,\ket{\psi_i})\}_{i=1}^n$ 
in a prepare and measure scheme is given, a preparation device can be based on a source of the entangled state 
\begin{equation}
  \ket{\psi}_{AA'} = \sum_{j=1}^n \sqrt{p_j} \ket{m_j}_A \otimes \ket{\psi_j}_{A'}, 
\end{equation}
where 
$\{\ket{m_j}\}_{j=1}^n$ 
are orthonormal.
Alice's measurement will then be simply a projective measurement, 
$A_i = \proj{m_i}$.
Note that while the $n$ nonorthogonal states 
$\ket{\psi_j}$ 
often span a Hilbert space $\mathcal{H}_{A'}$ of dimension less than $n$, 
the system $A$ is here $n$-dimensional.
It is possible to write $\ket{\psi}_{AA'}$ in the Schmidt decomposition,
\begin{equation}
  \ket{\psi}_{AA'} = \sum_{i=1}^{\dim(A')} = s_i \ket{u_i} \otimes \ket{v_i}, 
\end{equation}
thus identifying a subspace of $A$ of the same dimension as $A'$.
If we also project the original POVM elements $\proj{i}$ onto this subspace, 
we can get an entanglement based scheme where both systems are of dimension $\dim(A')$.

The main advantage of the equivalence between entanglement based and prepare and measure QKD is that it allows us to analyze a prepare and measure scheme by analyzing the equivalent entanglement based scheme. 
We will therefore use the data from prepare and measure schemes to reconstruct an effective bipartite state $\rho^{AB}$, which will be the state that had arrived at Alice and Bob's measurement devices if the state preparation had been based on a source of entangled states.
We will see in sections \ref{sec:parameterestimation} and \ref{sec:coherentanalysis} that we can use this state in the security analysis of the protocol.

\subsection{Attacks on QKD protocols}

As mentioned earlier, Eve is allowed to do anything allowed by quantum mechanics in her eavesdropping of the quantum channel.
We can divide Eve's attacks into three classes, depending how much of that power Eve actually uses.
\begin{description}
 \item[Individual attack.]\index{individual attack} Eve attaches a probe to each of the quantum signals and lets the probe interact with the signal in the way she chooses. Then she measures each probe with any POVM\footnote{In protocols that include \textindex{sifting}, Eve may be allowed to wait for the sifting information before measuring.}.
 \item[Collective attack.]\index{collective attack} Eve attaches a probe to each of the quantum signals and lets the probe interact with the signal in the way she chooses. She may then do a collective measurement on all the probes together (which may give here some advantage over measuring each individually). She may also postpone this measurement indefinitely, e.g. until the key is being used (which may reveal some the bits).
 \item[Coherent attack.]\index{coherent attack} Eve uses one big probe to interact coherently with all the quantum signals at the same time. Sometimes also called \emphindex{joint attack}.
\end{description}
Individual attacks are rarely analyzed anymore.
The main motivation for analyzing them was that there were no known ways of quantifying quantum information about classical data and people wanted to show that Eve's information about the key was small.

In a collective attack, Eve is allowed to keep quantum information about the measurement results. 
The simplifying assumption here is that since Eve uses individual probes, the signal pairs that Alice and Bob receive will be \textindex{independent and identically distributed} 
(i.i.d.).
That is, after receiving $n$ signals each, they have a state of the i.i.d.~form $(\rho^{AB})^{\otimes n}$.
Alice and Bob estimate $\rho^{AB}$ by estimating the joint probability distribution $q_{ij}$ from their data and using Eq.~\eqref{eq:entbasedstateestimation}.

In a coherent attack, the signals that Alice and Bob receive may be correlated in an arbitrary way. 
They simply receive one big state distributed on $2n$ quantum systems. 
For some protocols their structure makes it possible to reduce a coherent attack to a collective attack \cite{kraus05a}.
For other protocols using finite-dimensional signal states, 
the exponential \textindex{de Finetti theorem} (or \textindex{global representation theorem}) \cite{renner07a} can be applied.
It states that as long as there is a permutation symmetry between the signals, the state that Alice and Bob receive can be regarded as being i.d.d.~on all but a small subset of the signals.
This means that for many protocols, it is sufficient to show security against a collective attack.

In this thesis we do not attempt to show security for any QKD protocols. 
Rather we show in chapters \ref{ch:rcad}, \ref{ch:k2analytical}, and \ref{ch:numerical} that a collective attack is sufficient to break the security for some types of postprocessing when the estimated state is too noisy.
We will therefore always assume that Alice and Bob's data come from comes from measurements on i.i.d. states, and the only justification we need for that is that this is a possible strategy for Eve.

Eve can choose any state $\rho^{AB}$ and then distribute many copies to Alice and Bob.
While Alice and Bob will be able to estimate this state (or at least some of its parameters), they have to assume that Eve can be correlated with this state in an arbitrary way. 
The worst case is that Eve has the full purification of the state (since nothing is correlated with the random results of a measurement on a pure state).
Alice and Bob therefore always assume that their state is part of a pure state, 
$\rho^{AB} = \tr_E \proj{\Psi}_E$, 
where Eve is in possession of system $E$.

In an entanglement based scheme it is evident that Eve may indeed choose to give Alice and Bob any state $\rho^{AB}$ of her choosing, since she is assumed to have full control over its distribution.
In a prepare and measure scheme or---equivalently---an entanglement based scheme where the source is in Alice's lab, Eve only has access to one of the two particles from the source.
Alice's part $\rho^A$ of the final state $\rho^{AB}$ must therefore be the same as her part of the source's state $\proj{\psi}_{AA'}$,
\begin{equation}
 \rho^A = \tr_B [\rho^{AB}] = \tr_{A'} [\proj{\psi}_{AA'}].
\end{equation}
This additional restriction is useful for Alice and Bob in cases where their measurements do not allow them to estimate all parameters of $\rho^{A}$.
Apart from this restriction, Eve can transform the state to any state of her choosing by operating only on the system in transit from Alice to Bob \cite{curty04a}.

\subsection{The six-state QKD protocol}
\label{sec:qkdprotocols}

Two of the most celebrated QKD protocols are the \textindex{BB84 protocol} \cite{bennett84a} and the \textindex{six-state protocol} \cite{bruss98a,bechmann99a}. 
They are both originally qubit-based prepare and measure protocols, but as we saw in section \ref{sec:pmentequivalence} they can also be converted to entanglement based protocols.
The \textindex{BB84 protocol} is the one which is most well known and most used in experiment. 
The \textindex{six-state protocol} adds two signal states which makes theoretical analysis easier, since all parameters of the state in the entanglement based scheme can be estimated and it has a higher noise threshold.


\index{six-state protocol|(}
The six-state protocol uses the following six states as signal states,
\begin{subequations}
\begin{align}
 &\ket{+} := \tfrac{1}{\sqrt{2}}(\ket{0} + \ket{1}),\\
 \label{eq:6statestateminus}
 &\ket{-} := \tfrac{1}{\sqrt{2}}(\ket{0} - \ket{1}),\\
 &\ket{+\im} := \tfrac{1}{\sqrt{2}}(\ket{0} + \im \ket{1}),\\
 &\ket{-\im} := \tfrac{1}{\sqrt{2}}(\ket{0} - \im \ket{1}),\\
 \label{eq:6statestatezero}
 &\ket{0},\\
 &\ket{1}.
\end{align}
\end{subequations}
The two first states are eigenvector of $\sigma_x$ and are therefore called the $x$ basis.
The next two are eigenvectors of $\sigma_y$ and are called the $y$ basis.
The last two are eigenvectors of $\sigma_z$ and are called the $z$ basis or \textindex{computational basis}.
Each basis is chosen with equal probability and each state within a basis is chosen with equal probability, giving probability $1/6$ for each signal state.

The six-state measurement on Bob's side is described by a POVM with the elements
\begin{subequations}
\begin{align}
\label{eq:6statepmpovmfirst}
F_+ &= \tfrac{1}{3}\proj{+},\\
\label{eq:6statepmpovmminus}
F_- &= \tfrac{1}{3}\proj{-},\\
F_{+\im} &= \tfrac{1}{3}\proj{+\im},\\
F_{-\im} &= \tfrac{1}{3}\proj{-\im}\\
F_0 &= \tfrac{1}{3}\proj{0},\\
\label{eq:6statepmpovmlast}
F_1 &= \tfrac{1}{3}\proj{1}.
\end{align}
\end{subequations}
This can be implemented as a random basis choice followed by a projective measurement in that basis. 

The three bases that are used in the six-state protocol are 
\emph{mutually unbiased}\index{mutually unbiased bases}, 
which means that if a state from one basis is measured in one of the other bases, the outcome is uniformly random and therefore contains no information about the original state.
Alice therefore announces to which basis each of her signals belong, and Bob similarly announces which basis was used to measure each signal. 
They then discard their data for each signal where they used a different basis, since Bob's measurement result then is unlikely to be correlated with Alice's choice of signal.
This step is called \emphindex{sifting}.

On average the sifting step will throw away 2/3 of the signals. 
It is possible to throw away only a negligible fraction by choosing the $z$ basis with probability $1-\epsilon$ and the two others with probability $\epsilon/2$ for a small $\epsilon$ \cite{lo05b}.
Since we are only interested in whether a key can be distilled or not, we keep the probabilities equal.


The prepare and measure six-state scheme can be converted to an entanglement based scheme using the recipe in the previous section.
The entangled state is 
\begin{multline}
 \label{eq:6stateentstart}
 \ket{\psi}_{AA'} = \sqrt{\tfrac{1}{6}}\ket{m_0}_A\ket{+}_{A'} 
                    + \sqrt{\tfrac{1}{6}}\ket{m_1}_A\ket{-}_{A'}
                    + \sqrt{\tfrac{1}{6}}\ket{m_2}_A\ket{+\im}_{A'} \\
                    + \sqrt{\tfrac{1}{6}}\ket{m_3}_A\ket{-\im}_{A'}
                    + \sqrt{\tfrac{1}{6}}\ket{m_4}_A\ket{0}_{A'} 
                    + \sqrt{\tfrac{1}{6}}\ket{m_5}_A\ket{1}_{A'},
\end{multline}
and measuring it in the computational basis at $A$ effectively prepares the signal states with the correct probabilities.
The state in Eq.~\eqref{eq:6stateentstart} can be written as 
\begin{multline}
 \label{eq:6stateentschmidt}
 \ket{\psi}_{AA'} \\
         = 
                    \left(\sqrt{\tfrac{1}{12}}\ket{m_0}_A 
                    + \sqrt{\tfrac{1}{12}}\ket{m_1}_A 
                    + \sqrt{\tfrac{1}{12}}\ket{m_2}_A
                    + \sqrt{\tfrac{1}{12}}\ket{m_3}_A 
                    + \sqrt{\tfrac{1}{6}}\ket{m_4}_A\right)
                          \ket{0}_{A'}\\
         +
                    \left(\sqrt{\tfrac{1}{12}}\ket{m_0}_A 
                    - \sqrt{\tfrac{1}{12}}\ket{m_1}_A 
                    +\im \sqrt{\tfrac{1}{12}}\ket{m_2}_A
                    -\im \sqrt{\tfrac{1}{12}}\ket{m_3}_A 
                    + \sqrt{\tfrac{1}{6}}\ket{m_5}_A\right)
                          \ket{1}_{A'}
\end{multline}
which is a Schmidt decomposition since the two vectors,
\begin{subequations}
\begin{align}
 \tfrac{1}{\sqrt{2}} \ket{u_0} &= \sqrt{\tfrac{1}{12}}\ket{m_0}_A 
                    + \sqrt{\tfrac{1}{12}}\ket{m_1}_A 
                    + \sqrt{\tfrac{1}{12}}\ket{m_2}_A
                    + \sqrt{\tfrac{1}{12}}\ket{m_3}_A 
                    + \sqrt{\tfrac{1}{6}}\ket{m_4}_A,\\
 \tfrac{1}{\sqrt{2}} \ket{u_1} &= \sqrt{\tfrac{1}{12}}\ket{m_0}_A 
                    - \sqrt{\tfrac{1}{12}}\ket{m_1}_A 
                    +\im \sqrt{\tfrac{1}{12}}\ket{m_2}_A
                    -\im \sqrt{\tfrac{1}{12}}\ket{m_3}_A 
                    + \sqrt{\tfrac{1}{6}}\ket{m_5}_A
\end{align}
\end{subequations}
are orthogonal.
This means that we can look at the initial state as a two-qubit \textindex{maximally entangled} state
$\ket{\psi}_{AA'} = \frac{1}{\sqrt{2}}(\ket{u_0}\ket{0} + \ket{u_1}\ket{1})$.
We let $\ket{u_0}$ and $\ket{u_1}$ be the computational basis for the qubit space in the $A$ system and rename to the standard convention so that the form of the state is  
\begin{equation}
 \ket{\psi}_{AA'} = \frac{1}{\sqrt{2}}(\ket{00} + \ket{11}).
\end{equation}
To get the POVM elements on this 2-dimensional $A$ system, we project the vectors $\ket{m_i}$ down to the subspace spanned by $\ket{u_0}_A$ and $\ket{u_1}_A$ and rename correspondingly.
Using the projector $\pi = \ketbra{0}{u_0} + \ketbra{1}{u_1}$, we get
\begin{subequations}
\begin{align}
\pi \ket{m_0} &= \sqrt{\tfrac{1}{6}} \ket{0} + \sqrt{\tfrac{1}{6}} \ket{1} 
              = \sqrt{\tfrac{1}{3}} \ket{+},\\
\pi \ket{m_1} &= \sqrt{\tfrac{1}{6}} \ket{0} - \sqrt{\tfrac{1}{6}} \ket{1} 
              = \sqrt{\tfrac{1}{3}} \ket{-},\\
\pi \ket{m_2} &= \sqrt{\tfrac{1}{6}} \ket{0} - \im \sqrt{\tfrac{1}{6}} \ket{1} 
              = \sqrt{\tfrac{1}{3}} \ket{- \im},\\
\pi \ket{m_3} &= \sqrt{\tfrac{1}{6}} \ket{0} + \im \sqrt{\tfrac{1}{6}} \ket{1} 
              = \sqrt{\tfrac{1}{3}} \ket{+ \im},\\
\pi \ket{m_4} &= \sqrt{\tfrac{1}{3}} \ket{0}, \\
\pi \ket{m_5} &= \sqrt{\tfrac{3}{3}} \ket{1}.
\end{align}
\end{subequations}
The POVM elements are then $\left\{ \pi \proj{m_i} \pi^\dagger \right\}_{i=0}^5$.
This is the same measurement as Bob's, as given in Eqs.\eqref{eq:6statepmpovmfirst}-\eqref{eq:6statepmpovmlast}.
Note that from Eq.~\eqref{eq:6stateentstart} we can see that Alice's POVM element $\pi \proj{m_2} \pi^\dagger = \frac{1}{3} \proj{-\im}$ on the $A$ system projects the $A'$ system to the state $\proj{+\im}$. 
In general, measurements in the $y$ basis on $A$ projects the state to the orthogonal state on $A'$, while in the $x$- and $z$ bases, the states are projected to the same state in the two systems.

The \textindex{sifting} is done such that it is indistinguishable from the prepare and measure sifting. 
Instead of announcing in which basis her signal state was sent in, Alice really announces in which basis the measurement result was, since the measurement effectively prepares a signal state in the same basis.
\index{six-state protocol|)}

\subsection{The BB84 protocol}

\index{BB84 protocol|(}
The BB84 protocol is similar to the \textindex{six-state protocol} with the exceptions that the signal states $\ket{+\im}$ and $\ket{-\im}$ are not used, i.e the BB84 signal states are
\begin{subequations}
\begin{align}
 &\ket{+} := \tfrac{1}{\sqrt{2}}(\ket{0} + \ket{1}),\\
 &\ket{-} := \tfrac{1}{\sqrt{2}}(\ket{0} - \ket{1}),\\
 &\ket{0},\\
 &\ket{1}.
\end{align}
\end{subequations}
Like for the six-state protocol, the signal states are chosen with equal probability.
Similarly, only measurements in the $x$- and $z$ basis are used.
This means that instead of the six POVM elements in Eqs.~\eqref{eq:6statepmpovmfirst}-\eqref{eq:6statepmpovmlast} 
the POVM elements for the BB84 measurement are
\begin{subequations}
\begin{align}
\label{eq:BB84pmpovmfirst}
F_0 &= \frac{1}{2}\proj{0},\\
F_1 &= \frac{1}{2}\proj{1},\\
F_+ &= \frac{1}{2}\proj{+},\\
\label{eq:BB84pmpovmminus}
F_- &= \frac{1}{2}\proj{-}.
\end{align}
\end{subequations}
In the entanglement based BB84, both Alice and Bob use this measurement.

The sifting in the BB84 protocol is analogue to the six-state sifting. 
Alice and Bob keep their data if the signal and measurement (or the two measurements in the entanglement based picture) are in the same basis.
The BB84 sifting only discards 1/2 of the signals due to wrong basis instead of 2/3 for the six-state sifting.
Like for the six-state protocol, one basis can be chosen more often in order make this loss negligible, but we do not consider this here.
\index{BB84 protocol|)}

\subsection{Parameter estimation in the six-state and BB84 protocols}
\label{sec:parameterestimation}
\index{parameter estimation|(}
\index{BB84 protocol|(}
\index{six-state protocol|(}

\subsubsection{Full parameter estimation}

For the purpose of analyzing the protocols, we consider the entanglement based schemes, even if the actual implementation may be prepare and measure.
Let us first assume that the \textindex{parameter estimation} is done before the sifting step throws away the data where Alice and Bob used different basis.
We will relax this later so that they can do a meaningful parameter estimation also after sifting.

For the six-state protocol, both Alice and Bob receive a qubit which they measure using the POVM from Eqs.~\eqref{eq:6statepmpovmfirst}-\eqref{eq:6statepmpovmlast}.
From those measurement results that they announce publicly, they reconstruct the joint probability distribution $q_{ij}$.
This probability distribution is somewhat redundant in that the coarse grained measurement results
\begin{subequations}
\begin{align}
F_0 + F_1 &= \frac{1}{3} \calone\\
F_+ + F_- &= \frac{1}{3} \calone\\
F_{+\im} + F_{-\im} &= \frac{1}{3} \calone 
\end{align}
\end{subequations}
are proportional to $\calone$ and therefore their probability is independent of the measured state (it is simply the probability to choose a basis).
For this and other reasons it can be useful to first use the probabilities to compute the 
$R$-matrix\index{R-matrix@$R$-matrix} of the state (see section \ref{sec:qubitsandpaulis}), where each matrix element is given by the expectation value 
$r_{ij} := \tr[(\sigma_i \otimes \sigma_j)\rho^{AB}]$ where $\sigma_0 := \calone$ and $\sigma_1, \sigma_2, \sigma_3$ are $\sigma_x, \sigma_y, \sigma_z$, respectively. 
For this, they can use
\begin{subequations}
\begin{align}
\sigma_x &= \proj{+} - \proj{-} = 3 (F_+ - F_-),\\
\sigma_y &= \proj{+\im} - \proj{-\im} = 3 (F_{+\im} - F_{-\im}),\\
\sigma_z &= \proj{0} - \proj{1} = 3 (F_0 - F_1).
\end{align}
\end{subequations}
For example, 
the $r_{13}$ expectation value is
\begin{align}
 r_{13} &:= \tr[(\sigma_x \otimes \sigma_z)\rho^{AB}] \nonumber\\
        &= 
          \frac{
           \tr [
            \{(F_+ - F_-) \otimes (F_0 - F_1)\} \rho^{AB}
           ]
          }{
           \tr [
            \{(F_+ + F_-) \otimes (F_0 + F_1)\} \rho^{AB}
           ]
          }  \nonumber \\
         &= \frac{q_{+,0} - q_{+,1} - q_{-,0} + q_{-,1} }{q_{+,0} + q_{+,1} + q_{-,0} + q_{-,1}}.
\end{align}
The state can be reconstructed from the $R$-matrix\index{R-matrix@$R$-matrix} as
\begin{equation}
 \label{eq:Rmatrixstate}
 \rho^{AB} = \frac{1}{4} \sum_{i=0}^3 \sum_{j=0}^3 r_{ij} \sigma_i \otimes \sigma_j.
\end{equation}
Since the full $R$-matrix\index{R-matrix@$R$-matrix} is known from the measurements in the six-state protocol, the state is fully determined.

The state can also for the BB84 protocol be reconstructed via the $R$-matrix\index{R-matrix@$R$-matrix}.
In this case, however, the coefficients $r_{2i}$ and $r_{i2}$ in Eq.~\eqref{eq:Rmatrixstate} are unknown, so we get an equivalence class of possible states.

\subsubsection{Simplified parameter estimation}

As the parametrization via the $R$-matrix\index{R-matrix@$R$-matrix} shows, we need to estimate 15 parameters to estimate the two-qubit state (the $r_{00}$ element should be 1 for a normalized state).
By symmetrizing the state, we can estimate the state using fewer parameters.

Consider the following imaginary symmetrization step.
After Alice and Bob have received their qubits but before they measure them, Alice chooses two random bits $b_1$ and $b_2$ which she sends to Bob.
If $b_1 = 1$ they both apply a \textindex{bit-flip operator}, $\sigma_x$ to the qubit.
If $b_2 = 1$ they both apply a \textindex{phase-flip operator}, $\sigma_z$ to the qubit.
They do this to all their qubit pairs and then forget which bits received phase flips and bit flips.
This kind of symmetrization by randomly applying different but correlated unitary operators is called \emphindex{twirling}.
Their symmetrized state is now
\begin{equation}
 \rho_\text{sym}^{AB} = \frac{1}{4}
                        \sum_{b_1 = 0}^1 \sum_{b_2 = 0}^1 
                        (\sigma_z^{b_2} \sigma_x^{b_1} \otimes \sigma_z^{b_2} \sigma_x^{b_1}) 
                         \rho^{AB}
                        (\sigma_x^{b_1} \sigma_z^{b_2} \otimes \sigma_x^{b_1} \sigma_z^{b_2}) 
\end{equation}
This state is invariant under both $\sigma_x \otimes \sigma_x$ and $\sigma_z \otimes \sigma_z$.
We can write out the original state $\rho^{AB}$ using the $R$-matrix\index{R-matrix@$R$-matrix},
as in \eqref{eq:Rmatrixstate}.
By using the commutation and anticommutation rules for Pauli operators\index{Pauli operator} we see that most of the terms cancel and we end up with 
\begin{equation}
 \label{eq:belldiagsigma}
 \rho_\text{sym}^{AB} = \frac{1}{4} \sum_{i=0}^3 r_{ii} \sigma_i \otimes \sigma_i,
\end{equation}
i.e. the diagonal of the $R$-matrix\index{R-matrix@$R$-matrix} is the same, but the off-diagonal terms vanish.
Such a state is Bell diagonal\index{Bell-diagonal state} (see section \ref{sec:qubitsandpaulis}).
After symmetrizing the state this way, it is fully described by the three parameters $r_{11}, r_{22}$, and $r_{33}$, which are the same in the original state and the symmetrized state.

In practice, we can effectively apply the symmetrization step by updating the measurement results to what they would have been if we had done the symmetrization before the measurement.
Each qubit will be measured in the $x$, $y$, or $z$ basis.
For qubits that are measured in the $x$ basis, a $\sigma_x$ operation before the measurement does not change the result of the measurement, since it projects onto $\ket{+}$ and $\ket{-}$ which are eigenvectors of $\sigma_x$. 
A $\sigma_z$ before measurement will give the opposite result. 
Therefore, flipping the measurement result of and $x$-basis measurement whenever $b_1 = 1$ achieves the same effect as symmetrizing before that measurement.
Similarly, measurement results in the $y$ basis are flipped when $b_1 \ne b_2$ and results in the $z$ basis are flipped when $b_2 = 1$.
This process will decorrelate results in different bases since the each result then will be flipped with probability 1/2 independently.
The measurement results in the same basis will still be correlated, but any bias will be removed since both results are flipped together with probability 1/2.

Since the off-diagonal terms in the $R$-matrix\index{R-matrix@$R$-matrix} vanish for the symmetrized state, there is no point in estimating them in the first place.
Since we need only the measurement results from the same basis to estimate the diagonal elements, we can do the \textindex{parameter estimation} after the \textindex{sifting}.
The parameters $r_{11}$, $r_{22}$ and $r_{33}$ can be written as
\begin{subequations}
\begin{align}
 r_{11} &= 1 - 2Q_x\\
 r_{22} &= 2Q_y - 1\\
 r_{33} &= 1 - 2Q_z
\end{align}
\end{subequations}
where 
\begin{subequations}
\begin{align}
 Q_x &= \frac{q_{+,-} + q_{-,+}}{q_{+,+} + q_{+,-} + q_{-,+} + q_{-,-}}\\
 Q_y &= \frac{q_{+\im,+\im} + q_{-\im,-\im}}{q_{+\im,+\im} + q_{+\im,-\im} + q_{-\im,+\im} + q_{-\im,-\im}}\\
 Q_z &= \frac{q_{0,1} + q_{1,0}}{q_{0,0} + q_{0,1} + q_{1,0} + q_{1,1}}
\end{align}
\end{subequations}
are the \textindex{quantum bit error rate}s in each basis.
Note that we count equal outcomes as an error in the $y$ basis, since a perfect channel would give opposite outcomes.
The symmetrized state is one of the possible states that Eve could give Alice and Bob and the symmetrization process will not change the already symmetrized state.
When only the sifted bits are used in the parameter estimation, there is no way for Alice and Bob to tell the difference between a state and its symmetrized version, so the symmetrization step does not give any extra power to Eve in this case.

The \textindex{Bell-diagonal state} estimated from the QBER in the three bases can be written as a convex combination of \textindex{Bell state}s,
\begin{equation}
\rho^{AB}_\text{sym} = p_I \proj{\Phi^+} + p_x \proj{\Psi^+} + p_y \proj{\Psi^-} + p_z \proj{\Phi^-}, 
\end{equation}
where 
\begin{subequations}
\begin{align}
Q_x = p_y + p_z, \\
Q_y = p_x + p_z, \\
Q_z = p_x + p_y, 
\end{align}
\end{subequations}
and $p_I + p_x + p_y + p_z = 1$.

For the BB84 protocol the error rate in the $y$ basis, $Q_y$, is not known so we still get an equivalence class of possible states. 
Having symmetrized the state we only have one open parameter compared to seven before symmetrization.

\subsubsection{Identifying a state for each basis}

In the six-state protocol, the state is measured in three different bases.
This means that there can be different error rates in the \textindex{raw key} coming from different bases and also Eve's knowledge about the raw key will be different in the three bases.
We prefer to standardize the state so that the measured bits are in the computational basis.
We imagine that the measurements in the different bases is being performed by rotating the $x$ basis or $y$ basis to the $z$ basis before measuring in this basis. 
We then get three different states before measurement in the computational ($z$) basis.

For the basis rotation we use the unitary operator 
\begin{equation}
T := \ketbra{+}{0} 
    - \im \ketbra{-}{1} = \frac{1}{\sqrt{2}}
      \begin{bmatrix} 1 & -\im \\ 1 & \im \end{bmatrix}.
\end{equation}
This unitary operator has the property that 
$T \sigma_x T^\dagger = \sigma_y$, 
$T \sigma_y T^\dagger = \sigma_z$, and 
$T \sigma_z T^\dagger = \sigma_x$.
In the $y$ basis, a state with little noise will give anticorrelated results. 
We therefore want to flip the results in this basis. 
The most straightforward way to achieve this is to rotate using $T$ in Alice's system and the complex conjugate $T^* = \ketbra{+}{0} - \im \ketbra{-}{1}$ in Bob's system.
This rotation has the following effect on the non-identity terms of Eq.~\eqref{eq:belldiagsigma},
\begin{subequations}
\begin{align}
 (T \otimes T^*) (\sigma_x \otimes \sigma_x) (T \otimes T^*)^\dagger &= -\sigma_y \otimes \sigma_y,\\
 (T \otimes T^*) (\sigma_y \otimes \sigma_y) (T \otimes T^*)^\dagger &= -\sigma_z \otimes \sigma_z,\\
 (T \otimes T^*) (\sigma_z \otimes \sigma_z) (T \otimes T^*)^\dagger &= \phantom{-}\sigma_x \otimes \sigma_x.
\end{align}
\end{subequations}
This means that if the $R$-matrix\index{R-matrix@$R$-matrix} of the original symmetrized state is
$\diag(1, r_{11}, r_{22},$ $r_{33})$, 
the $R$-matrix after applying $T \otimes T^*$ once is 
$\diag(1,r_{33},-r_{11},-r_{22})$, and after applying it twice it is
$\diag(1,-r_{22},-r_{33},r_{11})$. 
In other words, the expectation value in the computational basis, 
$\langle \sigma_z \otimes \sigma_z \rangle$ is 
$r_{33}$ when measured on $\rho^{AB}_\text{sym}$, 
$-r_{22}$ when measured on $(T \otimes T^*)\rho^{AB}_\text{sym}(T \otimes T^*)^\dagger$, 
and
$r_{11}$ when measured on $(T \otimes T^*)^2\rho^{AB}_\text{sym}(T \otimes T^*)^{2\dagger}$.
We denote the rotated states by 
\begin{subequations}
\begin{align}
 \rho^{AB}_x &= (T \otimes T^*)^2 \rho^{AB}_\text{sym} (T \otimes T^*)^{2\dagger},\\
 \rho^{AB}_y &= (T \otimes T^*) \rho^{AB}_\text{sym} (T \otimes T^*)^\dagger,\\
 \rho^{AB}_z &= \rho^{AB}_\text{sym}.\\
\end{align}
\end{subequations}
These states will be the starting states which we try to distill a secret key from by using advantage distillation followed by other postprocessing procedures. 
It is entirely possible that a secret key can be distilled from only one or two of these states. 
In terms of the measured data, this means that the \textindex{raw key} from one basis may have too much error combined with too low secrecy to allow key distillation while the raw key from another basis may be distillable.
In the analysis we will not treat the three states separately, but rather give statements about key distillation from a specific \textindex{Bell-diagonal state}. 
After parameter estimation in the six-state protocol it is important to check all three Bell-diagonal states against any distillation criterion.

The Bell-diagonal state will often be given on the spectral decomposition instead of by the $R$-matrix.
The action of the rotation on the \textindex{Bell state}s are
\begin{subequations}
\begin{align}
(T \otimes T^*) \ket{\Phi^+} &= \phantom{-}\phantom{\im} \ket{\Phi^+}  
    &  (T \otimes T^*)^2 \ket{\Phi^+} &= \phantom{-}\phantom{\im} \ket{\Phi^+}, \\
(T \otimes T^*) \ket{\Psi^+} &=         {-}        {\im} \ket{\Psi^-}  
    &  (T \otimes T^*)^2 \ket{\Psi^+} &= \phantom{-}\phantom{\im} \ket{\Phi^-}, \\
(T \otimes T^*) \ket{\Psi^-} &= \phantom{-}        {\im} \ket{\Phi^-}  
    &  (T \otimes T^*)^2 \ket{\Psi^-} &= \phantom{-}        {\im} \ket{\Psi^+}, \\
(T \otimes T^*) \ket{\Phi^-} &= \phantom{-}\phantom{\im} \ket{\Psi^+}  
    &  (T \otimes T^*)^2 \ket{\Phi^-} &=         {-}        {\im} \ket{\Psi^-}. 
\end{align}
\end{subequations}
So if the estimated state is 
\begin{subequations}
\begin{align}
 \label{eq:belldiagonalZbasis}
 \rho^{AB}_\text{sym} = \rho^{AB}_z 
  &= p_I \proj{\Phi^+} + p_x \proj{\Psi^+} + p_y \proj{\Psi^-} + p_z \proj{\Phi^-},\\
\intertext{the state from the $y$ and $x$ basis are}
\rho^{AB}_y 
&= p_I \proj{\Phi^+} + p_z \proj{\Psi^+} + p_x \proj{\Psi^-} + p_y \proj{\Phi^-},\\
\label{eq:belldiagonalXbasis}
\rho^{AB}_x 
&= p_I \proj{\Phi^+} + p_y \proj{\Psi^+} + p_z \proj{\Psi^-} + p_x \proj{\Phi^-}.
\end{align}
\end{subequations}
An example state that can only be distilled in one basis according to condition \eqref{eq:chau-conditionINTRO} in chapter \ref{ch:rcad} is the Bell-diagonal state defined by 
$(p_I, p_x, p_y, p_z) = (\frac{4}{7}, \frac{3}{14}, 0 \frac{3}{14})$.
The state itself does not satisfy the condition, but when rotated with $T \otimes T^*$ so we get the previous $y$ basis in the computational basis, it is distillable.

In the BB84 protocol, only the $x$ and $z$ basis are used. 
Instead of imagining using the unitary operators $T^2$ and $T$ to turn the $x$ and $y$ bases into the computational basis, we imagine using the unitary and Hermitian Hadamard operator 
\begin{equation}
 H := \ketbra{+}{0} + \ketbra{-}{1} =  \frac{1}{\sqrt{2}}\begin{bmatrix} 1 & 1\\ 1 & -1\end{bmatrix},
\end{equation}
which interchanges the $x$ and $z$ bases\footnote{
  It would be possible to use $T^2$ to turn the $x$ basis into the computational basis and use the resulting state for checking if key distillation form the $x$-basis measurement results is possible.
  However, this would complicate the discussion when we soon only use the \emph{average} quantum bit error rate to estimate our state.
}.
The effect on Pauli operators is that 
$H \sigma_x H^\dagger =  \sigma_z$, 
$H \sigma_y H^\dagger = -\sigma_y$, and 
$H \sigma_z H^\dagger =  \sigma_x$.
This gives the following effect on the terms in Eq.~\eqref{eq:belldiagsigma} when both systems are rotated,
\begin{subequations}
\begin{align}
 (H \otimes H^*) (\sigma_x \otimes \sigma_x) (H \otimes H^*)^\dagger &= \sigma_z \otimes \sigma_z,\\
 (H \otimes H^*) (\sigma_y \otimes \sigma_y) (H \otimes H^*)^\dagger &= \sigma_y \otimes \sigma_y,\\
 (H \otimes H^*) (\sigma_z \otimes \sigma_z) (H \otimes H^*)^\dagger &= \sigma_x \otimes \sigma_x.
\end{align}
\end{subequations}
On the \textindex{Bell state}s the result is,
\begin{subequations}
\begin{align}
(H \otimes H) \ket{\Phi^+} &= \phantom{-} \ket{\Phi^+}  \\
(H \otimes H) \ket{\Psi^+} &= \phantom{-} \ket{\Phi^-}  \\
(H \otimes H) \ket{\Psi^-} &=         {-} \ket{\Psi^-}  \\
(H \otimes H) \ket{\Phi^-} &= \phantom{-} \ket{\Psi^+}  
\end{align}
\end{subequations}
If the original state parametrized by the $R$-matrix\index{R-matrix@$R$-matrix} is
\begin{align}
 \label{eq:rhozBB84r}
 \rho^{AB}_\text{sym} &= \frac{1}{4} \left( 
   \calone \otimes \calone
  + r_{11} \sigma_x \otimes \sigma_x 
  + r_{22} \sigma_y \otimes \sigma_y 
  + r_{33} \sigma_z \otimes \sigma_z 
\right),
\intertext{the state after a Hadamard rotation is}
 \label{eq:rhoxBB84r}
 (H \otimes H)\rho^{AB}_\text{sym}(H \otimes H)^\dagger &= \frac{1}{4} \left( 
   \calone \otimes \calone
  + r_{33} \sigma_x \otimes \sigma_x 
  + r_{22} \sigma_y \otimes \sigma_y 
  + r_{11} \sigma_z \otimes \sigma_z 
\right).
\end{align}
Since Alice and Bob cannot estimate $r_{22}$ from their BB84 measurements, this is an open parameter in both states. 
When written as a convex combination of Bell states\index{Bell state}, the states are
\begin{subequations}
\begin{align}
\label{eq:rhozBB84}
 \rho^{AB}_z &= p_I \proj{\Phi^+} + p_x \proj{\Psi^+} + p_y \proj{\Psi^-} + p_z \proj{\Phi^-},\\
 \label{eq:rhoxBB84}
 (H \otimes H)\rho^{AB}_\text{sym}(&H \otimes H)^\dagger  \nonumber\\
 &= p_I \proj{\Phi^+} + p_z \proj{\Psi^+} + p_y \proj{\Psi^-} + p_x \proj{\Phi^-},
\end{align}
\end{subequations}
where the eigenvalues are not fully determined due to their dependence on the unknown parameter $r_{22}$ given in Eqs.~\eqref{eq:belleigenvaluesfromRmatrixfirst}-\eqref{eq:belleigenvaluesfromRmatrixlast}.

\subsubsection{Fully simplified parameter estimation}

Sometimes it is useful to characterize the quality of Alice and Bob's state using only a single parameter, the average \textindex{quantum bit error rate} in the bases that are measured.
For the six-state protocol this average QBER is 
$Q = (Q_x + Q_y + Q_z)/3$.

By forgetting which basis each of the measured bits comes from, the average state is
\begin{equation}
 \rho^{AB}_\text{iso} = \frac{\rho^{AB}_x + \rho^{AB}_y + \rho^{AB}_z}{3}
\end{equation}
The $R$-matrix\index{R-matrix@$R$-matrix} can be written as 
$\diag(1,r,-r,r)$ 
where 
$r = \frac{1}{3} (r_{11} - r_{22} + r_{33})$.
When written as a mixture of the three Bell-diagonal states in Eqs.~\eqref{eq:belldiagonalZbasis}--\eqref{eq:belldiagonalXbasis},
this state becomes 
\begin{equation}
\rho^{AB}_\text{iso}
  = p_I \proj{\Phi^+} + p \proj{\Psi^+} + p \proj{\Psi^-} + p \proj{\Phi^-}
\end{equation}
where $p = (p_x + p_y + p_z)/3$ and $p_I = 1-3p$.
A state of this form is called an \emphindex{isotropic state}.
The parameter $p$ is half the average QBER, $p = Q/2$.
When we consider the six-state protocol with a given QBER, this is the starting state.

In the BB84 protocol, the average of the two states from 
Eqs. \eqref{eq:rhozBB84r} and\eqref{eq:rhoxBB84r} is
\begin{subequations}
\begin{align}
 \rho^{AB}_\text{BB84} &= \frac{\rho^{AB}_\text{sym} + (H \otimes H)\rho^{AB}_\text{sym}(H \otimes H)^\dagger}{2}\\
 &= \frac{1}{4} \left( 
   \calone \otimes \calone
  + r \sigma_x \otimes \sigma_x 
  + r_{22} \sigma_y \otimes \sigma_y 
  + r \sigma_z \otimes \sigma_z 
\right),\\
  &= p_I \proj{\Phi^+} + p \proj{\Psi^+} + p_y \proj{\Psi^-} + p \proj{\Phi^-},
\end{align}
\end{subequations}
where $r = (r_{11} + r_{33})/2 = 1-2Q$ and $p = (p_x + p_z)/2$ depends on the open parameter $r_{22}$.
We have here an equivalence class of starting states, parametrized by the open parameter $r_{22}$.
If a procedure is to succeed in distilling a secret key, it must distill key for all the states in this equivalence class.

\index{parameter estimation|)}
\index{BB84 protocol|)}
\index{six-state protocol|)}

\subsection{Analysis using coherent bipartite states}
\label{sec:coherentanalysis}

We justified the simplified parameter estimation by showing that even though Alice and Bob could not do the symmetrizing operations before measuring the qubits, they could update their measurement results to what they would have been if the quantum operations had been performed before the measurements.
The difference would not have any effect on the final measurement results or Eve's knowledge about them.
This may be seen as an application of the \emphindex{principle of deferred measurement} from quantum computing \cite{nielsen00a,griffiths96a}; 
any measurement during the intermediate stage of a computation can be moved to the end and if any of the measurement results are used to control further operations, they may be replaced with quantum controls.

The postprocessing that we will analyze in chapters \ref{ch:lad}--\ref{ch:numerical} in this thesis is purely classical and operates on blocks of the bits coming from these measurements. 
Again, we will imagine that we postpone the measurements as far as possible, but since the postprocessing involves classical communication we will need to measure in order to obtain those classical bits. 
We will, however, only make a minimal measurement to obtain those bits; 
when a postprocessing step calls for announcing a bit that depends on the bit value of several of the bits from the \textindex{raw key}, 
we imagine computing that bit using a quantum circuit and only measuring the qubit that contains the result of the computation in the computational basis. 

A similar approach is used when showing the security of QKD protocols by showing the equivalence to an entanglement purification protocol \cite{shor00a,gottesman03a}.
In that case the correctness and secrecy of the key follows from the fact that if the measurements are postponed until the end of the protocol and all processing is done coherently, the states right before the measurement would be maximally entangled states (or sometimes other private states \cite{renes07a}).
In our case we use show that the quantum state after the postprocessing steps involving two-way communication has finished has a \emphindex{symmetric extension}, and therefore no coherent or incoherent processing involving only one-way communication can result in a secret key.
One difference between the two cases is that in a security proof one may give Eve some extra power, since that does not weaken the security statement.
Since we use the existence of a symmetric extension to show that a key cannot be distilled, we must be careful to not give Eve access to anything she does not have.
In particular, steps like the P-step of \cite{gottesman03a} discard qubits which amount to giving them to Eve, since we always assume she has the purification of Alice and Bob's state.
If a protocol calls for discarding qubits, they must still be taken into account when checking for a symmetric extension, so we do not distinguish between the qubits.
We emphasize again that these quantum operations are purely fictional and only serve to show that classes of classical processing on classical bits cannot lead to a secret key when the Eve has the ability to manipulate quantum systems as she prefers.

It deserves mentioning that it is possible to take a different path after parameter estimation step.
Instead of continuing to postpone the measurement and imagine doing coherent processing, one can compute the classical-classical-quantum (ccq) state \index{ccq state} \cite{devetak04a,devetak05a,horodecki05a,kraus05a}, the state shared between Alice, Bob and Eve after Alice and Bob have done their measurements.
It is computed by first computing the purification of Alice and Bob's (possibly symmetrized) state and giving the purifying system to Eve. 
Then Alice and Bob measure their systems and index their classical result by a one-dimensional projector $\proj{i}$.
The ccq state is then of the form
\begin{equation}
 \label{eq:ccq}
 \rho_\text{ccq}^{ABE} = \sum_{ij} p_{ij} \proj{i}_A \otimes \proj{j}_B \otimes \rho_{ij}^E.
\end{equation}
The probability $p_{ij}$ is the probability for Alice getting measurement outcome $i$ and Bob measurement outcome $j$, and $\rho_{ij}^E$ is Eve's conditional state for each case.
Any processing of this state is now straightforward. 
Alice and Bob can add systems, pass messages to each other, and discard systems and Eve's state must be updated with the all communication from Alice to Bob. 
This class of processing is called \emph{LOPC} for \emphindex{Local operations and public communication}.
A detailed comparison between this class and bipartite LOCC \textindex{Local operations and classical communication} is included in \cite{horodecki09a}.

\section{Semidefinite programming}
\label{sec:prelim:sdp}

We will use \textindex{semidefinite programming} in several places in this thesis. 
We show in section \ref{sec:sdpimplementation} how to formulate a $2k$-qubit Bell-diagonal symmetric extension problem as a \textindex{semidefinite program}.
First we use this in section \ref{sec:belldiagsymext-sdp} to solve the symmetric extension problem analytically for (two-qubit) Bell-diagonal states.
Then in chapter \ref{ch:numerical} we need to test several $2k$-qubit Bell-diagonal states for a symmetric extension and we do this by formulating it as a semidefinite program which we can solve numerically.

Every symmetric extension problem can be formulated as as semidefinite program \cite{doherty02a,doherty04a} and also other subsystem compatibility problems \cite{hall07a} fall into this category.
Semidefinite programming also has applications in many areas of engineering \cite{vandenberghe96a,boyd}.

\subsection{Definition}
\label{sec:sdpdef}


A \emphindex{semidefinite program} \emph{(SDP)} is a convex optimization problem where we optimize over an affine subset of $n \times n$ positive semidefinite Hermitian\footnote{
Semidefinite programs are often formulated with real symmetric matrices, but in the context of quantum mechanics it is useful to allow complex Hermitian matrices. 
It turns out that the density matrices we want to check for symmetric extension in this thesis are all real, so we only really need the real symmetric formulation.
} matrices and the objective function is a linear function of the matrix.


Semidefinite programs can be cast in two forms \cite[pp.~168--169]{boyd}. The first is the \emph{standard form}\index{standard form of an SDP}
\begin{subequations}
\begin{align}
 \label{eq:standardformobj}
 \textrm{minimize }    \quad & \tr[C X]\\
 \textrm{subject to }  \quad & \tr[A_i X] = b_i, \quad i = 1, \ldots, p\\
 \label{eq:standardformpositivity}
                             & X \geq 0
\end{align}
\end{subequations}
where $C$ and $A_i$ are given Hermitian matrices\footnote{We also assume here that the $A_i$ are linearly independent. 
If not, either it is not possible to satisfy the linear constraints or some of the constraints are redundant and can be removed.}, $b_i$ are given fixed parameters, and $X$ is the Hermitian matrix to be optimized over.
The affine subset is here specified by the $p$ linear constraints on $X$, $\tr[A_i X] = b_i$. 

We can complete the matrices $A_i$ with other matrices $B_i$ such that they together form a Hermitian basis for the real vector space of Hermitian matrices.
We can choose these such that $\tr[A_i B_i] = 0$, i.e. they are orthogonal to $A_i$ with the matrix inner product so that $A_i$ and $B_i$ span orthogonal subspaces.
Then $\tr[B_i X] =: x_i$ are open parameters that can be changed freely without violating the linear constraints.
It is possible to write the matrix $X$ as a function of these open parameters,
\begin{equation}
\label{eq:sdpXasdualdum}
 X = \sum_{i=1}^p b_i A'_i + \sum_{i=1}^m x_i B'_i.
\end{equation}
where $p + m = n^2$ (or $p+m = n(n+1)/2$ for real symmetric matrices).
Here, $A'_i$ is a dual basis of $A_i$, i.e. a basis such that $\tr[A'_i A_j] = \delta_{ij}$ and similarly $B'_i$ is the dual basis of $B_i$, each on their own subspace.
The objective function \eqref{eq:standardformobj} can then also be written as a function of the open variables $x_i$.
The matrix $C$ that defines the objective function is in general a linear combination of the $A_i$ and $B_i$ that span the operator space, 
\begin{equation}
 C = \sum_{i=1}^p a_i A_i + \sum_{i=1}^{m} c_i B_i
\end{equation}
We then get 
\begin{equation}
 \tr[C X] = \sum_{i=1}^p a_i \tr[A_i X] + \sum_{i=1}^{m} c_i \tr[B_i X] 
            = \sum_{i=1}^p a_i b_i + \sum_{i=1}^{m} c_i x_i
\end{equation}
The first summation here is constant, so we only need to optimize the terms involving $x_i$.
We then get a semidefinite program on what is called the \emph{inequality form}\index{inequality form of an SDP},
\begin{subequations}
\begin{align}
 \label{eq:ineqformobj}
 \textrm{minimize }    \quad & \sum_{i=1}^{m} c_i x_i\\
 \label{eq:ineqformconstraints}
 \textrm{subject to }  \quad & F(x) := G + \sum_{i=1}^{m} x_i F_i \geq 0
\end{align}
\end{subequations}
Here, $G$ is the first sum in Eq.~\eqref{eq:sdpXasdualdum}, which is a constant matrix, and $F_i = B'_i$.
This semidefinite program is equivalent to the SDP in Eqs.~\eqref{eq:standardformobj}-\eqref{eq:standardformpositivity}, except that the objective function has been shifted by a constant.
An inequality of the form \eqref{eq:ineqformconstraints} is called a \emphindex{linear matrix inequality} \emph{(LMI)}\index{LMI|see{linear matrix inequality}}.

While the standard form of an SDP parametrizes the linear constraints and leaves the rest open, the inequality form specifies a set of open parameters and therefore needs no parametrization of the constraints.
Both define the same affine set of possible matrices, subject to an additional positivity constraint.
In both cases it is possible to use any operator basis that spans the same spaces as $A_i$ and $F_i$ in order to specify the affine set.

We denote any feasible $X$ that minimizes the objective function \eqref{eq:standardformobj} by $X^*$. 
Similarly, an $x$ that minimizes the objective function in Eq.~\eqref{eq:ineqformobj} is denoted by $x^*$. 

If we are only interested in whether a solution that satisfies the constraints exists or not, we can set $C = 0$ in the objective function \eqref{eq:standardformobj} or all $c_i = 0$ in the objective function \eqref{eq:ineqformobj}. 
The optimization is then trivial.
Such an SDP is called a \emphindex{feasibility problem}.

\subsection{Duality}
\label{sec:sdpduality}

To each semidefinite program we can associate a \emphindex{Lagrangian} \cite[p.~265]{boyd}. 
For a problem on inequality form 
\eqref{eq:ineqformobj}-\eqref{eq:ineqformconstraints}
this is given by 
\begin{subequations}
\begin{align}
\label{eq:lagrangiandef}
 L(x,Z) &:= \sum_{i=1}^{m} c_i x_i - \tr[F(x) Z]\\
\label{eq:lagrangianexpression}
        & = \sum_{i=1}^m  x_i(c_i - \tr[F_i Z]) - \tr[GZ],
\end{align}
\end{subequations}
where $Z$ is a new variable Hermitian matrix called a \textindex{Lagrange multiplier}.

From the Lagrangian we get the \emphindex{Lagrange dual function} (or just \textindex{dual function}) by taking the infimum over all $x$,
\begin{equation}
 g(Z) := \inf_{x} L(x,Z) = \left\{ 
 \begin{aligned}
  &-\tr[GZ] &\quad  &{}\text{for }\tr[F_i Z] = c_i, \\
  &- \infty &\quad  &{}\text{otherwise}.
 \end{aligned}
 \right. 
\end{equation}
The dual function gives us lower bounds on the objective function \eqref{eq:ineqformobj}.
For any positive semidefinite $Z$, we have that $\tr[F(x) Z] \geq 0$ so the Lagrangian \eqref{eq:lagrangiandef} is less than the objective function in Eq.~\eqref{eq:ineqformobj} for any $x$.
Since the dual function is the infimum of the Lagrangian it is even less and it is therefore a lower bound on the objective function which is independent of the open variables $x_i$.
It is therefore in particular a lower bound on the minimal value of the objective function.
It is of course a useless lower bound for any $Z$ that does not satisfy $\tr[F_i Z] = c_i$.

The \emphindex{Lagrange dual problem}\index{dual problem} is the problem of maximizing the dual function in order to get a lower bound which is as good as possible.
It is also a semidefinite program and is most easily expressed on standard form with a maximization instead of a minimization,
\begin{subequations}
\begin{align}
 \label{eq:dualstandardformobj}
 \textrm{maximize }    \quad & -\tr[G Z]\\
 \textrm{subject to }  \quad & \tr[F_i Z] = c_i, \quad i = 1, \ldots, p\\
 \label{eq:dualstandardformpositivity}
                             & Z \geq 0.
\end{align}
\end{subequations}
The dual problem is specified by the same parameters as the original problem (also called the \emph{primal problem} in duality settings)---the numbers $c_i$ and the matrices $F_i$ and $G$.

The property that the maximum of the dual problem is less than the minimum of the primal problem is called \emphindex{weak duality}. 
The difference between $\sum_{i=1}^m c_i x_i$ and $-Tr[F(x) Z]$ for some feasible $x$ and $Z$ is called the \emphindex{duality gap} and is always nonnegative.
It is equal to the matrix inner product between matrices from the primal and dual problem,
\begin{equation}
\label{eq:dualitygap}
 \tr[F(x) Z] = \sum_{i=1}^m x_i \tr[F_i Z] + \tr[GZ] = \sum_{i=1}^m c_i x_i  - (- \tr[GZ]). 
\end{equation}
For most semidefinite programs the optimal duality gap is zero, i.e. the optimal objective functions of the primal and dual problems are equal.
This property is called \emph{strong duality}.
A sufficient condition for strong duality is that either the primal or dual problem is \emphindex{strictly feasible}, i.e. there exists open parameters $x_i$ such that $F(x)$ is not only positive semidefinite but also positive definite or $Z$ such that $\tr[F_i Z] = c_i$ and $Z > 0$ \cite[Theorem 3.1]{vandenberghe96a}.

Strong duality does not exclude the possibility that one of the two SDPs are infeasible. 
For example, the primal problem may be strictly feasible, but the minimization gives $-\infty$ (i.e. there is no minimum).
In that case there is no maximum for the dual problem.
If both problems are strictly feasible, we are guaranteed that both optima exist and are equal \cite[Theorem 3.1]{vandenberghe96a}.
With a vanishing optimal duality gap, 
\begin{equation}
 \tr[F(x^*) Z^*] = 0
\end{equation}
the matrices $F(x^*)$ and $Z^*$ must reside on orthogonal subspaces,
\begin{equation}
 \label{eq:complementaryslackness}
 F(x^*) Z^* = 0.
\end{equation}
This property is called \emphindex{complementary slackness}.
It will be very useful in section \ref{sec:belldiagsymext-sdp} when we solve an SDP analytically.

\part{Symmetric extension and related properties}
\label{part:symext}

\chapter*{Prologue}
\addcontentsline{toc}{chapter}{Prologue} 
\markboth{Prologue}{} 

There has been an interest for how bipartite quantum states can be part of multi-partite quantum states for a long time.
A relevant question for computation of molecular energy levels fifty years ago was: given certain conditions on the state of $N$ particles, what are the possible two-particle density matrices \cite[(6) p. 175]{coulson60a}.
When the $N$ particles are fermions, this problem is known as the $N$-representability\index{N-representability@$N$-representability} problem
\cite{coleman63a}.
Different problems relating to the compatibility of density matrices on overlapping sets of subsystems with a global state has been studied by several authors over the years \cite{werner90a,jones05a,butterley06a,liu06a,hall07a}.

The symmetric extension problem is a special class of subsystem compatibility problem.
In general one may ask for a $(n,m)$-symmetric extension, meaning that we are asking if there exists a state on $n$ copies of system $A$ and $m$ copies of system $B$ such that any of the $A$ systems together with any of the $B$ systems has a density matrix equal to the given bipartite state. 
This kind of symmetric extension problem can be used to decide if a state is separable or entangled \cite{doherty02a,doherty04a} and also has implications for Bell-inequalities with $n$ settings at Alice or $m$ settings at Bob \cite{terhal03a}.

In this thesis we only study the (1,2)-symmetric extension problem where we are looking for tripartite extensions, on one copy of system $A$ and two copies of system $B$.
Chapters \ref{ch:symext} and \ref{ch:twoqubits} contain our results for states with a symmetric extension, chapter \ref{ch:bosonicfermionic} relates the symmetric extension problem to the bosonic and fermionic extension problems, and chapter \ref{ch:degradable} links the results to degradable and antidegradable quantum channels.

\chapter{Symmetric extension}
\label{ch:symext}

\section{Definition}

Most of this thesis is related to the concept of a \emphindex{symmetric extension}.
Let us first define what we mean by that.

\begin{defn}
 \label{def:symext}
 A bipartite quantum state $\rho^{AB}$ is said to have a \emph{symmetric extension} if there exists a tripartite quantum state $\rho^{ABB'}$ that satisfies
 \begin{subequations}
 \begin{gather}
  \label{eq:tracecondition}
  \tr_{B'} [\rho^{ABB'}] = \rho^{AB}\\
  \label{eq:swapcondition}
  (\calone_A \otimes \swap) \rho^{ABB'} (\calone_A \otimes \swap)^\dagger = \rho^{ABB'}
 \end{gather}
 \end{subequations}
where $\swap$ is the unitary operator that swaps systems $B$ and $B'$.
We may also say that a state with a symmetric extension is \emphindex{symmetric extendible}.
\end{defn}

In the definition, we have introduced a third quantum system $B'$. 
As the name suggests, this is of the same dimension as system $B$.
In addition, the swap operator $\swap$ implies that there is an \textindex{isometry} between $B$ and $B'$ (or rather between their corresponding Hilbert spaces $\mathcal{H}_B$ and $\mathcal{H}_{B'}$), so that for any vector in or operator on $\mathcal{H}_B$ there is a corresponding vector or operator on $\mathcal{H}_{B'}$.
The swap operator can be written out as 
\begin{equation}
 \swap = \sum_{ij} \ketbra{ij}{ji}_{BB'} = \sum_{ij} \ketbra{i}{j}_B \otimes \ketbra{j}{i}_{B'},
\end{equation}
where $\ket{i}_B$ and $\ket{i}_{B'}$ are corresponding vectors in the two Hilbert spaces.
When the swap operator swaps two of many systems, we will often omit the identity on the other systems in the notation, e.g. writing only $\swap\ket{\psi}_{ABB'}$ instead of $(\calone_A \otimes \swap)\ket{\psi}_{ABB'}$ when the operator acts on a tripartite system $ABB'$.

The first property, Eq.~\eqref{eq:tracecondition}, sometimes called the \emphindex{trace condition}, ensures that $\rho^{ABB'}$ is really an extension of $\rho^{AB}$ and not some unrelated quantum state.
The second property, Eq.~\eqref{eq:swapcondition}, the \emphindex{swap condition}, enforces that $B$ and $B'$ are on equal footing, since swapping the two will leave the tripartite state invariant.

One could imagine using a seemingly weaker condition instead of the swap condition, Eq.~\eqref{eq:swapcondition}. 
The condition
\begin{equation}
 \tr_{B'} [\rho^{ABB'}] = \tr_{B} [\rho^{ABB'}],
\end{equation}
which we can also write as $\rho^{AB} = \rho^{AB'}$, only ensures that having $A$ and $B$ together is equivalent to having $A$ and $B'$.
Note that here we also make implicit use of the \textindex{isometry} between $B$ and $B'$, since we are comparing density operators on different systems.
For our applications in QKD this is all we need, since it means that an eavesdropper in possession of the $B'$ system is in the same position as Bob with respect to Alice's system.

On closer inspection it turns out that the definitions with the two variants of the swap condition are equivalent. 
Given a $\rho^{AB}$, let $\sigma^{ABB'}$ be a state that satisfies $\tr_{B'} [\sigma^{ABB'}] = \rho^{AB}$ and $\tr_{B'} [\sigma^{ABB'}] = \tr_{B} [\sigma^{ABB}]$, but not necessarily $(\calone_A \otimes \swap) \sigma^{ABB'} (\calone_A \otimes \swap)^\dagger = \sigma^{ABB'}$.
The state $\sigma^{ABB'}$ would then be a valid symmetric extension according to the seemingly weaker definition, but not according to the stronger.
But a symmetric extension is a property of $\rho^{AB}$, not $\sigma^{ABB'}$, and a valid extension from both definitions is the symmetrized state $\rho^{ABB'} := (\sigma^{ABB'} + (\calone_A \otimes \swap) \sigma^{ABB'} (\calone_A \otimes \swap)^\dagger)/2$.
So the state $\rho^{AB}$ has a symmetric extension according to both definitions.
Since definition \ref{def:symext} puts more constraints on the tripartite state, that makes it easier to look for a symmetric extension or verify that none exists.

Note that a symmetric extension does not mean that the state is supported only on the \textindex{symmetric subspace} of $\mathcal{H}_B \otimes \mathcal{H}_{B'}$. 
One can also ask for such extensions and we call them bosonic extensions\index{bosonic extension}. 
A bosonic extension satisfies the stronger one-sided swap condition $\swap \rho^{ABB'} = \rho^{ABB'}$. 
Similarly there are fermionic extensions\index{fermionic extension} that satisfy $\swap \rho^{ABB'} = -\rho^{ABB'}$.
Bosonic and fermionic extensions are special cases of symmetric extensions and they are analyzed in further detail in chapter \ref{ch:bosonicfermionic}.

In this work, we always use the convention that a copy of system $B$ is added. 
Of course this is purely conventional and we could instead add a copy of system $A$, but our convention will be useful when we consider one-way communication from Alice to Bob in a QKD setting.
In general, it is possible to add more copies to both sides. 
If we look for a state with a total of $n$ copies of $A$ and $m$ copies of $B$, such that it is invariant under any permutation of the $A$ systems or $B$ systems, this is called an $(n,m)$ symmetric extension.
This concept has been used to derive an algorithm for checking if a state is separable \cite{doherty02a,doherty04a}.
In this setting, our symmetric extensions are $(1,2)$ symmetric extensions, but since we only consider those, we will not write that out.


\section{Properties}
\label{sec:symext-properties}

\subsection{Basic properties}
\label{sec:basic-symext-properties}

The set of states with a symmetric extension have some properties that are worth mentioning. 
First of all, the set is \emph{convex}\index{convexity}.
If $\rho_1^{AB}$ and $\rho_2^{AB}$ are states with the symmetric extensions $\rho_1^{ABB'}$ and $\rho_2^{ABB'}$, then the convex combination $\rho_p^{AB} := (1-p)\rho_1^{AB} + p \rho_2^{AB}$ also has a symmetric extension, namely 
$\rho_p^{ABB'} := (1-p)\rho_1^{ABB'} + p \rho_2^{ABB'}$.
The convex structure means that there are extremal states with a symmetric extension, and we will see in section~\ref{sec:pure-extendible-decomp} and section~\ref{sec:condition} that these states have some nice properties.

Another useful property is that the set is closed under 1-LOCC\index{1-LOCC}, Local Operation and one-way Classical Communication\index{Local Operation and one-way Classical Communication}, 
where the communication flows from Alice to Bob in our convention.
Even if the operation is allowed to be non-trace-preserving (sometimes called 1-SLOCC\index{1-SLOCC} for Stochastic Local Operation and one-way Classical Communication\index{Stochastic Local Operation and one-way Classical Communication}) a symmetric extension cannot be broken when communication from Bob to Alice is not allowed.
\begin{lemma}
 (Nowakowski and Horodecki \cite{nowakowski09a}) Let $\Lambda$ be a (not necessarily trace-preserving) quantum operation that can be realized with local operations assisted by one-way classical communication (1-LOCC), i.e., it is of the form 
\begin{equation}
  \Lambda(\rho) = \sum_{ij}(\calone \otimes B_{ij})(A_i \otimes \calone) \rho (A_i \otimes \calone)^\dagger (\calone \otimes B_{ij})^\dagger
\end{equation}
where $\sum_{i} A_i^\dagger A_i \le \calone$ and $\sum_j B_{ij}^\dagger B_{ij} = \calone$ for all $i$ since Bob cannot communicate the outcome of a probabilistic operation back to Alice. 

If $\rho^{AB}$ admits a symmetric extension, then so does $\Lambda(\rho^{AB})$.
\end{lemma}

An interesting special case is when Alice performs an invertible filter operation and Bob performs a unitary. Then the operation can be reversed with nonzero probability, so the output state admits a symmetric extension if \emph{and only if} the input state admits one. 

\subsection{Decomposition into pure-symmetric extendible states}
\label{sec:pure-extendible-decomp}

The convex structure of the set of states with a symmetric extension allows us to decompose nonextremal states into convex combination of states with special properties.
This is analogous to the convex set of separable states.
Separable quantum states are those states that can be written as convex combinations of product states $\rho_A \otimes \rho_B$ and they can even be decomposed further into convex combinations of pure product states. I.e.
\begin{equation}
  \rho_\text{sep} = \sum_j p_j \proj{\psi_j} \otimes \proj{\phi_j}.
\end{equation}
Although it can be difficult to determine whether or not a given state can be written on this form or not---and if it can, to find some $\ket{\psi_j}$ and $\ket{\phi_j}$ explicitly---the fact that all separable states can be written like this allows us to prove properties of separable states in general. 

Clearly, it is not true that any $\rho^{AB}$ that has a symmetric extension can be decomposed into pure states with the same property. 
This is because the only pure states that have a symmetric extension are the pure product states, and their convex hull is the set of separable states. 
But it turns out that if we consider the \emph{extended} states---the $\rho^{ABB^\prime}$ that are invariant under exchange of $B$ and $B^\prime$---they can be written as convex combinations of pure states with the same property. 
In fact, the pure states in the spectral decomposition can be chosen to have this property.

\begin{lemma}
\label{lem:symspecdecomp}
A tripartite state $\rho^{ABB^\prime}$ which is invariant under exchange of $B$ and $B^\prime$, $\rho^{ABB^\prime} = \swap \rho^{ABB^\prime} \swap^\dagger$, can be written in the spectral decomposition
\begin{equation}
  \rho^{ABB^\prime} = \sum_j \lambda_j \proj{\phi_j}
\end{equation} 
in such a way that $\proj{\phi_j} = \swap \proj{\phi_j} \swap^\dagger$, i.e.~$\swap \ket{\phi_j} = \pm \ket{\phi_j}$.
\end{lemma}
\begin{proof} 
  Since $\rho^{ABB^\prime} = \swap \rho^{ABB^\prime} \swap^\dagger$, $\rho^{ABB^\prime} \swap = \swap \rho^{ABB^\prime}$, so $\rho^{ABB^\prime}$ and $\swap$ (which is shorthand for $\calone_A \otimes \swap$) are commuting diagonalizable operators and therefore have a common set of eigenvectors. 
  Since $\swap^2 = I$, $\swap$ has eigenvalues $\pm 1$ and all its eigenvectors therefore satisfy $\swap \ket{\phi_j} = \pm \ket{\phi_j}$.
\end{proof}

The above lemma applies to the extended state $\rho^{ABB^\prime}$, but our main interest is for bipartite states $\rho^{AB}$ that \emph{admit} a symmetric extension. 
By tracing out the $B^\prime$ system we get
\begin{corollary}
\label{cor:pure-extendible-decomp}
  A bipartite quantum state $\rho^{AB}$ admits a symmetric extension if and only if it can be written as a convex combination 
  \begin{equation}
  \rho^{AB} = \sum_j p_j \rho_j^{AB}; \quad 0 \le p_j \le 1; \quad \sum_j p_j = 1,
  \end{equation}
  of states $\rho_j^{AB}$ which admit a \emph{pure} symmetric extension.
\end{corollary}

Hence, all the extremal states in the convex set of symmetric extendible states are extendible to pure states. We will call those states \emphindex{pure-extendible}. 
Note that not all states with a pure symmetric extension are extremal. 
There are states that admit either a pure or a mixed symmetric extension.
From lemma \ref{lem:symspecdecomp} we know that the mixed symmetric extension
is a convex combination of pure tripartite states that are invariant under swap of $B$ and $B'$.
The original state is then a convex combination of the pure-extendible reductions of those tripartite states, and is therefore not extremal.
Theorem \ref{thm:extremal} in section \ref{sec:necsufqubit} gives a full classification of the nonextremal pure-extendible states for the case of two qubits.

In the next section,
we give a simple necessary condition for a state to be pure-extendible. 
In section~\ref{sec:pure-extendible} we show that the condition is also sufficient if and only if it is a state of two qubits.


\subsection{The spectrum condition for pure-extendible states}
\label{sec:condition}

Let $\spec{\rho}$ denote the vector of nonzero eigenvalues of $\rho$ in nonincreasing order.

\begin{theorem}
  \label{thm:spectrumcondition}
  Let $\rho^{AB}$ be a state that has the pure symmetric extension $\proj{\psi}_{ABB^\prime}$. Then 
  \begin{equation}
    \label{eq:spectrumcondition}
    \spec{\rho^{AB}} = \spec{\rho^B}.
  \end{equation}
\end{theorem}
\begin{proof} 
  Using the Schmidt decomposition with the splitting $AB|B^\prime$, we can write the pure symmetric extension as
  \begin{equation}
    \ket{\psi}_{ABB^\prime} = \sum_j \sqrt{\lambda_j} \ket{\phi_j}_{AB} \ket{j}_{B^\prime}.
  \end{equation}
  The reduced density matrices of this state are
  \begin{equation}
    \rho^{AB} = \sum_j \lambda_j \proj{\phi_j}, \qquad \rho^{B^\prime} = \sum \lambda_j \proj{j},
  \end{equation}
  i.e.~the spectra of $\rho^{AB}$ and $\rho^{B^\prime}$ are equal. 
  By symmetry between $B$ and $B^\prime$, $\rho^B = \rho^{B^\prime}$ so $\spec{\rho^{AB}} = \spec{\rho^{B}}$.
\end{proof}

In general, we don't expect all states that satisfy condition \eqref{eq:spectrumcondition} to have a pure symmetric extension (but see Thm.~\ref{thm:necsufqubit} in section~\ref{sec:necsufqubit}). 
The following corollary provides a test that can rule out a pure-symmetric extension.
\begin{corollary}
\label{cor:purextfilter}
For any state $\rho^{AB}$ that has a pure symmetric extension and any operator $M$ on $\mathcal{H}_A$, the (unnormalized) state 
\begin{equation}
  \widetilde{\rho}^{AB} = (M \otimes \calone_B) \rho^{AB} (M \otimes \calone_B)^\dagger
\end{equation}
satisfies condition \eqref{eq:spectrumcondition}.
\end{corollary}
\begin{proof} 
Let $\ket{\psi}_{ABB'} = \pm \swap \ket{\psi}_{ABB'}$ be the pure symmetric extension of $\rho_{AB}$. 
The filter $M$ acts only on $\mathcal{H}_A$, so it commutes with $\swap$.
Therefore $M \ket{\psi}_{ABB'} = \pm M \swap \ket{\psi}_{ABB'} = \pm \swap M \ket{\psi}_{ABB'}$, 
so $M \ket{\psi}_{ABB'}$ is a symmetric extension of its reduced state $\widetilde{\rho}_{AB}$. 
Because of theorem \ref{thm:spectrumcondition}, $\widetilde{\rho}_{AB}$ then satisfies Eq.~\eqref{eq:spectrumcondition}.
\end{proof}

This condition is useful, since if given a state 
that is not pure-extendible
but satisfies condition \eqref{eq:spectrumcondition},
applying a random filter on system $A$ will usually break the condition and reveal that it is not pure-extendible.

\section{Symmetric extension for multi-qubit Bell-diagonal states}
\label{sec:simplifying}
\index{multi-qubit Bell-diagonal state}

In many cases the bipartite states that want to check for a symmetric extension come from a special class, 
and this may simplify the problem.
Usually, this comes about because of some symmetry on the class of states we are considering that induces a symmetry on the form of the extended state (should it exist).
The states that we consider here are states of $N$ qubit pairs in a generalized Bell-diagonal state.

\begin{defn}
A \emphindex{$2N$-qubit Bell-diagonal state} is a state of the form 
\begin{equation}
\label{eq:startform1}
\begin{aligned}
 \rho^{AB} = \rho^{A_1B_1 \ldots A_N B_N} 
&= \sum_{i_1 \cdots i_N} p_{i_1 \cdots i_N} \proj{\beta_{i_1 \cdots i_N}} \\
&= \sum_{i_1 \cdots i_N} p_{i_1 \cdots i_N} \proj{\beta_{i_1}}_{A_1 B_1} \otimes \cdots \otimes \proj{\beta_{i_N}}_{A_N B_N}
\end{aligned}
\end{equation}
where 
\begin{equation}
\label{eq:startform1expl}
 \ket{\beta_{i_1 \cdots i_N}} 
:= (\calone \otimes \sigma_{i_1}) \otimes \cdots \otimes (\calone \otimes \sigma_{i_N}) \ket{\Phi^+}^{\otimes N}, 
\quad \ket{\beta_i} := (\calone \otimes \sigma_{i}) \ket{\Phi^+}
\end{equation}
for $i \in \{0, 1, 2, 3\}$ or equivalently $i \in \{I, x, y, z\}$.
\end{defn}
From Eq.~\eqref{eq:startform1expl} we see that we may interpret $\ket{\Psi^+}, \ket{\Psi^-}$, and $\ket{\Phi_-}$ as $\sigma_x, \sigma_y,$ and $\sigma_z$ errors on one of the subsystems that originally started out in the state $\ket{\Phi^+}$.

We want to decide when states of this form have a symmetric extension. 
While any symmetric extension problem can be formulated as a semidefinite program that can be solved numerically for modest dimension \cite{doherty02a}, we can use the symmetry of the problem to simplify it considerably, and even solve it analytically when $N=1$ (see section~\ref{sec:belldiagsymext}).

\subsection{Symmetries of the extended state}
\label{sec:symofextstate}

A $2N$-qubit Bell-diagonal state is invariant under the following $2N + 1$ operations:
\begin{itemize}
 \item[(i)]   $\sigma_x^{A_i} \otimes \sigma_x^{B_i}$ on any subsystem $A_i B_i$,
 \item[(ii)]  $\sigma_z^{A_i} \otimes \sigma_z^{B_i}$ on any subsystem $A_i B_i$,
 \item[(iii)] Transposition of the whole quantum state.
\end{itemize}
From this, we can impose on a symmetric extension $\rho^{ABB'} = \rho^{A_1 B_1 B_1' \cdots A_N B_N B_N'}$---should it exist---that it be invariant under
\begin{itemize}
 \item[(i)]   $\sigma_x^{A_i} \otimes \sigma_x^{B_i} \otimes \sigma_x^{B_i{}'}$ on any subsystem $A_i B_i B_i'$,
 \item[(ii)]  $\sigma_z^{A_i} \otimes \sigma_z^{B_i} \otimes \sigma_z^{B_i{}'}$ on any subsystem $A_i B_i B_i'$,
 \item[(iii)] Transposition of the whole quantum state.
\end{itemize}
The reason we can impose this is the following.
Suppose that a state $\rho^{ABB'} = \rho^{A_1 B_1 B_1' \cdots A_N B_N B_N'}$ is a symmetric extension of the state $\rho^{AB} = \rho^{A_1 B_1 \cdots A_N B_N}$. 
Then, applying any of the $2N+1$ above operations to $\rho^{ABB'}$ will give a new (possibly different) state $\sigma^{ABB'}$. 
Since the operation on $ABB'$ reduces to one of the symmetry operations on $AB$, 
the new state $\sigma^{ABB'}$ will have the same reduction as $\rho^{ABB'}$, 
$\tr_{B'}[\sigma^{ABB'}] = \rho^{AB}$, 
and since the operations on $ABB'$ treat $B$ and $B'$ the same way, 
$\sigma^{ABB'}$ inherits the swap symmetry from $\rho^{ABB'}$.
It is therefore another symmetric extension of $\rho^{AB}$.
A third valid symmetric extension is then the state 
$\rho_\textrm{sym}^{ABB'} = (\rho^{ABB'} + \sigma^{ABB'})/2$ and since all the symmetry operations considered here are involutions, this state is invariant under the considered operation.
Hence, if there is a symmetric extension of a $2N$-qubit Bell-diagonal state, there is also a symmetric extension with the imposed symmetries.

Note that while $\rho^{AB}$ also is invariant under transposing only a subsystem $A_i B_i$, it will not always be possible to find a symmetric extension which is invariant under transposing $A_i B_i B_i'$. 
While the operator resulting from transposing a subsystem of $\rho^{ABB'}$ will have the correct marginal state and swap symmetry, it may have negative eigenvalues and will then not be a quantum state and therefore not a valid symmetric extension.

\subsection{Imposing the symmetries}

To impose the symmetries on the state $\rho^{ABB'}$ it is convenient to express it using the Pauli operator basis. Any $3N$-qubit state can be expanded as
\begin{equation}
\label{eq:pauliform}
 \rho^{ABB'} = 
 \frac{1}{2^{3N}}
 \sum_{\substack{i_1 \cdots i_N\\j_1 \cdots j_N\\k_1 \cdots k_N}}
 \alpha_{i_1 j_1 k_1 \cdots i_N j_N k_N}
 \sigma_{i_1}^{A_1} \otimes \sigma_{j_1}^{B_1} \otimes \sigma_{k_1}^{B_1{}'} 
 \otimes \cdots \otimes 
 \sigma_{i_N}^{A_N} \otimes \sigma_{j_N}^{B_N} \otimes \sigma_{k_N}^{B_N{}'}.
\end{equation}
Some of the coefficients $\alpha_{i_1 j_1 k_1 \cdots i_N j_N k_N}$ are fixed by the requirement that the state be compatible with the given $\rho^{AB}$ and that it be invariant under swap of $B$ and $B'$.
The imposed symmetry simplifies the problem in two ways. 
First, most of the coefficients vanishes. 
Second, we can reduce the dimension of the matrix involved from $2^{3N}$ to $2^{2N}$ by bringing it to block diagonal form and focusing on a single block. 

\subsubsection{Open and fixed parameters}
\label{sec:openandfixedparams}

Let us first see how the symmetry constraints on $\rho^{ABB'}$ forces most of the coefficients $\alpha_{i_1 j_1 k_1 \cdots i_N j_N k_N}$ to zero.
Consider first the imposed constraint that the state $\rho^{ABB'}$ be invariant if we apply $\sigma_x^{A_i} \otimes \sigma_x^{B_i} \otimes \sigma_x^{B_i{}'}$ on a given subsystem $i$.
Each term in the sum \eqref{eq:pauliform} will either be left alone or get a negative sign from this operation, depending on whether the Pauli operators on subsystem $i$ commute with $\sigma_x^{A_i} \otimes \sigma_x^{B_i} \otimes \sigma_x^{B_i{}'}$ or not.
For example with $N=2$, when $\sigma_x^{A_2} \otimes \sigma_x^{B_2} \otimes \sigma_x^{B_2{}'}$ is applied (i.e. $i = 2$),
the term 
$\sigma_x^{A_1} \otimes \calone_{B_1} \otimes \sigma_x^{B_1{}'} \otimes \sigma_y^{A_2} \otimes \sigma_y^{B_2} \otimes \sigma_z^{B_2{}'}$
gets a negative sign since
\begin{equation}
 (\sigma_x^{A_2} \otimes \sigma_x^{B_2} \otimes \sigma_x^{B_2{}'})
 (\sigma_y^{A_2} \otimes \sigma_y^{B_2} \otimes \sigma_z^{B_2{}'})
 (\sigma_x^{A_2} \otimes \sigma_x^{B_2} \otimes \sigma_x^{B_2{}'})^\dagger
 =
 - \sigma_y^{A_2} \otimes \sigma_y^{B_2} \otimes \sigma_z^{B_2{}'}.
\end{equation}
This is most easily seen as a consequence of the fact that 
$(\sigma_x^{A_2} \otimes \sigma_x^{B_2} \otimes \sigma_x^{B_2{}'})$
and 
$(\sigma_y^{A_2} \otimes \sigma_y^{B_2} \otimes \sigma_z^{B_2{}'})$
anticommute.

The terms that anticommute pick up a negative sign, and since the terms are all linearly independent, these terms will have to vanish for the state to be invariant.
The only combinations of Pauli operators on system $i$ that commute with both $\sigma_x^{A_i} \otimes \sigma_x^{B_i} \otimes \sigma_x^{B_i{}'}$ and
$\sigma_z^{A_i} \otimes \sigma_z^{B_i} \otimes \sigma_z^{B_i{}'}$ are
\begin{itemize}
\item $\calone \otimes \calone \otimes \calone$, 
\item all three permutations of $\calone \otimes \sigma_i \otimes \sigma_i$, where the index runs over only $x$, $y$ and $z$
\item all six permutations of 
$\sigma_x \otimes \sigma_y \otimes \sigma_z$.
\end{itemize} 
Therefore, all terms with nonzero coefficient in \eqref{eq:pauliform} will have one of these 16 combinations of Pauli operators in \emph{each} subsystem.
We label these tensor products of three Pauli matrices $P_i$, for $i \in \{1, \ldots 16\}$ as indicated in table \ref{tab:projectedpaulis}.
We can then rewrite the sum \eqref{eq:pauliform} by including only the terms that can be constructed from the Pauli combinations $P_i$,
\begin{equation}
  \label{eq:pauliformP}
  \rho^{ABB'} = \frac{1}{2^{3N}} \sum_{i_1 \cdots i_N = 1}^{16} \gamma_{i_1 \cdots i_N} P_{i_1} \otimes \cdots \otimes P_{i_N} .
\end{equation}
Each of the coefficients $\gamma_{i_1 \cdots i_N}$ is equal to a corresponding coefficient $\alpha_{i_1 j_1 k_1 \cdots i_N j_N k_N}$ from Eq.~\eqref{eq:pauliform}.
This reduces the number of terms from $4^{3N}$ to $16^N = 4^{2N}$.

The operators of the class $\sigma_x \otimes \sigma_y \otimes \sigma_z$ (i.e. $P_{11} \ldots P_{16}$) pick up a negative sign under transposition, so only terms with an even number of subsystems with this kind of Pauli combination can have a nonzero coefficient.
This ensures that like the original state $\rho^{AB}$, its symmetric extension $\rho^{ABB'}$ is invariant under transposition (i.e. it has only real matrix elements).

The above reduction in the number of open parameters was only due to the symmetries that a symmetric extension must possess.
Let us now see what additional constraints arises from the criterion that $\rho^{ABB'}$ be a symmetric extension of $\rho^{AB}$.
The first constraint is the \textindex{trace condition}, that $\tr_{B'}[\rho^{ABB'}] = \rho^{AB}$. 
Since the reduced state $\rho^{AB}$ is a multi-qubit Bell-diagonal state, it has the form
\begin{equation}
 \rho^{AB} = 
 \frac{1}{2^{2N}}
 \sum_{i_1 \cdots i_N} 
 \beta_{i_1 \cdots i_N}
 \sigma_{i_1}^{A_1} \otimes \sigma_{i_1}^{B_1} 
 \otimes \cdots \otimes
 \sigma_{i_N}^{A_N} \otimes \sigma_{i_N}^{B_N} ,
\end{equation}
where the coefficient $\beta_{0 \cdots 0}$ is 1 for a normalized state.
For a state of the form \eqref{eq:pauliform} to reduce to this, all coefficients that correspond to terms with $\calone$ on the $B'$ system, must be equal to the corresponding coefficient for the reduced state, 
\begin{equation}
 \label{eq:traceconditionalphabeta}
 \alpha_{i_1 i_1 0 i_2 i_2 0 \cdots i_N i_N 0} = \beta_{i_1 i_2 \cdots i_N}.
\end{equation}
The other nonzero coefficients are not constrained by this, since all terms which has a Pauli operator different from $\calone$ in one of the $B'_i$ subsystems vanish when we trace over $B'$.

The other type of constraints are those we get from $\rho^{ABB'}$ being a invariant under swap of $B$ and $B'$. 
This means that most of the coefficients have an equal partner, which is the one that is obtained by swapping the two last indices in every subsystem,
\begin{equation}
 \label{eq:swapconditionalpha}
 \alpha_{i_1 j_1 k_1 \cdots i_N j_N k_N}
 =
 \alpha_{i_1 k_1 j_1 \cdots i_N k_N j_N}.
\end{equation}

We can translate these conditions to the more economical state parametrization \eqref{eq:pauliformP}.
Only the Pauli type operators $P_1, P_5, P_6$, and $P_7$ survive tracing out the $B'$ system.
The trace condition \eqref{eq:traceconditionalphabeta} now becomes the following. 
For any $\gamma_{i_1 \cdots i_N}$ where all indices are $1,5,6,$ or $7$,
\begin{equation}
 \gamma_{i_1 \cdots i_N} = \beta_{j_1 \cdots j_N}
\end{equation}
where $j_k = i_k$ for $i_k =1$ and $j_k = i_k - 3$ for $i_k \in \{5,6,7\}$.
The symmetry constraint \eqref{eq:swapconditionalpha} gets a little more complicated when expressed in terms of $\gamma_{i_1 \cdots i_N}$, 
\begin{equation}
 \gamma_{i_1 \cdots i_N} = \gamma_{j_1 \cdots j_N} \quad \text{where} \quad 
 j_k = \left\{
 \begin{aligned}
  &i_k \text{ for } i_k \in \{ 1,2,3,4\}\\
  &i_k + 3 \text{ for } i_k \in \{ 5,6,7,11,12,13\}\\
  &i_k - 3 \text{ for } i_k \in \{ 8,9,10,14,15,16\}.
 \end{aligned}
 \right.
\end{equation}

These are all the constraints that we can put on the parameters, 
and those that are not explicitly constrained, can take any real value.
The symmetric extension problem is now to find values for these open parameters such that the resulting matrix is positive semidefinite.

\subsubsection{Using lower-dimensional matrices}
\label{sssec:lowerdimmatrices}

The state $\rho^{ABB'}$ is a $2^{3N} \times 2^{3N}$ matrix, and checking positive semidefiniteness for a generic big matrix is not trivial.
In this case, we can use the symmetry to simplify this part of the problem as well, by identifying a smaller matrix whose positivity is equivalent to that of $\rho^{ABB'}$.

Since the state $\rho^{ABB'}$ is invariant under 
$\sigma_z^{A_i} \otimes \sigma_z^{B_i} \otimes \sigma_z^{B_i{}'}$, 
each eigenvector of $\rho^{ABB'}$ is either in the $+1$ eigenspace or $-1$ eigenspace of this operator.
The same holds for any of the other subsystems $A_i B_i B'_i$, so each eigenvector has to be in one of the $2^N$ simultaneous eigenspaces of all the $N$ operators
$\sigma_z^{A_i} \otimes \sigma_z^{B_i} \otimes \sigma_z^{B_i{}'}$.
Therefore, the state can be written as a direct sum of positive operators on the different eigenspaces,
\begin{equation}
  \rho^{ABB'} = \bigoplus_{s \in \{+,-\}^N} \rho_s.
\end{equation}
The $i$th symbol in the $N$-symbol string $s$ indicates whether the operator is the positive ($+$) or negative ($-$) eigenspace on system $i$.
The operator $\rho_{+--}$, for example, has its support on the intersection of the  $+1$ eigenspace of $\sigma_z^{A_1} \otimes \sigma_z^{B_1} \otimes \sigma_z^{B_1{}'}$ and the $-1$ eigenspaces of $\sigma_z^{A_2} \otimes \sigma_z^{B_2} \otimes \sigma_z^{B_2{}'}$ and $\sigma_z^{A_3} \otimes \sigma_z^{B_3} \otimes \sigma_z^{B_3{}'}$.

Moreover, the state is invariant under 
$\sigma_x^{A_i} \otimes \sigma_x^{B_i} \otimes \sigma_x^{B_i{}'}$
for all $i \in \{1, \ldots, N\}$.
Recall that the $+1$ eigenspace of 
$\sigma_z \otimes \sigma_z \otimes \sigma_z$
is spanned by 
$\ket{000}, \ket{011}, \ket{101}, \ket{110}$ 
and the $-1$ eigenspace by
$\ket{111}, \ket{100}, \ket{010}, \ket{001}$, 
i.e.~the computational basis vectors with even and odd Hamming weight, respectively.
Applying a 
$\sigma_x \otimes \sigma_x \otimes \sigma_x$ 
operator to any vector in the $+1$ or $-1$  subspace will therefore flip it to the other subspace.
This means, for example, that $\rho_{++++}$ can be taken to the $+--+$ subspace by applying $\sigma_x \otimes \sigma_x \otimes \sigma_x$ to the second and third subsystems. 
Since the state $\rho^{ABB'}$ is supposed to be invariant under any such operation, 
all the $\rho_s$ have to be isomorphic, so if one of them is given, the whole density matrix $\rho^{ABB'}$ can be reconstructed through the symmetry operations.
To verify positivity, it is sufficient to check any one of the eigenspaces, since the eigenvalues will be the same on each eigenspace. 

To do this in practice, we project the state onto the intersection of all the $+1$ eigenspaces of the 
$\sigma_z^{A_i} \otimes \sigma_z^{B_i} \otimes \sigma_z^{B_i{}'}$ operators.
For each of the 16 Pauli combinations $P_1 \ldots P_{16}$, we can compute its projection onto the $+1$ eigenspace of $\sigma_z \otimes \sigma_z \otimes \sigma_z$.
For example, we can write the Pauli combination $P_2$ out in the computational basis,
\begin{equation}
  P_2
 = 
  \calone \otimes \sigma_x \otimes \sigma_x
 = 
  \begin{bmatrix}
   0 & 0 & 0 & 1  &  0 & 0 & 0 & 0\\ 
   0 & 0 & 1 & 0  &  0 & 0 & 0 & 0\\ 
   0 & 1 & 0 & 0  &  0 & 0 & 0 & 0\\ 
   1 & 0 & 0 & 0  &  0 & 0 & 0 & 0\\ 

   0 & 0 & 0 & 0  &  0 & 0 & 0 & 1\\ 
   0 & 0 & 0 & 0  &  0 & 0 & 1 & 0\\ 
   0 & 0 & 0 & 0  &  0 & 1 & 0 & 0\\ 
   0 & 0 & 0 & 0  &  1 & 0 & 0 & 0 
  \end{bmatrix}.
\end{equation}
The positive subspace is spanned by $\ket{000}$, $\ket{011}$, $\ket{101}$, and $\ket{110}$, which corresponds to the rows and columns $1$, $4$, $6$ and $7$.
A projection matrix that projects onto the positive subspace is therefore
\begin{equation}
 P_+ = 
 \begin{bmatrix}
  1 & 0 & 0 & 0  &  0 & 0 & 0 & 0\\
  0 & 0 & 0 & 1  &  0 & 0 & 0 & 0\\
  0 & 0 & 0 & 0  &  0 & 1 & 0 & 0\\
  0 & 0 & 0 & 0  &  0 & 0 & 1 & 0
 \end{bmatrix},
\end{equation}
and we can again write the resulting matrix as a tensor product of Pauli operators,
\begin{equation}
 Q_2 := P_+ P_2 P_+^\dagger
 =
 \begin{bmatrix}
   0 & 1 & 0 & 0\\ 
   1 & 0 & 0 & 0\\ 
   0 & 0 & 0 & 1\\ 
   0 & 0 & 1 & 0 
 \end{bmatrix}
 = \calone \otimes \sigma_x.
\end{equation}
When we do this for all 16 allowed combinations of Pauli operators, we get a projected operator $Q_i := P_+ P_i P_+$ for each original operator $P_i$ and
the new operators that are shown in table \ref{tab:projectedpaulis}.
\begin{table}[hbt]
\caption{The independent Pauli terms and their projections to the $+1$ eigenspace of $\sigma_z \otimes \sigma_z \otimes \sigma_z$.}
\label{tab:projectedpaulis}
\begin{tabular}{ccc}
\hline
\hline
Original & Projected\\
\hline
\hline
$P_1 = \calone  \otimes \calone  \otimes \calone $ & $ Q_{1} = \calone  \otimes \calone $ \\
\hline
$P_2 = \calone  \otimes \sigma_x \otimes \sigma_x$ & $ Q_{2} = \calone  \otimes \sigma_x$ \\
$P_3 = \calone  \otimes \sigma_y \otimes \sigma_y$ & $ Q_{3} =-\sigma_z \otimes \sigma_x$ \\
$P_4 = \calone  \otimes \sigma_z \otimes \sigma_z$ & $ Q_{4} =  \sigma_z \otimes \calone $ \\
\hline
$P_5 = \sigma_x \otimes \sigma_x \otimes \calone $ & $ Q_{5} = \sigma_x \otimes \sigma_x$ \\
$P_6 = \sigma_y \otimes \sigma_y \otimes \calone $ & $ Q_{6} = \sigma_y \otimes \sigma_y$ \\
$P_7 = \sigma_z \otimes \sigma_z \otimes \calone $ & $ Q_{7} = \sigma_z \otimes \sigma_z$ \\
\hline
$P_8 = \sigma_x \otimes \calone  \otimes \sigma_x$ & $ Q_{8} = \sigma_x \otimes \calone $ \\
$P_9 = \sigma_y \otimes \calone  \otimes \sigma_y$ & $ Q_{9} =-\sigma_x \otimes \sigma_z$ \\
$P_{10} = \sigma_z \otimes \calone  \otimes \sigma_z$ & $ Q_{10} = \calone  \otimes \sigma_z$ \\
\hline
$P_{11} = \sigma_x \otimes \sigma_y \otimes \sigma_z$ & $ Q_{11} = \sigma_y \otimes \sigma_x$ \\
$P_{12} = \sigma_y \otimes \sigma_z \otimes \sigma_x$ & $ Q_{12} = \sigma_y \otimes \sigma_z$ \\
$P_{13} = \sigma_z \otimes \sigma_x \otimes \sigma_y$ & $ Q_{13} = \calone  \otimes \sigma_y$ \\
$P_{14} = \sigma_x \otimes \sigma_z \otimes \sigma_y$ & $ Q_{14} = \sigma_y \otimes \calone $ \\
$P_{15} = \sigma_y \otimes \sigma_x \otimes \sigma_z$ & $ Q_{15} = \sigma_x \otimes \sigma_y$ \\
$P_{16} = \sigma_z \otimes \sigma_y \otimes \sigma_x$ & $ Q_{16} = \sigma_z \otimes \sigma_y$ \\
\hline
\hline
\end{tabular}
\end{table}

The projected operators $Q_i$ can be substituted directly for the original operators $P_i$ in each term of the expansion \eqref{eq:pauliformP} in order to obtain $\rho_{++ \cdots +}$,
\begin{equation}
  \label{eq:pauliformQ}
  \rho_{++ \cdots +} =  \frac{1}{2^{3N}} \sum_{i_1 \cdots i_N = 1}^{16} \gamma_{i_1 \cdots i_N} Q_{i_1} \otimes \cdots \otimes Q_{i_N} .
\end{equation}
If we can assign values to the open parameters such that the projected matrix $\rho_{++\cdots +}$ is positive semidefinite, the same values will give a valid symmetric extension of $\rho^{AB}$ when the original three Pauli operators $P_i$ are used in each term of Eq.~\eqref{eq:pauliformQ}.

Having used the symmetry to reduce the number of open parameters and dimension of the matrix for which we need to ensure positivity, we list the number of open parameters and matrix dimension with and without symmetry in table \ref{tab:numberofparameters}.

\begin{table}[hbt]
\caption{Number of open parameters and dimension with and without symmetry.}
\label{tab:numberofparameters}
\begin{tabular}{c|cc|cc}
\hline
\hline
                  & without & symmetry   & with &symmetry \\
\hline
qubit pairs & \# of open params & dimension & \# of open params & dimension\\
\hline
\hline
1 &     48   &   8 &  3    & 4\\
2 &   3840   &  64 &  60   & 16\\
3 & 258048   & 512 & 1008  & 64\\
4 & 16711680 & 4096& 16320 & 256\\
\hline
\hline
\end{tabular}
\end{table}

\subsection{Using additional \texorpdfstring{$S^\dagger \otimes S$}{S\032\201 \327 S} symmetry of the reduced state}
\label{sec:additionalsymmetry}

We will be particularly interested in finding a symmetric extension for $2N$-qubit Bell-diagonal states that have an additional symmetry, namely that swapping $\sigma_x$ errors and $\sigma_y$ errors in any of the subsystems leaves the state invariant.
In other words, for any qubit with a \textindex{bit error}, the probability of having a \textindex{phase error} is 1/2.
The symmetry operations related to this is $S^\dagger \otimes S$ where the unitary operator $S := \proj{0} + \im \proj{1}$ is called the phase gate.
Since $S^2 = (S^\dagger)^2 = \sigma_z$, this symmetry implies the $\sigma_z \otimes \sigma_z$ symmetry that we have for all Bell-diagonal states.
Since Alice and Bob's state is invariant under the operation
\begin{itemize}
 \item[(iv)]   $S^\dagger_{A_i} \otimes S_{B_i}$ on any subsystem $A_i B_i$,
\end{itemize}
we can impose that the extended state be invariant under the operation
\begin{itemize}
 \item[(iv)]   $S^\dagger_{A_i} \otimes S_{B_i} \otimes S_{B'_i}$ on any subsystem $A_i B_i B'_i$,.
\end{itemize}

Since $S$ is diagonal, it commutes with both the diagonal matrices, $\calone$ and $\sigma_z$.
For the $\sigma_x$ and $\sigma_y$ matrices, the relevant relations are  
\begin{subequations}
 \begin{align}
  S \sigma_x S^\dagger &= \sigma_y,\\
  S \sigma_y S^\dagger &= -\sigma_x,\\
  S^\dagger \sigma_x S &= -\sigma_y,\\
  S^\dagger \sigma_y S &= \sigma_x.
 \end{align}
\end{subequations}
This means that some of the coefficients in \eqref{eq:pauliform} and \eqref{eq:pauliformP} have to be equal.
The operation $S^\dagger \otimes S \otimes S$ takes 
\begin{align*}
P_2 = \calone  \otimes \sigma_x \otimes \sigma_x &\leftrightarrow \phantom{-} \calone  \otimes \sigma_y \otimes \sigma_y = P_3, \\
P_5 = \sigma_x \otimes \sigma_x \otimes \calone &\leftrightarrow - \sigma_y \otimes \sigma_y \otimes \calone = -P_6, \\
P_8 =\sigma_x \otimes \calone  \otimes \sigma_x &\leftrightarrow - \sigma_y \otimes \calone  \otimes \sigma_y = -P_9,\\
P_{11} = \sigma_x \otimes \sigma_y \otimes \sigma_z &\leftrightarrow \phantom{-} \sigma_y \otimes \sigma_x \otimes \sigma_z = P_{15},\\
P_{12} = \sigma_y \otimes \sigma_z \otimes \sigma_x &\leftrightarrow \phantom{-} \sigma_x \otimes \sigma_z \otimes \sigma_y = P_{14},\\
P_{13} = \sigma_z \otimes \sigma_x \otimes \sigma_y &\leftrightarrow - \sigma_z \otimes \sigma_y \otimes \sigma_x = -P_{16}.
\end{align*}
and the other combinations $P_i$ are invariant
(note that $S^\dagger \otimes S \otimes S$ is not an involution, but its square is $\sigma_z \otimes \sigma_z \otimes \sigma_z$ and all the $P_i$ are invariant under that).
This means that instead of building the extended state from the 16 Pauli combinations $P_i$ listed in table \ref{tab:projectedpaulis}, it is sufficient to use the 10 combinations $R_i$ from table \ref{tab:projectedpaulismoresymmetry}, so that we get a sum of the form
\begin{equation}
  \label{eq:pauliformR}
  \rho^{ABB'} =  \frac{1}{2^{3N}} \sum_{i_1 \cdots i_N = 1}^{10} \xi_{i_1 \cdots i_N} R_{i_1} \otimes \cdots \otimes R_{i_N} .
\end{equation}
This takes the number of terms to consider from $16^N$ without the $S^\dagger \otimes S \otimes S$ symmetry to $10^N$ when we include it.

\begin{table}[hbt]
\caption{Independent Pauli terms with $S^\dagger \otimes S \otimes S$ symmetry, both in original form and projected to the $+1$ eigenspace of $\sigma_z \otimes \sigma_z$.}
\label{tab:projectedpaulismoresymmetry}
\begin{tabular}{cc}
\hline
\hline
Original & Projected\\
\hline
\hline
$R_1 = \calone  \otimes \calone  \otimes \calone $ & $T_1 = \calone  \otimes \calone $ \\
\hline
$R_2 = \calone  \otimes \sigma_x \otimes \sigma_x + \calone  \otimes \sigma_y \otimes \sigma_y$ 
& $T_2 = \calone  \otimes \sigma_x - \sigma_z \otimes \sigma_x$ \\
$ R_3 = \calone  \otimes \sigma_z \otimes \sigma_z$ & $T_3 = \sigma_z \otimes \calone $ \\
\hline
$R_4 = \sigma_x \otimes \sigma_x \otimes \calone - \sigma_y \otimes \sigma_y \otimes \calone $ 
& $T_4 = \sigma_x \otimes \sigma_x -  \sigma_y \otimes \sigma_y$ \\
$R_5 = \sigma_z \otimes \sigma_z \otimes \calone $ & $T_5 =  \sigma_z \otimes \sigma_z$ \\
\hline
$R_6 = \sigma_x \otimes \calone  \otimes \sigma_x - \sigma_y \otimes \calone  \otimes \sigma_y$ 
& $T_6 =  \sigma_x \otimes \calone + \sigma_x \otimes \sigma_z$ \\
$R_7 = \sigma_z \otimes \calone  \otimes \sigma_z$ & $T_7 =  \calone  \otimes \sigma_z$ \\
\hline
$R_8 = \sigma_x \otimes \sigma_y \otimes \sigma_z + \sigma_y \otimes \sigma_x \otimes \sigma_z$ 
& $T_8 =  \sigma_y \otimes \sigma_x + \sigma_x \otimes \sigma_y$ \\
$R_9 = \sigma_y \otimes \sigma_z \otimes \sigma_x + \sigma_x \otimes \sigma_z \otimes \sigma_y$ 
& $T_9 =  \sigma_y \otimes \sigma_z + \sigma_y \otimes \calone $ \\
$R_{10} = \sigma_z \otimes \sigma_x \otimes \sigma_y - \sigma_z \otimes \sigma_y \otimes \sigma_x$ 
& $T_{10} =  \calone  \otimes \sigma_y - \sigma_z \otimes \sigma_y$\\
\hline
\hline
\end{tabular}
\end{table}

All the operators $R_i$ in table \ref{tab:projectedpaulismoresymmetry} are invariant under $S^\dagger \otimes S \otimes S$. 

Under swapping the last two qubits, $R_1, R_2, R_3, R_5,$ and $R_7$ are invariant. 
The operator $R_4$ becomes $R_6$ and vice versa, and the same holds for $R_8$ and $R_9$.
The last operator, $R_{10}$ acquires a negative sign when the two last qubits are swapped.
These relations define pairs of coefficients $\xi_{i_1 \cdots i_N}$ that have to be equal in the sum \eqref{eq:pauliformR}.

As before, we can substitute the operators $R_i$ on the full Hilbert space of $ABB'$ with the projected operators $T_i$ when we check for positivity,
\begin{equation}
  \label{eq:pauliformS}
  \rho_{++ \cdots +} =  \frac{1}{2^{3N}} \sum_{i_1 \cdots i_N = 1}^{10} \xi_{i_1 \cdots i_N} T_{i_1} \otimes \cdots \otimes T_{i_N} .
\end{equation}

\section{The symmetric extension problem as a semidefinite program}
\label{sec:sdpimplementation}

The symmetric extension problem can be cast in the form of a \textindex{semidefinite program} (SDP) \cite{doherty02a,doherty04a}. 
By determining the open and fixed parameters of the extended state $\rho^{ABB'}$ in section \ref{sec:openandfixedparams} we have already informally formulated the symmetric extension problem as a \textindex{feasibility problem} (see section \ref{sec:sdpdef})---an SDP on the inequality form \eqref{eq:ineqformobj}-\eqref{eq:ineqformconstraints} with all $c_i = 0$.
With the minimization gone we are only left with finding open parameters $x_i$ such that the \textindex{linear matrix inequality}
from Eq.~\eqref{eq:ineqformconstraints},
\begin{equation}
\label{eq:symextsdpLMI}
 F(x) := G + \sum_{i=1}^{m} x_i F_i \geq 0
\end{equation}
is satisfied.

The matrix $F(x)$ in represents the extended quantum state $\rho^{ABB'}$, parametrized as in Eq.~\eqref{eq:pauliform}, except that we omit the normalization factor $1/2^{3N}$ so that $\tr[F(x)] = \tr[\calone] = 2^{3N}$.
At this point, the normalization factor does not make a difference, since it does not affect the semidefiniteness of $F(x)$.
There is one open parameter $x_i$ for every independent open coefficient 
(some of the $\alpha_{i_1 j_1 k_1 \cdots i_N j_N k_N}$ have to be equal because of the swap symmetry), 
and the corresponding $F_i$ is simply the sum of those
$\sigma_{i_1}^{A_1} \otimes \sigma_{j_1}^{B_1} \otimes \sigma_{k_1}^{B_1{}'} 
 \otimes \cdots \otimes 
 \sigma_{i_N}^{A_N} \otimes \sigma_{j_N}^{B_N} \otimes \sigma_{k_N}^{B_N{}'}$
that correspond to that open coefficient.
Those of the $\alpha_{i_1 j_1 k_1 \cdots i_N j_N k_N}$ that are fixed by the requirement that the state reduce to $\rho^{AB}$ when $B'$ is traced out are summed with the corresponding Pauli operators in the fixed matrix $G$.

This approach works even if there are no symmetries, but as we saw in section \ref{sec:openandfixedparams} the symmetry requirements fix a lot of the otherwise open parameters so that the search space gets smaller.
When we have the additional $S^\dagger \otimes S$ symmetry we can also lump together many of the previously independent open parameters into one open parameter to further reduce the search space as described in section \ref{sec:additionalsymmetry}.
This means that each $F_i$ will have a greater number of Pauli terms.

As described in section \ref{sssec:lowerdimmatrices}, positivity of the matrix on a subspace is sufficient to ensure positivity of the whole matrix due to the Bell-diagonal form of the state.
In the semidefinite program this means that all the $F_i$ can be substituted for smaller matrices generated by using the projected forms of the Pauli operators from table \ref{tab:projectedpaulis} (without $S^\dagger \otimes S$ symmetry) or table \ref{tab:projectedpaulismoresymmetry} (with $S^\dagger \otimes S$ symmetry).

The feasibility problem only tells us whether or not the problem is feasible, i.e.~if the state has a symmetric extension. 
By modifying the SDP slightly we can introduce an objective function whose optimal value will give us an indication of how far away the state is from losing or gaining a symmetric extension.
For this, instead of the fixed parameter $\alpha_{000\cdots 000}$ we use $t + \alpha_{000\cdots 000}$ where $t$ is a new open parameter.
$\alpha_{000\cdots 000} \calone^{ABB'}$ is still a part of $G$ while $t \calone^{ABB'}$ is one of the $x_i F_i$ terms in Eq. \eqref{eq:symextsdpLMI}.
Let this be $x_1 F_1$.
We then also let $c_1 = 1$ (with all other $c_i = 0$ still), so that the objective function becomes simply $t$.

This makes sure that the problem is always feasible---if $e_\textrm{min}$ is the (negative) minimum eigenvalue of some matrix $F(x)$, 
then adding $|e_\textrm{min}|$ to $t$ will make the matrix positive semidefinite.
Moreover, if the minimum value of $t$ is greater that $0$, it means that the state cannot have a symmetric extension, since a valid quantum state would have $t = 0$ and satisfy the positivity constraint. 
If the minimum is less than $0$,  we can get a symmetric extension by setting $t = 0$ and leaving the other open parameters unchanged. 
This adds some fraction of $\calone$ to a positive matrix so the resulting matrix that represents a trace-1 quantum state (when $\alpha_{000\cdots000}=1$ as it should be for a normalized quantum state) is still positive.

The semidefinite program now takes the form
\begin{align}
 \textrm{minimize }    \quad & t\\
 \textrm{subject to }  \quad & 2^{3N}\rho(\alpha) := 
\sum_{\alpha \in \Gamma_\textrm{fixed}} \alpha \sum_{\sigma \in S_\alpha} \sigma
+  \sum_{\alpha \in \Gamma_\textrm{open}} \alpha \sum_{\sigma \in S_\alpha} \sigma
+ t \calone
\geq 0.
\end{align}
Here, $\Gamma_\textrm{fixed}$ is the set of $\alpha_{i_1 \cdots k_N}$ that are fixed by the partial trace condition, $\Gamma_\textrm{open}$ is the set of independent open parameters, and $S_\alpha$ is the set of 
$\sigma_{i_1}^{A_1} \otimes \sigma_{j_1}^{B_1} \otimes \sigma_{k_1}^{B_1{}'} 
 \otimes \cdots \otimes 
 \sigma_{i_N}^{A_N} \otimes \sigma_{j_N}^{B_N} \otimes \sigma_{k_N}^{B_N{}'}$ 
that correspond to the independent parameter $\alpha$.

\chapter{Symmetric extension of two-qubit states}
\label{ch:twoqubits}

\section{Conjectured formula}
\label{sec:conjformula}

In the previous chapter, we have found some properties of the states with a symmetric extension, 
but none of them will tell us directly if a state has a symmetric extension or not.
If we restrict our attention to the simplest possible bipartite quantum system---two qubits---we can get stronger results. 
For many classes of two-qubit states we can derive formulas for when they have a symmetric extension. 

The situation is analogous to the separability\index{separability problem} problem. 
Like the set of states with a symmetric extension, the set of separable states is convex, and the extremal states (pure product states) are easier to deal with than general separable states.
For systems of dimension $2 \times 2$ and $2 \times 3$, the PPT\index{PPT criterion} criterion [ref Peres] is a necessary and sufficient condition for separability [ref Horodeckis].
If all the eigenvalues of $(\rho^{AB})^\Gamma$ (where $\Gamma$ denotes a partial transpose) are nonnegative, then we know that the state is separable, otherwise it is entangled.
We know that looking only at the eigenvalues of $\rho^{AB}$ together with its reductions $\rho^A$ and $\rho^B$ is insufficient to determine separability\cite{nielsen01a}, even if we restrict our attention to two qubits..

For symmetric extension of two-qubit states we think the situation is a little different.
Like for the separability problem, a simple formula is enough to determine whether a two-qubit state has a symmetric extension or not, but looking at the eigenvalues of $\rho^{AB}$ and $\rho^B$ seems to be sufficient. 
We propose a formula based only on these eigenvalues that we conjecture determines if a state has a symmetric extension.
We can prove the formula only in special cases and this will be done in the rest of this chapter after strengthening some of the results from the previous chapter for the case of two qubits.

\begin{conjecture}
\label{conj:twoqubitextension}
A two-qubit state $\rho^{AB}$ with reduced state $\rho^{B}$ has a symmetric extension if and only if
\begin{equation}
\label{eq:twoqubitextension}
  \tr[(\rho^B)^2] \geq \tr[(\rho^{AB})^2] - 4\sqrt{\det(\rho^{AB})}.
\end{equation}
\end{conjecture}

The conjecture has received extensive numerical testing and no counterexamples were found. 

In the rest of this chapter we first specialize and strengthen the results from section~\ref{sec:condition} for states with a pure symmetric extension.
Then we use this and other results from the previous chapter to derive conditions that are equivalent to Eq.~\eqref{eq:twoqubitextension} for Bell-diagonal states, rank-2 states, a subset of the states invariant under $\sigma_z \otimes \sigma_z$, and states with support only on the symmetric subspace.

\section{Pure-extendible states of two qubits}
\label{sec:pure-extendible}
\index{pure-extendible}

\subsection{Spectrum condition is sufficient for two qubits}
\label{sec:necsufqubit}

In this section it is shown that if $\rho^{AB}$ is a two-qubit state and satisfies $\spec{\rho^{AB}} = \spec{\rho^{B}}$, then there exists a pure state $\ket{\psi}_{ABB^\prime}$ such that $\ket{\psi}_{ABB^\prime} = \swap \ket{\psi}_{ABB^\prime}$. 
We first start by giving an equivalent condition to the spectrum condition.
\begin{lemma}
  \label{Bsymmetric}
Given a bipartite state $\rho^{AB}$. Then $\spec{\rho^{AB}} = \spec{\rho^{B}}$ if and only if there exists a pure tripartite state $\ket{\psi}_{ABB^\prime}$ with reductions  $\rho^{AB}$, $\rho^B$ and $\rho^{B^\prime}$ where $\rho^B = \rho^{B^\prime}$.
\end{lemma}
\begin{proof} 
  Assume $\spec{\rho^{AB}} = \spec{\rho^{B}} = (\lambda_j)$. 
  We can write the states in the spectral decomposition, 
  $\rho^{AB} = \sum_j \lambda_j \proj{\varphi_j}$, 
  $\rho^{B} = \sum_j \lambda_j \proj{b_j}$. 
  Then, a purification of $\rho^{AB}$ is 
\begin{equation}
  \label{eq:pureABBprime}
  \ket{\psi}_{ABB^\prime} = \sum_j \sqrt{\lambda_j} \ket{\varphi_j}_{AB} \ket{b_j}_{B^\prime}.
\end{equation}
Tracing out the $AB$ system we get $\rho^{B^\prime} = \sum_j \lambda_j \proj{b_j} = \rho^{B}$.

Conversely, assume that there exists a pure tripartite state $\ket{\psi}_{ABB^\prime}$ which has the given $\rho^{AB}$ as a reduction and that the reductions of $\ket{\psi}_{ABB^\prime}$ to $B$ and $B'$ satisfies $\rho^B = \rho^{B^\prime}$.
The Schmidt decomposition with the $AB|B'$ splitting of the state $\ket{\psi}_{ABB^\prime}$ takes the form \eqref{eq:pureABBprime}, where $\lambda_j$ and $\sum_j \lambda_j \proj{\varphi_j}$ must be the eigenvalues and eigenvectors of the given $\rho^{AB}$.
The reduction of $\ket{\psi}_{ABB^\prime}$ to $B'$ is 
$\rho^{B'} = \sum_j \lambda_j \proj{b_j}$.
Since $\rho^B = \rho^{B'}$ this is also the reduction to $B$, so $\rho^B$ has the same spectrum as $\rho^{AB}$.
\end{proof}

In other words, the spectrum condition $\spec{\rho^{AB}} = \spec{\rho^{B}}$ means that there exists a pure extension of $\rho^{AB}$ to $\ket{\psi}_{ABB'}$ such that the reduced states on $B$ and $B'$ are equal.
Note that it is not guaranteed that the reduced state on $AB$ is equal to the reduced state on $AB'$ which would be  necessary for a pure \emph{symmetric} extension.
The following theorem says that for a two-qubit system, we can also find a proper pure symmetric extension whenever the spectrum condition is satisfied.

\begin{theorem}
\label{thm:necsufqubit}
For a two-qubit state, $\spec{\rho^{AB}} = \spec{\rho^{B}}$ is a necessary and sufficient condition for it to have a pure symmetric extension.
\end{theorem}
\begin{proof} 
The condition is necessary for any dimension and this is already dealt with in section \ref{sec:condition}. Here, we only prove sufficiency for two qubits. 
By lemma \ref{Bsymmetric}, the condition implies that there exists a purification $\ket{\psi}_{ABB^\prime}$ of $\rho^{AB}$ which is such that $\rho^B = \rho^{B^\prime}$.
We will prove that for such a pure state, there is always a unitary operator on the $B^\prime$ system alone that will make it symmetric between $B$ and $B^\prime$. 

First, we prove the special case when $\rho^B$ is completely mixed. 
Then $\rho^{BB^\prime} = \tr_A [\proj{\psi}_{ABB^\prime}]$ is a state with maximally mixed subsystems. 
For such a state, there exist local unitary operators $U_B, V_{B^\prime}$ such that $(U_B \otimes V_{B^\prime})\rho^{BB^\prime}(U_B \otimes V_{B^\prime})^\dagger$ is Bell-diagonal \cite{horodecki96c}. Moreover, since $A$ is a qubit, $\rho_{BB^\prime}$ is of rank-2 and we have
\begin{equation}
\label{eq:belldiagform}
(U_B \otimes V_{B^\prime})\rho_{BB^\prime}(U_B \otimes V_{B^\prime})^\dagger
= 
p \proj{\psi_1} + (1-p) \proj{\psi_2},
\end{equation}
with $\ket{\psi_1}$ and $\ket{\psi_2}$ two of the four Bell-diagonal states $\ket{\Phi^\pm} = (\ket{00} \pm \ket{11})/\sqrt{2}$, $\ket{\Psi^\pm} = (\ket{01} \pm \ket{10})/\sqrt{2}$. 
Since the Bell-basis can be permuted arbitrarily with local unitary operators \cite{bennett96b}, we can choose $U_B$ and $V_{B^\prime}$ such that $\ket{\psi_1} = \ket{\Phi^+}$ and $\ket{\psi_2} = \ket{\Phi^-}$, so that we avoid the antisymmetric state $\ket{\Psi^-}$. 
The state in Eq.~\eqref{eq:belldiagform} can now be purified to $\sqrt{p} \ket{0}_A \ket{\Phi^+}_{BB^\prime} + \sqrt{1-p} \ket{1}_A \ket{\Phi^-}_{BB^\prime}$. 
Since all purifications of a state are equivalent up to a local unitary on the purifying system---in this case $A$---this is related to the pure state that we started out with as
\begin{multline}
(T_A \otimes U_B \otimes V_{B^\prime}) \ket{\psi}_{ABB^\prime}
\\
=
\sqrt{p} \ket{0}_A \otimes \ket{\Phi^+}_{BB^\prime} + \sqrt{1-p} \ket{1}_A \otimes \ket{\Phi^-}_{BB^\prime},
\end{multline}
where $T_A$ is the unitary operator on $A$ that relates this purification to the one where $A$ is left unchanged. 
We now perform the unitary $T_A^\dagger \otimes U_B^\dagger \otimes U_{B^\prime}^\dagger$ on the state, and a unitary of this form will not change the symmetry between $B$ and $B^\prime$. 
This gives
\begin{multline}
(I_A \otimes I_B \otimes U_B^\dagger V_{B^\prime}) \ket{\psi}_{ABB^\prime}
=
(T_A^\dagger \otimes U_B^\dagger \otimes U_{B^\prime}^\dagger) \times \\
(\sqrt{p} \ket{0}_A \otimes \ket{\Phi^+}_{BB^\prime} + \sqrt{1-p} \ket{1}_A \otimes \ket{\Phi^-}_{BB^\prime}).
\end{multline}
From this we can conclude that performing the unitary $U^\dagger V$ on system $B^\prime$ will take the starting state $\ket{\psi}_{ABB^\prime}$ to a symmetric one, so the state $\rho^{AB}$ has a pure symmetric extension.

We now consider the generic case when the reduced state $\rho^B$ is \emph{not} maximally mixed. 
In this case, the two nondegenerate eigenvectors of $\rho^B$ provide a preferred basis for $\mathcal{H}_B$ and the corresponding basis in $\mathcal{H}_{B'}$ is an eigenbasis for $\rho^{B'}$.
By choosing the bases in this way, we make sure that $\rho^{B} = \rho^{B'}$ are diagonal.

An arbitrary state vector of the system $ABB^\prime$ can be written as 
\begin{equation}
\label{eq:tripartitegeneric}
a \ket{000} + b\ket{001} + c\ket{010} + d\ket{011} + e\ket{100} + f\ket{101} + g\ket{110} + h\ket{111}, 
\end{equation}
 where $a,\ldots,h$ are complex numbers whose absolute square sum to 1. 
It is symmetric under permutation of $B$ and $B^\prime$ if $b = c$ and $f = g$. 


The two reduced density matrices of the generic state are in the computational basis 
\begin{align}
  \rho^B &= 
    \begin{bmatrix}
	|a|^2 + |b|^2 + |e|^2 + |f|^2  & a c^* + b d^* + e g^* + f h^* \\
        a^* c + b^* d + e^* g + f^* h  &  |c|^2 + |d|^2 + |g|^2 + |h|^2
    \end{bmatrix},\\
  \rho^{B^\prime} &= 
    \begin{bmatrix}
	|a|^2 + |c|^2 + |e|^2 + |g|^2  & a b^* + c d^* + e f^* + g h^* \\
        a^* b + c^* d + e^* f + g^* h  &  |b|^2 + |d|^2 + |f|^2 + |h|^2
    \end{bmatrix}.
\end{align}
On these matrices we will impose that they are equal and diagonal, since we can achieve that by choosing the eigenvectors of $\rho^{B}$ and $\rho^{B'}$ as bases.
In the next paragraphs, we show that this implies that 
\begin{gather}
 |b| = |c|, \\
 |f| = |g|, \\ 
 |c||g| \left(\e^{\im(\phi_b - \phi_c)} - \e^{\im(\phi_f - \phi_g)} \right) = 0.
\end{gather}
We then show that a unitary operator from Eq.~\eqref{eq:firstcorrectingunitary} or Eq.~\eqref{eq:lastcorrectingunitary} can be applied on $B'$ to give a pure symmetric extension, 
i.e. a state vector with $b = c$ and $f = g$.

The equations we get from imposing that $\rho^{B}$ and $\rho^{B'}$ are equal and diagonal are
\begin{subequations}
\begin{align}
  |b|^2 + |f|^2 = |c|^2 + |g|^2,  \label{eq:cond1}\\
   a c^* + b d^* + e g^* + f h^* &= 0, \label{eq:cond2}\\
   a b^* + c d^* + e f^* + g h^* &= 0, \label{eq:cond3}
\end{align}
\end{subequations}
where the first is from the diagonal entries of $\rho^B$ being equal to those of $\rho^{B'}$ and the others from the off-diagonal elements being 0.

First we show that unless $|b| = |c|$ and $|f| = |g|$, $\rho_B$ has to be maximally mixed. 
For a contradiction, assume that $|b| \ne |c|$. 
Then by \eqref{eq:cond1} also $|f| \ne |g|$. 
From \eqref{eq:cond2} and \eqref{eq:cond3} one can then isolate $e$ and $h^*$:
\begin{subequations}
\label{subeq:eh}
\begin{align}
  e   &= \frac{a(b^*f   - c^*g)   + d^*(cf   - bg)  }{|g|^2 - |f|^2},\\
  h^* &= \frac{a(c^*f^* - b^*g^*) + d^*(bf^* - cg^*)}{|g|^2 - |f|^2}.
\end{align}
\end{subequations}
From this one can compute $|e|^2 - |h|^2$ and by using \eqref{eq:cond1} this simplifies to
\begin{equation}
  \label{eq:mixedcondition}
  |e|^2 - |h|^2 = |d|^2 - |a|^2 .
\end{equation}
Taken together with \eqref{eq:cond1}, this is exactly the condition that the two diagonal elements in $\rho^B$ and $\rho^{B^\prime}$ are equal, so the the subsystems must be completely mixed.
Since the subsystems are not completely mixed, we have a contradiction and we must have $|b| = |c|$ and $|f| = |g|$.
This shows that the absolute values of the relevant amplitudes are equal.

Now we want to find the relations between the complex phases of $b$, $c$, $f$ and $g$. 
Denote $b = |b|\e^{\im\phi_b}$, $c = |c|\e^{\im\phi_c}$, $f = |f|\e^{\im\phi_f}$ and $g = |g|\e^{\im\phi_g}$. 
Multiplying \eqref{eq:cond2} by $g$, \eqref{eq:cond3} by $f$, taking the difference, and using $|f| = |g|$, we obtain
\begin{equation}
  a(c^*g - b^*f) + d^*(bg - cf) = 0 .
\end{equation} 
Since $|c||g| = |c||f| = |b||f| = |b||g|$, this becomes
\begin{equation}
  \e^{\im \phi_g}|c||g|(a \e^{-\im \phi_b} + d^*\e^{\im \phi_c})(\e^{\im(\phi_b - \phi_c)} - \e^{\im(\phi_f - \phi_g)}) = 0.
\end{equation} 
At least one of the following two equations must then hold. 
Either 
\begin{equation}
\label{eq:phaserelation}
  |c||g| \left(\e^{\im(\phi_b - \phi_c)} - \e^{\im(\phi_f - \phi_g)} \right) = 0,
\end{equation}
which will be a useful relation, \emph{or}
\begin{equation}
\label{eq:phasealternative}
  d^*\e^{\im \phi_c} = - a \e^{-\im \phi_b}.
\end{equation}
In the case that \eqref{eq:phaserelation} does not hold, \eqref{eq:phasealternative} must hold, and
we will first see that this leads to a contradiction, since it implies that subsystem $B$ is completely mixed. 

If we insert \eqref{eq:phasealternative} into \eqref{eq:cond3} and use $|b| = |c|$ and $|f| = |g|$, we obtain
\begin{equation}
  \label{eq:phasealternativeconsequence}
  h^* g = -e f^*.
\end{equation}
Since \eqref{eq:phaserelation} does not hold, $|f| = |g| \ne 0$ and therefore \eqref{eq:phasealternativeconsequence} implies $|e| = |h|$. 
The condition \eqref{eq:phasealternative} already means that $|a| = |d|$, so again we have that \eqref{eq:mixedcondition} holds so the diagonal terms in $\rho^B$ are equal and we are in the maximally mixed case.

Since it leads to a contradiction if \eqref{eq:phaserelation} does not hold, we have that \eqref{eq:phaserelation}  must hold.
So while the absolute values of the relevant amplitudes are equal, the complex phases might be different. 
This can be corrected with a phase gate on $B'$ as follows. 
If $b = c = 0$, the unitary operator on $B'$,
\begin{equation}
  \label{eq:firstcorrectingunitary}
  U_{B^\prime} = \proj{0} + \e^{-\im(\phi_f - \phi_g)} \proj{1}
\end{equation}
will correct the phases so that $f = g$, and if $f = g = 0$
\begin{equation}
  \label{eq:lastcorrectingunitary}
  U_{B^\prime} = \proj{0} + \e^{-\im(\phi_b - \phi_c)} \proj{1}
\end{equation}
will make sure that $b = c$.
If none of the relevant amplitudes are zero, Eq.~\eqref{eq:phaserelation} implies that the two expressions are equal, so the same unitary operator will correct both amplitude relations. 
We can therefore always apply a unitary operator to $B'$ to obtain pure tripartite state of the general form \eqref{eq:tripartitegeneric} with $b = c$ and $f = g$.

Hence, for two-qubit states $\rho^{AB}$ that satisfy the spectrum condition \eqref{eq:spectrumcondition}, we have shown that there exists a pure state vector $\ket{\psi}_{ABB^\prime}$ which is symmetric, $\ket{\psi}_{ABB^\prime} =  \swap \ket{\psi}_{ABB^\prime}$.
\end{proof}

This theorem, together with corollary \ref{cor:pure-extendible-decomp} (section~\ref{sec:pure-extendible-decomp}), fully characterizes the set of two-qubit states with a symmetric extension. 
It is the convex hull of the set of states that satisfies condition \eqref{eq:spectrumcondition}. 
Not all the states that satisfy Eq.~\eqref{eq:spectrumcondition} are extremal for the set of states with symmetric extension, however. 
While any pure-extendible states that is itself pure (i.e.~a product state) is extremal for both the set of states and the subset of extendible states, 
there are some mixed pure-extendible states that are not extremal.
The following theorem characterizes the mixed nonextremal pure-extendible states of two qubits.

\begin{theorem}
\label{thm:extremal}
For a two-qubit state $\rho^{AB}$ which is mixed and pure-extendible the following are equivalent: 
\begin{itemize}
\item[1.] $\rho^{AB}$ can be written as a convex combination of other pure-extendible states (i.e. it is not extremal)
\item[2.] $\rho^{AB}$ is separable
\item[3.] $\rho^{AB}$ is of the form
\end{itemize} 
\begin{equation}
\label{eq:purextsepform}
\rho^{AB} = \lambda \proj{\psi_0 0} + (1-\lambda) \proj{\psi_1 1},
\end{equation}
where $\braket{0}{1} = 0$, $\braket{\psi_0}{\psi_1}$ is arbitrary and $0 < \lambda < 1$.
\end{theorem}
\begin{proof} 
3 $\Rightarrow$ 2 is trivial as the state in Eq.~\eqref{eq:purextsepform} is a convex combination of two product states. 2 $\Rightarrow$ 1 is also trivial, since any mixed separable state can be decomposed into a convex combination of pure product states $\rho_{AB} = \sum_j p_j \proj{\psi_j \phi_j}$ and the product states have the pure symmetric extension $\ket{\psi_j}_A\ket{\phi_j}_B\ket{\phi_j}_{B^\prime}$.

The only nontrivial part is 1 $\Rightarrow$ 3. For this part, assume that $\rho^{AB}$ can be written as a convex combination of other pure-extendible states,
\begin{equation}
  \label{eq:pureABdecomp}
  \rho^{AB} = \sum_j p_j \rho_j^{AB},
\end{equation}
where $\rho^{AB}$ and all $\rho_j^{AB}$ satisfy the spectrum condition \eqref{eq:spectrumcondition}. Tracing out $A$ gives 
\begin{equation}
  \label{eq:pureBdecomp}
  \rho^{B} = \sum_j p_j \rho_j^{B}.
\end{equation}
Since $\rho^{AB}$ has support on a 2-dimensional subspace, the support of the $\rho_j^{AB}$ must be on that same subspace. 
We can parametrize the states on $AB$ by Pauli operators $\calone_S, \Sigma_x, \Sigma_y, \Sigma_z$ on this 2-dimensional subspace, 
\begin{equation}
  \label{eq:paulidecompAB}
  \rho_j^{AB} = \frac{1}{2} (\calone_S + X_j \Sigma_x + Y_j \Sigma_y + Z_j \Sigma_z) = \frac{1}{2} ( \calone_S + \vec{R} \cdot \vec{\Sigma}).
\end{equation}
Note that the $\calone_S$ here is not the identity on the 4-dimensional Hilbert space of the system $AB$, but a projector to the 2-dimensional support of $\rho^{AB}$. 
The reduced states on system $B$ can be written as
\begin{equation}
   \label{eq:paulidecompB}
  \rho_j^{B} = \frac{1}{2} (\calone_B + x_j \sigma_x + y_j \sigma_y + z_j \sigma_z) = \frac{1}{2} ( \calone_B + \vec{r} \cdot \vec{\sigma}),
\end{equation}
where $\sigma_x, \sigma_y, \sigma_z$ are the Pauli operators on the qubit $B$. 
Similarly, we can write $\rho_j^{AB} = (\calone_S + \vec{R}_j \cdot \vec{\Sigma})/2$ and $\rho_j^{B} = (\calone_B + \vec{r}_j \cdot \vec{\sigma})/2$. 
In this representation, Eq.~\eqref{eq:pureABdecomp} and Eq.~\eqref{eq:pureBdecomp} becomes $\vec{R} = \sum_j p_j \vec{R}_j$ and $\vec{r} = \sum_j p_j \vec{r}_j$. 

The eigenvalues of $\rho^{AB}$ and $\rho^B$ are determined by the length of the vectors
$\vec{R}$ and $\vec{r}$,
\begin{align}
 \label{eq:r_ev_AB}
  \spec{\rho^{AB}} &= \frac{1}{2}(1 + |\vec{R}|,1 - |\vec{R}|), \\
 \label{eq:r_ev_B}
  \spec{\rho^B}    &= \frac{1}{2}(1 + |\vec{r}|,1 - |\vec{r}|).
\end{align}
The $\rho_j^{AB}$ and $\rho^{AB}$ are pure-extendible, so they satisfy condition \eqref{eq:spectrumcondition}. In terms of the above parametrization, this means that $|\vec{R}_j| = |\vec{r}_j|$ and $|\vec{R}| = |\vec{r}|$. 

Since tracing out a part of a quantum system never can increase the trace distance between the states \cite{ruskai94a}, we have 
\begin{equation}
\| \rho_j^{AB} - \rho_k^{AB}\|_1 \geq \| \rho_j^{B} - \rho_k^{B}\|_1.
\end{equation}
The trace distance can be written in terms of $\vec{R}_j$ and $\vec{r}_j$ as $\| \rho_j^{AB} - \rho_k^{AB}\|_1 = |\vec{R}_j - \vec{R}_k|$ and  $\| \rho_j^{B} - \rho_k^{B}\|_1 = |\vec{r}_j - \vec{r}_k|$. 
From $|\vec{R}_j - \vec{R}_k|^2 \geq |\vec{r}_j - \vec{r}_k|^2$ we get
\begin{equation*}
   |\vec{R}_j|^2 - 2\vec{R}_j \cdot \vec{R}_k + |\vec{R}_k|^2 \geq |\vec{r}_j|^2 - 2\vec{r}_j \cdot \vec{r}_k + |\vec{r}_k|^2,
\end{equation*}
and since $|\vec{R}_j| = |\vec{r}_j|$, this gives
\begin{equation}
  \label{eq:ip_ineq_sepproof}
  \vec{R}_j \cdot \vec{R}_k \leq \vec{r}_j \cdot \vec{r}_k.
\end{equation}

Now we can use $|\vec{R}| = |\vec{r}|$ and Eq.~\eqref{eq:ip_ineq_sepproof} to show that when $\rho^{AB}$ is a pure-extendible state, the trace distance between the $\rho_j^{AB}$ does not decrease when system $A$ is traced out.
\begin{align}
 |\vec{R}|^2 &= \biggl( \sum_j p_j \vec{R}_j \biggr) \biggl( \sum_k p_k \vec{R}_k \biggr) \\
             &= \sum_j p_j^2 |\vec{R}_j|^2 + 2 \sum_{j < k} p_j p_k \vec{R}_j \cdot \vec{R}_k \\
 |\vec{r}|^2 &= \sum_j p_j^2 |\vec{r}_j|^2 + 2 \sum_{j < k} p_j p_k \vec{r}_j \cdot \vec{r}_k 
\end{align}
so by demanding $|\vec{R}| = |\vec{r}|$ and using $|\vec{R}_j| = |\vec{r}_j|$, we get
\begin{equation}
 \sum_{j < k} p_j p_k \vec{R}_j \cdot \vec{R}_k = \sum_{j < k} p_j p_k \vec{r}_j \cdot \vec{r}_k.
\end{equation}
By Eq.~\eqref{eq:ip_ineq_sepproof} none of the terms on the LHS can be greater than the corresponding term on the RHS. The only way for this to be satisfied is that 
\begin{equation}
 \vec{R}_j \cdot \vec{R}_k = \vec{r}_j \cdot \vec{r}_k,
\end{equation}
for all pairs $(j,k)$.
By reversing the calculation leading to \eqref{eq:ip_ineq_sepproof} we get that $|\vec{R}_j - \vec{R}_k|^2 = |\vec{r}_j - \vec{r}_k|^2$ and 
\begin{equation}
\label{eq:equal_trnorm}
\| \rho_j^{AB} - \rho_k^{AB}\|_1 = \| \rho_j^{B} - \rho_k^{B}\|_1.
\end{equation}

The next step is to use Eq.~\eqref{eq:equal_trnorm} to find the structure of the support of $\rho^{AB}$.
The difference $\rho_j^{AB} - \rho_k^{AB}$ must be on the same two-dimensional subspace that all the $\rho_j^{AB}$ are confined to. 
Being the difference between two operators with trace one, it is also traceless, so in the spectral decomposition it can be written as
\begin{equation}
\label{eq:orth_decomp}
 \rho_j^{AB} - \rho_k^{AB} = r \proj{\psi_+} - r \proj{\psi_-}
\end{equation}
for some $r \geq 0$. 
The orthogonal vectors $\ket{\psi_+}$ and $\ket{\psi_-}$ define the two-dimensional support of $\rho_j^{AB}$ and $\rho^{AB}$. 
From Eq.~\eqref{eq:equal_trnorm} and taking the trace norm of both sides of Eq.~\eqref{eq:orth_decomp} it is clear that $\|  \rho_j^{B} - \rho_k^{B} \|_1 = 2r$. 

Let $\rho_+^B = \tr_A \proj{\psi_+}$ and $\rho_-^B = \tr_A \proj{\psi_-}$. 
Tracing out the $A$ system in Eq.~\eqref{eq:orth_decomp} and taking the trace norm gives 
$r \| \rho_+^B - \rho_-^B \|_1 = \|  \rho_j^{B} - \rho_k^{B} \|_1 = 2r$, or 
\begin{equation}
 \| \rho_+^B - \rho_-^B \|_1 = 2.
\end{equation}
This is the maximal distance between two states in trace norm, and it means that $\rho_+^B$ and $\rho_-^B$ have support on orthogonal subspaces. 
Since $B$ is a qubit, $\rho_+^B$ and $\rho_-^B$ must be orthogonal pure states which we denote $\rho_+^B = \proj{0}$, $\rho_-^B = \proj{1}$. 
This also means that 
$\ket{\psi_+}$ and $\ket{\psi_-}$ are product states,
\begin{align}
 \ket{\psi_+} &= \ket{\psi_0} \otimes \ket{0}, \\
 \ket{\psi_-} &= \ket{\psi_1} \otimes \ket{1},
\end{align}
where $\ket{\psi_0}$ and $\ket{\psi_1}$ are arbitrary.

Any state on the subspace spanned by $\ket{\psi_+}$ and $\ket{\psi_-}$ can be expressed as 
\begin{equation}
 \rho^{AB} = \sum_{m,n = 0}^1 \rho_{mn} \ket{\psi_m}\bra{\psi_n} \otimes \ket{m}\bra{n},
\end{equation}
with the reduced state being
\begin{equation}
 \rho^{B} = \sum_{m,n = 0}^1 \rho_{mn} \braket{\psi_n}{\psi_m} \otimes \ket{m}\bra{n}.
\end{equation}
Since $\rho^{AB}$ is pure-extendible, it satisfies Eq.~\eqref{eq:spectrumcondition} and for qubits this is equivalent to the condition that the purities of the global and reduced states are equal, $\tr[(\rho^{AB})^2] = \tr[(\rho^{B})^2]$. The purities are
\begin{align}
 \tr[(\rho^{AB})^2] &= \rho_{00}^2 + \rho_{11}^2 + |\rho_{01}|^2 \\
 \tr[(\rho^{B})^2] &= \rho_{00}^2 + \rho_{11}^2 + |\rho_{01}|^2 |\braket{\psi_0}{\psi_1}|^2.
\end{align}
For the purities to be equal, either $\rho_{01}=0$ or $|\braket{\psi_0}{\psi_1}| = 1$. In the first case, the state would be 
\begin{equation}
 \label{eq:sepform2}
 \rho^{AB} = \rho_{00} \proj{\psi_0 0} + \rho_{11} \proj{\psi_1 1}
\end{equation}
which is the sought separable form. In the other case $\ket{\psi_0}$ and $\ket{\psi_1}$ only differ by a phase, so all states in the subspace are product states of the form $\proj{\psi_0} \otimes \rho^B$ which is the special case of Eq.~\eqref{eq:sepform2} where $\ket{\psi_0} = \ket{\psi_1}$.
\end{proof}

\subsection{Spectrum condition is not sufficient for any higher dimension}
\label{sec:counterex}

In the previous section, we have seen that the spectrum condition \eqref{eq:spectrumcondition} is not only necessary but also sufficient for the state to have a pure symmetric extension when the system considered is a pair of qubits. 
One may ask if the same might be true for any higher dimensional system. We show some counterexamples that exclude this possibility for any dimension greater than $2 \times 2$.

\begin{figure}
 \resizebox{0.7\textwidth}{!}{\includegraphics{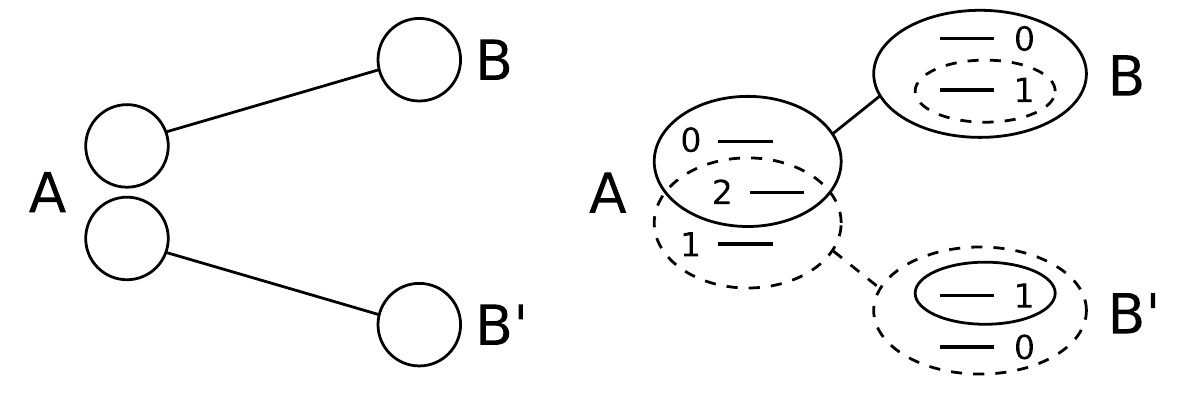}}
  \caption[Examples of tripartite states states where $\rho_{AB}$ satisfies the spectrum condition \eqref{eq:spectrumcondition}, but does not have a symmetric extension.]{\label{fig:counterexNx2} Examples of tripartite states states where $\rho_{AB}$ satisfies the spectrum condition \eqref{eq:spectrumcondition}, but does not have a symmetric extension. The $4 \times 2$ state from example \ref{ex:4x2} on the left and the $3\times 2$ state from example \ref{ex:3x2} on the right.}
\end{figure}

\begin{example}
\label{ex:4x2}
($4 \times 2$). 
The simplest example is when Alice holds two qubits and Bob one. 
One of Alice's qubits is maximally mixed, while the other is maximally entangled with Bob's qubit.
\begin{equation}
  \rho^{A_1 A_2 B} = \frac{\calone_{A_1}}{2} \otimes \proj{\Phi^+}_{A_2 B}
\end{equation}
The global density matrix $\rho^{A_1 A_2 B}$ has nonzero eigenvalues $\{1/2,1/2\}$, and so has the local one $\rho^{B}$. 
The state therefore satisfies the spectrum condition $\spec{\rho^{A_1 A_2 B}} = \spec{\rho^B}$, but does not have a symmetric extension, since by tracing out $A_1$, Alice can make a pure maximally entangled state.
The purification of the state is illustrated in Fig.~\ref{fig:counterexNx2}.
\end{example}

While the above example is conceptually simple, it does not exclude that the spectrum condition could be sufficient when Alice holds a qutrit. The following example is similar in spirit to the above, and shows that for system of size $3 \times 2$ and higher, the spectrum condition cannot be sufficient. 
\begin{example}
\label{ex:3x2}
($3 \times 2$). Consider the (unnormalized) vectors of a tripartite system 
\begin{align}
  \ket{v_1} &= \ket{001} + \ket{211}, \\
  \ket{v_2} &= \ket{110} + \ket{211}, 
\end{align}
where the registers are $A$, $B$ and $B^\prime$. The vectors are illustrated in Fig.~\ref{fig:counterexNx2}, the solid line corresponds to $\ket{v_1}$ and the dashed line to $\ket{v_2}$. The vector $\ket{v_1}$ is entangled between $A$ and $B$, while $\ket{v_2}$ is entangled between $A$ and $B^\prime$. Interchanging 0 and 1 at $A$ and swapping $B$ and $B^\prime$, takes $\ket{v_1}$ to $\ket{v_2}$ and \emph{vice versa}. 
Adding the two vectors and normalizing gives the state
\begin{equation}
  \ket{\psi} = \frac{1}{\sqrt{6}} \ket{001} + \frac{1}{\sqrt{6}} \ket{110} + \sqrt{\frac{2}{3}} \ket{211}.
\end{equation}
The reduced states are 
\begin{align}
 \rho^{AB} &= \frac{5}{6} \proj{\psi_{1/5}} + \frac{1}{6} \proj{11}, \\
 \rho^B &= \frac{5}{6} \proj{1} + \frac{1}{6} \proj{0},
\end{align}
where 
\begin{equation}
  \ket{\psi_{1/5}} = \frac{1}{\sqrt{5}} \ket{00} + \sqrt{\frac{4}{5}} \ket{21}.
\end{equation}
The nonzero eigenvalues are the same for $\rho^{AB}$ and $\rho^{B}$, so $\rho^{AB}$ satisfies the spectrum condition. 
However, it does not have a symmetric extension. 
This is most easily seen by applying the filter $F = \proj{0} + \proj{2}$ to $A$. 
This succeeds with probability 5/6 and the state after the filter is the pure entangled state $\ket{\psi_{1/5}}$, which has no symmetric extension.
\end{example}

Both examples above are states that can be extended to states that are invariant under some $U_A \otimes \swap$, where $U_A$ is a unitary on $A$, but not under $\calone_A \otimes \swap$. 
For the $4 \times 2$ case, $U_A$ was the unitary swapping $A_1$ and $A_2$, while in the $3 \times 2$ example it was ${\ket{0}\bra{1}} + {\ket{1}\bra{0}} + {\ket{2}\bra{2}}$, a unitary swapping the basis vectors $\ket{0}$ and $\ket{1}$. 
One can use the same arguments as in the proof of theorem \ref{thm:spectrumcondition} to show that any pure state that has a symmetry of the type $U_A \otimes \swap$ has a reduction to $AB$ that satisfies the condition \eqref{eq:spectrumcondition}.

The above examples show that the condition \eqref{eq:spectrumcondition} cannot be sufficient for pure extendibility for $M \times N$ systems where $M \ge 3$ and $N \ge 2$. This leaves open the question whether it is sufficient for $2\times N$ for any $N > 2$. We therefore now give an example of a class of states with system dimension $2 \times 3$ that satisfies condition \eqref{eq:spectrumcondition}, but has no symmetric extension. 

\begin{example}
($2 \times 3$). Consider states with spectral decomposition
\begin{equation}
  \rho^{AB} = \sum_{j=0}^2 \lambda_j \proj{\psi_j},
\end{equation}
where the eigenvectors are $\ket{\psi_0} = \ket{12}$, $\ket{\psi_1} = \ket{02}$ and $\ket{\psi_2} = \sqrt{s}\ket{00} + \sqrt{1-s}\ket{11}$. 
For such a state to satisfy the spectrum condition \eqref{eq:spectrumcondition}, the eigenvalues must be $\lambda_0 = s/2$, $\lambda_1 = (1-s)/2$ and $\lambda_2 = 1/2$. 
To $\rho^{AB}$ we now apply a filter operation in the standard basis in the $A$ system, $F = \sqrt{p} \proj{0} + \proj{1}$. This is a 1-SLOCC operation and cannot break a symmetric extension. 
After a successful filter, the global and local eigenvalues $\lambda_j(p)$ and $\lambda_j^B(p)$ are 
\begin{equation*}
  \begin{aligned}
    \lambda_0(p) &= \frac{s}{1+p},\\
    \lambda_1(p) &= \frac{(1-s)p}{1+p},\\
    \lambda_2(p) &= \frac{1-s(1-p)}{1+p},
  \end{aligned}
\qquad
  \begin{aligned}
    \lambda_0^B(p) &= \frac{sp}{1+p},\\
    \lambda_1^B(p) &= \frac{1-s}{1+p},\\
    \lambda_2^B(p) &= \frac{1-(1-s)(1-p)}{1+p}.
  \end{aligned}
\end{equation*}
Except when $s \in \{0,1/2,1\}$ or $p \in \{0,1\}$, the spectra of the local and global density matrices are different. 
Since a filtering like this will keep a pure symmetric extension if the original state had one (Corollary \ref{cor:purextfilter}, section~\ref{sec:condition}), $\rho^{AB}$ cannot have a pure symmetric extension. 
For $1/2 < s < 1$ and $0 < p < 1$, the state has no symmetric extension at all. 
This is because in this regime the coherent information $I(A\rangle B) := S(\rho^{B}) - S(\rho^{AB})$, where $S(\cdot)$ is the von Neumann entropy, is positive. 
This is a lower bound to the distillable entanglement with one-way communication from $A$ to $B$ \cite{devetak05a}. 
By monogamy of entanglement, $\rho^{AB}$ cannot have a symmetric extension.
\end{example}

\section{Rank-2 states}

In this section, we prove conjecture \ref{conj:twoqubitextension} from section~\ref{sec:conjformula} for two-qubit states of rank 2.
When $\rho^{AB}$ has rank 2, the determinant in Eq.~\eqref{eq:twoqubitextension} vanishes, 
and since the remaining inequality only compares the purity of the states, 
we can as well use the maximum eigenvalues to compare it.

\begin{theorem}
\label{thm:rank2}
A two-qubit state $\rho^{AB}$ of rank 2 has a symmetric extension if and only if 
\begin{equation}
\label{eq:rank2cond}
\lambda_\text{max}(\rho^{AB}) \leq \lambda_\text{max}(\rho^B)
\end{equation}
\end{theorem}
\begin{figure}
\includegraphics{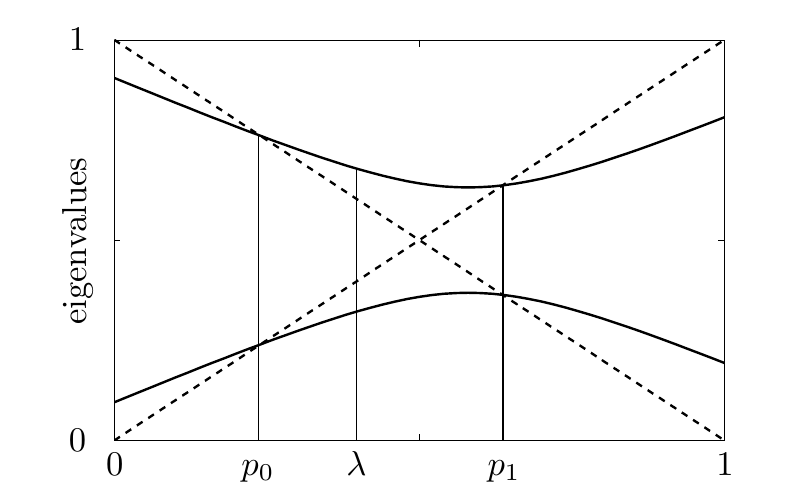}
\caption[Decomposition into pure-extendible states for two qubit states of rank~2.]{\label{fig:rank2} Decomposition into pure-extendible states for two qubit states of rank 2. The dashed lines are global eigenvalues as parametrized on the $x$-axis. The solid lines are the local eigenvalues. The state with global eigenvalues $(1-\lambda,\lambda)$ is a convex combination of states with global eigenvalues $(1-p_0,p_0)$ and $(p_1,1-p_1)$, which have the same local eigenvalues and therefore a pure symmetric extension.} 
\end{figure}
\begin{proof} 
We first prove the ``if'' part. 
Assume that $\rho^{AB}$ is a two-qubit state of rank 2 that satisfies Eq.~\eqref{eq:rank2cond}. 
We can write it in the spectral decomposition
\begin{equation}
  \rho^{AB} = (1-\lambda) \proj{\psi_0} + \lambda \proj{\psi_1}.
\end{equation}
Consider the class of states with the same eigenvectors as above, parametrized by  $p$, $\rho_p^{AB} = (1-p) \proj{\psi_0} + p \proj{\psi_1}$. 
Now, $\rho^{AB} = \rho_\lambda^{AB}$. 
For $p=0$ and $p=1$, the corresponding pure states satisfy
$\lambda_\text{max}(\rho_p^{B}) \leq \lambda_\text{max}(\rho_p^{AB}) = 1$. 
Since at $p = \lambda$, $\lambda_\text{max}(\rho_p^{B}) \geq \lambda_\text{max}(\rho_p^{AB})$ by assumption and $\lambda_\text{max}$ is a continuous function of the parameter $p$, 
there must exist parameters $p_0 \in [0,\lambda]$, $p_1 \in [\lambda,1]$ such that 
$\lambda_\text{max}(\rho_{p_0}^{B}) = \lambda_\text{max}(\rho_{p_0}^{AB})$ 
and
$\lambda_\text{max}(\rho_{p_1}^{B}) = \lambda_\text{max}(\rho_{p_1}^{AB})$ (see Fig.~\ref{fig:rank2}).  
From theorem \ref{thm:necsufqubit}, we know that $\rho_{p_0}^{AB}$ and $\rho_{p_1}^{AB}$ have pure symmetric extensions, $\ket{\psi_{p_0}}_{ABB^\prime}$ and $\ket{\psi_{p_1}}_{ABB^\prime}$. 
Since $\rho_\lambda^{AB}$ is a convex combination $\rho_\lambda^{AB} = (1-q) \rho_{p_0}^{AB} + q \rho_{p_1}^{AB}$, where $q = (\lambda - p_0)/(p_1 - p_0)$, a symmetric extension of $\rho_\lambda^{AB}$ is $\rho^{ABB^\prime} = (1-q) \proj{\psi_{p_0}}_{ABB^\prime} + q \proj{\psi_{p_1}}_{ABB^\prime}$. 

Now, for the ``only if'' part, assume that $\rho^{AB}$ is a bipartite state of rank 2 that has a symmetric extension to two copies of the qubit $B$ (in this part we do not use that $A$ is a qubit). 
Then by corollary \ref{cor:pure-extendible-decomp} it can be written as a convex combination of pure-extendible states
\begin{equation}
  \label{eq:ABdecomp}
  \rho^{AB} = \sum_j p_j \rho_j^{AB},
\end{equation}
and tracing out $A$,
\begin{equation}
  \label{eq:Bdecomp}
  \rho^{B} = \sum_j p_j \rho_j^{B}.
\end{equation}
Like in the proof of theorem \ref{thm:extremal} in section~\ref{sec:necsufqubit} we can use the fact that 
$\rho^{AB}$ has support on a 2-dimensional subspace to parametrize it using 
Pauli operators as in Eq.~\eqref{eq:paulidecompAB}. 
Likewise we expand $\rho^{B}$ as in Eq.~\eqref{eq:paulidecompB}, so 
Eq.~\eqref{eq:ABdecomp} and Eq.~\eqref{eq:Bdecomp} become
$\vec{R} = \sum_j p_j \vec{R}_j$ and $\vec{r} = \sum_j p_j \vec{r}_j$.
We can proceed exactly as in the previous proof to arrive at
Eq.~\eqref{eq:ip_ineq_sepproof} which says that
$\vec{R}_j \cdot \vec{R}_k \leq \vec{r}_j \cdot \vec{r}_k$.
for all $j$ and $k$.

Since $\rho_j^{AB}$ are pure-extendible states, 
they have the same eigenvalues as the corresponding $\rho_j^{B}$ and therefore 
$|\vec{R}_j| = |\vec{r}_j|$.
Now we can use this and Eq.~\eqref{eq:ip_ineq_sepproof} to compare $|\vec{R}|$ and $|\vec{r}|$,
\begin{align}
 |\vec{R}|^2 &= \biggl( \sum_j p_j \vec{R}_j \biggr) \biggl( \sum_k p_k \vec{R}_k \biggr) \\
             &= \sum_j p_j^2 |\vec{R}_j|^2 + 2 \sum_{j < k} p_j p_k \vec{R}_j \cdot \vec{R}_k \\
          &\leq \sum_j p_j^2 |\vec{r}_j|^2 + 2 \sum_{j < k} p_j p_k \vec{r}_j \cdot \vec{r}_k \\ 
             &= |\vec{r}|^2.
\end{align}
From $|\vec{R}| \leq |\vec{r}|$ and the relations to eigenvalues Eq.~\eqref{eq:r_ev_AB} and Eq.~\eqref{eq:r_ev_B} we can conclude that $\lambda_\text{max}(\rho^{AB}) \leq \lambda_\text{max}(\rho^B)$ which completes the proof.
\end{proof}

\begin{remark}
\label{remark:rank2necessary}
 The assumption that system $A$ is a qubit was only needed in the ``if'' part of the proof 
to conclude that states that satisfy the spectrum condition $\lambda_\text{max}(\rho^{AB}) = \lambda_\text{max}(\rho^{B})$ have a symmetric extension.
The rest of the proof, in particular the ``only if'' part, is independent of this assumption.
Therefore, no $N \times 2$ state of rank 2 that satisfies $\lambda_\text{max}(\rho^{AB}) > \lambda_\text{max}(\rho^{B})$ can have a symmetric extension.
\end{remark}

\section{Symmetric subspace}

In this section, we prove conjecture \ref{conj:twoqubitextension} from section~\ref{sec:conjformula} for two-qubit states that have their support only on the symmetric subspace, i.e. the subspace spanned by the three symmetric vectors $\ket{\Phi^+}$, $\ket{\Phi^-}$, and $\ket{\Psi^+}$.
We will make use of the Pauli operator basis to expand the density operator,
\begin{equation}
\label{eq:rhopauliexpansion}
\rho^{AB} = \frac{1}{4} \sum_{i=0}^3 r_{ij} \sigma_i \otimes \sigma_j.
\end{equation}
The $4 \times 4$ matrix of the indices $r_{ij}$ is called the $R$-matrix\index{R-matrix@$R$-matrix} of the state, and its elements are explicitly given by $r_{ij} := \tr[\rho^{AB}(\sigma_i \otimes \sigma_j)]$, $i,j \in \{0,1,2,3\}$.
Let us first see which $R$-matrices correspond to states with support on the symmetric subspace.

\begin{lemma}
  \label{lem:symsubspaceRmatrix}
   The two-qubit Hermitian operator expanded in the local Pauli basis
   \begin{equation}
   \label{eq:pauliexpansion}
   M := \frac{1}{4} \sum_{i=0}^3 r_{ij} \sigma_i \otimes \sigma_j
   \end{equation}
   has support only on the symmetric subspace 
   if and only if 
   \begin{itemize}
   \item[(i)] $r_{ij} = r_{ji}$ and 
   \item[(ii)] $r_{00} = r_{11} + r_{22} + r_{33}$.
   \end{itemize}
\end{lemma}
\begin{proof}
 Condition (i) means that $M$ is invariant under exchange of $A$ and $B$, which is certainly a necessary condition for being on the symmetric subspace.
 Condition (ii) comes from imposing that the state having no overlap with the 1-dimensional antisymmetric subspace, spanned by the single vector $\ket{\Psi^-}$. 
 Expanding $\proj{\Psi^-} = \frac{1}{4}(\calone \otimes \calone - \sum_{k=1}^3 \sigma_k \otimes \sigma_k)$, we impose
 \begin{gather}
  0 = \bra{\Psi^-}\rho\ket{\Psi^-} \\
  = \tr[\rho \proj{\Psi^-}] \\
  = \frac{1}{16} \Big[\sum_{ij} r_{ij} \tr[(\sigma_i \otimes \sigma_j)(\sigma_0 \otimes \sigma_0)] - \sum_{ij}\sum_{k=1}^3 r_{ij} \tr[(\sigma_i \otimes \sigma_j)(\sigma_k \otimes \sigma_k)] \Big]\\
  = \frac{1}{16} \Big[\sum_{ij} r_{ij} 4\delta_{i0}\delta_{j0} - \sum_{ij}\sum_{k=1}^3 r_{ij} 4 \delta_{ik} \delta_{jk} \Big]\\
  = \frac{1}{4} [ r_{00} - \sum_{k=1}^3 r_{kk} ]
 \end{gather}
 which is the same as condition (ii).
 The other part is to show that any Hermitian operator with $r_{ij}$ that satisfies conditions (i) and (ii) has support only on the symmetric subspace. 
 This follows since any operator that satisfies (i) is a direct sum of operators on the symmetric and antisymmetric subspace. 
 Condition (ii) then ensures that there is no support on the 1-dimensional antisymmetric subspace.
\end{proof}

\begin{theorem}
 A two-qubit state $\rho^{AB}$ whose support is the symmetric subspace has a symmetric extension if and only if Eq.~\eqref{eq:twoqubitextension} holds. It here reduces to 
\begin{equation}
 \label{eq:twoqubitextensionsymsubspace}
 \tr[(\rho^B)^2] \geq \tr[(\rho^{AB})^2].
\end{equation}
\end{theorem}
Note that the condition also is necessary and sufficient if the support of $\rho^{AB}$ is a proper subspace of the symmetric subspace. 
In that case, it is either of rank 2 and covered by theorem \ref{thm:rank2} or it is of rank 1 and a pure product state.
\begin{proof}
For the ``only if'' part, assume that $\rho^{AB}$ has the symmetric subspace as its support and has a symmetric extension. 
The condition \eqref{eq:twoqubitextensionsymsubspace} can be written as 
$f(\rho^{AB}) := \tr[(\rho^B)^2] - \tr[(\rho^{AB})^2] \geq 0$.
First, we want to show that when restricted to states that are invariant under swap of the $A$ and $B$ systems (of which states on the symmetric subspace is a subclass), $f$ is a concave function. 

When we expand in the Pauli basis, the state and its reduced state are given by
\begin{gather}
 \rho^{AB} = \frac{1}{4} \sum_{ij} r_{ij} \sigma_i \otimes \sigma_j,\\
 \rho^{B} = \frac{1}{2} \sum_{j} r_{0j} \sigma_j.
\end{gather}
By using that $\tr[\sigma_i \sigma_j] = 2\delta_{ij}$ and that 
$\tr[(\sigma_i \otimes \sigma_j)(\sigma_k \otimes \sigma_l)] = 4 \delta_{ik} \delta{jl}$ we get
\begin{gather}
 \tr[\rho_{AB}^2] = \frac{1}{4} \sum_{ij} r_{ij}^2,\\
 \tr[\rho_{B}^2]  = \frac{1}{2} \sum_{j}  r_{0j}^2,\\
 \tr[\rho_{B}^2] - \tr[\rho_{AB}^2] = 
 \frac{1}{4} \sum_{j=0}^3 r_{0j}^2 - \frac{1}{4} \sum_{i=1}^3 \sum_{j=0}^3 r_{ij}^2.
\end{gather}
This is neither concave nor convex, since the first sum is of convex terms and the negative sign makes the second part concave.
However, for states that are invariant when swapping $A$ and $B$, $r_{ij} = r_{ji}$, so the three terms in the first sum corresponding to $j \ne 0$ are cancelled by a corresponding term in the second sum, leaving that for these states
\begin{equation}
 \tr[\rho_{B}^2] - \tr[\rho_{AB}^2] = \frac{1}{4} \left( r_{00}^2 - \sum_{i=1}^3 \sum_{j=1}^3 r_{ij}^2\right).
\end{equation}
Since $r_{00} = 1$ is constant, this is a concave function.

The state $\rho^{AB}$ has a symmetric extension, so it can be written as a convex combination of states with a pure symmetric extension (corollary \ref{cor:pure-extendible-decomp}),
\begin{equation}
 \rho^{AB} = \sum_i p_i \rho_i^{AB}.
\end{equation}
Since $\rho^{AB}$ only has support on the symmetric subspace, the pure-extendible states must also have their support on this subspace.
The pure-extendible states satisfy the spectrum condition $\vec{\lambda}(\rho_{AB}) = \vec{\lambda}(\rho_{B})$ (theorem \ref{thm:spectrumcondition}), 
so $f(\rho_i^{AB}) = 0$ for all $i$. 
Since $f$ is concave on the symmetric subspace, 
 \begin{equation}
  f(\rho^{AB}) = f\Big(\sum_i p_i \rho_i^{AB}\Big) \geq \sum_i p_i f(\rho_i^{AB}) = 0,
 \end{equation}
so condition \eqref{eq:twoqubitextensionsymsubspace} is satisfied.

For the ``if'' part, assume that $\rho^{AB}$ has the symmetric subspace as its support and satisfies condition \eqref{eq:twoqubitextensionsymsubspace}.
In this part of the proof, we will show how any such state can be decomposed into a convex combination of other states that we know have a symmetric extension. 
We will first split off a separable part if the inequality is not tight, and then deal with part where the condition is satisfied with equality.

If the inequality in Eq.~\eqref{eq:twoqubitextensionsymsubspace} is strict, 
we can write $f(\rho^{AB}) = \tr[(\rho^B)^2] - \tr[(\rho^{AB})^2] > 0$.
We can then decompose the state into the maximally mixed state on the symmetric subspace and another state $\rho_0$, 
\begin{equation}
 \rho^{AB} = p \frac{P_\textrm{sym}}{3} + (1-p) \rho_0,
\end{equation}
where $P_\textrm{sym}$ is the projector onto the symmetric subspace and $P_\textrm{sym}/3$ is a separable state and  therefore has a symmetric extension.
We may increase $p$ gradually until either $\rank(\rho_0) < 3$ or $f(\rho_{0})  = 0$.
In the first case, we already know from theorem \ref{thm:rank2} that $\rho_0$ has a symmetric extension, since it is a rank-2 or rank-1 state with $f(\rho_0) > 0$. 
We therefore have that our state is a convex combination of two states with symmetric extension and therefore has symmetric extension itself.
In the second case, we have decomposed our original state into a state with symmetric extension ($P_\textrm{sym}/3$) and a state $\rho_0$ which satisfies $f(\rho_0) = 0$.
We therefore now assume that $f(\rho^{AB}) = 0$ and we will show that such a state can be decomposed into two states with a pure symmetric extension.

From $f(\rho^{AB}) = 0$, the $R$-matrix of the state $\rho^{AB}$ gets another constraint in addition to (i) and (ii) from lemma \ref{lem:symsubspaceRmatrix}. 
This constraint is 
\begin{equation}
 \label{eq:Rmatrixf0}
 r_{00}^2 = \sum_{i=1}^3 \sum_{j=1}^3 r_{ij}^2.
\end{equation}
Now, choose an arbitrary vector $\vec{k} := (k_1, k_2, k_3)$ and consider the one-parameter family of Hermitian operators defined by 
\begin{equation}
 \rho(t) = \rho^{AB} + t \sum_{j=1}^3 k_j (\calone \otimes \sigma_j + \sigma_j \otimes \calone).
\end{equation}
The value of $t$ will only influence $R$-matrix elements of the form $r_{i0}$ and $r_{0i}$.
The state will therefore remain on the symmetric subspace and still satisfy Eq.~\eqref{eq:Rmatrixf0} and $f(\rho(t)) = 0$ for any value of $t$.
Since $\rho^{AB}$ has full rank on the symmetric subspace, $\rho(t)$ will be a state for $|t|$ small enough.
There will be a $t_+ > 0$ and a $t_- < 0$ such that $\rho(t)$ is a state for $t_- \leq t \leq t_+$, but it will have negative eigenvalues for $t < t_-$ and $t > t_+$. 
Since the spectrum is a continuous function of the operator, the states $\rho(t_-)$ and $\rho(t_+)$ will have maximum two nonzero eigenvalues.
Since they also satisfy $\tr[\rho^{AB}(t_\pm)] = \tr[\rho^B(t_\pm)]$ and $\tr[\rho^{AB}(t_\pm)^2] = \tr[\rho^B(t_\pm)^2]$, 
it means that $\vec{\lambda}(\rho^{AB}(t_\pm)) = \vec{\lambda}(\rho^{B}(t_\pm))$. 
This is equivalent to $\rho^{AB}(t_\pm)$ having pure symmetric extensions by theorem \ref{thm:necsufqubit} (section \ref{sec:necsufqubit}).
The original state is a mixture of these two pure-extendible states:
\begin{equation}
 \rho^{AB} = \frac{-t_-}{t_+ - t_-} \rho(t_+) + \frac{t_+}{t_+ - t_-} \rho(t_-),
\end{equation}
and therefore has a symmetric extension.
\end{proof}

\section{Bell-diagonal states}
\label{sec:belldiagsymext}

Bell-diagonal states are most commonly defined as two-qubit states of the form
\begin{equation}
 \rho^{AB} = p_I \proj{\Phi^+} + p_x \proj{\Psi^+} + p_y \proj{\Psi^-} + p_z \proj{\Phi^-}.
\end{equation}
Since all the maximally entangled states have maximally mixed subsystems, we also have $\rho^B = \tr[\rho^{AB}]\calone/2$ for Bell-diagonal states. 
The conjectured condition \eqref{eq:twoqubitextension} for two-qubit symmetric extension can therefore be written as
\begin{equation}
\label{eq:twoqubitextensionbelldiag}
 4\sqrt{\det(\rho^{AB})} \geq \tr[(\rho^{AB})^2] - \frac{1}{2} (\tr[\rho^{AB}])^2.
\end{equation}
In this section, we will follow the procedure of section~\ref{sec:simplifying} for the special case of one qubit pair ($N = 1$) in order to use the symmetry of this class of states to simplify the problem (section \ref{sec:belldiagsymext-symmetry}).
We will then in section \ref{sec:belldiagsymext-sdp} formulate the simplified problem as a semidefinite program and solve it analytically. 
The solution to the semidefinite program is not exactly Eq.~\eqref{eq:twoqubitextensionbelldiag}, but rather the condition specified in 
Eqs.~\eqref{eq:rankoneXsufcondsum}--\eqref{eq:rank2Xposconstraint2sum}.
We will find this formulation of the condition useful in section \ref{sec:cross-section-symext}.
In section \ref{sec:belldiagsymext-equivalence} we show that the two conditions are equivalent.

\subsection{Using symmetry to simplify the Bell-diagonal symmetric extension problem}
\label{sec:belldiagsymext-symmetry}

Due to the symmetry of the state (as listed in section~\ref{sec:symofextstate}) it can also be parametrized as 
\begin{equation}
 \rho^{AB} = \frac{1}{4}(\alpha_0 \calone \otimes \calone + \alpha_x \sigma_x \otimes \sigma_x 
            + \alpha_y \sigma_y \otimes \sigma_y + \alpha_z \sigma_z \otimes \sigma_z),
\end{equation}
where
\begin{subequations}
\begin{align}
\label{eq:xyzalphafirst}
\alpha_0 &=  \phantom{-} p_I + p_x + p_y + p_z\\
\alpha_x &=  \phantom{-} p_I + p_x - p_y - p_z\\
\alpha_y &=    -         p_I + p_x - p_y + p_z\\
\label{eq:xyzalphalast}
\alpha_z &=  \phantom{-} p_I - p_x - p_y + p_z.
\end{align}
\end{subequations}
When we impose the corresponding symmetries on the extended state only terms of the forms $\calone \otimes \calone \otimes \calone$, $\sigma_i \otimes \sigma_i \otimes \calone$, $\sigma_i \otimes \calone \otimes \sigma_i$, and $\calone \otimes \sigma_i \otimes \sigma_i$ for $i \in \{x,y,z\}$ survive.
No terms of the form $\sigma_x \otimes \sigma_y \otimes \sigma_z$ and permutations appear, since we have only one qubit pair and these terms appear only for two and more qubit pairs.
The extended state therefore has to be of the form
\begin{align}
 \label{eq:belldiagextendedform}
 \rho^{ABB'} = \frac{1}{8} &( \alpha_0 \calone \otimes \calone \otimes \calone \nonumber \\
  & + \alpha_x \sigma_x \otimes \sigma_x \otimes \calone + \alpha_y \sigma_y \otimes \sigma_y \otimes \calone + \alpha_z \sigma_z \otimes \sigma_z \otimes \calone \nonumber \\ 
  & + \alpha_x \sigma_x \otimes \calone \otimes \sigma_x + \alpha_y \sigma_y \otimes \calone \otimes \sigma_y + \alpha_z \sigma_z \otimes \calone \otimes \sigma_z \nonumber \\
  & + \beta_x \calone \otimes \sigma_x \otimes \sigma_x + \beta_y \calone \otimes \sigma_y \otimes \sigma_y + \beta_z \calone \otimes \sigma_z \otimes \sigma_z ),
\end{align}
where the $\alpha_i$ parameters are fixed by $\rho^{AB}$ and the three parameters $\beta_i$ are open.
This is a $8 \times 8$ matrix, but we can reduce the relevant dimension to by focusing on the $+1$ eigenspace of $\sigma_z \otimes \sigma_z \otimes \sigma_z$ and replace the terms in Eq.~\eqref{eq:belldiagextendedform} according to table \ref{tab:projectedpaulis}.
This gives the projected extended states
\begin{align}
 \label{eq:belldiagextendedformprojected}
 P_+ \rho^{ABB'} P_+ &= \frac{1}{8} ( \alpha_0 \calone \otimes \calone \nonumber \\
  & \qquad + \alpha_x \sigma_x \otimes \sigma_x  + \alpha_y \sigma_y \otimes \sigma_y + \alpha_z \sigma_z \otimes \sigma_z \nonumber \\ 
  & \qquad + \alpha_x \sigma_x \otimes \calone - \alpha_y \sigma_x \otimes \sigma_z + \alpha_z \calone \otimes \sigma_z \nonumber\\
  & \qquad+ \beta_x \calone \otimes \sigma_x  - \beta_y \sigma_z \otimes \sigma_x + \beta_z \sigma_z \otimes \calone )\\
  & = \frac{1}{8}
      \begin{bmatrix}
       \alpha_0 + 2\alpha_z + \beta_z & \beta_x - \beta_y & \alpha_x - \alpha_y & \alpha_x - \alpha_y \\
       \beta_x - \beta_y & \alpha_0 - 2\alpha_z + \beta_z & \alpha_x + \alpha_y & \alpha_x + \alpha_y \\
       \alpha_x - \alpha_y & \alpha_x + \alpha_y & \alpha_0 - \beta_z & \beta_x + \beta_y \\
       \alpha_x - \alpha_y & \alpha_x + \alpha_y & \beta_x + \beta_y & \alpha_0 - \beta_z 
      \end{bmatrix}\\
 & = \frac{1}{8}
      \begin{bmatrix}
       \alpha_0 + 2\alpha_1 + \beta_1 & \sqrt{2} \beta_2 & \sqrt{2} \alpha_2 & \sqrt{2} \alpha_2 \\
       \sqrt{2} \beta_2 & \alpha_0 - 2\alpha_1 + \beta_1 & \sqrt{2} \alpha_3 & \sqrt{2} \alpha_3 \\
       \sqrt{2} \alpha_2 & \sqrt{2} \alpha_3 & \alpha_0 - \beta_1 & \sqrt{2} \beta_3 \\
       \sqrt{2} \alpha_2 & \sqrt{2} \alpha_3 & \sqrt{2} \beta_3 & \alpha_0 - \beta_1 
      \end{bmatrix},
\end{align}
where we in the last line have defined $\alpha_1 := \alpha_z$, $\alpha_2 := (\alpha_x - \alpha_y)/\sqrt{2}$, and $\alpha_3 := (\alpha_x + \alpha_y)/\sqrt{2}$ and similarly for $\beta_1$, $\beta_2$, and $\beta_3$.
These parameters will be useful later, so we include their relation to the eigenvalues of the Bell-diagonal state,
\begin{subequations}
\label{eq:alpha-transformation}
\begin{align}
 \label{eq:alpha-transformation-first}
 \alpha_0 &= p_I + p_x + p_y + p_z = 1,\\
 \label{eq:alpha-transformation-alpha1}
 \alpha_1 &= p_I - p_x - p_y + p_z,\\
 \label{eq:alpha-transformation-alpha2}
 \alpha_2 &= \sqrt{2} ( p_I - p_z ),\\
 \label{eq:alpha-transformation-last}
 \alpha_3 &= \sqrt{2} ( p_x - p_y ).
\end{align}
\end{subequations}
The new open variables $\beta_2$ and $\beta_3$ are independent, so we can choose 
$\beta_3 = (\alpha_0 - \beta_z)/\sqrt{2}$.
With that choice, the last two rows and columns are equal so we only need to check when we can find values for the two remaining open parameters $\beta_1$ and $\beta_2$ such that the upper-left $3 \times 3$ matrix is positive semidefinite.
If we can, the state defined by the $\alpha_i$ has a symmetric extension, otherwise it has no symmetric extension.

\subsection{Using semidefinite programming to solve the simplified problem}
\label{sec:belldiagsymext-sdp}

To solve this simplified problem, we will use techniques from semidefinite programming (see section \ref{sec:prelim:sdp}). 
As it stands, the problem is the following \textindex{feasibility problem} on the standard form  \eqref{eq:standardformobj}-\eqref{eq:standardformpositivity},
\begin{align}
 \textrm{minimize }    \quad & \tr[C X]\\
 \textrm{subject to }  \quad & \tr[A_i X] = b_i, \quad i = 1, \ldots, 4\\
                             & X \geq 0
\end{align}
where 
\begin{align*}
 C &=  \begin{bmatrix} 0 & 0 & 0 \\ 0 & 0 & 0 \\ 0 & 0 & 0 \end{bmatrix},\\
 A_1 &= \begin{bmatrix} 1 & 0 & 0 \\ 0 & -1& 0 \\ 0 & 0 & 0 \end{bmatrix}, &\quad  b_1 &= 4\alpha_1,\\
 A_2 &= \begin{bmatrix} 0 & 0 & 1 \\ 0 & 0 & 0 \\ 1 & 0 & 0 \end{bmatrix}, &\quad  b_2 &= 2\sqrt{2}\alpha_2,\\
 A_3 &= \begin{bmatrix} 0 & 0 & 0 \\ 0 & 0 & 1 \\ 0 & 1 & 0 \end{bmatrix}, &\quad  b_3 &= 2\sqrt{2}\alpha_3,\\
 A_4 &= \begin{bmatrix} 1 & 0 & 0 \\ 0 & 1 & 0 \\ 0 & 0 & 2 \end{bmatrix}, &\quad  b_4 &= 4\alpha_0,
\end{align*}
and $X$ is the upper-left $3 \times 3$ part of $P_+\rho^{ABB'} P_+$ (without the factor $1/8$) and that is the variable to be optimized. 
Since $C = 0$, there is no optimization to be done if a feasible solution is found.
We can get an SDP that involves optimization by removing the last constraint 
and minimizing the corresponding variable instead of fixing it.
By taking $C = \diag(1,1,2)$, we minimize the trace of the density operator.
This gives us an SDP which is similar to the original one:
\begin{align}
 \label{eq:primalobjective}
 \textrm{minimize }    \quad & \tr[C X]\\
 \label{eq:primallinearconstraints}
 \textrm{subject to }  \quad & \tr[A_i X] = b_i, \quad i = 1, \ldots, 3\\
 \label{eq:primalpositivityconstraint}
                             & X \geq 0
\end{align}
where 
\begin{align*}
 C &= \begin{bmatrix} 1 & 0 & 0 \\ 0 & 1 & 0 \\ 0 & 0 & 2 \end{bmatrix},\\
 A_1 &= \begin{bmatrix} 1 & 0 & 0 \\ 0 & -1& 0 \\ 0 & 0 & 0 \end{bmatrix}, &\quad b_1 &= 4\alpha_1,\\
 A_2 &= \begin{bmatrix} 0 & 0 & 1 \\ 0 & 0 & 0 \\ 1 & 0 & 0 \end{bmatrix}, &\quad b_2 &= 2\sqrt{2}\alpha_2,\\
 A_3 &= \begin{bmatrix} 0 & 0 & 0 \\ 0 & 0 & 1 \\ 0 & 1 & 0 \end{bmatrix}, &\quad b_3 &= 2\sqrt{2}\alpha_3.
\end{align*}
A value of the objective function of $4 \alpha_0$ corresponds to a normalized density matrix. 
If the optimum of the objective function is less than that, a multiple of $C$ can be added in order to obtain a suboptimal feasible $X$ that represents a normalized symmetric extension. 
If the optimum is greater than $4 \alpha_0$, a symmetric extension cannot exist, since the symmetric extension would give a feasible $X$ with $\tr[C X] = 4 \alpha_0$.

The advantage of expressing the SDP like this is that it is now always strictly feasible; 
by adding a large enough multiple of $C$ or $\calone$ to $X$ it is always possible to make it satisfy $X > 0$.
This gives us an advantage in that \textindex{strong duality} holds and we can use \textindex{complementary slackness} in analyzing the optima (see section \ref{sec:sdpduality}).
We express the dual of the above SDP on inequality form,
\begin{align}
 \label{eq:dualobjective}
 \textrm{minimize }    \quad & \sum_{i=1}^3 b_i x_i\\
  \label{eq:dualconstraint}
 \textrm{subject to }  \quad & A(x) := C + x_1 A_1 + x_2 A_2 + x_3 A_3 \geq 0
\end{align}
where $x_i$ are the variables that are to be optimized and $A_i$ and $b_i$ are taken from the primal SDP.
The form of $A(x)$ is then 
\begin{equation}
\label{eq:Axform}
 A(x) = \begin{bmatrix}
         1 + x_1 & 0 & x_2 \\ 0 & 1 - x_1 & x_3 \\ x_2 & x_3 & 2
        \end{bmatrix}.
\end{equation}
The state has a symmetric extension if and only if $\sum_{i=1}^3 b_ix_i^* \geq -4\alpha_0$\footnote{
If we had maximized $- \sum_{i=1}^3 b_i x_i$ instead of minimizing $\sum_{i=1}^3 b_i x_i$, the limit would have been the same as for the primal problem. The dual of a minimization problem is a maximization problem, but here the minimization is conceptually simpler, so we express the problem this way.
}. 
If a feasible $x$ is found that gives an objective function below $-4\alpha_0$, 
we know that the minimum will not be greater and therefore the corresponding state cannot have a symmetric extension.

The complementary slackness condition \eqref{eq:complementaryslackness} says that for the optimal $X^*$ of the primal problem and the optimal $x^*$ of the dual problem, we have
\begin{equation}
 \label{eq:slackness}
 A(x^*) X^* = 0.
\end{equation}
The first simplification we get from condition \eqref{eq:slackness} is that 
$\rank[A(x^*)] + \rank(X^*) \linebreak[1]\leq 3$
since $A(x^*)$ and $X^*$ must have support on orthogonal subspaces.
Since both $A(x^*) = 0$ and $X^* = 0$ are excluded by the constraints, 
at least one of $F(x^*)$ and $Z^*$ must have rank one.

In order to solve the semidefinite problem, we proceed as follows.
We first consider $X$ of rank one.
This will give us a sufficient condition for a symmetric extension.
We then consider the case when this condition is not satisfied.
Under the assumption that the state still has a symmetric extension, we use complementary slackness to show that there can only be four possible $X^*$ for any given $\rho^{AB}$. 
If none of these candidates satisfy $X \geq 0$, we get a contradiction and the state 
cannot have a symmetric extension.
It turns out that the candidates all satisfy $\tr[C X] \leq 4\alpha_0$, though, 
so if one of them also is positive semidefinite, it also proves that the state has a symmetric extension.
We end up with the conditions for symmetric extension listed in 
Eqs.~\eqref{eq:rankoneXsufcondsum}--\eqref{eq:rank2Xposconstraint2sum}.

Start by finding the possible values for the objective function when $X$ is rank one.
From the constraints $\tr[A_iX]={b_i}$ of the primal problem and filling in with new open parameters $t_i$, $X$ gets the form 
\begin{equation}
\label{eq:Xform}
X = 
      \begin{bmatrix}
       4 \alpha_1 + t_1 & t_3 & \sqrt{2} \alpha_2\\
       t_3 & t_1 & \sqrt{2} \alpha_3\\
       \sqrt{2} \alpha_2 & \sqrt{2} \alpha_3 & t_2
      \end{bmatrix}.
\end{equation}
The objective function only depends on the diagonal of this matrix, so we want to determine $t_1$ and $t_2$ from the rank-one condition.
Since $X$ is real and symmetric, we can parametrize its eigenvector
with three real numbers $a_i$. This gives
an alternative characterization of $X$, 
\begin{equation}
\label{eq:rankone}
  X = 
\begin{bmatrix}
  a_1^2 & a_1 a_2 & a_1 a_3  \\
  a_1 a_2 & a_2^2 & a_2 a_3  \\
  a_1 a_3 & a_2 a_3 & a_3^2
\end{bmatrix},
\end{equation}
and we can solve the problem by equating these.  
Taking the ratio of the 1,3 and 2,3 elements, we get
$a_1/a_2 = \alpha_2 /
\alpha_3$ when $a_2, a_3, \alpha_3\neq 0$. 
The ratio of the 1,1 and 2,2 elements is the square of this, 
which implies that $t_1=4\alpha_1\alpha_3^2/(\alpha_2^2-\alpha_3^2)$.
Now use the fact that the square of the 2,3 element equals $t_1 t_2$ to find $t_2=(\alpha_2^2-\alpha_3^2)/2\alpha_{1}$. 
The objective function is then 
\begin{equation}
\label{eq:objfunXrankone}
        \tr[C X]=\frac{ (\alpha_2^2 - \alpha_3^2)^2 + 4 \alpha_1^2 (\alpha_2^2 + \alpha_3^2)}{\alpha_1 (\alpha_2^2-\alpha_3^2)},
\end{equation}
and since it is fixed by the state (and the requirement that $X$ be rank one), no minimization is required.
If this expression is less than or equal to $4\alpha_0$, the state has a symmetric extension. 
If the value is greater than $4\alpha_0$, we cannot conclude yet since it could be that $X^*$ is of rank two and has trace less than or equal to $4 \alpha_0$. 
Thus,
\begin{equation}
\label{eq:rankoneXsufcond}
 4 \alpha_0 \alpha_1 (\alpha_2^2-\alpha_3^2) - (\alpha_2^2 - \alpha_3^2)^2 - 4 \alpha_1^2 (\alpha_2^2 + \alpha_3^2) \geq 0
\end{equation}
is a sufficient but not necessary condition for the state to have a symmetric extension.
While this expression will be useful in chapter \ref{ch:rcad}, it is worth noting that it looks nicer when converted back to the parameters defined in Eqs. \eqref{eq:xyzalphafirst}-\eqref{eq:xyzalphalast},
\begin{equation}
 \label{eq:rankoneXsufcondxyz}
 2 \alpha_0 \alpha_x \alpha_y \alpha_z + \alpha_x^2 \alpha_y^2 + \alpha_x^2 \alpha_z^2 + \alpha_y^2 \alpha_z^2 \leq 0.
\end{equation}

If Eq.~\eqref{eq:rankoneXsufcond} is satisfied, we know that the state has a symmetric extension, so from now on we assume that it is not satisfied.
For a contradiction (in some cases), we now assume that the state has a symmetric extension.
This means that $\rank(X^*) = 2$ and because of complementary slackness 
$\rank[A(x^*)] = 1$.
We therefore want to find out for what $x_1$, $x_2$, and $x_3$ has to satisfy for $A(x)$ to be rank one.
Like before, we equate $A(x)$ as given by Eq.~\eqref{eq:Axform} to the generic rank-one matrix in Eq.~\eqref{eq:rankone}.
From the 1,2 element, it is clear that either $a_1$ or $a_2$ must be zero.
This zeroes out the first or second column and row, and we immediately obtain $x_1=\pm 1$ and for the plus and minus cases we get $x_3=0$ and $x_2=0$, respectively. 
This leaves a matrix with a nonzero $2 \times 2$ block, which must have determinant zero. 
From this we get $x_2^2 = 4$ and $x_3^2 = 4$ in the two cases, so $x = (x_1,x_2,x_3)$ can only take one of the four values
$(1, \pm 2, 0), (-1, 0, \pm 2)$.
The corresponding four values of the objective function \eqref{eq:dualobjective} are
\begin{equation}
 4\alpha_1 \pm 4\sqrt{2} \alpha_2 \quad \text{and} \quad -4\alpha_1 \pm 4\sqrt{2} \alpha_3.
\end{equation}
If any of these would be less than $-4\alpha_0$, we would be able to exclude the possibility of a symmetric extension at this point. 
However, this is not possible for any states, 
since each of the four inequalities translate into $p_i \leq 0$ for one of the four eigenvalues $p_I, p_x, p_y$, and $p_z$.

The four possible candidates for $x^*$ cannot by themselves contradict our assumption of a symmetric extension for possible values of $\alpha_i$.
However, under this assumption one of these candidates must be optimal. 
There must, therefore, be a complementary optimal $X^*$ of the primal problem for which
the complementary slackness condition \eqref{eq:slackness} is satisfied.
For each of the four possible $x^*$, we can impose the complementary slackness condition $A(x) X = 0$ to an $X$ of the form \eqref{eq:Xform}, 
and check if the resulting $X$ can be positive semidefinite as required by the SDP conditions. 
For the two vectors $x = (1,\pm 2, 0)$, this gives these values for the open parameters in Eq.~\eqref{eq:Xform},
\begin{subequations}
\begin{align}
t_1 &= \mp \sqrt{2} \alpha_2 - 4 \alpha_1 \\
t_2 &= \mp \sqrt{2} \alpha_2\\
t_3 &= \mp \sqrt{2} \alpha_3
\end{align}
\end{subequations}
which gives the two possible matrices
\begin{equation}
\label{eq:rank2X}
        X=\sqrt{2}
          \begin{pmatrix}
           \mp \alpha_2 & \mp \alpha_3 & \alpha_2\\
           \mp \alpha_3 & \mp \alpha_2 - 2\sqrt{2} \alpha_1 & \alpha_3\\
           \alpha_2     &    \alpha_3  & \mp \alpha_2
          \end{pmatrix}.
\end{equation}
Since the third column is proportional to the first, 
the matrix is positive semidefinite if and only if the upper left $2 \times 2$ block is. 
This is positive semidefinite if and only if both the determinant and the two diagonal elements are nonnegative. 
The determinant is in this case proportional to $\alpha_2^2-\alpha_3^2 \pm 2\sqrt{2}\alpha_1\alpha_2$, 
so the matrix is positive semidefinite if and only if 
$\mp \alpha_2 \geq 0$,
$\mp \alpha_2 -2\sqrt{2}\alpha_1 \geq 0$, 
and 
$\alpha_2^2-\alpha_3^2 \pm 2\sqrt{2}\alpha_1\alpha_2 \geq 0$.
The possible matrices for $x = (-1,0, \pm 2)$ are
\begin{equation}
        X=\sqrt{2}
          \begin{pmatrix}
           \mp \alpha_3 + 2\sqrt{2}\alpha_1 & \mp \alpha_2 & \alpha_2\\
           \mp \alpha_2 & \mp \alpha_3 & \alpha_3\\
           \alpha_2     &    \alpha_3  & \mp \alpha_3
          \end{pmatrix}.
\end{equation}
which are the matrices we get from Eq.~\eqref{eq:rank2X} by interchanging the first and second rows and columns and making the substitutions 
$\alpha_2 \leftrightarrow \alpha_3$, $\alpha_1 \leftrightarrow - \alpha_1$. 
The positivity conditions are 
$\mp \alpha_3 \geq 0$,
$\mp \alpha_3 + 2\sqrt{2}\alpha_1 \geq 0$, 
and $\alpha_3^2-\alpha_2^2 \mp 2\sqrt{2}\alpha_1\alpha_3 \geq 0$. 
Thus, if the state does not satisfy condition \eqref{eq:rankoneXsufcond}, 
and also none of the four positivity constraints,
\begin{subequations}
\label{eq:rank2Xposconstraints}
\begin{align}
\label{eq:rank2Xposconstraint1}
 \alpha_2^2-\alpha_3^2  \pm 2\sqrt{2}\alpha_1 \alpha_2 \geq 0 
 \quad & \text{and} \quad  
 \mp \alpha_2 \geq 0
 \quad & \text{and} \quad  
 \mp \alpha_2 - 2\sqrt{2} \alpha_1 \geq 0, \\
\label{eq:rank2Xposconstraint2}
 \alpha_3^2-\alpha_2^2  \mp 2\sqrt{2}\alpha_1 \alpha_3 \geq 0 
 \quad & \text{and} \quad  
 \mp \alpha_3 \geq 0
 \quad & \text{and} \quad  
 \mp \alpha_3 + 2\sqrt{2} \alpha_1 \geq 0,
\end{align}
\end{subequations}
we cannot have $\rank[A(x^*) = 1]$, so our assumption that the state has a symmetric extension is contradicted.

If on the other hand, one or more of the positivity constraints in Eq.~\eqref{eq:rank2Xposconstraint1} and Eq.~\eqref{eq:rank2Xposconstraint2} are satisfied,
there is no contradiction and the state could have a symmetric extension. 
Actually, we can use the $X$ which satisfies $X \geq 0$ to prove that a symmetric extension exists.
Computing the value of the objective function with $X$ from Eq.~\eqref{eq:rank2X} gives 
$\tr[C X] = -4\alpha_1 \mp 4\sqrt{2} \alpha_2 \leq 4 \alpha_0$, 
where the inequality is simply the two positivity conditions $p_I \geq 0$ and $p_z \geq 0$, for the plus and minus cases, respectively.
For the $X$ corresponding to $x = (-1, 0, \pm 2)$, we can show similarly that the objective function $\tr[C X]$ is smaller than $4 \alpha_0$ by using $p_x \geq 0$ and $p_y \geq 0$.
The $X$ which is positive semidefinite will therefore satisfy all the constraints \eqref{eq:primallinearconstraints} and \eqref{eq:primalpositivityconstraint}, and since it gives a value of the objective function which is less than or equal to $4\alpha_0$, the state must have a symmetric extension.

\begin{figure*}
  \resizebox{\textwidth}{!}{\includegraphics{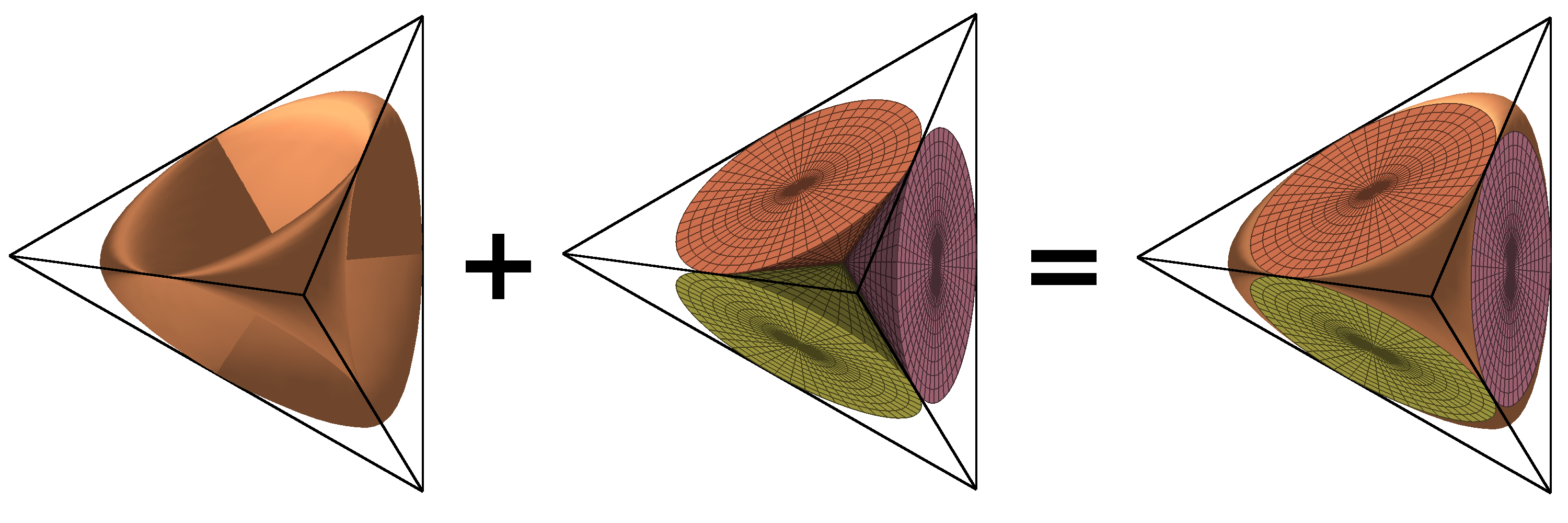}}
\caption[The set of Bell-diagonal states with a symmetric extension.]{\label{fig:extstates}The set of Bell-diagonal states that satisfies the rank-one $X$ condition \eqref{eq:rankoneXsufcondsum} (left), 
rank-two $X$ conditions \eqref{eq:rank2Xposconstraint1sum} or \eqref{eq:rank2Xposconstraint2sum} (center), and the union of the two (right). 
The figures have a maximally entangled state on each vertex and the surfaces have the symmetry of the tetrahedron.}
\end{figure*}

Altogether, we have shown that if any of the conditions 
\eqref{eq:rankoneXsufcond}, \eqref{eq:rank2Xposconstraint1}, or \eqref{eq:rank2Xposconstraint2} 
are satisfied, the state has a symmetric extension, otherwise it does not. 
Since at least one of $\mp \alpha_2 \geq 0$ always holds, 
we can combine the two options in Eq.~\eqref{eq:rank2Xposconstraint1} into one at the cost of adding an absolute value. 
We can do the same for $\mp \alpha_3 \geq 0$ in Eq.~\eqref{eq:rank2Xposconstraint2}
and combining everything we get that a state has symmetric extension if and only if one or more of the following three positivity conditions hold:
\begin{subequations}
\label{eq:conditionsummary}
\begin{gather}
\label{eq:rankoneXsufcondsum}
 4 \alpha_1 (\alpha_2^2-\alpha_3^2) - (\alpha_2^2 - \alpha_3^2)^2 - 4 \alpha_1^2 (\alpha_2^2 + \alpha_3^2) \geq 0, \\
\label{eq:rank2Xposconstraint1sum}
 \alpha_2^2-\alpha_3^2 - 2\sqrt{2}\alpha_1|\alpha_2| \geq 0
  \quad \text {and} \quad
 |\alpha_2| - 2\sqrt{2} \alpha_1 \geq 0, \\
\label{eq:rank2Xposconstraint2sum}
 \alpha_3^2-\alpha_2^2 + 2\sqrt{2}\alpha_1|\alpha_3| \geq 0
 \quad \text {and} \quad
 |\alpha_3| + 2\sqrt{2} \alpha_1 \geq 0.
\end{gather}
\end{subequations}
The set of Bell-diagonal states with symmetric extension is pictured in Fig.~\ref{fig:extstates}.
Condition \eqref{eq:rankoneXsufcondsum} describes a body that includes the 
symmetric extendible states closest to the maximally entangled states. 
It is, however, not convex. 
The conditions \eqref{eq:rank2Xposconstraint1sum} and \eqref{eq:rank2Xposconstraint2sum} 
describe four cones with vertex at the maximally mixed state 
and a maximal circular base on each face of the tetrahedron. 
The cones fill in the convex hull of the first body, 
so that the body of symmetric extendible states is just 
the convex hull of the body from condition \eqref{eq:rankoneXsufcondsum}.

\subsection{Equivalence between SDP solution and conjecture}
\label{sec:belldiagsymext-equivalence}

Instead of having a condition which is a single inequality, the solution of the semidefinite program is three positivity conditions Eqs.~\eqref{eq:rankoneXsufcondsum}-\eqref{eq:rank2Xposconstraint2sum} where the state has a symmetric extension if any one of them holds. 
We can split the conjectured condition \eqref{eq:twoqubitextensionbelldiag} up in a similar way.
If the RHS is not positive, the inequality is always satisfied, since the LHS cannot be negative.
If the RHS is positive, we can square both sides and get rid of the square root. 
This gives the two inequalities
\begin{subequations}
\begin{gather}
 \label{eq:puritydifference}
 \tr[(\rho^{AB})^2] - \frac{1}{2} (\tr[\rho^{AB}])^2 \leq 0\\
 \label{eq:rankoneXsufcondtraceanddet}
 4^2\det(\rho^{AB}) \geq \big\{ \tr[(\rho^{AB})^2] - \frac{1}{2} (\tr[\rho^{AB}])^2 \big\}^2.
\end{gather}
\end{subequations}
If any of the two inequalities are satisfied, the original inequality \eqref{eq:twoqubitextensionbelldiag} is satisfied.
Here, we want to prove that the set of positivity conditions Eqs.~\eqref{eq:rankoneXsufcondsum}-\eqref{eq:rank2Xposconstraint2sum} is equivalent to the set of inequalities Eqs.~\eqref{eq:puritydifference}-\eqref{eq:rankoneXsufcondtraceanddet}.
For the two sets of conditions to be equivalent, \emph{each} of the conditions in one set must imply \emph{at least one} of the conditions in the other.

The two inequalities Eq.~\eqref{eq:rankoneXsufcondsum} and Eq.~\eqref{eq:rankoneXsufcondtraceanddet} are equivalent, so they therefore imply each other. 
The equivalence can be shown by inserting $p_I$, $p_x$, $p_y$, and $p_z$ into both inequalities, but that is a very tedious approach.
We here show how to do it using the parameters $\alpha_0$, $\alpha_x$, $\alpha_y$, and $\alpha_z$ from Eqs.~\eqref{eq:xyzalphafirst}-\eqref{eq:xyzalphalast}.
Multiplying both sides of Eq.~\eqref{eq:rankoneXsufcondtraceanddet} with $4^2$ gives 
\begin{equation}
 \label{eq:rankoneXsufcondtraceanddet44}
 \det(4\rho^{AB}) \geq \big\{ 4\tr[(\rho^{AB})^2] - 2(\tr[\rho^{AB}])^2 \big\}^2.
\end{equation}
The LHS can be computed by writing out the density matrix as a function of $\alpha_0$, $\alpha_x$, $\alpha_y$, and $\alpha_z$,
\begin{align}
 \det(4\rho^{AB}) &= 
  \begin{vmatrix}
   \alpha_0 + \alpha_z & 0 & 0 & \alpha_x - \alpha_y \\
   0 & \alpha_0 - \alpha_z & \alpha_x + \alpha_y & 0 \\
   0 & \alpha_x + \alpha_y & \alpha_0 - \alpha_z & 0 \\
   \alpha_x - \alpha_y & 0 & 0 & \alpha_0 + \alpha_z \\
  \end{vmatrix}\\
  & = 
  \alpha_0^4 + \alpha_x^4 + \alpha_y^4 + \alpha_z^4 
   - 2 \alpha_0^2 (\alpha_x^2 + \alpha_y^2 + \alpha_z^2) \nonumber \\
   & \quad
  - 2 \alpha_x^2 \alpha_y^2 - 2 \alpha_x^2 \alpha_z^2 - 2 \alpha_y^2 \alpha_z^2
  - 8 \alpha_0 \alpha_x \alpha_y \alpha_z.
\end{align}
For the RHS, we can use that $\tr[(\rho^{AB})^2] = \frac{1}{4} (\alpha_0^2 + \alpha_x^2 + \alpha_y^2 + \alpha_z^2)$ and that $\tr[\rho^{AB}] = \alpha_0$ to get 
\begin{align}
 \big\{ 4\tr[(\rho^{AB})^2] - 2(\tr[\rho^{AB}])^2 \big\}^2
 &= \alpha_0^4 + \alpha_x^4 + \alpha_y^4 + \alpha_z^4
  - 2 \alpha_0^2 (\alpha_x^2 + \alpha_y^2 + \alpha_z^2) \nonumber \\
  & \quad
   + 2 \alpha_x^2 \alpha_y^2 + 2 \alpha_x^2 \alpha_z^2 + 2 \alpha_y^2 \alpha_z^2.
\end{align}
Inserting the expressions for the LHS and the RHS into Eq.~\eqref{eq:rankoneXsufcondtraceanddet44}, dividing by 4, and moving everything to one side then gives Eq.~\eqref{eq:rankoneXsufcondxyz} which is equivalent to Eq.~\eqref{eq:rankoneXsufcondsum}.

For the other inequalities, the relations are more involved.
The inequalities \eqref{eq:rank2Xposconstraint1sum} and \eqref{eq:rank2Xposconstraint2sum} are similar in that they describe cones from the maximally mixed state to each face of the state space.
If any one of these inequalities are satisfied, it implies that the inequality \eqref{eq:puritydifference} holds, or in more visual terms, the four cones defined by the inequalities \eqref{eq:rank2Xposconstraint1sum} and
 \eqref{eq:rank2Xposconstraint2sum} are contained inside the ball defined by the inequality \eqref{eq:puritydifference}.
If \eqref{eq:puritydifference} holds, we can only show that the state must have a symmetric extension, and therefore one of the inequalities 
\eqref{eq:rankoneXsufcondsum}-\eqref{eq:rank2Xposconstraint2sum}
must hold, but which one(s) may depend on the exact state.

The inequality \eqref{eq:puritydifference} can be converted to the parameters $\alpha_0, \alpha_1, \alpha_2$, and $\alpha_3$,
\begin{equation}
\label{eq:twoqubitextensionbellpurityalpha}
\alpha_1^2 + \alpha_2^2 + \alpha_3^2 \leq \alpha_0^2,
\end{equation}
where it is evident that it describes a ball with radius $\alpha_0$.
For the inequalities \eqref{eq:rank2Xposconstraint1sum} and \eqref{eq:rank2Xposconstraint2sum}, we can split up the absolute value to go back to the form given in Eqs.~\eqref{eq:rank2Xposconstraint1} and \eqref{eq:rank2Xposconstraint2},
\begin{align*}
 \alpha_2^2-\alpha_3^2  \pm 2\sqrt{2}\alpha_1 \alpha_2 \geq 0 
 &\quad &\text{and} &\quad  
 \mp \alpha_2 \geq 0
 &\quad &\text{and} &\quad  
 \mp \alpha_2 - 2\sqrt{2} \alpha_1 \geq 0, \\
 \alpha_3^2-\alpha_2^2  \mp 2\sqrt{2}\alpha_1 \alpha_3 \geq 0 
 &\quad &\text{and} &\quad  
 \mp \alpha_3 \geq 0
 &\quad &\text{and} &\quad  
 \mp \alpha_3 + 2\sqrt{2} \alpha_1 \geq 0.
\end{align*}
For each of the four cases, we can apply an orthogonal coordinate transformation that takes it to a simpler form.
The transformations for \eqref{eq:rank2Xposconstraint1} are 
\begin{align*}
\alpha_1 &= \sqrt{2/3} x - \sqrt{1/3} y, \\
\alpha_2 &= \mp \sqrt{1/3} x \mp \sqrt{2/3} y.\\
\alpha_3 &= z,
\end{align*}
 The transformations for \eqref{eq:rank2Xposconstraint2} are obtained by interchanging $\alpha_2$ with $\alpha_3$ and $\alpha_1$ with $-\alpha_1$. 
All four positivity conditions then become simply 
\begin{equation}
\label{eq:rank2Xposconstraintsxyz}
 x^2 + z^2 \leq 2y^2 \quad \text{and} \quad -x \leq \sqrt{2} y \quad \text{and} \quad x \leq \sqrt{2}y,
\end{equation}
where the first inequality describes a double cone with a circular base and the other two cut away one part such that only the cone where $y \geq 0$ remains.
The purity condition \eqref{eq:twoqubitextensionbellpurityalpha} that we want to prove becomes 
$x^2 + y^2 + z^2 \leq \alpha_0$
for all the four coordinate transformations.
For each transformation, one of the positivity conditions for the eigenvalues $p_i$ translates into 
$y \leq \alpha_0/\sqrt{3}$, and since we have only nonnegative $y$ we also have $y^2 \leq \alpha_0^2/3$.
When we combine this with Eq.~\eqref{eq:rank2Xposconstraintsxyz}
we get $x^2 + y^2 + z^2 \leq 3 y^2 \leq \alpha_0$, which proves that Eq.~\eqref{eq:puritydifference} holds if any of Eq.~\eqref{eq:rank2Xposconstraint1sum} and Eq.~\eqref{eq:rank2Xposconstraint2sum} holds.

The last implication we need to show is that any state that satisfies
\eqref{eq:puritydifference},  $\tr(\rho_{AB}^2) \leq 1/2$, 
or equivalently \eqref{eq:twoqubitextensionbellpurityalpha}, 
also satisfies at least one of \eqref{eq:rankoneXsufcondsum}--\eqref{eq:rank2Xposconstraint2sum}.
For this we use the proven fact that these inequalities are necessary and sufficient conditions for the state to have a symmetric extension, and therefore the set must be convex. 
Any Bell-diagonal state that satisfies inequality \eqref{eq:puritydifference} can be written as a convex combination of states that saturates it. 
One way to achieve this for a normalized state is to choose the four states $\rho_I := (1-q_I)\proj{\Phi^+} + q_I \rho_{AB}$, $\rho_X := (1-q_X)\proj{\Psi^+} + q_X \rho_{AB}$, etc.~with the choice of $q_j$ that gives $\tr[\rho_{j}^2] = 1/2$, 
and write $\rho_{AB}$ as a convex combination of these.
Since the determinant of a state always is nonnegative, 
all these extremal states satisfy \eqref{eq:rankoneXsufcondtraceanddet}, 
and therefore also \eqref{eq:rankoneXsufcondsum}, 
so they must have a symmetric extension.
Convex combinations of states with symmetric extension also have symmetric extension, 
so any state with $\tr(\rho_{AB}^2) \leq 1/2$ has symmetric extension and therefore satisfies one of 
Eqs.~\eqref{eq:rankoneXsufcondsum}--\eqref{eq:rank2Xposconstraint2sum}.

\section{ZZ-invariant states}

Now, we consider states that are invariant under $\sigma_z \otimes \sigma_z$. 
This class includes the Bell-diagonal states, which are those states that in addition are invariant under $\sigma_x \otimes \sigma_x$.
The states invariant under $\sigma_z \otimes \sigma_z$ are of the form
\begin{equation}
\label{eq:Zcorrform}
  \rho^{AB} = 
\begin{bmatrix}
p_1 & 0 & 0 & x \\
0 & p_2 & y & 0 \\ 
0 & y & p_3 & 0 \\ 
x & 0 & 0 & p_4 
\end{bmatrix}
\end{equation}
in the product basis $\ket{00}, \ket{01}, \ket{10}, \ket{11}$.
The Bell-diagonal states is the special case 
where $p_1 = p_4$ and $p_2 = p_3$.
In this section, we will show that the conjectured condition \eqref{eq:twoqubitextension} is necessary and sufficient in another special case of this class, namely when $y = 0$.
This special class is the set of states that are invariant under $S^\dagger \otimes S$, where the qubit unitary operator $S := \proj{0} + \im \proj{1}$ is often called the \textindex{phase gate}.
Invariance under $S^\dagger \otimes S$ implies invariance under $\sigma_z \otimes \sigma_z$, 
since $S^2 = (S^\dagger)^2 = \sigma_z$.

Let us first, however, simplify the problem for the whole class of states invariant under $\sigma_z \otimes \sigma_z$.
Without loss of generality, we can assume that $p_1 \geq p_2, p_3, p_4$ and that $x$ and $y$ are both real and nonnegative, since this can be accomplished by changing the local basis.
The following lemma gives a necessary and sufficient condition for a state of the form \eqref{eq:Zcorrform} to have a symmetric extension.

\begin{lemma}
\label{lem:Zcorrnecsuf}
 A state of the form \eqref{eq:Zcorrform} has symmetric extension if and only if
there exist $s \in [0,p_2]$ and $t \in [0,\min(p_3,p_4)]$ such that 
\begin{subequations}
\begin{align}
  \label{eq:xcondst}
  x &\leq \sqrt{s}\sqrt{p_1-t} + \sqrt{t}\sqrt{p_4-s},\\
  \label{eq:ycondst}
  y &\leq \sqrt{s}\sqrt{p_2-t} + \sqrt{t}\sqrt{p_3-s}.
\end{align}
\end{subequations}
\end{lemma}
\begin{proof} 
 For the ``if'' part, we give an explicit symmetric extension of the state for the case when the inequalities are saturated. 
The extended state is then the rank-2 state 
$\rho_{ABB'} = p \proj{\psi_1} + (1-p) \proj{\psi_2}$ where
\begin{equation}
\label{eq:stextension}
\begin{aligned}
\sqrt{p}\ket{\psi_1} &= \sqrt{p_1-t}\ket{000} + \sqrt{p_2-t}\ket{011} \\
                     & \quad + \sqrt{s}\ket{101} + \sqrt{s}\ket{110},\\
\sqrt{1-p}\ket{\psi_2} &= \sqrt{t}\ket{001} + \sqrt{t}\ket{010}  \\
                     & \quad + \sqrt{p_3-s}\ket{100} + \sqrt{p_4-s}\ket{111}.
\end{aligned}
\end{equation}
If a state has symmetric extension for a given $x$ and $y$, then also states with smaller $x$ or $y$ have symmetric extension.
This is because local unitary operators can change the sign of either $x$ or $y$,  
$S \otimes S$ will change the sign of $x$ while $S^\dagger \otimes S$ does the same for $y$.
The resulting states will also have a symmetric extension.
Mixing the original state with one of these states will reduce either $x$ or $y$ of the original state, and convex combinations of extendible states also have a symmetric extension. 
Hence, we can have inequality instead of equality in Eq.~\eqref{eq:xcondst} and Eq.~\eqref{eq:ycondst}.

For the ``only if'' part, a generic symmetric operator on $ABB'$ that reduces to 
the form \eqref{eq:Zcorrform} when $B'$ is traced out has the form 
\begin{equation}
 \begin{bmatrix}
  p_1 - t & \times & \times & \times & \times & k_1 & k_1 & \times \\
  \times & t      & \times & \times & l_1     & \times & \times & k_2\\
  \times & \times & t      & \times & l_1     & \times & \times & k_2\\
  \times & \times & \times & p_2 - t & \times & l_2    & l_2    & \times\\
  \times & l_1^*  & l_1^*  & \times & p_3 - s & \times & \times & \times\\
  k_1^*  & \times & \times & l_2^*  & \times  & s      & \times & \times\\
  k_1^*  & \times & \times & l_2^*  & \times  & \times & s       & \times\\
  \times & k_2^*  & k_2^*  & \times & \times  & \times & \times & p_4-s\\
 \end{bmatrix}.
\end{equation}
Here, $k_1 + k_2 = x$ and $l_1 + l_2 = y$. 
For this to be positive semidefinite, all subdeterminants must be positive. 
From positivity of the subdeterminants 
\begin{equation*}
 \begin{vmatrix}
  p_1 -t & k_1 \\
  k_1^*  & s
 \end{vmatrix}
 \text{ and }
 \begin{vmatrix}
  t      & k_2 \\
  k_2^*  & p_4 -s
 \end{vmatrix},
\end{equation*}
we get that $x = k_1 + k_2 \leq |k_1| + |k_2| \leq \sqrt{s}\sqrt{p_1-t} + \sqrt{t}\sqrt{p_4-s}$. 
From the subdeterminants involving $l_1$ and $l_2$ we get
$y = l_1 + l_2 \leq |l_1| + |l_2| \leq \sqrt{t}\sqrt{p_3-s} + \sqrt{s}\sqrt{p_2-t}$.
\end{proof}

Since $p_1 \geq p_2$ the possible values for $t$ in \eqref{eq:xcondst} and \eqref{eq:ycondst} are between 0 and $p_2$.
The parameter $s$, however, is bounded from above by both $p_3$ and $p_4$. 
Before we go to the special case $y = 0$ we treat the case $p_3 \geq p_4$ separately, 
since knowing which of the two bounds applies will simplify the analysis.
When $p_3 \geq p_4$, the state has a symmetric extension for any $x$ and $y$, 
since even the rank-2 state we get by taking the maximum 
$x = \sqrt{p_1 p_4}$ and $y = \sqrt{p_2 p_3}$ has symmetric extension 
by theorem \ref{thm:rank2}. 
It is also easy to verify that in this case $\tr[(\rho^{B})^2] \geq \tr[(\rho^{AB})^2]$, 
so the condition \eqref{eq:twoqubitextension} is always satisfied.

We now prove conjecture \ref{conj:twoqubitextension} for the special case $y=0$ (i.e. $S^\dagger \otimes S$ invariant states).

\begin{theorem}
 A state $\rho^{AB}$ of the form \eqref{eq:Zcorrform} with $y=0$ has a symmetric extension if and only if 
\begin{equation}
  \label{eq:twoqubitextensionZZ}
  \tr[(\rho^B)^2] \geq \tr[(\rho^{AB})^2] - 4\sqrt{\det(\rho^{AB})}.
\end{equation}
\end{theorem}
\begin{proof}
Since $y=0$, Eq.~\eqref{eq:ycondst} is satisfied for any $s \in [0,p_3]$ and $t \in [0,p_2]$.
Our only objective is therefore to maximize the right hand side of \eqref{eq:xcondst},  
\begin{equation}
f(s,t) := \sqrt{s}\sqrt{p_1-t} + \sqrt{t}\sqrt{p_4-s},
\end{equation} 
on this domain, 
since the maximum will be the maximum $x$ that allows for a symmetric extension.
Without the constraints on $s$ and $t$, $f(s,t)$ reaches its maximum value of $\sqrt{p_1 p_4}$ for any value of $(s,t)$ that satisfies
$p_1 s + p_4 t = p_1 p_4$.
Since $s \leq p_3$ and $t \leq p_2$ this maximum value may or may not be obtainable.
The maximum value of $p_1 s + p_4 t$ is $p_1 p_3 + p_2 p_4$, 
so if $p_1 p_3 + p_2 p_4 \geq p_1 p_4$, then $x = \sqrt{p_1 p_4}$ can be obtained by choosing $s = p_3$ and $t = (p_1 p_4 - p_1 p_3)/p_4 \leq p_2$.
When $p_1 p_3 + p_2 p_4 < p_1 p_4$, however, we will have 
$f(s,t) < \sqrt{p_1 p_4}$
for all possible $(s,t)$.
In this case, the optimal choice of $(s,t)$ is $(p_3, p_2)$, 
since in the region where
$p_1 p_3 + p_2 p_4 < p_1 p_4$, 
the function $f(s,t)$ increases both when $s$ and $t$ increases.
The maximum value for $x$ is then $\sqrt{p_3}\sqrt{p_1-p_2} + \sqrt{p_2}\sqrt{p_4-p_3}$.
Summing up, a state of the form \eqref{eq:Zcorrform} with $y = 0$ 
has a symmetric extension if and only if
\begin{equation}
\label{eq:Zcorrboundwithyzeroapp}
 x \leq \left\{ 
\begin{aligned}
 &\sqrt{p_1 p_4} 
    \qquad \text{for} \quad p_1 p_3 + p_2 p_4 \geq p_1 p_4 \\
 &\sqrt{p_3}\sqrt{p_1-p_2} + \sqrt{p_2}\sqrt{p_4-p_3}  
    \quad \text{otherwise.}
\end{aligned}
\right.
\end{equation}

The remaining part is to show that the condition \eqref{eq:twoqubitextensionZZ} is equivalent to this. 
The condition is equivalent to at least one of the following two inequalities holding
\begin{subequations}
\begin{gather}
  \label{eq:twoqubitextensionsplit1}
  \tr[(\rho^{AB})^2] - \tr[(\rho^{B})^2] \leq 0 \\
  \label{eq:twoqubitextensionsplit2}
  4 \sqrt{\det(\rho^{AB})} \geq |\tr[(\rho^{AB})^2] - \tr[(\rho^{B})^2]|.
\end{gather}
\end{subequations}
Since $y=0$, we get $\det(\rho^{AB}) = p_2 p_3(p_1 p_4 - x^2)$
and $\tr[(\rho^{AB})^2] - \tr[(\rho^B)^2] = 2(x^2 - p_1 p_3 - p_2 p_4)$.
Inserting this into \eqref{eq:twoqubitextensionsplit1} and \eqref{eq:twoqubitextensionsplit2} and solving for $x$ gives
\begin{subequations}
\begin{align}
  \label{eq:twoqubitextensionsplit1x}
  x &\leq \sqrt{p_1 p_3 + p_2 p_4} \\
  \label{eq:twoqubitextensionsplit2x}
  x &\leq \sqrt{p_3}\sqrt{p_1-p_2} + \sqrt{p_2}\sqrt{p_4-p_3}.
\end{align}
\end{subequations}
Only one of these inequalities have to be satisfied for a state with symmetric extension, so the upper bound on $x$ is the maximum of the two.
By comparing the two bounds, we find in which region each of the two is valid, and get
\begin{equation}
 x \leq \left\{ 
\begin{aligned}
 &\sqrt{p_1 p_3 + p_2 p_4} 
    \qquad \text{for} \quad p_1 p_3 + p_2 p_4 \geq p_1 p_4 \\
 &\sqrt{p_3}\sqrt{p_1-p_2} + \sqrt{p_2}\sqrt{p_4-p_3}  
    \quad \text{otherwise.}
\end{aligned}
\right.
\end{equation}
The only region where $\sqrt{p_1 p_3 + p_2 p_4}$ is the valid upper bound is when it is greater than $\sqrt{p_1 p_4}$. 
Since $x$ never can exceed $\sqrt{p_1 p_4}$ for any state, this is the same as \eqref{eq:Zcorrboundwithyzeroapp}. 
\end{proof}

Any two-qubit state with three degenerate eigenvalues will be of this class. 
In this case, $\ket{00}$ and $\ket{11}$ can be taken as the Schmidt basis vectors 
of the nondegenerate eigenvector. 
We can then write the state as 
$(\lambda_1 - \lambda) \proj{\psi} + \lambda/4 \calone$, where $\lambda_1$ is the nondegenerate eigenvalue, $\lambda$ the degenerate eigenvalue, and $\ket{\psi}$ the nondegenerate eigenvector.
Since $\calone$ is diagonal and $\proj{\psi}$ only has an off-diagonal entry in the $x$ position, the state is of the form \eqref{eq:Zcorrform} with $y=0$.

\chapter{Bosonic and fermionic extensions}
\label{ch:bosonicfermionic}

In this thesis we use the term ``symmetric extension'' for extensions that are invariant under exchange of two systems, without considering if its support is on the symmetric or antisymmetric subspace or both. 
An extension that resides only on the symmetric subspace of 
$\mathcal{H}_B \otimes \mathcal{H}_{B'}$
we call a \emphindex{bosonic extension}, while one that resides on the antisymmetric subspace is 
a \emphindex{fermionic extension}.
Generic symmetric extensions are mixtures of bosonic and fermionic extensions.
Bosonic ($+$) and fermionic ($-$) extensions satisfy 
$\pi_\pm \rho^{ABB^\prime} \pi_\pm^\dagger$ for $\pi_\pm := \frac{1}{2}(\calone \pm \swap)$ the projector onto the symmetric/antisymmetric subspace,
and 
$\swap \rho^{ABB'} = \pm \rho^{ABB'}$. 
Here we show that when the subsystem to be extended is a qubit, 
the states with symmetric and bosonic extension coincides, and that this is not true in general.

\begin{proposition}
  If a quantum state $\rho^{AB}$ of dimension $N \times 2$ has a symmetric extension to $\rho^{ABB^\prime}$ it also has a bosonic extension $\sigma^{ABB^\prime}$, i.e.~that satisfies also 
\begin{equation}
 \sigma^{ABB^\prime} = \frac{1}{2}(\calone + \swap) \sigma^{ABB^\prime} \frac{1}{2}(\calone + \swap)
\end{equation}
\end{proposition}
\begin{proof} 
 Decompose the extended state $\rho^{ABB^\prime}$ with the spectral decomposition as in lemma \ref{lem:symspecdecomp},
\begin{equation}
 \rho^{ABB^\prime} = \sum_j \lambda_j^+ \proj{\phi_j^+} + \sum_k \lambda_k^- \proj{\phi_k^-}
\end{equation}
where $\ket{\phi_j^+} = \swap \ket{\phi_j^+}$ and $\ket{\phi_k^-} = - \swap \ket{\phi_k^-}$ are symmetric and antisymmetric, respectively. 
The vectors are of the form 
$\ket{\phi_j^\pm} = \sum_k \alpha_{jk} \ket{\psi_{jk}}_A \ket{\psi_{k}^\pm}_{BB^\prime}$ where $\ket{\psi_k^\pm}_{BB^\prime}$ are in the symmetric and antisymmetric subspace of $\mathcal{H}_B \otimes \mathcal{H}_{B^\prime}$. 
When $B$ and $B^\prime$ are qubits, the antisymmetric subspace is one-dimensional and is spanned by the vector $\ket{\Psi^-} = 1/\sqrt{2}(\ket{01} - \ket{10})$. 
The antisymmetric vectors are therefore of the product form 
$\ket{\phi_k^-} = \ket{\psi_k}_A \ket{\Psi^-}_{BB^\prime}$. 
Replacing them with symmetric vectors of the form 
$\ket{\xi_k^+} = \ket{\psi_k}_A \ket{\Psi^+}_{BB^\prime}$ 
where $\ket{\Psi^+} = 1/\sqrt{2}(\ket{01} + \ket{10})$ yields a state 
\begin{equation}
 \sigma^{ABB^\prime} = \sum_j \lambda_j^+ \proj{\phi_j^+} + \sum_k \lambda_k^- \proj{\xi_k^+}
\end{equation}
which has support on the symmetric subspace. 
Note that $\lambda_j^+$ and $\lambda_k^-$ are no longer eigenvalues of this state. 
But since the reduced states of $\ket{\xi_k^+}$ are the same as for $\ket{\phi_k^-}$, we have that $\rho^{AB} := \tr_{B^\prime} [\rho^{ABB^\prime}] = \tr_{B^\prime} [\sigma^{ABB^\prime}]$, so $\sigma^{ABB^\prime}$ is a valid bosonic extension of $\rho^{AB}$. 
\end{proof}

To show that this is an effect of the low dimension of the $B$ system, we give an example of a state of two qutrits that has a fermionic but not a bosonic extension.

\begin{example}
  Consider a tripartite pure state on $ABB^\prime$ of the form
\begin{equation}
\ket{\psi} = \alpha (\ket{012} - \ket{021}) + \beta (\ket{120} - \ket{102}) + \gamma (\ket{201} - \ket{210})
\end{equation}
where $\alpha, \beta, \gamma \ne 0$ are all different. 
This is a fermionic extension of the reduced state $\rho^{AB} = \tr_{B^\prime} [\proj{\psi}]$. 
If $\rho^{AB}$ had a bosonic extension, a trace preserving and completely positive (TPCP) map on a purifying system (here $B'$) would be able to convert any purification of $\rho^{AB}$ into this bosonic extension. 
If the TPCP map is given by its \textindex{Kraus operators} ${K_j}$ which satisfy $\sum_j K_j^\dagger K_j = \calone_{B'}$, the output state when applied to $\ket{\psi}$ would be
\begin{equation}
 \sigma^{ABB'} = \sum_j (\calone_A \otimes \calone_B \otimes K_j) \proj{\psi} (\calone_A \otimes \calone_B \otimes K_j)^\dagger .
\end{equation}
If $\sigma^{ABB'}$ is a bosonic extension, all the terms in this sum must be on the symmetric subspace.
Consider one of the Kraus operators, $K$.
Applying it to $\ket{\psi}$ gives 
\begin{equation}
 (\calone_A \otimes \calone_B \otimes K) \ket{\psi} = \alpha  \ket{0}_A \ket{\psi_0}_{BB'}
                                        + \beta   \ket{1}_A \ket{\psi_1}_{BB'}
                                        + \gamma  \ket{2}_A \ket{\psi_2}_{BB'},
\end{equation}
where 
$\ket{\psi_0} = \ket{1} \otimes K \ket{2} - \ket{2} \otimes K \ket{1}$, 
$\ket{\psi_1} = \ket{2} \otimes K \ket{0} - \ket{0} \otimes K \ket{2}$, and 
$\ket{\psi_2} = \ket{0} \otimes K \ket{1} - \ket{1} \otimes K \ket{0}$.
Each of the $\ket{\psi_j}$ needs to be on the symmetric subspace of 
$\mathcal{H}_B \otimes \mathcal{H}_{B'}$.
Expressing $K$ as $\sum_{jk} k_{jk}\ketbra{j}{k}$ and
imposing $\swap \ket{\psi_1} = \ket{\psi_1}$ gives us that 
$k_{01} = k_{02} = 0$ and $k_{22} = -k_{11}$.
Doing the same with the other vectors we get that $k_{jk} = 0$ for any $j \ne k$, 
$k_{00} = -k_{22}$ and $k_{11} = -k_{00}$. 
The only possible solution to this is that $K$ vanishes, 
so no nonzero $K$ applied on $B'$ can give a vector which is on the symmetric subspace.
Hence, the state $\rho^{AB}$ cannot have a bosonic extension.
\end{example}

This means that there are states $\rho^{AB}$ with a symmetric extension that cannot be extended 
to a pure state on four systems $\ket{\psi}_{ABB^\prime R}$ in such a way that $\ket{\psi}_{ABB^\prime R} = \pm \swap \ket{\psi}_{ABB^\prime R}$. 
This condition means that the extension is bosonic ($+$) or fermionic ($-$), 
but some states with symmetric extension admit neither. 
One example is if $\rho^{AB}$ does not admit a fermionic extension and 
$\sigma^{AB}$ does not admit a bosonic extension.
Then the state $(\proj{0}_{A'} \otimes \rho^{AB} + \proj{1}_{A'} \otimes \sigma^{AB})/2$
cannot admit a bosonic nor a fermionic extensions.

The above proposition and example prove that while for dimension $N \times 2$, a symmetric extension is equivalent to a bosonic extension, the equivalence is broken already at dimension $3 \times 3$. 
This leaves the question open for dimension $2 \times N$. 
Preliminary tests suggest that we can convert symmetric extensions to bosonic extensions for dimension $2 \times 4$, but not for $2 \times 5$. 
We therefore conjecture that symmetric extensions and bosonic extensions are equivalent up to dimension $2 \times 4$, but not higher.

\chapter{Degradable and antidegradable channels}
\label{ch:degradable}

So far we have been interested in quantum states and whether they have a symmetric extension. 
We make the connection to \emph{degradable}\index{degradable channel} \cite{devetak05b} and \emph{antidegradable}\index{antidegradable channel} \cite{wolf07a} quantum channels which are related concepts in quantum channel theory.
If a channel is degradable or antidegradable this greatly simplifies the evaluation of the quantum capacity of the channel.

A quantum channel can be represented by a unitary operator acting jointly on the system and the environment---where the environment starts out in a pure state---followed by tracing out the environment. Given a channel $\mathcal{N}: \mathcal{N}(\rho) = \tr_E (U(\rho \otimes \proj{0}_E)U^\dagger)$, the \emphindex{complementary channel} is the channel to the environment, where the system is traced out, $\mathcal{N}^C (\rho) = \tr_S (U(\rho \otimes \proj{0}_E)U^\dagger)$.
The complementary channel is only defined up to a unitary on the output system, and the channel itself
is a complementary channel of its complementary channel.
A channel $\mathcal{N}$ is called degradable if there exists another channel $\mathcal{D}$, that will degrade the channel to the complementary channel when applied on the output, $\mathcal{N}^C = \mathcal{D} \circ \mathcal{N}$. Similarly, the channel is called antidegradable if the complementary channel is degradable, $\mathcal{N} = \mathcal{D} \circ \mathcal{N}^C$ for some channel $\mathcal{D}$.

Using the Choi-Jamio{\l}kowski isomorphism \cite{choi75a,jamiolkowski72a}, we can represent any channel by the bipartite quantum state 
resulting from the channel acting on
one half of a maximally entangled state. 
We use the convention where Alice prepares a maximally entangled state and sends the second subsystem to Bob through the channel, a procedure that leaves the first subsystem maximally mixed \footnote{A more common convention is to let the channel act on the first subsystem of the operator $M = \sum_{j,k=0}^{d-1} \ket{jj}\bra{kk}$ which is a maximally entangled state, normalized such that $\tr_A(M) = \calone_B$.},
\begin{equation}
 \rho_\mathcal{N} = \frac{1}{d} \sum_{j,k=0}^{d-1} \ket{i}\bra{j} \otimes \mathcal{N}(\ket{i}\bra{j}).
\end{equation}

Like in the rest of this thesis, we always consider symmetric extensions to two copies of the second subsystem, which in the Choi-Jamio{\l}kowski representation represents the output system.
\begin{lemma}
\label{lem:symextantidegradable}
 A channel $\mathcal{N}$ is antidegradable if and only if its Choi-Jamio{\l}kowski representation $\rho_\mathcal{N}$ has a symmetric extension.
\end{lemma}
\begin{proof} 
 Let the channel $\mathcal{N}$ be antidegradable, and let $\mathcal{D}$ be the channel that degrades the complementary channel, $\mathcal{N} = \mathcal{D} \circ \mathcal{N}^C$.
Applying $\mathcal{N}$ on the second half of a maximally entangled state and applying $\mathcal{D}$ to the environment produces a tripartite state $\rho^{ABE}$ where the reduced states satisfy $\rho^{AB} = \rho^{AE} = \rho_\mathcal{N}$, but it does not need to be invariant under $\swapBE$. 
The state $(\rho^{ABE} + \swapBE \rho^{ABE} \swapBE^\dagger)/2$ has the same reduced states and is also invariant under exchange of $B$ and $E$. 
It is therefore a symmetric extension of $\rho_\mathcal{N} = \rho^{AB}$.

Conversely, let the Choi-Jamio{\l}kowski representation $\rho_\mathcal{N}$ have a symmetric extension $\rho^{ABB^\prime}$. 
This satisfies $\rho^{AB} = \rho^{AB^\prime} = \rho_\mathcal{N}$ and has a purification $\ket{\psi}_{ABB^{\prime}R}$. 
The Choi-Jamio{\l}kowski representation of the complementary channel is then
$\rho^{AB'R}$ where $B'R$ is the output system.
Clearly, a degrading channel is then
$\mathcal{D}(\rho^{B^\prime R}) = \tr_R(\rho^{B^\prime R})$.
\end{proof}

This means that all necessary or sufficient conditions derived for symmetric extension are also necessary or sufficient conditions for the Choi-Jamio{\l}kowski representation of an antidegradable channel. In particular, if conjecture \ref{conj:twoqubitextension} is true, it will also characterize the antidegradable qubit channels.
 
By interchanging the roles of the output and the environment, we can reduce to problem of deciding whether a channel is degradable to deciding whether the Choi-Jamio{\l}kowski representation of the complementary channel has a symmetric extension. 
A channel $\mathcal{N}$ with $d_A$-dimensional input, $d_B$-dimensional output, and environment dimension of $d_E$ is degradable if and only if $\rho_{\mathcal{N}^C}$ of dimension $d_A \times d_E$ and rank $d_B$ has symmetric extension.
Wolf and P{\'e}rez-Garc{\'i}a \cite{wolf07a} found that when $d_E = 2$, a qubit channel is either degradable, antidegradable or both. 
This also follows from our theorem \ref{thm:rank2} about symmetric extension of rank-2 two-qubit states. 
For qubit channels with larger environment there are examples of channels that are neither, even close to the identity channel \cite{smith07a}. 
Using the following theorem, we can show that no qubit channels with $d_E \geq 2$ can be degradable.

\begin{theorem}
\label{thm:limitedBrank}
 Any bipartite state $\rho^{AB}$ of rank 2 with a symmetric extension has a reduced state that satisfies $\rank(\rho^B) \leq 2$.
\end{theorem}
\begin{proof} 
 By corollary \ref{cor:pure-extendible-decomp}, $\rho^{AB}$ can be decomposed into pure-extendible states
\begin{equation}
 \rho^{AB} = \sum_j p_j \rho_j^{AB},
\end{equation}
where the $\rho_j^{AB}$ all satisfy the spectrum condition \eqref{eq:spectrumcondition}. 
Since $\rho^{AB}$ is of rank 2, 
$\rank(\rho_j^{AB}) \leq 2$ for all $j$. 

If $\max_j \rank(\rho_j^{AB}) = 1$, all the pure-extendible states are pure product states $\rho_j^{AB} = \proj{\psi_j \otimes \phi_j}$ by condition \eqref{eq:spectrumcondition}. 
Because the rank of $\rho^{AB}$ is 2, there can only be two independent product vectors, say $\ket{\phi_1 \otimes \psi_1}$ and $\ket{\phi_2 \otimes \psi_2}$, so the support of $\rho^B$ is spanned by $\psi_1$ and $\psi_2$ and therefore at most two-dimensional.

If there is at least one $j$ such that $\rank(\rho_j^{AB}) = 2$, this defines a 2-dimensional subspace where all other $\rho_j^{AB}$ must have their support. 
Let $\rho_1^{AB}$ be one of the $\rho_j^{AB}$ with rank 2. 
Let the spectral decomposition for it and its reduction to $B$ be 
$\rho_1^{AB} = \gamma \proj{\phi_0} + (1-\gamma) \proj{\phi_1}$ and 
$\rho_1^B = \gamma \proj{0} + (1 - \gamma) \proj{1}$, respectively, 
in accordance with the spectrum condition \eqref{eq:spectrumcondition}. 
The eigenvectors of $\rho_1^{AB}$ can be decomposed as 
$\ket{\phi_k} = \ket{\widetilde{\psi}_{k0}}_A \ket{0}_B + \ket{\widetilde{\psi}_{k1}}_A \ket{1}_B$, 
where maximum one of the four unnormalized $\ket{\widetilde{\psi}_{kl}}_A$ can be the zero vector. 
Since all the other $\rho_j^{AB}$ have to have support within $\mathspan\{\ket{\phi_1},\ket{\phi_2}\}$, 
they can only ever have reduced states $\rho_j^B$ that are supported on $\mathspan\{\ket{0},\ket{1} \}$. 
Therefore, also $\rho^B$ is supported on $\mathspan\{\ket{0},\ket{1} \}$ and has $\rank(\rho^{B}) \leq 2$.
\end{proof}

This reduces the $N \times M$ symmetric extension problem for states of rank 2 to $N \times 2$. 
From remark \ref{remark:rank2necessary}, we already have a necessary condition for this case, 
namely that $\lambda_\text{max}(\rho^{B}) \geq \lambda_\text{max}(\rho^{AB})$. 
This also generalizes theorem \ref{thm:rank2} to give necessary and sufficient conditions for symmetric extension of a $2 \times N$ state of rank 2.
Such a state has symmetric extension if and only if  $\lambda_\text{max}(\rho^{B}) \geq \lambda_\text{max}(\rho^{AB})$ and $\rank(\rho^{B}) \leq 2$.

From the connection between symmetric extension and antidegradable channels in lemma \ref{lem:symextantidegradable}, the following corollary automatically follows.

\begin{corollary}
 Any antidegradable channel $\mathcal{N}$ with qubit environment has output of rank 2. 
 If $\rho_\mathcal{N}$ is the Choi-Jamio{\l}kowski state representing the channel, $\lambda_\text{max}(\rho_\mathcal{N}) \leq \lambda_\text{max}(\tr_A [\rho_\mathcal{N}])$. 
\end{corollary}
Exchanging the output and the environment changes antidegradability into degradability:
\begin{corollary}
\label{cor:degradableenvlimit}
 Any degradable channel with qubit output has $d_E \leq 2$. 
 If $\rho_\mathcal{N}$ is the Choi-Jamio{\l}kowski state representing the channel, $\lambda_\text{max}(\tr_A [\rho_\mathcal{N}]) \leq \lambda_\text{max}(\rho_\mathcal{N})$.
\end{corollary}

This result has also been independently obtained by Cubitt \textit{et~al.}~\cite{cubitt08a} by other methods.
One could imagine that theorem \ref{thm:limitedBrank} would generalize to higher rank
so that the rank of the $\rho^B$ system always would be bounded by the rank of $\rho^{AB}$ for symmetric extendible states. 
This would mean that the dimension of the environment always would be bounded by the output rank for degradable channels.
However, Cubitt \textit{et~al.}~\cite{cubitt08a} has proved that this only holds for channels with qubit and qutrit outputs. 
If the rank of a symmetric extendible state is $R$, 
the technique from the proof of theorem \ref{thm:limitedBrank} can fail only if $1 < \max_j \rank(\rho_j^{AB}) < R$.
This gives the following corollary:
\begin{corollary}
\label{cor:highBrankstructure}
If $\rho^{AB}$ has a $(1,2)$-symmetric extension and $\rank(\rho^B) > \rank(\rho^{AB})$, 
then for any decomposition into pure-extendible states
\begin{equation*}
 \rho^{AB} = \sum_j p_j \rho_j^{AB},
\end{equation*}
$\rank(\rho_j^{AB}) < \rank(\rho^{AB})$ for all $j$.
\end{corollary}
\begin{proof} 
Assume that $\max_j \rank(\rho_j^{AB}) = \rank(\rho_1^{AB}) = \rank(\rho^{AB}) =: R$.
Let the spectral decomposition of $\rho_1^{AB}$ and its reduced state be
$\rho_1^{AB} = \sum_{k=1}^R \gamma_k \proj{\phi_k}$ 
and $\rho_1^{B} = \sum_{k=1}^R \gamma_k \proj{k}$. 
The eigenvectors of $\rho_1^{AB}$ can then be written as 
$\ket{\phi_k} = \sum_{m=1}^R \ket{\widetilde{\psi}_{km}}\ket{m}$.
Since $\rho_1^{AB}$ has the full rank of $\rho^{AB}$, 
the support of $\rho^{AB}$ must be the space spanned by the eigenvectors of $\rho_1^{AB}$.
This means that $\rho^{B}$ has support on $\mathspan\{ \{\ket{m}\}_{m=1}^R \}$ 
and therefore has rank $R$. 
Therefore, if $\rank(\rho^B) > R$ we cannot have $\max_j \rank(\rho_j^{AB}) = \rank(\rho^{AB})$.
\end{proof}

\part{Quantum key distribution with linear advantage distillation}
\label{part:qkd}

\chapter*{Prologue}
\addcontentsline{toc}{chapter}{Prologue} 
\markboth{Prologue}{} 


The first upper bounds on the error threshold for the BB84 and six-state protocols with one-way postprocessing were based on attacks using optimal \textindex{approximate cloning} \cite{cirac97a,fuchs97a,bechmann99a}.
While perfect cloning of a quantum state is impossible \cite{wootters82a}, it is possible to construct a cloning machine that from one input state produces two or more approximate copies of the original state\cite{buzek96a}.
The threshold kicks in when Eve produces two approximate clones of the signal states that are of equal quality.
By forwarding one of the clones to Bob, she makes sure that by performing the same measurement as Bob on her clone, she gets an estimate of Alice's signal state which is as good as Bob's estimate.

If Alice instead of signal states sends one part of an entangled pair, such an approximate cloning machine will produce a tripartite state which is invariant under swapping the two cloned subsystems.
In other words, the state that Alice and Bob share before measurement has a symmetric extension to the clone that Eve kept.

In this part we use the symmetric extension concept while analyzing two-way advantage distillation steps. 
When considering the alternative implementation where the key bits are only measured at the end of the protocol (see section \ref{sec:coherentanalysis}), a symmetric extension at the end of the two-way part of the postprocessing means that the remaining one-way part will not be able to distill a secret key.
The common theme in chapters \ref{ch:rcad}--\ref{ch:numerical} is that we start with a state which is on the threshold for being distillable and show that after some kind of advantage distillation step, it still has a symmetric extension.
Therefore, the considered advantage distillation step cannot give an improved threshold.
Chapter \ref{ch:lad} prepares the ground for this by defining and analyzing the classes of advantage distillation that we will analyze.

\chapter{Linear advantage distillation (LAD)}
\label{ch:lad}

\section{Definition}

We consider all possible advantage distillation steps based on announcements of parity bits from blocks of \textindex{raw key}. 
The block size is $n$, and from that $m$ independent parity bits are announced.
The blocks of bits are described as binary column vectors. 
Alice's raw data block is denoted $\mathbf{d_A}$ and Bob's raw data block is denoted $\mathbf{d_B}$ and they are both of length $n$.
The procedure then goes as follows.
\begin{defn}
 In \emphindex{linear advantage distillation} \emph{(LAD)} Alice announces 
 \begin{equation}
  \mathbf{a} := \mathbf{Hd_A}
 \end{equation}
 where $\mathbf{H}$ is an arbitrary $m \times n$ binary matrix called a \emphindex{parity check matrix}.
 Bob computes $\mathbf{b} := \mathbf{Hd_B}$ and accepts if $\mathbf{a} = \mathbf{b}$ and rejects otherwise.
\end{defn}

The entries that are $1$ in row $i$ in the parity check matrix $\mathbf{H}$ defines which bits the announced bit $a_i$ is a parity of.
Note that this procedure requires two-way communication since both parties need to know whether the block should be discarded.

This procedure is a special case of Maurer's original proposal for advantage distillation \cite{maurer93a}.
The original procedure involves randomly choosing a codeword from an error correcting code and when the code is linear, the matrix $\mathbf{H}$ is the parity check matrix of that code. 
The exact correspondence between the two formulations is shown in appendix \ref{app:adcorrespondence}.
In practice, the term advantage distillation has often been limited to the case when a repetition code is used \cite{chabanne06a,acin06a,bae07a,kraus07a}.
In this case, the parity check matrix is an $(n-1) \times n$ matrix of the form
\begin{equation}
\label{eq:repetitionparitycheck}
 \mathbf{H}_\text{RCAD} = 
 \begin{bmatrix}
    1   &    1   &   0 & \cdots & 0\\
    1   &    0   &   1 &        & 0 \\
 \vdots & \vdots &     & \ddots & \\
    1   &    0   &   0 &        & 1
 \end{bmatrix}.
\end{equation}
We call this kind of postprocessing \emph{repetition code advantage distillation\index{repetition code advantage distillation} (RCAD)} and we will come back to that in chapter \ref{ch:rcad}.

\section{Equivalent announcements}

We want to classify all the different parity check matrices that could be used in linear advantage distillation to potentially help distill a secret key from QKD setups with higher error rate than what is known to be possible with repetition code advantage distillation.
Since Alice and Bob compute the same parity bits and only keep the block if they are all the same, the advantage distillation step is symmetric between Alice and Bob.
In the following sections we only focus on Alice's announcement, keeping in mind that Bob effectively announces the same parity bits from his \textindex{raw key} when he announces that they should accept this block.

Many different parity check matrices will reveal exactly the same information about Alice's data $\mathbf{d_A}$.
For example announcing $a_i$ and $a_j$ reveals the same information as announcing $a_i$ and $a_i \oplus a_j$ (where $\oplus$ denotes addition modulo 2), since $a_j$ can be reconstructed from the latter when $a_i$ is known.
In general, the information in an announcement remains the same if two of the announced bits are swapped or if one announced bit is added to another (modulo 2).
This means that standard Gauss-Jordan row operations on the parity check matrix $\mathbf{H}$, gives an equivalent parity check matrix.

By Gaussian elimination any parity check matrix can therefore be brought to the unique \emphindex{row-reduced echelon form}, 
i.e.~such that
(i) all rows of all zeros are at the bottom,  
(ii) the first 1 on each row (called a pivot) appears strictly to the right of the first 1 on the row above it (if any), and 
(iii) the pivot is the only nonzero entry in its column.
For example, the matrix 
\begin{equation}
 \label{eq:equivexample}
 M = \begin{bmatrix}
      1 & 0 & 1 & 0 & 0 & 0 \\
      1 & 0 & 0 & 1 & 1 & 1 \\
      1 & 0 & 0 & 0 & 0 & 1 \\
      1 & 0 & 1 & 1 & 1 & 0 
     \end{bmatrix}
\end{equation}
has row-reduced echelon form
\begin{equation}
 \label{eq:equivexamplereduced}
 M_\text{rref} = \begin{bmatrix}
                 1 & 0 & 0 & 0 & 0 & 1 \\
                 0 & 0 & 1 & 0 & 0 & 1 \\
                 0 & 0 & 0 & 1 & 1 & 0 \\
                 0 & 0 & 0 & 0 & 0 & 0
                \end{bmatrix}.
\end{equation}
Any rows of all zeros correspond to dependent announcements that may be removed. 
An announcement of $m$ independent parity bits on a block of $n$ bits therefore corresponds to a unique parity check matrix on row-reduced echelon form without any zero rows.

\section{Equivalent announcements on permutation invariant blocks}
\label{sec:permequivalent}


The blocks of $n$ bits we will consider will come from measurements on blocks of $n$ copies of a two-qubit state shared by Alice and Bob. 
The blocks we are interested in therefore has a permutation symmetry, and for announcements on these blocks, parity check matrices that are equal up to a permutation of the columns are equivalent. 
This means that the parity check matrix now is equivalent under the following three operations\footnote{
  Operation (ii) is not really independent from (i), since any two rows $i$ and $j$ can be swapped by adding $i$ to $j$, then $j$ to $i$, and finally adding $i$ to $j$ again. Any permutation can then be built of from swaps implemented this way. Since we will use permutations of rows, it is convenient to have it listed as a separate operation.
}
\begin{itemize}
 \item[(i)] add one row to another (modulo 2),
 \item[(ii)] any permutation of the rows,
 \item[(iii)] any permutations of the columns.
\end{itemize}

By using the equivalence under permutation of columns we can move all the columns with pivots to the right and the others to the left, so that the parity check matrix takes the form 
\begin{equation}
\label{eq:standardannouncement}
\mathbf{H} = [ \mathbf{P} | \calone_m ]. 
\end{equation}
A parity check matrix of this type is said to be on \emphindex{systematic form}.
The matrix $\mathbf{P}$ is of size $m \times k$ where $k := n - m$ and is called the \emphindex{parity matrix}.
The parity matrix is sufficient to fully specify the announcement.

\begin{figure}
  \resizebox{\textwidth}{!}{
  \includegraphics{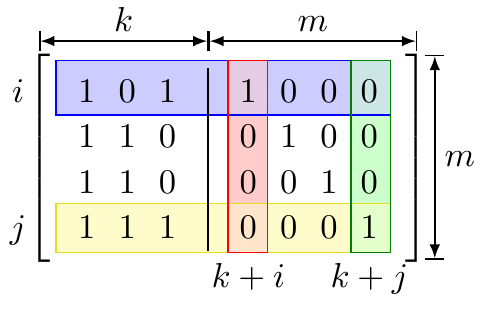}
  \includegraphics{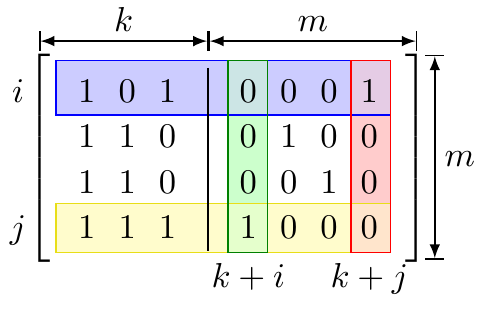}
  \includegraphics{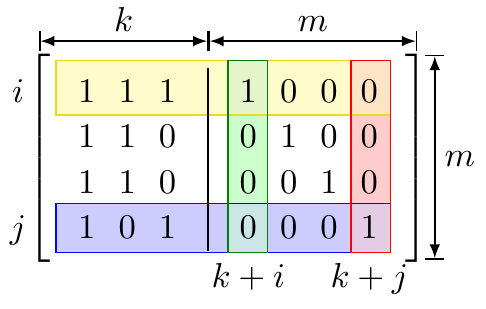}
  }
 \caption[
Equivalence operation resulting from swapping two columns from the $\calone$ part of a systematic parity check matrix.
]{
Illustration of the equivalence operation resulting from swapping two columns from the $\calone$ part (columns $k+i$ and $k+j$) of a parity check matrix on systematic form $[\mathbf{P} | \calone]$. 
On the left, the original parity check matrix $\mathbf{H}$ on systematic form with columns $k+i$ and $k+j$ and rows $i$ and $j$ highlighted. 
In the center, the resulting matrix after swapping columns $k+i$ and $k+j$. Note that at this point, the matrix is not of the systematic form. 
To restore the systematic form, rows $i$ and $j$ are swapped, resulting in the matrix on the right. 
The end result is that the rows $i$ and $j$ in the parity matrix are swapped.
}
 \label{fig:columnswapping1}
\end{figure}

The parity matrix $\mathbf{P}$ is not unique, though. 
The most obvious symmetry is that its columns may be permuted in any order, 
since this was the symmetry that allowed us to collect them at the left side of $\mathbf{H}$ in the first place.
Another straightforward symmetry is the equivalence under permuting the rows of $\mathbf{P}$. 
This comes from the symmetry of permuting columns in the $\calone_m$ part of $\mathbf{H}$.
If columns $k+i$ and $k+j$ in $\mathbf{H}$ are swapped, the resulting matrix can be brought back to the systematic form \eqref{eq:standardannouncement} by swapping rows $i$ and $j$ in $\mathbf{H}$ and therefore the rows of $\mathbf{P}$ also get swapped (see Fig.~\ref{fig:columnswapping1}).

\begin{figure}
  \resizebox{\textwidth}{!}{
  \includegraphics{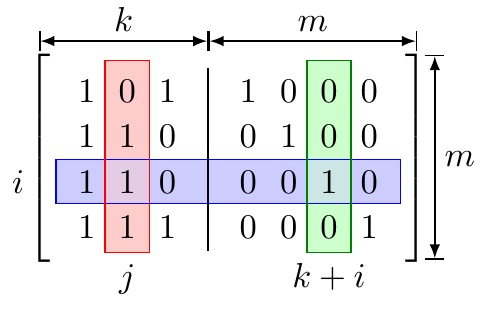}
  \includegraphics{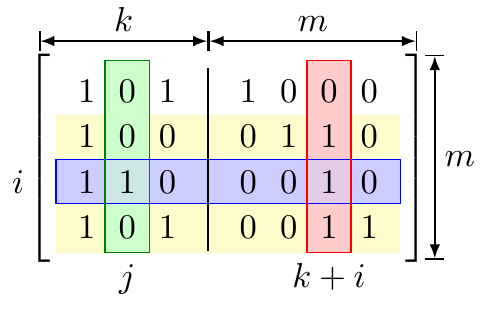}
  \includegraphics{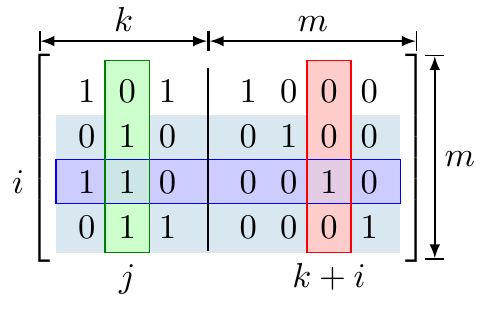}
  }
 \caption[Equivalence operation resulting from swapping a column from the parity matrix part with a column from the $\calone$ part of a systematic parity check matrix.]{Illustration of the equivalence operation resulting from swapping a column from the parity matrix part (column $j$) with a column from the $\calone$ part (column $k+i$) of a parity check matrix on systematic form $[\mathbf{P} | \calone]$. On the left, the original parity check matrix $\mathbf{H}$ on systematic form with row $i$ and columns $j$ and $k+i$ highlighted. In the center, the resulting matrix after swapping columns $j$ and $k+i$. Note that at this point, the matrix is not of the systematic form due to the additional 1 entries in column $k+i$. To restore the systematic form, row $i$ is added to all other rows that has a $1$ in column $k+i$, resulting in the matrix on the right.}
 \label{fig:columnswapping}
\end{figure}

The last kind of equivalence we get from the permutation symmetry of the block is not so trivial.
This is what we get from swapping columns in $\mathbf{P}$ (index from 1 to $k$) with columns in $\calone_m$ (index from $k+1$ to $n$).
Suppose entry $(i,j)$ with $j \leq k$ in $\mathbf{H}$ (and $\mathbf{P}$) is a $1$ and we swap column $j$ with column $k+i$ (see Fig.~\ref{fig:columnswapping}). 
The right $m \times m$ part of this matrix now has a $1$ on each diagonal entry and to take it back to the systematic form \eqref{eq:standardannouncement} it is sufficient to add (modulo 2) row $i$ to all the other rows where column $k+i$ (which came from column $j$) now has a $1$.
On $\mathbf{P}$ this means that row $i$ and column $j$ are unchanged. Also rows that have a $0$ in column $j$ remain the same, while those rows that have a $1$ in column $j$ have row $i$ added to them (modulo 2), in all columns but column $j$.
Note that if column $j$ in $\mathbf{P}$ is the only place that row $i$ has a $1$, this operation will leave $\mathbf{P}$ unchanged.

The above symmetry operation that came from swapping columns $j$ and $k+i$ in $\mathbf{H}$ was defined when entry $(i,j)$ of $\mathbf{P}$ was a $1$.
If entry $(i,j)$ of $\mathbf{P}$ (and $\mathbf{H}$) is $0$, the right $m \times m$ part of $\mathbf{H}$ will get a row of all zeros when columns $j$ and $k+i$ are swapped, and can therefore not be brought to standard form with only row operations.
Therefore, the swap of columns $j$ and $k+i$ of $\mathbf{H}$ will in this case not give rise to a symmetry operation on $\mathbf{P}$.

Any permutation of the columns of the matrix $\mathbf{H}$ can be achieved by a finite number of swaps. 
When $\mathbf{H}$ is on systematic form, the permutations of columns in the parity matrix part  $\mathbf{P}$ can be generated by swaps of columns in this part and the same goes for the $\calone_m$ part. 
Therefore, any permutation of the rows and columns of $\mathbf{P}$ leads to an equivalent parity matrix, since they are generated by swaps of rows and columns.
A general permutation of columns will mix the two parts. 
It will always be such that $l$ ($\leq \min(m,k)$) columns in the $\mathbf{P}$ part end up in the $\calone_m$ part and vice versa.
A general permutation can therefore be broken down into three parts: first a permutation which do not mix the two parts, second swapping columns $1\ldots l$ with $k+1\ldots k+l$, and third another permutation which do not mix the two parts.
The first and the third permutation correspond to permutations of rows and columns in the parity matrix $\mathbf{P}$. 
The second permutation can be done by $l$ consecutive swaps, each of which has to have a $1$ at the right place in the left part $\mathbf{P}$ in order to avoid an all-zero row in the right $m \times m$ part of the parity check matrix $\mathbf{H}$. 
Therefore, the only column permutations that allow for the systematic form to be reconstructed from row operations are the 3 mentioned types. 
This means that any two equivalent parity matrices $\mathbf{P}_1$ and $\mathbf{P}_2$ can be taken from one to the other with the symmetry operations
\begin{enumerate}
 \item[(i)] Swap two columns
 \item[(ii)] Swap two rows
 \item[(iii)] For any entry $(i,j)$ with a $1$, keep row $i$ and row $j$ unchanged. 
Keep rows which has a $0$ in column $j$ unchanged. 
For the rows that have a $1$ in column $j$, add (modulo 2) row $i$ to that row, except for column $j$.
\end{enumerate}

\section{Reducible announcements}
\label{sec:reducibleannouncements}

The two previous sections show how different announcements of $m$ bits on block size $n$ reveal the same information about the bits in that block. 
When testing a set of announcements it is therefore sufficient to test using one parity check matrix from each equivalence class.
Some announcement are \emph{reducible}\index{reducible announcement} in that they are equivalent to one or more announcements with fewer announced bits or on smaller block size. 

The first kind of reduction is when the rows in the parity check matrix $H$ are linearly dependent. 
If that is the case, the announced bits are redundant and some of them can be removed. 
On row-reduced echelon form, the parity check matrix then has one or more rows of all zeros, so the corresponding announced parity bit will always be zero. 
It is therefore equivalent to use a parity check matrix where the rows of all zeros are removed, reducing $m$.

The second kind of reduction is when a column in the parity check matrix consists of only zeros.
This means that the corresponding bit in the raw data does not influence the announced data, so no information about that bit will be revealed.
We can then as well remove that bit from the data and the zero column from the parity check matrix, reducing $n$.
The example $4 \times 6$ matrix in Eq.~\eqref{eq:equivexample} has both a zero column and linearly dependent rows, so it can be reduced to a $3 \times 5$ matrix by removing the second column and last row in Eq.~\eqref{eq:equivexamplereduced}.

A third kind of reduction is when a row in the parity check matrix only has a single $1$ and the other entries in that row are zero.
In the parity matrix, this will show up as a row of all zeros. 
If row $i$ is such a row and the $1$ is in column $j$ it means that the $i$th announced bit will be equal to the $j$th raw bit, so the $j$th bit is no longer secret after the announcement.
In this case, we can remove the bit from both the announcement and the raw data in order to have an equivalent announcement.
This means that we remove row $i$ and column $j$ from the parity check matrix. 
On a parity matrix, this corresponds to removing the (all zero) row $i$.

The two last kinds of announcements are both examples of announcements where a single data bit is treated separately from the rest. 
In general we can have a subblock of bits that are treated independently from the other bits.
For example, an announcement of $m_1$ parity bits from a block of $n_1$ data bits and an announcement of $m_2$ parity bits from a block of $n_2$ data bits can be interpreted as an announcement of $m_1 + m_2$ bits from a block of $n_1 + n_2$ bits. 
The parity check matrix will then be a direct sum of the two parity check matrices, 
$\mathbf{H} = \mathbf{H_1} \oplus \mathbf{H_2}$.
If we can bring the a parity check matrix to this form using equivalence operations, we can reduce it two two independent announcements. 
For example, the $3 \times 6$ parity check matrix on systematic form
\begin{equation}
 \begin{bmatrix}
   1 & 0 & 1 & 1 & 0 & 0\\
   0 & 1 & 0 & 0 & 1 & 0\\
   1 & 0 & 0 & 0 & 0 & 1
 \end{bmatrix}
\end{equation}
can be written as 
\begin{equation}
 \begin{bmatrix}
   1 & 1 & 1 & 0 & 0 & 0\\
   1 & 0 & 0 & 1 & 0 & 0\\
   0 & 0 & 0 & 0 & 1 & 1
 \end{bmatrix}
\end{equation}
by moving column 2 and 5 to the right and row 2 to the bottom.
We see that the latter form is a direct sum of a $2 \times 4$ and a $1 \times 2$ parity check matrix on systematic form.
From the example, we see that if we have parity check matrices on systematic form, it is sufficient to bring the parity matrix $\mathbf{P}$ to the form $\mathbf{P} = \mathbf{P_1} \oplus \mathbf{P_2}$, since the $\calone$ part can always be split up and added such that the parity check matrix also takes the form of a direct sum.

\section{Quantum implementation of systematic announcements}
\label{sec:quantumimplementation}

In linear advantage distillation (LAD), both Alice and Bob announce parity bits from their data blocks and postselect on equal outcome. 
This section describes how we implement this announcement in order to keep track of how the underlying quantum state changes.

The announcement is performed on a block of bits that comes from measurement of qubits in the computational basis.
We can imagine that instead of measuring all qubits and then announcing parities of the obtained data, we measure those parities directly on the quantum state and announce them (this is justified in section \ref{sec:coherentanalysis}).
Say that one of the rows in a parity check matrix (not necessarily on systematic form \eqref{eq:standardannouncement}) is $[ 1 0 0 1 1]$, i.e.~we are supposed to announce the parity of the first, fourth, and fifth bit. 
Instead of measuring $\sigma_z$ (giving results $\pm 1$, where $+1$ maps to $0$ and $-1$ maps to $1$) on each of those qubits and announcing the result, we can measure the operator 
$\sigma_z \otimes \calone \otimes \calone \otimes \sigma_z \otimes \sigma_z$. 
In practice this can be done by attaching an ancilla in the state $\ket{0}$ and performing three CNOT gates, with the ancilla as target and the first, fourth, and fifth qubit as source, followed by a measurement of $\sigma_z$ on the ancilla.
By following this procedure, we only measure the information we are announcing and the rest is left as quantum information that can be measured later. 
If the qubits that we start out with in the block are entangled with another block, we can assess the secrecy of the bits after the announcement by keeping track of the quantum state that is left after the measurement.

When we cast the announcement into systematic form \eqref{eq:standardannouncement}, the last $m$ qubits only affects the outcome of one of the announced bits.
Because of this we can avoid the use of ancilla qubits in the measurement and instead measure on the last $m$ qubits.
The way to do this is the following: 
For each $1$ in the $\mathbf{P}$ part of $\mathbf{H}$ there will be one CNOT with one of the $k$ first qubits as source and one of the $m$ last qubits as target.
If there is a 1 in entry $(i,j)$ where $j \leq k$, the corresponding CNOT will have qubit $j$ as source and qubit $k+i$ as target. 
After this unitary operation, parity bit $i$ can be measured simply by measuring qubit $k+i$.
After measurement and announcement of the $m$ parity bits, we are left with a quantum state on $k$ qubits.

A quantum circuit for this operation can be constructed directly from the parity check matrix by going through one row at a time.
If we let the time flow downwards, such that each line that denotes a qubit is directly below the corresponding column in the matrix, the diagonal of the $\calone_m$ part of $\mathbf{H}$ will show up as a series of targets on the corresponding qubits.
On the first $k$ qubits there will be a control dot for each 1 in the $\mathbf{P}$ part of the matrix. 
For example, the parity check matrix
\begin{equation}
 \mathbf{H} = \left[
\begin{array}{cc|ccc}
 0 & 1 & 1 & 0 & 0 \\
 0 & 1 & 0 & 1 & 0 \\
 1 & 1 & 0 & 0 & 1 
\end{array}
\right]
\end{equation}
has the corresponding quantum circuit,
\begin{center}
\rotatebox{-90}{
\Qcircuit @C=1em @R=1em {
& \qw      & \qw & \qw      & \qw & \targ    & \targ   & \measureD{\vphantom{M}} \\
& \qw      & \qw & \targ    & \qw & \qw      & \qw     & \measureD{\vphantom{M}} \\
& \targ    & \qw & \qw      & \qw & \qw      & \qw     & \measureD{\vphantom{M}} \\
{\ar @{--} [0,7]}& &  &     &     &          &         &      \\ 
& \ctrl{-2}& \qw & \ctrl{-3}& \qw & \qw      &\ctrl{-4}& \qw  \\
& \qw      & \qw &\qw       & \qw & \ctrl{-5}& \qw     & \qw 
}
}
\end{center}
The first two CNOTs correspond to the first two rows, while the last two CNOTs correspond to the last row.

In linear advantage distillation, we always use the same circuit on Alice's part and Bob's part of the system. 
When we have two (possibly different) \textindex{Bell state}s and apply a CNOT from the first to the second qubit on each side, the result is still two Bell states \cite{bennett96b}.
The final measurement only serves to distinguish whether the measured pair was in one of\footnote{
The actual outcome of each individual measurement contains no information, since for any multi-qubit \textindex{Bell-diagonal state}, a measurement on any single qubit will have a completely random outcome. Only the correlations between qubits in a qubit pair contain information.
}
$\ket{\Phi^\pm} := (\ket{00} \pm \ket{11})/\sqrt{2}$ or one of $\ket{\Psi^\pm} := (\ket{01} \pm \ket{10})/\sqrt{2}$.
So since in our case the system before announcement and postselection is in a \textindex{multi-qubit Bell-diagonal state} (i.e. a probability distribution over different possible strings of Bell states, see section \ref{sec:simplifying}), it will remain so after afterwards.

With the chosen implementation of the measurement of the parity bits, the state after measurement is unique given the starting state.
We could also have chosen a different measurement of the same parity bits, for example by adding a CNOT in the end on the first two qubits in the example circuit and a few controlled phases between some pairs.
This would correspond to a measurement with the same POVM as the chosen one (since the probability for each outcome is the same) but different Kraus operators (since the post-measurement state is different).
Since the Kraus operators that correspond to the same POVM are all related by a unitary operator, the different possible post-LAD states would be related by local unitary operators.
They are therefore all equivalent when it comes to symmetric extension and many other relevant properties.

\section{Postselection vs. communication in reverse direction}

The advantage distillation step is symmetric upon switching the roles of Alice and Bob, 
since by accepting, Bob effectively announces that he has the same parity bits as Alice has announced.
For the purpose of breaking a symmetric extension, one could argue that it would have been sufficient if Bob announced this data, making the initial announcement in the reverse direction from the one-way communication that comes later during privacy amplification and error correction.
If this does not break the symmetric extension, Alice announcing some data will not help since this does not change the symmetry between $B$ and $B'$. 
If, on the other hand, Bob's announcement does break the symmetric extension, we could risk that Alice's measurement and announcement would actually reintroduce a symmetric extension. 

The main reason for letting Alice release the same information as Bob is to make the problem tractable. 
If Alice does nothing, the system would be $2^n \times 2^k$ dimensional instead of $2^k \times 2^k$.
Moreover, we would not be able to use the nice properties of the multi-qubit Bell-diagonal states. 
Also, as we will argue in the following, with the Bell-diagonal starting states that we have, it is unlikely that Alice's announcement and the following postselection does any damage.

Initially, Alice and Bob have many copies of a Bell-diagonal state (see section \ref{sec:parameterestimation})
\begin{equation}
 \label{eq:discbelldiag}
 \rho^{AB} = p_I \proj{\Phi^+} + p_x \proj{\Psi^+} + p_y \proj{\Psi^-} + p_z \proj{\Phi^-}.
\end{equation}
The purification of this state can be written as
\begin{align}
 \ket{\psi}_{ABE} 
 &= \sqrt{p_I} \ket{\Phi^+}_{AB}\ket{I}_{E} + \sqrt{p_x} \ket{\Psi^+}_{AB}\ket{x}_{E} \nonumber \\
 & \quad + \sqrt{p_y} \ket{\Psi^-}_{AB}\ket{y}_{E} + \sqrt{p_z} \ket{\Phi^-}_{AB}\ket{z}_{E}\\
 &= \ket{00}\frac{1}{\sqrt{2}}(\sqrt{p_I}\ket{I} + \sqrt{p_z}\ket{z})
 + \ket{11}\frac{1}{\sqrt{2}}(\sqrt{p_I}\ket{I} - \sqrt{p_z}\ket{z}) \nonumber \\
 & \quad + \ket{01}\frac{1}{\sqrt{2}}(\sqrt{p_x}\ket{x} + \sqrt{p_y}\ket{y})
 + \ket{10}\frac{1}{\sqrt{2}}(\sqrt{p_x}\ket{x} - \sqrt{p_y}\ket{y}),\\
 &= \ket{00}_{AB}\ket{\psi_{Iz}^+}_E
 + \ket{11}_{AB}\ket{\psi_{Iz}^-}_E 
 + \ket{01}_{AB}\ket{\psi_{xy}^+}_E
 + \ket{10}_{AB}\ket{\psi_{xy}^-}_E,
\end{align}
where 
$\ket{\psi_{Iz}^\pm} := \frac{1}{\sqrt{2}}(\sqrt{p_I}\ket{I} \pm \sqrt{p_z}\ket{z})$
and 
$\ket{\psi_{xy}^\pm} : = \frac{1}{\sqrt{2}}(\sqrt{p_x}\ket{x} \pm \sqrt{p_y}\ket{y})$
are unnormalized vectors that satisfy 
$\braket{\psi_{Iz}^\pm}{\psi_{xy}^\pm} = \braket{\psi_{Iz}^\mp}{\psi_{xy}^\pm} = 0$. 
From the latter form we easily get the 
classical-classical-quantum\index{ccq state} (ccq) state \cite{devetak04a,horodecki09a}
\begin{align}
 \rho_\textrm{ccq}^{ABE} 
 &= \proj{00}_{AB} \proj{\psi_{Iz}^+}_E
 + \proj{11}_{AB} \proj{\psi_{Iz}^-}_E\nonumber \\
 & \quad + \proj{01}_{AB} \proj{\psi_{xy}^+}_E
 + \proj{10}_{AB} \proj{\psi_{xy}^-}_E
\end{align}
This state summarizes Eve's quantum knowledge about Alice and Bob's classical bits.
Eve's states corresponding to $00$ and $11$ are orthogonal to the ones that correspond to $01$ and $10$, so by measuring if her state is in the subspace spanned by $\ket{I}$ and $\ket{z}$ or $\ket{x}$ and $\ket{y}$, she can tell whether Alice's and Bob's bits are equal \emph{without disturbing the ccq-state}.
So if Bob announces the parity of some of her bits, Eve will know which of Alice's corresponding bits are the same and different, and therefore she knows Alice's value for the corresponding parity bit. 
Alice's announcement that she has the same parity bits is therefore of no help to Eve.
Even if Alice announced all the parity bits---regardless of whether they were equal or not---this
information would already be known to Eve, but would tell Bob exactly which parity bits were different. 

The next step is to discard the block if any of the announced bits are different.
At this point, all the parties have classical information telling them what the error pattern in the announced bits are, so Alice and Bob's state is an orthogonal mixture of states with the different error patterns. 
If this state does not have a symmetric extension, at least one of the error patterns must correspond to a state without a symmetric extension. 
We have here focused on the case where there are no detected errors. 
While it may be possible to increase the key rate by also attempting to distill key from blocks with few errors in addition to those with no errors, we doubt that states with detected errors will be able to not have a symmetric extension whilst the state with no detected errors has one.
The fact that both Alice and Bob reveal their parities and postselect on equal result is therefore very unlikely to give an advantage to Eve over the situation where only Bob announces the parity bits.

\chapter[LAD with one encoded bit does not improve threshold]{Linear advantage distillation with one encoded bit does not improve threshold}
\label{ch:rcad}
\markboth{Chapter \thechapter. LAD with one encoded bit \ldots}{}


In this chapter we consider the special case of \textindex{linear advantage distillation} (LAD) with one encoded bit.
For blocksize $n$, this means that $m = n-1$ and $k = 1$.
The \textindex{parity check matrix} $\mathbf{H}$ therefore has dimension $(n-1) \times n$ and the \textindex{parity matrix} $\mathbf{P}$ has dimension $(n-1) \times 1$.
We show that for the \textindex{Bell-diagonal state}s outside the set that was shown by Chau to be distillable with this kind of advantage distillation \cite{chau02a}, any kind of advantage distillation with one encoded bit will result in a state with a symmetric extension. 
The current thresholds for the six-state \index{six-state protocol}\cite{bruss98a,bechmann99a} and BB84 \index{BB84 protocol}\cite{bennett84a} protocols will therefore not be improved with this kind of postprocessing.
This is similar to the analysis by Ac\'in \etal \cite{acin06a,bae07a}
but since we know when a state has a symmetric extension, 
we do not need to construct an explicit attack. 
The approach described here has been generalized to qudit versions of the six-state and BB84 protocols by Ranade et al.\cite{ranade09a,ranade09b}.


\section{Repetition code advantage distillation (RCAD)}

When the parity check matrix on systematic form $\mathbf{H} = [\mathbf{P}|\calone]$ defines an error correcting code with only one encoded bit, the parity matrix $\mathbf{P}$ is a $(n-1) \times 1$ matrix.
We have seen in section~\ref{sec:reducibleannouncements} that if a row in the parity matrix consists of only zeros, it is equivalent to one where that row is removed (reducing the block size), so for codes where the advantage distillation cannot be reduced to smaller block sizes, all rows of the parity matrix must contain at least one $1$.
Since each row in $\mathbf{P}$ now consists of only one element, that element has to be $1$.
Therefore the only irreducible announcement of this type with blocksize $n$ is 
\begin{equation}
\label{eq:repetitionparitycheckLAD1}
 \mathbf{H}_\text{RCAD} = 
 \begin{bmatrix}
    1   &    1   &   0 & \cdots & 0\\
    1   &    0   &   1 &        & 0 \\
 \vdots & \vdots &     & \ddots & \\
    1   &    0   &   0 &        & 1
 \end{bmatrix}.
\end{equation}
This is the parity check matrix for a repetition code and we therefore call linear advantage distillation with one encoded bit \emphindex{repetition code advantage distillation} \emph{(RCAD)}.

A 2-way procedure to distill secret key from quantum correlations in a prepare and measure scheme invented by Gottesman and Lo \cite{gottesman03a}, 
was proven by Chau \cite{chau02a} to work for Bell-diagonal states with error rates that satisfy
\begin{equation} 
\label{eq:chau-conditionINTRO}
(p_I - p_z)^2 > (p_I + p_z)(p_x + p_y). 
\end{equation}
No known procedure can distill a secret key if the underlying Bell-diagonal state is outside this set.
The border of this set corresponds to a QBER ($p_x + p_y$) of 27.64~\% for the six-state scheme ($p_x = p_y = p_z$) and 20~\% for BB84 ($p_x = p_z$, $p_y = 0$), which are therefore the best error thresholds for these protocols.

The procedure works by first applying a number of so-called \textindex{B-steps} (for bit error detection), 
then \textindex{P-steps} (for phase error correction) and in the end a one-way quantum error correcting code. 
The B-step works on two bit pairs. 
On each side the parity of the bits is computed and compared to the other side. 
If the parity differs, there must have been an error and both pairs are discarded. 
If the parity is equal, the first pair is kept. 
This step requires two-way communication since both parties need to know if they should keep the first pair.
The P-step works on three bit pairs. 
The output bit on each side is the parity of the three bits. 
This does not require any communication at all, 
but it simulates a phase error correction step where two qubits are measured to give a phase error syndrome which is sent from Alice to Bob for comparison.
Alternatively, we can look at it as keeping the two extra qubits on each side in a
\emph{shield} system which limits Eve's knowledge about the \emph{key} system \cite{horodecki05a}.
Irrespective of how we look at it, a P-step does not require communication from Bob to Alice, and can therefore not break a symmetric extension.
If states with symmetric extension are to be distilled into secret key, the B-steps must break the symmetric extension.

The B-steps are the advantage distillation part of this procedure.
They come from the \emphindex{bit pair iteration} protocol introduced by Gander and Maurer \cite{gander94a}.
Instead of using a large blocksize to approach the threshold, it uses many rounds.
While each B-step only uses a blocksize 2 (with parity check matrix $\mathbf{H} = \begin{bmatrix} 1 & 1 \end{bmatrix}$), 
we can think about $N$ rounds as an announcement on blocksize $n = 2^N$.
If no error is detected, the announcement reveals the parity of any two of the $2^N$ data bits.
This is the same information that is revealed by RCAD on the same block.
The bit pair iteration is more efficient than RCAD, since if an error is detected in some of the subblocks in an early round, 
only that subblock will need to be discarded, while with RCAD we will discard all $2^N$ bit pairs if an error is detected.
Here, we do not care about how many bit pairs is needed  on average to distill a key, only if it is possible or not.
What is important is that the state  after successful RCAD is the same as after successful bit pair iteration, so for our purposes RCAD and B-steps are equivalent.

Similarly, the formulation of RCAD where Alice chooses a bit at random and announces the data if the bit is 0 and the inverted data if the bit is 1, is equivalent to our formulation.
This formulation is used by Ac{\'in} and Bae \cite{acin06a,bae07a} and is based on the original formulation of advantage distillation by Maurer \cite{maurer93a}. 
In appendix \ref{app:adcorrespondence} we show the equivalence in the general setting of linear advantage distillation.
Actually, any kind of advantage distillation where we after the announcement know that the data can only have come from two possible strings, is equivalent to RCAD.
If the two possible $n$-bit strings are $a_1$ and $a_2$, they will in general differ at some $m \geq 1$ positions. 
In those positions where they do not differ, the raw data is revealed and is therefore useless for key distillation and can be removed. 
This leaves two possible $m$-bit strings which differ in all positions, 
which is exactly same situation as if RCAD is performed on those $m$ bits.

\section{State after RCAD}
\label{sec:stateafterrcad}

As shown in section~\ref{sec:parameterestimation}, we can regard the quantum state between Alice and Bob as many copies of a \textindex{Bell-diagonal state}, 
 \begin{equation}
  \label{eq:belldiagstateRCAD}
  \rho^{AB} = p_I \proj{\Phi^+} + p_x \proj{\Psi^+} + p_y \proj{\Psi^-} + p_z \proj{\Phi^-}.
\end{equation}
We can think of this as a maximally entangled state $\ket{\Phi^+}$ which has a probability $p_i$ for having suffered a $\sigma_i$ error on Bob's qubit.
We then perform RCAD on a block of $n$ copies of this state.
Bob's output qubit has a \textindex{bit error} (either $\sigma_x$ or $\sigma_y$) iff all the qubits in the block had a bit error and 
it has no bit error iff no qubit in the block had a bit error. 
The other bit error patterns are detected and the block is discarded.
The output qubit has a \textindex{phase error} ($\sigma_y$ or $\sigma_z$) if an odd number of input qubits had a phase error, 
and no phase error if an even number of input qubits had a phase error.
So the output qubit is error free iff an even number of input qubits had a $\sigma_z$ error and the rest were error free. 
The probability for this to happen given the state of the input qubits is 
$p_I^\text{RCAD} = p_I^n + \binom{n}{2} p_I^{n-2} p_z^2 + \binom{n}{4} p_I^{n-4} p_z^4 + \cdots = \sum_{j=0}^{\lfloor n/2 \rfloor} \binom{n}{2j} p_I^{n-2j} p_z^{2j}$.
This is every second term in the binomial expansion of $(p_I \pm p_Z)^n$, 
and by taking the average of the $\pm$ cases we get the terms we want, 
$p_I^\text{RCAD} = \frac{1}{2}\left( (p_I + p_z)^n + (p_I - p_z)^n\right)$.
By making similar arguments we get $p_z^\text{RCAD}$ from the terms with an odd number of $\sigma_z$ errors, and $p_x^\text{CAD}$ and $p_y^\text{RCAD}$ from the cases where there are $\sigma_x$ and $\sigma_y$ errors instead of $\mathbbm{1}$ and $\sigma_z$.
This gives the following eigenvalues for the state after RCAD:
\begin{subequations}
\begin{align}
\label{eq:RCADstatefirst}
p_I^\text{RCAD} &= \sum_{j=0}^{\lfloor n/2 \rfloor} \binom{n}{2j} p_I^{n-2j} p_z^{2j} \nonumber \\
&= \frac{1}{2}\left( (p_I + p_z)^n + (p_I - p_z)^n\right)\\ 
p_z^\text{RCAD} &= \sum_{j=0}^{\lfloor (n-1)/2 \rfloor} \binom{n}{2j+1} p_I^{n-(2j+1)} p_z^{2j+1} \nonumber \\
&= \frac{1}{2}\left( (p_I + p_z)^n - (p_I - p_z)^n\right)\\ 
p_x^\text{RCAD} &= \frac{1}{2}\left( (p_x + p_y)^n + (p_x - p_y)^n\right)\\ 
\label{eq:RCADstatelast}
p_y^\text{RCAD} &= \frac{1}{2}\left( (p_x + p_y)^n - (p_x - p_y)^n\right)
\end{align}
\end{subequations}
where the sum of these probabilities gives the probability for RCAD to succeed.

To quantify how the procedure improves or deteriorates the ability of a state to produce a key, 
as defined by \eqref{eq:chau-conditionINTRO}, 
we define the quantity\footnote{We do not care about the rate, so we expect no relation between $D_C$ and the key rate. It is possible to have arbitrarily high $D_C$ and at the same time arbitrarily low key rate.}
\index{D@$D_C$}
\begin{equation}
\label{eq:chau-distillability}
 D_C := \log_2 \left(  \frac{(p_I - p_z)^2}{(p_I + p_z)(p_x + p_y)} \right).
\end{equation}
This quantity is positive on all distillable states, negative on states where $(p_I - p_z)^2 < (p_I + p_z)(p_x + p_y)$ and zero on the border. 
By inserting the recursion relations \eqref{eq:RCADstatefirst}-\eqref{eq:RCADstatelast} into \eqref{eq:chau-distillability}, 
we see that $D_C^\text{RCAD} = n D_C$. 
Thus, if the state starts out with negative $D_C$, it will remain negative, 
if it starts out being zero it will remain so, 
and if it starts out being positive it can reach an arbitrary positive value by choosing $n$ large enough. 
We will next show that this allows the procedure to break the symmetric extension when $D_C > 0$ and not otherwise. 
More precisely, we will show that reaching $D_C \geq 2$ is sufficient for breaking symmetric extension, 
whereas all states with $D_C \leq 0$ have a symmetric extension.

\begin{figure}
\scalebox{0.85}{\includegraphics{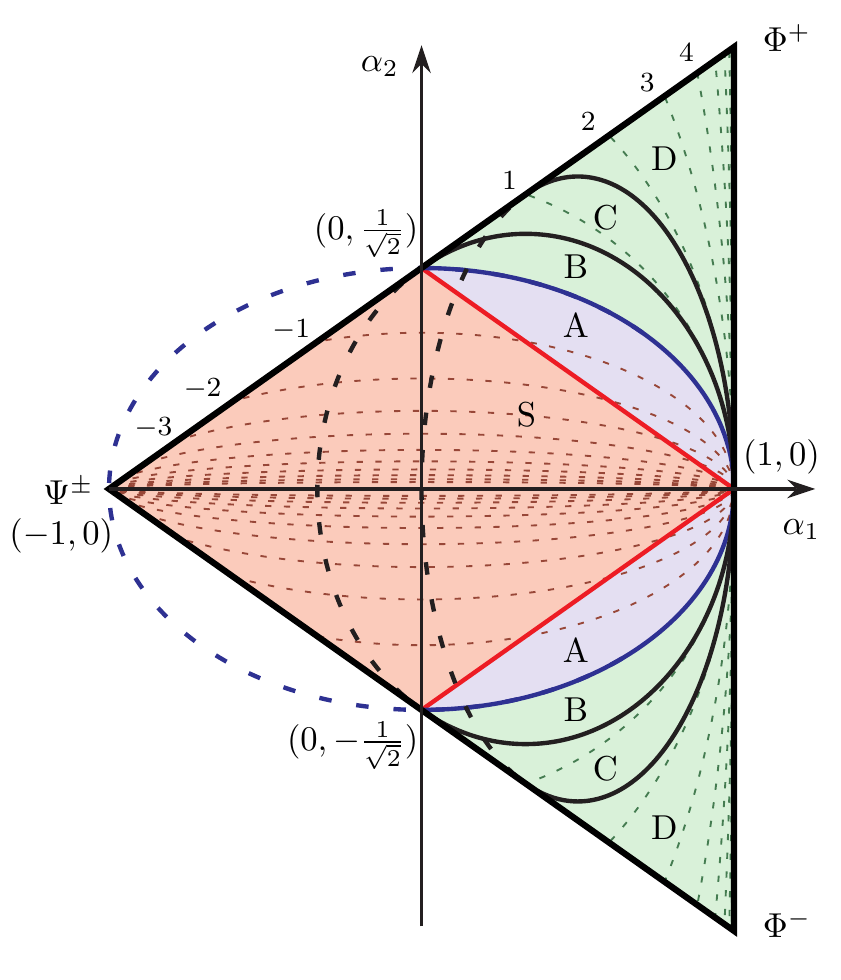}}
\caption[Symmetric extension and key distillability for Bell-diagonal states]{\label{fig:statespace} 
Plot of the Bell-diagonal state space as a function of $\alpha_1$ and $\alpha_2$, with $\alpha_3$ projected out.
The thin dashed lines indicates the value of $D_C$.
The shaded region S corresponds to separable states (for at least some $\alpha_3$). 
Region A is the set of entangled states for which the B-steps fail to break a symmetric extension. 
The border between A and B corresponds to $D_C = 0$.
Regions A and B together are the entangled states with symmetric extension for all possible values of $\alpha_3$, 
while in region C all the states with $\alpha_3 = 0$ has symmetric extension, 
but some states with other $\alpha_3$ do not.
In region D no states have symmetric extension.
The borders between regions B and C and regions C and D both have the shape of ellipses.
The former is described by Eq. \eqref{eq:surfaceellipse}, 
while the latter is described by Eq. \eqref{eq:sixstateellipse}.}
\end{figure}

To show this, we describe the states by the same parameters $\alpha_i$
that we used in the symmetric extension calculation 
and defined in Eqs.~\eqref{eq:alpha-transformation-first}-\eqref{eq:alpha-transformation-last}.
In these coordinates, 
 $D_C = \log_2 (  2 \alpha_2^2/(\alpha_0^2-\alpha_1^2) )$
does not depend on $\alpha_3$ at all. 
The equations for the surfaces of constant $D_C$ are then
\begin{equation}
\label{eq:chauconst}
 \alpha_1^2 + 2\cdot 2^{-D_C} \alpha_2^2 = \alpha_0 = 1
\end{equation}
(we assume normalized states in this chapter, so $\alpha_0 = 1$).
These are the equations for ellipses with center in the origin, 
constant $\alpha_1$-semiaxis $1$ and $D_C$-dependent $\alpha_2$-semiaxis $2^{(D_C -1)/2}$. 
The surfaces are plotted in figure \ref{fig:statespace}. 
In the figure the ellipse that extends outside the state space and separates region A and B is the surface where $D_C = 0$. 
Inside that ellipse are thin dashed lines indicating $D_C = -1, -2, \ldots$, 
and outside are similar lines indicating $D_C = 1, 2, \ldots$.
The two other curves relate to symmetric extension which we will deal with next.

\section{Symmetric extension for cross sections}
\label{sec:cross-section-symext}

In section~\ref{sec:belldiagsymext}, we derived conditions \eqref{eq:rankoneXsufcondsum}-\eqref{eq:rank2Xposconstraint2sum} for when a state has a symmetric extension,
\begin{subequations}
\label{eq:conditionsummaryRCAD}
\begin{gather}
\label{eq:rankoneXsufcondRCAD}
 4 \alpha_1 (\alpha_2^2-\alpha_3^2) - (\alpha_2^2 - \alpha_3^2)^2 - 4 \alpha_1^2 (\alpha_2^2 + \alpha_3^2) \geq 0, \\
\label{eq:rank2Xposconstraint1RCAD}
 \alpha_2^2-\alpha_3^2 - 2\sqrt{2}\alpha_1|\alpha_2| \geq 0
  \quad \text {and} \quad
 |\alpha_2| - 2\sqrt{2} \alpha_1 \geq 0, \\
\label{eq:rank2Xposconstraint2RCAD}
 \alpha_3^2-\alpha_2^2 + 2\sqrt{2}\alpha_1|\alpha_3| \geq 0
 \quad \text {and} \quad
 |\alpha_3| + 2\sqrt{2} \alpha_1 \geq 0,
\end{gather}
\end{subequations}
where the state has a symmetric extension if and only if at least one of the conditions are satisfied.

Unlike the surfaces for constant $D_C$, 
the surface of the set of extendible states is dependent on $\alpha_3$. 
In comparing symmetric extension to the $D_C$ surfaces 
we will be particularly interested in three cross sections 
through the symmetric extension surface. 
One is through the center of the tetrahedron that defines the state space, 
where $\alpha_3 = 0$ ($p_x = p_y$). 
The two others are the two faces of the tetrahedron where
$\alpha_3 = \pm (1 - \alpha_1)/\sqrt{2}$ ($p_y = 0$ and $p_x = 0$).

\begin{figure}
\scalebox{0.85}{\includegraphics{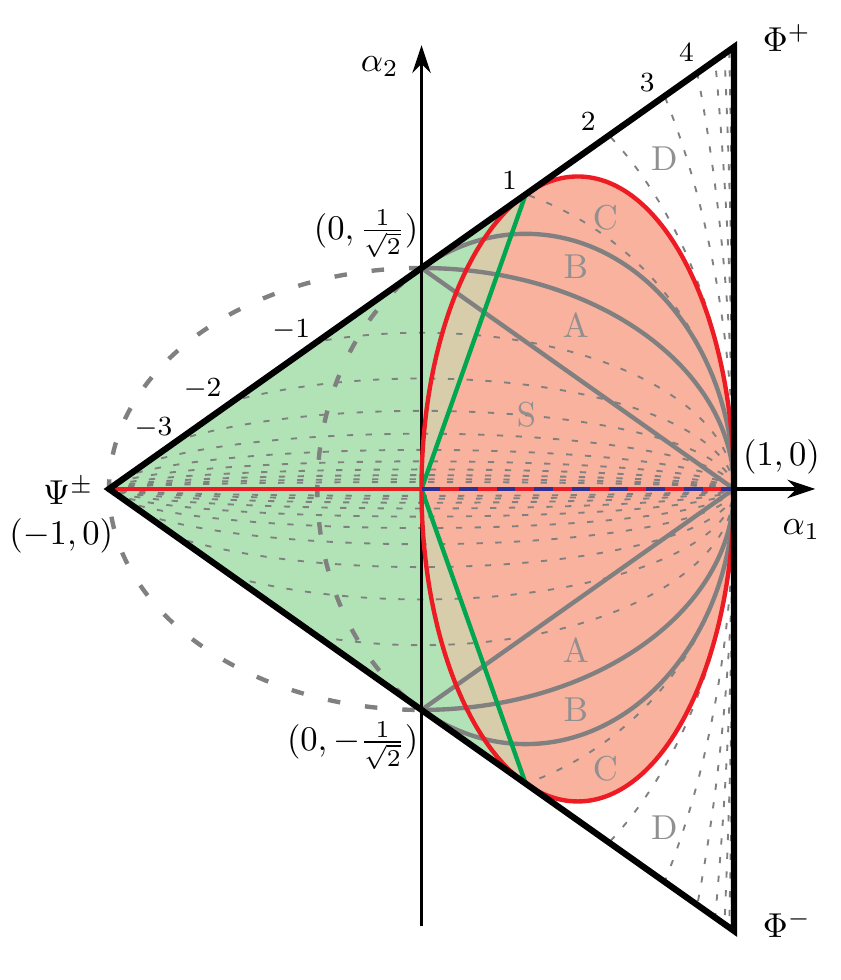}}
\caption[States with a symmetric extension at the central plane of the tetrahedal state space.]{\label{fig:statespace-center} 
The colored area shows which states have a symmetric extension at the cross section $\alpha_3 = 0$.
It is the union of the three sets defined by Eqs.~\eqref{eq:rankoneXsufcondRCAD}-\eqref{eq:rank2Xposconstraint2RCAD}.
The red set consisting of the ellipse to the right of the $\alpha_2$ axis and the whole $\alpha_2 = 0$ line is the defined by Eq.~\eqref{eq:sixstatesymext}.
The green set that partially overlaps with the ellipse is defined by Eq.~\eqref{eq:rank2Xposconstraint1center}. 
The blue set defined by Eq.~\eqref{eq:rank2Xposconstraint2center} consists of only the positive $\alpha_1$ axis.
}
\end{figure}

For the cross section where $p_x = p_y$ we set $\alpha_3 = 0$ in 
\eqref{eq:rankoneXsufcondRCAD} to get the inequality
\begin{equation}
\label{eq:sixstatesymext}
  -{\alpha_2^2} \left( 4\left(\alpha_1 - \frac{1}{2}\right)^2 + \alpha_2^2 -1\right) \geq 0.
\end{equation}
This tells us that any state with $\alpha_2 = 0$ (and at the same time $\alpha_3 = 0$, that is, on the $\alpha_1$ axis) has a symmetric extension, 
and they also happen to be separable. 
When $\alpha_2 \ne 0$, we get
\begin{equation}
\label{eq:sixstateellipse}
 4\left(\alpha_1 - \frac{1}{2}\right)^2 + \alpha_2^2  \leq 1
\end{equation}
which describes an ellipse with center in $(\alpha_1,\alpha_2) = (1/2,0)$, $\alpha_1$-semiaxis $1/2$ and $\alpha_2$-semiaxis $1$.
In figure \ref{fig:statespace} this is the solid curve separating regions C and D and is also shown as a red ellipse in figure \ref{fig:statespace-center}.
Setting $\alpha_3 = 0$ in Eqs.~\eqref{eq:rank2Xposconstraint1RCAD} and \eqref{eq:rank2Xposconstraint2RCAD} gives the two conditions
\begin{subequations}
\begin{gather}
\label{eq:rank2Xposconstraint1center}
|\alpha_2| \geq 2\sqrt{2} \alpha_1 \\
\label{eq:rank2Xposconstraint2center}
|\alpha_2| = 0 \quad \text {and} \quad \alpha_1 \geq 0.
\end{gather}
\end{subequations}
The first of these two lines describes the area above and below lines going from the origin to the point where the ellipse touches the border of the state space (shown in green in figure \ref{fig:statespace-center}).
The second only describes the the $\alpha_1$ axis for positive $\alpha_1$ (shown in blue in figure \ref{fig:statespace-center}) , which is already part of the ellipse.
From this it is clear that in the cross section $\alpha_3 = 0$, only states in region D in figure \ref{fig:statespace} do not have a symmetric extension.

\begin{figure}
\scalebox{0.85}{\includegraphics{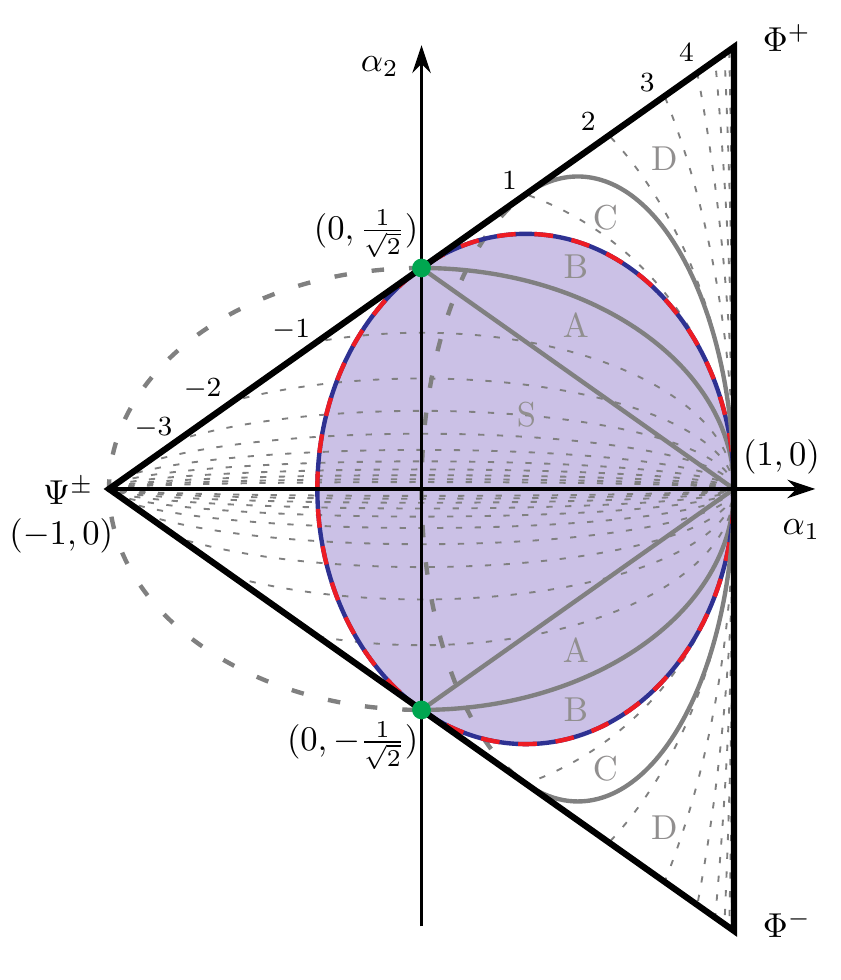}} 
\caption[States with a symmetric extension at two faces of the tetrahedral state space.]{\label{fig:statespace-face} 
The colored area shows which states have a symmetric extension at the cross sections $\alpha_3 = \pm (1 - \alpha_1)/\sqrt{2}$ which corresponds to two of the faces of the tetrahedral state space.
It is the union of the three sets defined by Eqs.~\eqref{eq:rankoneXsufcondRCAD}-\eqref{eq:rank2Xposconstraint2RCAD}.
The red set consisting of the perimeter of the blue ellipse is defined by Eq.~\eqref{eq:surfaceellipse}.
The green set consists of only the two points with $\alpha_1 = 0$ and $\alpha_2 = \pm 1/\sqrt{2}$. 
The blue set is the elliptic disc defined by Eq.~\eqref{eq:surfaceellipse2}.
}
\end{figure}

For the cases $p_y = 0$ and $p_x = 0$ we insert $\alpha_3 = \pm (1 - \alpha_1)/\sqrt{2}$ into \eqref{eq:rankoneXsufcondRCAD}
to get 
\begin{equation}
  -\frac{1}{36}\left(\frac{9}{4}\left(\alpha_1-\frac{1}{3}\right)^2 + \frac{3}{2}\alpha_2^2 -1\right)^2 \geq 0.
\end{equation}
which simplifies to 
\begin{equation}
 \label{eq:surfaceellipse}
 \frac{9}{4}\left(\alpha_1-\frac{1}{3}\right)^2 + \frac{3}{2}\alpha_2^2 = 1.
\end{equation}
This describes another ellipse, with center in $(1/3,0)$, $\alpha_1$-semiaxis $2/3$ and $\alpha_2$-semiaxis $\sqrt{2/3}$. 
This is the solid line separating regions B and C in figure \ref{fig:statespace}.
Only the perimeter of the ellipse is a solution. 
This can also be seen from the 3-dimensional plot of Eq.~\eqref{eq:rankoneXsufcondRCAD} on the left in figure \ref{fig:extstates} where the solution set only touches the face of the tetrahedron along a line.
Only two points inside the state space satisfies Eq.~\eqref{eq:rank2Xposconstraint1RCAD} with $\alpha_3 = \pm (1 - \alpha_1)/\sqrt{2}$. 
This is for $\alpha_1 = 0$ and $|\alpha_2| = 1/\sqrt{2}$ which is where the two cones with faces up and down (in the center plot in figure \ref{fig:extstates}) touches the edge of the tetrahedron.
Finally, Eq.~\eqref{eq:rank2Xposconstraint2RCAD} becomes 
\begin{equation}
 \label{eq:surfaceellipse2}
 \frac{9}{4}\left(\alpha_1-\frac{1}{3}\right)^2 + \frac{3}{2}\alpha_2^2 \leq 1.
\end{equation}
This fills the ellipse from Eq.~\eqref{eq:surfaceellipse} and corresponds to the circular faces of the cones that faces to the sides in the center figure in figure \ref{fig:extstates}.
Like in figure \ref{fig:statespace-center}, the solution to the three conditions \eqref{eq:rankoneXsufcondRCAD}-\eqref{eq:rank2Xposconstraint2RCAD} are plotted in figure \ref{fig:statespace-face} in red, green, and blue.
The union of these (which is equal to the blue set) is the set of states with symmetric extension in this cross section.

The area labeled D in the figures \ref{fig:statespace}, \ref{fig:statespace-center}, and \ref{fig:statespace-face} is the only area where there are no states with a symmetric extension, for any $\alpha_3$.
This is because if a state defined by $(\alpha_1,\alpha_2,\alpha_3)$ has a symmetric extension, 
so does the state defined by $(\alpha_1,\alpha_2,-\alpha_3)$ since the states are related by local unitary operators. 
Then the convex combination $(\alpha_1,\alpha_2,0)$ would also have a symmetric extension. 
Similarly, the areas S, A, and B (colored in figure \ref{fig:statespace-face}) are the only areas where the corresponding state has a symmetric extension for any $\alpha_3$.
In this area, even the states on the surface of the tetrahedron has a symmetric extension, and by mixing the two states on each side (the $\pm$ cases in $\alpha_3 = \pm (1 - \alpha_1)/\sqrt{2}$) we get a state with symmetric extension for any $\alpha_3$.

\section{Distillability vs.~symmetric extension}


We are now in a position to relate $D_C$\index{D@$D_C$} to symmetric extension.
From figure \ref{fig:statespace} it is evident that in most of the state space, 
the surface $D_C = 2$ lies strictly outside the outer border for symmetric extension (i.e. it lies inside area D). 
Towards the the point $(1,0)$, however, all the lines for constant $D_C$, symmetric extension border, and separability border converge. 
This is also the state towards which the sequence of states converges for the most relevant starting states (e.g.~all states  with $p_I \geq 0.5, p_z > 0$) when the blocksize in RCAD increases.
Even though this is a separable state, in any neighborhood around it there will be states without a symmetric extension. 
By inserting $D_C = 2$ in equation \eqref{eq:chauconst}, the $D_C = 2$ border can be described by $\alpha_2^2 = 2 - 2 \alpha_1^2 =: f(\alpha_1)$. 
Similarly, the outer border for symmetric extension from Eq.~\eqref{eq:sixstateellipse} can be expressed as 
$\alpha_2^2 = 1 - 4(\alpha_1-1/2)^2 =: g(\alpha_1)$. 
Taking the difference, we get $\Delta(\alpha_2^2) = f(\alpha_1) - g(\alpha_1) = 2(\alpha_1-1)^2 \geq 0$, 
so the $D_C = 2$ surface is always outside the symmetric extension surface, 
except for the point $(1,0)$ where $D_C$ is not defined. 
Thus, no states for which $D_C \geq 2$, have a symmetric extension.
RCAD can therefore break a symmetric extension whenever $D_C > 0$ by using a blocksize greater than $2/D_C$.

In a similar fashion one can show that the Chau-border $D_C = 0$ never is outside the inner symmetric extension border (between regions B and C) in the interval $\alpha_1 \in [0,1]$, 
which is the region where the Chau-border is contained in the state space. 
They coincide at the points $(0,1/\sqrt{2})$ and $(1,0)$. 
Thus, any state with $D_C \leq 0$ has a symmetric extension.
Since RCAD cannot change the sign of $D_C$, no key can be distilled with RCAD followed by one-way postprocessing if the initial states satisfies $D_C \leq 0$. 

To apply the $D_C = 0$ threshold to the standard \index{six-state protocol}six-state and BB84\index{BB84 protocol} scenario, 
let us assume that Alice and Bob discard the data specifying which bits come from which bases, 
meaning the error rates in the different bases are identical.
For the six-state scheme, only one possible state is consistent with the observed 
error rate $Q$, namely $p_x = p_y = p_z = Q/2$. 
This immediately yields 
$Q_\textrm{max} = (5-\sqrt{5})/10 \approx 27.64 \%$ for $D_C = 0$.

From the BB84 measurements, only the error rates in the $x$ and $z$ basis can be observed, 
meaning the state is not completely determined, 
only that $Q=Q_x=Q_z$ for $Q_x := p_y + p_z$ and $Q_z := p_x + p_y$.
If $Q$ is below $1/2$ the possible eigenvalues are
$(1-2Q, Q, 0, Q) + t(1,-1,1,-1)$ for $t \in [0,Q/2]$.
When expressed in terms of $(\alpha_1,\alpha_2,\alpha_3)$, 
this becomes  $(1-2Q,\sqrt{2}(1-3Q),\sqrt{2}Q)+t(0,2\sqrt{2},-2\sqrt{2})$.
To determine if $D_C \leq 0$ for any of the possible states, 
we minimize $2^{D_C} = 2\alpha_2^2/(1-\alpha_1^2)$. 
This amounts to minimizing $|\alpha_2|$,
since $\alpha_1$ is fixed by $Q$, 
and it is obvious by inspection that $t=0$ gives the minimum. 
Solving $D_C = 0$, we find $Q_\textrm{max} = 1/5 = 20 \%$.

\chapter[LAD with two encoded bits does not improve threshold]{Linear advantage distillation with two encoded bits does not improve threshold}
\label{ch:k2analytical}
\markboth{Chapter \thechapter. LAD with two encoded bits \ldots}{}

In the previous chapter we analyzed RCAD (i.e. LAD with $k=1$) on bits that could come from any Bell-diagonal state and showed that for the set of states for which $D_C \geq 0$ the states did not get rid of their symmetric extension.
This showed that the well known QBER thresholds of $20.00 \%$ for the BB84 protocol and $27.64 \%$ for the six-state protocol cannot be improved without changing the advantage distillation step.
In this chapter we analyze the next logical step---LAD with $k=2$. 
Where we for $k=1$ had only one irreducible parity check matrix $\mathbf{H}$ for each blocksize, we now have several inequivalent parity check matrices for the same block size. 
At the same time we reduce the scope to include only the six-state protocol and show that no LAD scheme with $k=2$ can break the symmetric extension at or above the known RCAD QBER error threshold $27.64\%$.

In a six-state QKD setup with an average quantum bit error rate $Q$ in the three bases, 
one possible state that an attacker may give to Alice and Bob is many identical copies of the isotropic state
\begin{equation}
  \label{eq:6stateisotropic}
  \rho^{AB} = (1- 3p) \proj{\Phi^+} + p \proj{\Psi^+} + p \proj{\Psi^-} + p \proj{\Phi^-}
\end{equation}
with $Q = 2p$.
In order to show that the RCAD threshold cannot be improved, we start with the the lowest possible QBER (i.e. the best possible state) from which RCAD cannot produce a secret key. 
We get this with $p = (5-\sqrt{5})/20$ in Eq.~\eqref{eq:6stateisotropic} which corresponds to $Q = 2p \approx 27.64 \%$.
We show that for any  $(n-2) \times n$ parity check matrix $\mathbf{H}$ used in LAD on this state, the output state has a specific form. 
We then find an explicit symmetric extension for all states of that form.

\section{Form of state after LAD with \texorpdfstring{$k=2$}{k=2}}

The first step in showing that states on the threshold keep their symmetric extension is to show that any linear advantage distillation with $k=2$ can be broken down into RCAD on separate blocks, possibly followed by announcing the parity of the resulting bits. 
The following lemma makes this statement more precise.
\begin{lemma}
\label{lem:twoencodedsimp}
 Any $(n,2)$ linear advantage distillation announcement is equivalent to one of the following procedures.
 \begin{itemize}
 \item[1.] Divide the block into two subblocks of $n_1$ and $n_2$ bits ($n_i \geq 1, n_1 + n_2 = n$). Perform repetition code advantage distillation on each block separately.
 \item[2.] Divide the block into three subblocks of $n_1$, $n_2$, and $n_3$ $(n_i \geq 1, n_1 + n_2 + n_3 = n$). Perform repetition code advantage distillation on each block separately. 
 Then announce the parity of the three remaining bits.
  \end{itemize}
If a block contains only one bit, the repetition code advantage distillation step announces nothing and leaves the bit untouched.
\end{lemma}
The proof for this lemma can be found in appendix \ref{proof:twoencodedsimp}.

The breakdown into repetition code advantage distillation followed by a single announced bit is helpful since the former has been well studied before in the context of QKD \cite{gottesman03a, chau02a, acin06a,bae07a,myhr09a} and the latter operates only on three bits and is the same for all parity check matrices.
We are here interested in whether $k=2$ linear advantage distillation could break the symmetric extension where repetition code advantage distillation ($k=1$) cannot, 
so from here on we only consider announcements of the generic type 2 in lemma \ref{lem:twoencodedsimp}.

Alice and Bob start out with $n$ copies of $\rho^{AB}$ from Eq.\eqref{eq:6stateisotropic} with $p = (5 - \sqrt{5})/20$, the best state in terms of error rate that cannot be distilled with repetition code advantage distillation.
They then divide the block into three subblocks of $\{n_i\}_{i=1}^3$ and perform repetition code advantage distillation on each subblock. 
The following lemma gives the form of the output state of such a subblock after this step.
\begin{lemma}
\label{lem:RCADoutform}
 After performing successful repetition code advantage distillation on a block of $n$ copies of the state \eqref{eq:6stateisotropic} with $p = (5-\sqrt{5})/20$, the output state is a 
 Bell-diagonal state 
 \begin{equation}
  \label{eq:belldiagstate}
  \rho = p_I \proj{\Phi^+} + p_x \proj{\Psi^+} + p_y \proj{\Psi^-} + p_z \proj{\Phi^-}
\end{equation}
with
\begin{subequations}
\label{eq:RCADoutparametrization}
 \begin{align}
\label{eq:RCADoutparametrizationI}
  p_I &= \tfrac{1}{4}(1 + \cos \theta + \sin \theta)\\
\label{eq:RCADoutparametrizationX}
  p_x &= \tfrac{1}{4}(1 - \cos \theta)\\
\label{eq:RCADoutparametrizationY}
  p_y &= \tfrac{1}{4}(1 - \cos \theta)\\
\label{eq:RCADoutparametrizationZ}
  p_z &= \tfrac{1}{4}(1 + \cos \theta - \sin \theta)
 \end{align}
\end{subequations}
for $\theta$ such that $\cos \theta = \frac{1}{\sqrt{5}}$ for $n=1$ (i.e. nothing is announced),
$\cos \theta = \frac{\sqrt{5}}{3}$ for $n = 2$, 
and $\cos \theta \geq \frac{2}{\sqrt{5}}$ for $n \geq 3$. 
\end{lemma}
The proof for this lemma can be found in appendix \ref{proof:RCADoutform}.

The next step is that Alice and Bob announce the parity of the three remaining bits and postselect on equal outcome.
The following lemma gives the form of the $2 \times 2$-qubit Bell-diagonal state after such a procedure on a general $2 \times 3$-qubit Bell-diagonal state.


\begin{lemma}
\label{lem:threeparityoutput}
 Consider a $2 \times 3$-qubit Bell-diagonal state 
 \begin{equation}
\label{eq:twotimestwobelldiag}
 \sigma^{A_1 B_1 A_2 B_2 A_3 B_3} 
= \sum_{i \cdots j} p_{ijk} \proj{\beta_{i}}_{A_1 B_1} \otimes \proj{\beta_{j}}_{A_2 B_2} \otimes \proj{\beta_{k}}_{A_3 B_3}
\end{equation}
where $i,j,k \in \{I,x,y,z\}$.
After successful linear advantage distillation with the parity check matrix 
 \begin{equation}
  \mathbf{H} = \begin{bmatrix} 1 & 1 & 1 \end{bmatrix}
 \end{equation}
implemented as described in section \ref{sec:quantumimplementation} the resulting $2\times2$-qubit state is characterized by the new eigenvalues
\begin{subequations}
\begin{align}
 \label{eq:firstthreeparityoutput}
 p_{II}^\textrm{out} = p_{III} + p_{zzz}\\
 p_{Ix}^\textrm{out} = p_{Ixx} + p_{zyy}\\
 p_{Iy}^\textrm{out} = p_{Iyx} + p_{zxy}\\
 p_{Iz}^\textrm{out} = p_{IzI} + p_{zIz}\\
 p_{xI}^\textrm{out} = p_{xIx} + p_{yzy}\\
 p_{xx}^\textrm{out} = p_{xxI} + p_{yyz}\\
 p_{xy}^\textrm{out} = p_{xyI} + p_{yxz}\\
 p_{xz}^\textrm{out} = p_{xzx} + p_{yIy}\\
 p_{yI}^\textrm{out} = p_{yIx} + p_{xzy}\\
 p_{yx}^\textrm{out} = p_{yxI} + p_{xyz}\\
 p_{yy}^\textrm{out} = p_{yyI} + p_{xxz}\\
 p_{yz}^\textrm{out} = p_{yzx} + p_{xIy}\\
 p_{zI}^\textrm{out} = p_{zII} + p_{Izz}\\
 p_{zx}^\textrm{out} = p_{zxx} + p_{Iyy}\\
 p_{zy}^\textrm{out} = p_{zyx} + p_{Ixy}\\
 \label{eq:lastthreeparityoutput}
 p_{zz}^\textrm{out} = p_{zzI} + p_{IIz}
\end{align}
\end{subequations}
which are normalized such that their sum is the probability of success.
\end{lemma}

The proof for this lemma can be found in appendix \ref{proof:threeparityoutput}.

In our case, the $2 \times 3$ qubit Bell-diagonal state has a special form.
It consists of three independent Bell-diagonal states, so the probability distribution is a product distribution, $p_{ijk} = q_i r_j s_k$.
The distributions $q_i$, $r_j$, and $s_k$ are of the form given by Eqs. \eqref{eq:RCADoutparametrizationI}-\eqref{eq:RCADoutparametrizationZ}, each specified by one angle.
We let the three angles be $\theta$, $\phi$, and $\alpha$, respectively.
After the final part of the announcement we end up with a $2 \times 2$-qubit state, specified by the three parameters $\theta, \phi,$ and $\alpha$.
The following lemma characterizes the state after the final announcement of the final parity bit.

\begin{lemma}
\label{lem:LAD2outform}
Given three two-qubit systems, each in a state of the form of lemma \ref{lem:RCADoutform}, but with three different angles, $\theta$, $\phi$, and $\alpha$.
 After successful linear advantage distillation with the parity check matrix 
 \begin{equation}
  \mathbf{H} = \begin{bmatrix} 1 & 1 & 1 \end{bmatrix}
 \end{equation}
 the resulting state on $\mathbb{C}^4 \otimes \mathbb{C}^4$ has nonzero matrix elements
 \begin{subequations}
 \begin{align}
  \label{eq:firstABstate}
  \rho_{00,00} = \rho_{11,11} = \rho_{22,22} = \rho_{33,33} 
  &= \frac{1}{32} (1+\cos \theta)(1+\cos \phi)(1+\cos \alpha)\\
  \rho_{00,11} = \rho_{11,00} = \rho_{22,33} = \rho_{33,22} 
  &= \frac{1}{32} (1+\cos \theta) \sin \phi \sin \alpha\\
  \rho_{00,22} = \rho_{11,33} = \rho_{22,00} = \rho_{33,11} 
  &= \frac{1}{32} \sin \theta (1+\cos \phi) \sin \alpha\\
  \rho_{00,33} = \rho_{11,22} = \rho_{22,11} = \rho_{33,00} 
  &= \frac{1}{32} \sin \theta \sin \phi (1+\cos \alpha)\\
  \rho_{01,01} = \rho_{10,10} = \rho_{23,23} = \rho_{32,32} 
  &= \frac{1}{32} (1+\cos \theta)(1-\cos \phi)(1-\cos \alpha)\\
  \rho_{02,02} = \rho_{13,13} = \rho_{20,20} = \rho_{31,31} 
  &= \frac{1}{32} (1-\cos \theta)(1+\cos \phi)(1-\cos \alpha)\\
  \label{eq:lastABstate}
  \rho_{03,03} = \rho_{12,12} = \rho_{21,21} = \rho_{30,30} 
  &= \frac{1}{32} (1-\cos \theta)(1-\cos \phi)(1+\cos \alpha)
 \end{align}
 \end{subequations}
before normalization.
We have here labeled the basis of each four-dimensional subsystem in the usual way, 
$\ket{0}_A := \ket{00}_{A_1 A_2}$, $\ket{1}_A := \ket{01}_{A_1 A_2}$, $\ket{2}_A := \ket{10}_{A_1 A_2}$, and $\ket{3}_A := \ket{11}_{A_1 A_2}$, and similarly for $B$.
\end{lemma}

The proof for this lemma can be found in appendix \ref{proof:LAD2outform}.

\section{Simplifying the symmetric extension problem}
\label{sec:simplifying-k2analytical}

We now want to simplify the symmetric extension problem for a state with nonzero matrix elements of the form \eqref{eq:firstABstate}-\eqref{eq:lastABstate}.
Our ultimate goal in this chapter is to find a symmetric extension for all possible values of the parameters $\theta, \phi,$ and $\alpha$. 
We therefore only need a simplified condition that implies a symmetric extension, but not necessarily that a symmetric extension would imply that the simplified condition holds (as long as we can show the simplified condition to be satisfied).
Nevertheless, in this section we show that a state of this form has a symmetric extension \emph{if and only if} a $4 \times 4$ matrix satisfying certain constraints can be made positive semidefinite.

The $2\times 2$-qubit state is 16-dimensional and its symmetric extension is a $3 \times 2$-qubit state which is 64-dimensional and therefore represented by a $64 \times 64$ matrix. 
As argued in section \ref{sssec:lowerdimmatrices} we only need to consider one simultaneous eigenspace of $\sigma_z^{A_1} \otimes \sigma_z^{B_1} \otimes \sigma_z^{B'_1}$ 
and 
$\sigma_z^{A_2} \otimes \sigma_z^{B_2} \otimes \sigma_z^{B'_2}$
because of the Bell-diagonal symmetry. 
The intersection of the positive eigenspaces (which we call the $++$ subspace) is 16-dimensional and is spanned by the vectors
\begin{gather}
 \ket{000}, \nonumber \\
 \ket{110}, \ket{220}, \ket{330},\nonumber \\
 \ket{101}, \ket{202}, \ket{303}, \nonumber \\
 \ket{011}, \ket{022}, \ket{033}, \nonumber \\
 \ket{123}, \ket{231}, \ket{312}, \nonumber \\
 \ket{132}, \ket{213}, \ket{321}. \label{eq:basisorder}
\end{gather}
The subspace is closed under swapping $B$ and $B'$, so the symmetry constraint can be enforced within this subspace.
The marginal constraints---that $\rho^{ABB'}$ reduces to the correct $\rho^{AB}$ upon tracing out $B'$---involves all four subspaces, so we will now translate those elements that are outside this subspace into equal elements from this subspace.

Each of the seven equations \eqref{eq:firstABstate}-\eqref{eq:lastABstate} defines a marginal constraint. 
Let us first look at \eqref{eq:firstABstate}.
This says that 
\begin{multline}
 \rho_{000,000} + \rho_{001,001} + \rho_{002,002} + \rho_{003,003} = \rho_{00,00} \\
 = \frac{1}{32} (1+\cos \theta)(1+\cos \phi)(1+\cos \alpha).
\end{multline}
Of the four matrix elements on the left hand side, only $\rho_{000,000}$ is in the $++$ subspace. We can use symmetry operations to turn the other elements into elements on this subspace. 
Applying $\sigma_x^{A_2} \otimes \sigma_x^{B_2} \otimes \sigma_x^{B'_2}$ will swap the elements $\rho_{001,001}$ and $\rho_{110,110}$ and since the state is invariant under this operation the matrix elements must be equal.
By applying $\sigma_x$ on the other three qubits, we get that 
$\rho_{002,002} = \rho_{220,220}$ and $\sigma_x$ on all six qubits gives 
$\rho_{003,003} = \rho_{330,330}$.
Therefore, $\rho_{00,00} = \rho_{000,000} + \rho_{110,110} + \rho_{220,220} + \rho_{330,330}$.
It makes no difference if we start with other matrix elements of $\rho^{AB}$ that are equal to $\rho_{00}$, 
they end up with it being the sum of the same four matrix elements of the $++$ eigenspace part of $\rho^{ABB'}$.
The pattern here is that $\sigma_x$ on the three qubits in the second subsystem will swap indices $0 \leftrightarrow 1$ and $2 \leftrightarrow 3$, $\sigma_x$ on the three qubits in the first subsystem swap $0 \leftrightarrow 2$ and $1 \leftrightarrow 3$, and on all six qubits $0 \leftrightarrow 3$ and $1 \leftrightarrow 2$.
By doing this for all the seven marginal constraints \eqref{eq:firstABstate}-\eqref{eq:lastABstate}, we end up with the following constraints on $\rho^{ABB'}$.
\begin{align*}
 \label{eq:firstABBconstraint}
 \rho_{000,000} + \rho_{110,110} + \rho_{220,220} + \rho_{330,330} 
 &= \frac{1}{32} (1+\cos \theta)(1+\cos \phi)(1+\cos \alpha)\tag{$\diamondsuit$}\\
 \rho_{000,110} + \rho_{110,000} + \rho_{220,330} + \rho_{330,220} 
 &= \frac{1}{32} (1+\cos \theta)\sin \phi \sin \alpha\tag{$\heartsuit$}\\
 \rho_{000,220} + \rho_{110,330} + \rho_{220,000} + \rho_{330,110} 
 &= \frac{1}{32} \sin \theta(1+\cos \phi)\sin \alpha\tag{$\spadesuit$}\\
 \rho_{000,330} + \rho_{110,220} + \rho_{220,110} + \rho_{330,000} 
 &= \frac{1}{32} \sin \theta \sin \phi (1+\cos \alpha)\tag{$\clubsuit$}\\
 \rho_{011,011} + \rho_{101,101} + \rho_{231,231} + \rho_{321,321} 
 &= \frac{1}{32} (1+\cos \theta)(1-\cos \phi)(1-\cos \alpha)\tag{$\vartriangle$}\\
 \rho_{022,022} + \rho_{132,132} + \rho_{202,202} + \rho_{312,312} 
 &= \frac{1}{32} (1-\cos \theta)(1+\cos \phi)(1-\cos \alpha)\tag{$\triangledown$}\\
 \label{eq:lastABBconstraint}
 \rho_{033,033} + \rho_{123,123} + \rho_{213,213} + \rho_{303,303} 
 &= \frac{1}{32} (1-\cos \theta)(1-\cos \phi)(1+\cos \alpha)\tag{$\Box$}
\end{align*}
In addition, some sums are constrained to be zero (those corresponding to the 0 elements of $\rho^{AB}$). 
The labels are symbols from the matrix in figure \ref{fig:16x16constraints}, where each element from the $++$ subspace of $\rho^{ABB'}$ that enter in a marginal constraint is marked with a symbol identifying the constraint.
\begin{figure}[t]
\newcommand{\llabel}[1]{\makebox[0cm][r]{\scriptsize \ensuremath{\ket{#1}} \hspace*{1em}}}
\newcommand{\alabel}[1]{\makebox[0cm][l]{\rotatebox{90}{\makebox[0cm][l]{\hspace*{3ex}\scriptsize\ensuremath{\bra{#1}}}}}}
\vspace{3ex} 
\begin{equation*}
\label{page:constraintillustration}
\left[
\begin{array}{l|lll|lll|lll|lll|lll}
  \llabel{000}\alabel{000}\diamondsuit  &  \alabel{110}\heartsuit  &  \alabel{220}\spadesuit  &  \alabel{330}\clubsuit  &
  \alabel{101} & \alabel{202} & \alabel{303} &  
  \alabel{011} & \alabel{022} & \alabel{033} & 
  \alabel{123} & \alabel{231} & \alabel{312} &
  \alabel{132} & \alabel{213} & \alabel{321}  \\
 \hline
  \llabel{110}\heartsuit  &  \diamondsuit  &  \clubsuit  &  \spadesuit  & & & & & & & & & &    &     & \\
  \llabel{220}\spadesuit  &  \clubsuit  &  \diamondsuit  &  \heartsuit  & & & & & & & & & &    &     & \\
  \llabel{330}\clubsuit  &  \spadesuit  &  \heartsuit  &  \diamondsuit  & & & & & & & & & &    &     & \\
 \hline
   \llabel{101}& & & &          \vartriangle    & & & \oa &     &     &     & \oh &     &     &     & \og \\
   \llabel{202}& & & & &        \triangledown     & &     & \od &     &     &     & \oc & \oi &     &     \\
   \llabel{303}& & & & & &      \Box                &     &     & \oj & \of &     &     &     & \ob &     \\
   \hline
   \llabel{011}& & & & \oa &     &     &  \vartriangle    & & &     & \og &     &     &     & \oh \\
   \llabel{022}& & & &     & \od &     & &  \triangledown   & &     &     & \oi & \oc &     &     \\
   \llabel{033}& & & &     &     & \oj & & &  \Box            & \ob &     &     &     & \of&     \\
   \hline
   \llabel{123}& & & &     &     & \of &     &     & \ob &  \Box                & &     & & \oj &     \\
   \llabel{231}& & & & \oh &     &     & \og &     &     &     &  \vartriangle   & &      &     & \oa \\
   \llabel{312}& & & &     & \oc &     &     & \oi &     &     &     &  \triangledown  & \od &  &     \\
   \hline
   \llabel{132}& & & &     & \oi &     &     & \oc &     &     & & \od &  \triangledown  &     &     \\
   \llabel{213}& & & &     &     & \ob &     &     & \of & \oj &     &       & &  \Box    &     \\
   \llabel{321}& & & & \og &     &     & \oh &     &     &     & \oa & &     & &  \vartriangle        
\end{array}
\right]
\end{equation*}
\caption[
Illustration of the constraints on the $++$ subspace of $\rho^{ABB'}$.
]{
Illustration of the constraints on the $++$ subspace of $\rho^{ABB'}$. 
Each matrix element that enter in a constraint is marked by a symbol corresponding to that constraint.
}
\label{fig:16x16constraints}
\end{figure}
We also need the matrix to be invariant under swapping systems $B$ and $B'$. 
This corresponds to swapping rows and columns 2--4 with 5--7 and swapping 11--13 with 14--16. 

The circular symbols in figure \ref{fig:16x16constraints} (i.e. the off-diagonal symbols in rows/columns 5--16) are for constraints where the sum of the matrix elements are zero.
If a positive semidefinite matrix that satisfies all the constraints can be found, also the matrix where all these elements and all the off-diagonal elements in columns 8--16 (together with their transpose) are set to zero will be positive and fulfill all constraints. 
We therefore only need to look for a matrix where rows and columns 8--16 has nonzero elements only on the diagonal\footnote{
The $S^\dagger \otimes S \otimes S$ symmetry discussed in section \ref{sec:additionalsymmetry} can also be used to argue that some of these elements can be set to zero, but this symmetry argument is not sufficient for all the elements.
}.

If such a matrix can be found, also the matrix we get by setting the diagonal elements in columns 11--16  to zero will be positive semidefinite. 
It will also satisfy the swap symmetry, but we have to add the removed elements to the diagonal elements in columns 8--10 (which are not constrained by swap symmetry) in order to fulfill the constraints $\vartriangle, \triangledown$, and $\Box$ (remember that the swap symmetry demands that the diagonal elements 11-13 are equal to the diagonal elements 14-16, which is of course fulfilled when all are set to zero).
We can therefore limit the search to symmetric matrices where columns 11--16 are zero and columns 8--10 are nonzero only on the diagonal.

Now, the diagonal constraints represented by $\vartriangle$, $\triangledown$, and $\Box$ are only that the sum of two elements should have a given value. 
Since the positivity of the matrix is unaffected as long as the diagonal elements 8--10 are nonnegative, the constraints translate into an upper bounds on the diagonal elements 5--7.
With these upper bounds, only the upper left $7 \times 7$ part of the matrix is relevant.
 
Because of the swap symmetry, the two remaining $3\times 3$ blocks on the diagonal (2--4 and 5--7) have to be equal. 
With the swap symmetry constraint, the upper left $4 \times 4$ matrix can be represented as
\begin{equation}
\label{eq:4x4constraints}
\left[
\begin{array}{c|ccc}
 \diamondsuit & \heartsuit & \spadesuit & \clubsuit \\
 \hline
 \heartsuit & (\diamondsuit\vartriangle) & \clubsuit & \spadesuit \\
 \spadesuit & \clubsuit & (\diamondsuit\triangledown) & \heartsuit \\
 \clubsuit & \spadesuit & \heartsuit & (\diamondsuit\Box) \\
\end{array}
\right].
\end{equation}
The parenthesis on the diagonal means that these elements are subject to two constraints, 
one being the sum of all four diagonal elements and in addition each of the individual elements except $(1,1)$ has an upper bound.
We can now write out the constraints on this subspace spanned by $\ket{000}$, $\ket{110}$, $\ket{220}$, and $\ket{330}$.
\begin{gather*}
 \label{eq:firstABBconstraint4x4}
 \rho_{000,000} + \rho_{110,110} + \rho_{220,220} + \rho_{330,330} 
 = \frac{1}{32} (1+\cos \theta)(1+\cos \phi)(1+\cos \alpha)\tag{$\diamondsuit$}\\
 \rho_{000,110} + \rho_{110,000} + \rho_{220,330} + \rho_{330,220} 
 = \frac{1}{32} (1+\cos \theta)\sin \phi \sin \alpha\tag{$\heartsuit$}\\
 \rho_{000,220} + \rho_{110,330} + \rho_{220,000} + \rho_{330,110} 
 = \frac{1}{32} \sin \theta(1+\cos \phi)\sin \alpha\tag{$\spadesuit$}\\
 \rho_{000,330} + \rho_{110,220} + \rho_{220,110} + \rho_{330,000} 
 = \frac{1}{32} \sin \theta \sin \phi (1+\cos \alpha)\tag{$\clubsuit$}\\
 \label{eq:firstABBineqconstraint}
 \rho_{110,110}  
 \leq \frac{1}{32} (1+\cos \theta)(1-\cos \phi)(1-\cos \alpha)\tag{$\vartriangle'$}\\
 \rho_{220,220} 
 \leq \frac{1}{32} (1-\cos \theta)(1+\cos \phi)(1-\cos \alpha)\tag{$\triangledown'$}\\
 \label{eq:lastABBconstraint4x4}
 \rho_{330,330} 
 \leq \frac{1}{32} (1-\cos \theta)(1-\cos \phi)(1+\cos \alpha)\tag{$\Box'$}
\end{gather*}
If these constraints can be satisfied by a positive semidefinite real symmetric $4 \times 4$ matrix
\begin{equation}
 \mathbf{M_4} = \left[
 \begin{array}{c|ccc}
   B & a & b & c\\
   \hline
   a &   &   &  \\
   b &   & \mathbf{M_3} & \\
   c &   &   &  \\
 \end{array}
 \right],
\end{equation}
a positive semidefinite matrix on the 7-dimensional subspace spanned by $\ket{000}$, $\ket{110}$, $\ket{220}$, $\ket{330}$, $\ket{101}$, $\ket{202}$, and $\ket{303}$ is 
\begin{equation}
 \mathbf{M_7} = \left[
 \begin{array}{c|ccc|ccc}
   B & a & b & c & a & b & c\\
   \hline
   a &   &                &  &  &                & \\
   b &   & \mathbf{M_3} &  &  & \mathbf{M_3} & \\
   c &   &                &  &  &                & \\
   \hline
   a &   &                &  &  &                & \\
   b &   & \mathbf{M_3} &  &  & \mathbf{M_3} & \\
   c &   &                &  &  &                & \\
 \end{array}
 \right].
\end{equation}
This satisfies both the marginal constraints and the swap symmetry constraint.
From this we can go back and reconstruct the $16 \times 16$ matrix by padding with zeros, except for the diagonal elements 8--10, which may have to be positive to fulfill the constraints. 
From this matrix, we can again reconstruct the full 64-dimensional (unnormalized) density matrix $\rho^{ABB'}$.
Therefore, a symmetric extension exists if and only if a real symmetric $4 \times 4$ matrix with the constraints given by \eqref{eq:4x4constraints} and $\diamondsuit$--$\Box'$ can be found.

\section{Finding an explicit symmetric extension}
\label{sec:explicitextension}


We now go on to show that for any output state after linear advantage distillation with $k=2$, we can find a positive semidefinite $4 \times 4$ matrix $\mathbf{M_4}$ that satisfies the constraints, 
and therefore the state has a symmetric extension.
We present two different forms of this matrix, which work in different regimes. 
The first one that turns out to work whenever the blocksize is $n \geq 5$  is 
\begin{equation}
\label{eq:lowerdiagform}
 \mathbf{M_4^{(1)}} = 
 \frac{1}{32}\left[
 \begin{array}{c|ccc}
   B & a & b & c\\
   \hline
   a & x & 0 & 0\\
   b & 0 & y & 0\\
   c & 0 & 0 & z\\
 \end{array}
 \right],
\end{equation}
where the constraints define the values of the parameters. 
The only additional choice we make is that the constraints \eqref{eq:firstABBineqconstraint}-\eqref{eq:lastABBconstraint4x4} be saturated.
The parameters then have to be
 \begin{align}
 B &= (1+\cos \theta)(1+ \cos \phi)(1 + \cos \alpha) \nonumber \\
   & \quad - (1 - \cos \theta)(1 + \cos \phi)(1 - \cos \alpha) \nonumber \\
   & \quad - (1 - \cos \theta)(1 - \cos \phi)(1 + \cos \alpha) \nonumber\\
   & \quad - (1 + \cos \theta)(1 - \cos \phi)(1 - \cos \alpha), \\
 a &= \frac{1}{2} (1 + \cos \theta) \sin \phi \sin \alpha, \\
 b &= \frac{1}{2} \sin \theta (1 + \cos \phi) \sin \alpha, \\
 c &= \frac{1}{2} \sin \theta \sin \phi (1 + \cos \alpha), \\
 x &= (1 + \cos \theta)(1 - \cos \phi)(1 - \cos \alpha),\\
 y &= (1 - \cos \theta)(1 + \cos \phi)(1 - \cos \alpha),\\
 z &= (1 - \cos \theta)(1 - \cos \phi)(1 + \cos \alpha).
 \end{align}
 Since $x$, $y$, and $z$ always are positive, the matrix is positive definite whenever its determinant is positive.
 The determinant of this matrix (without $\frac{1}{32}$) is 
 \begin{align}
  \det(32 \mathbf{M_4^{(1)}}) 
  &= Bxyz - a^2 yz - b^2 xz - c^2 xy \nonumber \\
  &= (1+\cos \theta)(1+\cos \phi)(1+\cos \alpha) \nonumber\\
  &\quad \times (1-\cos \theta)^2(1-\cos \phi)^2(1-\cos \alpha)^2 \nonumber\\
  &\quad \times \Big[ \frac{1}{4} (1 +\cos \theta)(1 + \cos \phi)(1 + \cos \alpha) \nonumber\\
                 &\quad \quad - (1 - \cos \theta)(1 + \cos \phi)(1 - \cos \alpha) \nonumber\\
                 &\quad \quad - (1 - \cos \theta)(1 - \cos \phi)(1 + \cos \alpha) \nonumber\\
                 &\quad \quad - (1 + \cos \theta)(1 - \cos \phi)(1 - \cos \alpha)
                \Big]
 \end{align}
So the matrix is positive definite whenever
\begin{align}
\label{eq:diagonalpositivity}
 f(\cos\theta,\cos \phi, \cos \alpha) &:= \frac{1}{4} (1 +\cos \theta)(1 + \cos \phi)(1 + \cos \alpha) \nonumber \\
                  &\quad - (1 - \cos \theta)(1 + \cos \phi)(1 - \cos \alpha) \nonumber\\
                  &\quad - (1 - \cos \theta)(1 - \cos \phi)(1 + \cos \alpha) \nonumber\\
                  &\quad - (1 + \cos \theta)(1 - \cos \phi)(1 - \cos \alpha) > 0.
\end{align}

On blocksize 5 there are two possible announcements of type 2 from lemma \ref{lem:twoencodedsimp}. 
The five qubits can be split into three subblocks like (2,2,1) or (3,1,1).
In the first case, two of the angles satisfy $\cos \theta = \frac{\sqrt{5}}{3}$ and one $\cos \theta = \frac{1}{\sqrt{5}}$ (see lemma \ref{lem:RCADoutform}).
In that case, the function in Eq. \eqref{eq:diagonalpositivity} evaluates to 
$f(\frac{\sqrt{5}}{3},\frac{\sqrt{5}}{3}, \frac{1}{\sqrt{5}}) = (140-44\sqrt{5})/36\sqrt{5} > 0$.
For the case (3,1,1), $f(\frac{2}{\sqrt{5}},\frac{1}{\sqrt{5}},\frac{1}{\sqrt{5}}) = 78 - 30\sqrt{5} > 0$.
Any announcements on larger blocks can be represented by increasing the size of one or more subblocks from these announcements. 
Lemma \ref{lem:RCADoutform} states that this will lead to larger $\cos \theta$, $\cos \phi$, or $\cos \alpha$.
If $f$ is monotonically increasing in each of its inputs, then the matrix of this form is positive for any higher blocksize.
We can compute the partial derivative of $f(x,y,z)$
\begin{equation}
 \frac{\partial f}{\partial x} = \frac{1}{4}(1+y)(1+z) + (1+y)(1-z) + 2(1-y)z
\end{equation}
and this is positive since $y,z \in (0,1)$.
Since $f$ is invariant under any permutation of its inputs, the same holds for the other inputs.
Therefore, for any output state after linear advantage distillation with $k=2$ and blocks of size 5 or larger, we can construct a symmetric extension from a matrix of the form \eqref{eq:lowerdiagform}.

The only remaining cases to cover now are blocksize 3 and 4. 
For blocksize 3, the parity of the three bits are announced directly, and for blocksize 4, two of the bits first go through a round of repetition code advantage distillation. 
For these cases, the form  \eqref{eq:lowerdiagform} is not positive semidefinite.
But we can use the fact that in both cases two of the subblocks are of size 1.
Let this correspond to the two angles $\phi$ and $\alpha$ which now satisfy
$\cos \phi = \cos \alpha = \frac{1}{\sqrt{5}}$.
We can now use a matrix of the form
\begin{equation}
\label{eq:equalangleform}
 \mathbf{M_4^{(2)}} = 
 \frac{1}{32}\left[
 \begin{array}{c|ccc}
   B & a & b & b\\
   \hline
   a & x & 0 & 0\\
   b & 0 & y & y\\
   b & 0 & y & y\\
 \end{array}
 \right].
\end{equation}
where the parameters are
\begin{align}
 B &= \frac{4}{5}[\sqrt{5}(1+\cos\theta) - 2(1-\cos\theta)], \\
 a &= \frac{2}{5} (-1 + 3\cos \theta), \\
 b &= \frac{1}{5} (\sqrt{5}+1)\sin \theta, \\
 x &= \frac{2}{5}(3-\sqrt{5})(1 + \cos \theta),\\
 y &= \frac{4}{5}(1-\cos\theta).
\end{align}
Since the third and fourth row and column are equal, we only need to check positivity of the matrix
\begin{equation}
\label{eq:equalangleformreduced}
 \mathbf{M_3^{(2)}} = 
 \frac{1}{32}\left[
 \begin{array}{c|cc}
   B & a & b\\
   \hline
   a & x & 0\\
   b & 0 & y\\
 \end{array}
 \right].
\end{equation}
Again, the matrix is positive definite if and only if its determinant is positive.
We get 
\begin{multline}
  \det(32 \mathbf{M_3^{(1)}}) 
  = Bxy - a^2 y - b^2 x \\
  = \frac{32}{5^3}(1-\cos\theta)[(\sqrt{5}-4)\cos^2\theta + (6\sqrt{5}-8)\cos\theta + (5\sqrt{5}-12)] 
\end{multline}
This is positive for 
\begin{equation}
 \cos \theta > \frac{3\sqrt{5} - 4 - 2 \sqrt{2\sqrt{5}-3}}{4-\sqrt{5}} \approx 0.16
\end{equation}
which is more than good enough, since we always have $\cos \theta \geq \frac{1}{\sqrt{5}} \approx 0.45$.
Hence, we can construct a symmetric extension also in the case of blocksize 3 and 4.

\chapter{Numerical results for 3 and 4 encoded bits}
\label{ch:numerical}

In this chapter we report on numerical results indicating that using a linear code with few encoded bits and moderate block size instead of the repetition code normally used in advantage distillation will not increase the tolerable error rate in the six-state protocol.
More precisely, we start with a block of $n$ qubit pairs in the isotropic state 
\begin{equation}
  \label{eq:6stateisotropicnumerical}
  \rho = (1- 3p) \proj{\Phi^+} + p \proj{\Psi^+} + p \proj{\Psi^-} + p \proj{\Phi^-}
\end{equation}
where $p$ depends on $n$ and is chosen such that it gives a useful comparison to repetition code advantage distillation on the same blocksize.
For given $n$ and $m$ we go through all inequivalent parity check matrices of size $m \times n$ and compute the $2k$-qubit state ($k = n-m$) after using each parity check matrix for advantage distillation.
For the resulting state, we use a numerical SDP solver to verify that it has a symmetric extension.

The starting state---as parametrized by $p$---is chosen to be less noisy than in the analytical calculations in the previous chapter.
In that case we chose the best possible isotropic state such that repetition code advantage distillation could not break the symmetric extension \emph{for any blocksize}.
Then we showed analytically that no linear advantage distillation with $k=2$ could break the symmetric extension for this state, again for any blocksize.
In this case we are dealing with a finite blocksize $n$.
It would be of very limited value to have a result that said that given a state for which RCAD can almost break the symmetric extension in the limit of infinite block size, no parity check matrix on a small block is able to break the symmetric extension.
We therefore choose the starting state such that RCAD almost breaks symmetric extension on blocksize $n$. 
More precisely we choose the minimum $p$ such that if we perform RCAD on $n$ copies of $\rho$ in Eq.~\eqref{eq:6stateisotropicnumerical} the resulting state has a symmetric extension.
This means that the state after RCAD on blocksize $n$ is on the surface of the set of states with a symmetric extension.
The results show that after linear advantage distillation with all other parity check matrices on $n$ copies of that starting state, the resulting state is in the interior of the set of states with a symmetric extension.
We therefore conclude that for the tested blocksizes, no other linear advantage distillation is as good as RCAD.

Since we are dealing with numerical calculations, they are limited by numerical precision and also run-time and memory limitations.
As the blocksize increases, the distance between states after advantage distillation gets smaller and approaches the numerical precision, while the number of inequivalent parity check matrices and the run-time required to compute the state after advantage distillation increases.
We therefore only do computations up to blocksize 9.

We use the equivalence relations from section~\ref{sec:permequivalent} to skip many parity check matrices that are equivalent to one we already checked. 
However, we do not take care to only check one single matrix from each equivalence class, since  the overhead of checking a few equivalent matrices
is not big compared to 
the extra work of implementing thorough equivalence checking.

The resulting state after advantage distillation is computed using the implementation from section \ref{sec:quantumimplementation}.
After each \textindex{BXOR} (bilateral exclusive OR, i.e. CNOT on each side) the state remains in a $2n$-qubit Bell-diagonal state and also after measuring out some qubits and postselecting on equal outcomes on Bob's side, the state remains Bell-diagonal \cite{bennett96b}.

For the isotropic starting state that we consider, another symmetry is useful.
The starting state satisfies 
$\rho = (S^\dagger \otimes S)\rho(S^\dagger \otimes S)^\dagger$ for $S := \proj{0} + \im \proj{1}$ the phase gate, which rotates the Bloch sphere an angle $\pi/2$ around the $z$ axis.
The effect of this operation on a Bell-diagonal state is to swap the eigenvalues corresponding to $\ket{\Psi^+}$ and $\ket{\Psi^-}$.
We have that both $S \otimes \calone$ and $S^\dagger \otimes \calone$ commute with CNOT (where the first qubit is the source and the second is the target).
So if we start with two qubit-pairs in a state that satisfies 
\begin{equation}
 \rho^{A_1A_2 B_1B_2} = (S^\dagger \otimes \calone \otimes S \otimes \calone) \rho^{A_1A_2 B_1B_2} (S^\dagger \otimes \calone \otimes S \otimes \calone)^\dagger
\end{equation}
then the state after applying a CNOT from the first to the second subsystem on each side
\begin{multline}
 \sigma^{A_1A_2 B_1B_2} = \\ 
 (\textrm{CNOT}_{A_1 \rightarrow A_2} \otimes \textrm{CNOT}_{B_1 \rightarrow B_2}) \rho^{A_1A_2 B_1B_2} (\textrm{CNOT}_{A_1 \rightarrow A_2} \otimes \textrm{CNOT}_{B_1 \rightarrow B_2})^\dagger
\end{multline}
also satisfies 
\begin{equation}
 \sigma^{A_1A_2 B_1B_2} = (S^\dagger \otimes \calone \otimes S \otimes \calone) \sigma^{A_1A_2 B_1B_2} (S^\dagger \otimes \calone \otimes S \otimes \calone)^\dagger.
\end{equation}
In the implementation of the linear advantage distillation from section \ref{sec:quantumimplementation}, the qubit pairs that will remain after measurement are only used as source in the CNOTs, so they keep the $(S^\dagger \otimes S)$-symmetry, even after the BXOR part.
The only remaining part of LAD is the measurement and postselection based on the measurement result. 
This can be seen as a non-trace-preserving, completely positive map on the qubits that are measured, and since the symmetry operation only acts on the qubits that are not measured, the two operations also commute, and the final state after LAD is symmetric under $S^\dagger \otimes S$ on any of the remaining qubit pairs.

Having established the form of the state after LAD, we want to solve the symmetric extension problem for each state.
Since the state is a $2k$-qubit Bell-diagonal state with the additional $S^\dagger \otimes S$ symmetry, the symmetric extension problem can be simplified using the techniques in section \ref{sec:simplifying} and implemented as an SDP as described in section \ref{sec:sdpimplementation}. 
The SDP is then solved using Yalmip \cite{yalmip} on Matlab with the SDPT3 solver \cite{sdpt3}.

In the tables that follow are the results of the SDP calculations for $k = 3$ which means that the parity matrix $\mathbf{P}$ is a $(n-3) \times 3$ matrix. 
The first seven columns of the tables specify how many of the $n-3$ rows that are equal to each of the seven possible nonzero 3-bit strings (remember that the order of the rows in the parity matrix is irrelevant).
For example, row 4 in table \ref{tab:k3n6} represents a parity matrix $\mathbf{P}$ where one row is $\begin{bmatrix}1 & 0 & 0\end{bmatrix}$ and two rows are $\begin{bmatrix}0 & 1 & 1\end{bmatrix}$.
So the parity check matrix is
\begin{equation}
 \mathbf{H} = [\mathbf{P}|\calone] = 
\begin{bmatrix} 1 & 0 & 0 & 1 & 0 & 0\\ 0 & 1 & 1 & 0 & 1 & 0 \\ 0 & 1 & 1 & 0 & 0 & 1\end{bmatrix},
\end{equation}
which is actually a \textindex{reducible announcement} (see section \ref{sec:reducibleannouncements}), since it acts independently on two subblocks of two and four bits.
We find it useful to include reducible announcements in the table for comparison with the irreducible announcements.

The 8th row is the value of the parameter that is minimized in the semidefinite program.
It can be interpreted as being the minimum $t$ such that $t\calone^{ABB'}/2^{3N} + \rho^{ABB'}$ is positive semidefinite or equivalently such that the state $t\calone^{AB}/2^{2N} + \rho^{AB}$ (normalized to trace $1+t$) has a symmetric extension.
A negative value means that the state is in the interior of the set of states with symmetric extension, 
a positive value means that the state does not have a symmetric extension, and a value of zero means that the state is on the surface of the set of extendible states (see section \ref{sec:sdpimplementation} for more details). 
The results are ordered with an increasing $t$ as we go down the table.
\\[\intextsep]
\begin{minipage}{\linewidth}
\begin{center}
\tabcaption{Results for $k = 3$ and $n = 4$}
\begin{tabular}{cccccccc}
 \hline
 100 & 010 & 001 & 111 & 110 & 101 & 011 & $t$\\
 \hline
     &     &     &     &     &     &  1  & -0.1452\\
     &     &     &  1  &     &     &     & -0.0537\\
  1  &     &     &     &     &     &     & -0.0455\\
  \hline
\end{tabular}
\end{center}
\end{minipage}
\\[\intextsep]
\\[\intextsep]
\begin{minipage}{\linewidth}
\begin{center}
\tabcaption{Results for $k = 3$ and $n = 5$}
\begin{tabular}{cccccccc}
 \hline
 100 & 010 & 001 & 111 & 110 & 101 & 011 & $t$\\
 \hline
     &     &     &     &     &  1  &  1  & -0.0728\\
  1  &     &     &     &     &     &  1  & -0.0655\\
     &     &     &     &     &     &  2  & -0.0584\\
  1  &     &     &  1  &     &     &     & -0.0233\\
  1  &  1  &     &     &     &     &     & -0.0205\\
  2  &     &     &     &     &     &     & -0.0183\\
  \hline
\end{tabular}
\end{center}
\end{minipage}
\\[\intextsep]
\\[\intextsep]
\begin{minipage}{\linewidth}
\begin{center}
\tabcaption{Results for $k = 3$ and $n = 6$\label{tab:k3n6}}
\begin{tabular}{cccccccc}
 \hline
 100 & 010 & 001 & 111 & 110 & 101 & 011 & $t$\\
 \hline
     &     &     &     &  1  &  1  &  1  & -0.0821\\
  1  &     &     &     &  1  &  1  &     & -0.0577\\
  1  &     &     &     &     &  1  &  1  & -0.0298\\
  1  &     &     &     &     &     &  2  & -0.0283\\
  1  &     &     &     &     &  2  &     & -0.0258\\
  2  &     &     &     &     &     &  1  & -0.0254\\
  2  &     &     &     &     &  1  &     & -0.0221\\
  1  &  1  &     &  1  &     &     &     & -0.0100\\
  1  &  1  &  1  &     &     &     &     & -0.0093\\
  2  &  1  &     &     &     &     &     & -0.0083\\
  3  &     &     &     &     &     &     & -0.0071\\
  \hline
\end{tabular}
\end{center}
\end{minipage}
\\[\intextsep]
\\[\intextsep]
\begin{minipage}{\linewidth}
\begin{center}
\tabcaption{Some results for $k = 3$ and $n = 7$}
\begin{tabular}{cccccccc}
 \hline
 100 & 010 & 001 & 111 & 110 & 101 & 011 & $t$\\
 \hline
     &     &     &  1  &  1  &  1  &  1  & -0.0659\\
  1  &     &     &     &  1  &  1  &  1  & -0.0329\\
  1  &  1  &     &     &  1  &  1  &     & -0.0275\\
  1  &  1  &     &     &  2  &     &     & -0.0252\\
  2  &     &     &     &  1  &  1  &     & -0.0213\\
$\vdots$&$\vdots$&$\vdots$&$\vdots$&$\vdots$&$\vdots$&$\vdots$& $\vdots$\\
  1  &  1  &  1  &  1  &     &     &     & -0.0043\\
  2  &  1  &     &  1  &     &     &     & -0.0040\\
  2  &  1  &  1  &     &     &     &     & -0.0038\\
  3  &     &     &  1  &     &     &     & -0.0036\\
  2  &  2  &     &     &     &     &     & -0.0034\\
  3  &  1  &     &     &     &     &     & -0.0032\\
  4  &     &     &     &     &     &     & -0.0026\\
  \hline
\end{tabular}
\end{center}
\end{minipage}
\\[\intextsep]
\\[\intextsep]
\begin{minipage}{\linewidth}
\begin{center}
\tabcaption{Some results for $k = 3$ and $n = 8$}
\begin{tabular}{cccccccc}
 \hline
 100 & 010 & 001 & 111 & 110 & 101 & 011 & $t$\\
 \hline
  1  &     &     &  1  &  1  &  1  &  1  & -0.0245\\
  1  &  1  &     &  1  &  1  &     &  1  & -0.0136\\
  1  &  1  &     &  1  &     &  1  &  1  & -0.0136\\
  1  &  1  &     &  1  &  2  &     &     & -0.0132\\
  1  &  1  &  1  &     &  2  &     &     & -0.0131\\
  1  &  1  &  1  &     &  1  &  1  &     & -0.0128\\
  2  &     &     &     &  1  &  1  &  1  & -0.0124\\
$\vdots$&$\vdots$&$\vdots$&$\vdots$&$\vdots$&$\vdots$&$\vdots$& $\vdots$\\
  2  &  2  &  1  &     &     &     &     & -0.0016\\
  3  &  1  &  1  &     &     &     &     & -0.0015\\
  4  &     &     &  1  &     &     &     & -0.0014\\
  3  &  2  &     &     &     &     &     & -0.0013\\
  4  &  1  &     &     &     &     &     & -0.0012\\
  5  &     &     &     &     &     &     & -0.0009\\
  \hline
\end{tabular}
\end{center}
\end{minipage}
\\[\intextsep]
\\[\intextsep]
\begin{minipage}{\linewidth}
\begin{center}
\tabcaption{Some results for $k = 3$ and $n = 9$}
\begin{tabular}{cccccccc}
 \hline
 100 & 010 & 001 & 111 & 110 & 101 & 011 & $t$\\
 \hline
  1  &  1  &     &     &  2  &  1  &  1  & -0.01245\\
  1  &  1  &     &  1  &  1  &  1  &  1  & -0.01059\\
  1  &  1  &  1  &     &  1  &  1  &  1  & -0.00997\\
  2  &     &     &  1  &  1  &  1  &  1  & -0.00892\\
  2  &  1  &     &     &  2  &  1  &     & -0.00869\\
  1  &  1  &  1  &  1  &  1  &  1  &     & -0.00542\\
$\vdots$&$\vdots$&$\vdots$&$\vdots$&$\vdots$&$\vdots$&$\vdots$& $\vdots$\\
  4  &  1  &  1  &     &     &     &     & -0.00057\\
  5  &     &     &  1  &     &     &     & -0.00053\\
  3  &  3  &     &     &     &     &     & -0.00052\\
  4  &  2  &     &     &     &     &     & -0.00051\\
  5  &  1  &     &     &     &     &     & -0.00048\\
  6  &     &     &     &     &     &     & -0.00033\\
  \hline
\end{tabular}
\end{center}
\end{minipage}
\\[\intextsep]

For $k=4$, we only test blocksizes $n = 5$ and $n = 6$. 
In the former case there is only one \textindex{irreducible announcement}, namely the one that announces the parity of all bits on each side. 
Three other parity matrices are included for comparison.
In the latter case there are three equivalence classes that take into account the whole block (those that do not have a columns of only zeros) and do not reduce to two announcements on separate blocks (those that have at least one column of two ones).
Five other parity matrices are included for comparison in that case.
In the tables, the parity matrix and the minimum value of $t$ is listed.
\\[\intextsep]
\begin{minipage}{\linewidth}
\begin{center}
\tabcaption{Results for $k = 4$ and $n = 5$}
\begin{tabular}{cc}
 \hline
 Parity matrix & $t$\\
 \hline
  $\begin{bmatrix}
    1 & 1 & 1 & 1
   \end{bmatrix}$
  & -0.0924\\
  $\begin{bmatrix}
    1 & 1 & 0 & 0
   \end{bmatrix}$
  & -0.0847\\
  $\begin{bmatrix}
    1 & 1 & 1 & 0
   \end{bmatrix}$
  & -0.0301\\
  $\begin{bmatrix}
    1 & 0 & 0 & 0
   \end{bmatrix}$
  & -0.0256\\
  \hline
\end{tabular}
\end{center}
\end{minipage}
\\[\intextsep]
\\[\intextsep]
\begin{minipage}{\linewidth}
\begin{center}
\tabcaption{Results for $k = 4$ and $n = 6$}
\begin{tabular}{cc}
 \hline
 Parity matrix & $t$\\
 \hline
  $\begin{bmatrix}
    1 & 1 & 1 & 1\\ 0 & 0 & 1 & 1
   \end{bmatrix}$
  & -0.0430\\
  $\begin{bmatrix}
    1 & 0 & 0 & 0\\ 1 & 1 & 1 & 1
   \end{bmatrix}$
  & -0.0399\\
  $\begin{bmatrix}
    1 & 0 & 0 & 0\\ 0 & 1 & 1 & 0
   \end{bmatrix}$
  & -0.0373\\
  $\begin{bmatrix}
    1 & 0 & 0 & 0\\ 1 & 1 & 0 & 0
   \end{bmatrix}$
  & -0.0351\\
  $\begin{bmatrix}
    0 & 1 & 1 & 1\\ 1 & 1 & 1 & 0
   \end{bmatrix}$
  & -0.0151\\
  $\begin{bmatrix}
    1 & 0 & 0 & 0\\ 0 & 1 & 1 & 1
   \end{bmatrix}$
  & -0.0135\\
  $\begin{bmatrix}
    1 & 0 & 0 & 0\\ 0 & 1 & 0 & 0
   \end{bmatrix}$
  & -0.0115\\
  $\begin{bmatrix}
    1 & 0 & 0 & 0\\ 1 & 0 & 0 & 0
   \end{bmatrix}$
  & -0.0103\\
  \hline
\end{tabular}
\end{center}
\end{minipage}
\\[\intextsep]

The fact that $t < 0$ for all these parity matrices shows that on the tested blocksizes, RCAD is strictly better at breaking symmetric extension than any other LAD scheme. 
We can also make some further observations from the tables.

\begin{itemize}
 \item The worst and best parity matrix both get a value of $t$ closer to zero when the blocksize increases. Remember that that in this calculation, the starting state gets more and more noisy for increasing block size, so this is not merely due to larger blocksize detecting more errors.
 \item The worst equivalence class is usually the one which contains a parity matrix where all the rows are different (exception for $n = 9$).
 \item Some of the parity matrices are reducible to RCAD on separate subblocks (those where all rows are of Hamming weight one) and we therefore know from previous results that they cannot break the symmetric extension. We include those for comparison.
 \item The best parity matrix is the one that reduces to RCAD on a subset of the bit pairs that is as large as possible and the rest are ignored (i.e. RCAD with blocksize 1).
 \item The parity matrices that reduce to RCAD on $k$ distinct blocks are among the best ones.
 \item The best parity matrix that does not reduce to RCAD on distinct blocks has one single row of only ones and otherwise RCAD on distinct blocks (usually on a single block).
\end{itemize}

The pattern seems to be that the more similar a linear advantage distillation scheme is to repetition code advantage distillation, the closer it is to breaking the symmetric extension.

\chapter*{Concluding remarks}
\addcontentsline{toc}{chapter}{Concluding remarks} 
\markboth{Concluding remarks}{} 
\label{ch:concluding}

The main contributions of this thesis can be divided into two areas, corresponding to part \ref{part:symext} and part \ref{part:qkd}.
The first part is concerned with the classification of bipartite quantum states related to the concept of a symmetric extension. 
This part is independent of any applications.
In the second part, the six-state quantum key distribution protocol---and to some extent the BB84 protocol---is analyzed using symmetric extension as a tool to show that certain classes of two-way postprocessing cannot possibly result in a secret key beyond a certain noise threshold.

In chapter \ref{ch:symext} of part \ref{part:symext} we started by deriving some basic properties of states with a symmetric extension. 
One key property is, from corollary \ref{cor:pure-extendible-decomp}, that any state with a symmetric extension can be written as a convex combination of states with a pure symmetric extension.
We then showed in theorem \ref{thm:spectrumcondition} that for any state with a pure symmetric extension, its spectrum is equal to the spectrum of the reduced density operator on the $B$ system.
In the rest of this chapter we considered a special class of states called 
\textindex{multi-qubit Bell-diagonal state}s. 
For this class of states, the symmetry simplifies the symmetric extension problem considerably.

In chapter \ref{ch:twoqubits}, we specialized to look at systems consisting of two qubits. 
Everything in this chapter revolves around conjecture \ref{conj:twoqubitextension}, a conjectured formula to characterize two-qubit states with a symmetric extension.
The characterization of pure-extendible states simplify for two qubits, so that in this case a state has a pure symmetric extension if and only if its local and global spectrum is equal (theorem \ref{thm:necsufqubit}).
We also presented examples that show that this only happens for two qubits and not for any other dimension.
Then follows a series of proofs of conjecture \ref{conj:twoqubitextension} for various classes of states.

Chapters \ref{ch:bosonicfermionic} and \ref{ch:degradable} deal with concepts related to symmetric extension. 
In chapter \ref{ch:bosonicfermionic} the relation to the more specialized concepts of \textindex{bosonic extension} and \textindex{fermionic extension} was analyzed. 
In particular, symmetric extension and bosonic extension is equivalent for systems of dimension $N \times 2$.
In chapter \ref{ch:degradable} we converted our results about states with a symmetric extension to results about degradable and antidegradable channels.
The main result here was corollary \ref{cor:degradableenvlimit} which states that a degradable channel with qubit output has an environment dimension less than or equal to 2.

With chapter \ref{ch:lad} we enter part \ref{part:qkd} which is the QKD part of this thesis.
This chapter defined the class of two-way postprocessing called \textindex{linear advantage distillation} as advantage distillation schemes based on a linear error correcting code and we used the parity check matrix to specify a particular scheme.
We characterized which parity check matrices are equivalent and some ways that a parity check matrix could be reduced to simpler parity check matrices.
The final section specified how a parity check matrix on systematic form is implemented on quantum bits when the measurement of the final key bits are deferred to the end of the protocol.

The final three chapters have a common structure.
They all started with the best possible states for which we do not know how to distill a secret key.
Then the form of the state after advantage distillation was computed.
Finally, it was shown that such a state has a symmetric extension.
This means that the advantage distillation step has failed to produce a state without a symmetric extension, making it impossible for the remaining one-way processing to distill a secret key.

In chapter \ref{ch:rcad} we started by showing that all advantage distillation schemes which announce so much information that only two strings remain as compatible raw strings, are all equivalent, and we call this 
\textindex{repetition code advantage distillation} (RCAD).
We here defined the quantity $D_C$\index{D@$D_C$} such that the states that we do not know how to distill are characterized by $D_C \leq 0$.
We then showed that RCAD on a block of states with $D_C \leq 0$ cannot produce a state with $D_C \geq 0$.
Finally, we showed that those states with $D_C \leq 0$ all have a symmetric extension.

The next logical step after analyzing RCAD, which uses a linear error correcting code with one encoded bit, was to check linear advantage distillation with two encoded bits, which we did in chapter \ref{ch:k2analytical}.
While key distillation from any Bell-diagonal starting state was considered in chapter \ref{ch:rcad}, we specialized to isotropic starting states in chapter \ref{ch:k2analytical}.
This means that the results here are not directly applicable to the BB84 protocol or the six-state protocol where we record the QBER of each basis, only the six-state protocol where we record the average QBER.
The chapter starts by identifying the form of the state after advantage distillation. 
Even though there are many parity check matrices to consider, they have enough common structure that the resulting state can be parametrized using three angles (lemma \ref{lem:LAD2outform}).
Then the symmetric extension problem was simplified for this class of states to only use a matrix of dimension $4 \times 4$.
Finally, an explicit $4 \times 4$ matrix was constructed.
From this matrix a full symmetric extension could be reconstructed.

After concluding that linear advantage distillation with two encoded bits cannot break a symmetric extension beyond the current threshold, the next steps were 3 and 4 encoded bits, which we considered in chapter \ref{ch:numerical}.
We started with an isotropic state like in chapter \ref{ch:k2analytical}, but in this case the different parity check matrices do not have enough common structure to allow us to parametrize the state after advantage distillation.
Instead, we let a computer run through all possible inequivalent parity check matrices for modest blocksizes.
For each parity check matrix the state after advantage distillation was computed and checked numerically for a symmetric extension.
The results showed that even though RCAD on the considered blocksize would be able to break a symmetric extension, none of the other parity check matrices could do that.
This indicates that of all the possible parity check matrices one may use, RCAD is the best option for advantage distillation, so other parity check matrices will not improve the known threshold.

The main conclusion that can be drawn from part \ref{part:qkd} is that the natural generalization of the currently most successful advantage distillation technique does not improve the QBER threshold.
The results cannot exclude that linear advantage distillation can succeed in increasing the threshold for more than 2 encoded bits with large blocks, but the details of the numerical results give strong hints that RCAD is the best linear advantage distillation scheme.

\appendix

\cleardoublepage
\phantomsection
\addcontentsline{toc}{part}{Appendices}
\part*{Appendices}

\chapter{Connection with Maurer's original advantage distillation}
\label{app:adcorrespondence}
\markboth{Appendix \thechapter. Connection with Maurer's advantage distillation}{}

In the original advantage distillation proposal by Maurer \cite{maurer93a} Alice randomly selects a codeword $\mathbf{c_s}$ from an error correcting code (ECC), and sends $\mathbf{m} := \mathbf{c} \oplus \mathbf{d_A}$ to Bob (and Eve) over the public channel, where the addition is bitwise modulo 2. Bob accepts the received word only if he can make a very reliable decision about the codeword sent my Alice.
In practice, advantage distillation has often been limited to the case when the ECC is a repetition code and Bob only accepts if $\mathbf{m} \oplus \mathbf{d_B}$ is a valid codeword without errors  \cite{chabanne06a,acin06a,bae07a,kraus07a}.
In our approach we consider the more general case where the ECC is a binary linear code and Bob still accepts only if he detects no errors.

We now want to show that announcing $\mathbf{m} := \mathbf{c} \oplus \mathbf{d_A}$ reveals exactly the same information as announcing $\mathbf{a} = \mathbf{H} \mathbf{d_A}$, the parities of the data as defined by the parity check matrix $\mathbf{H}$.
The equivalence can be seen as a specialization of the following lemma.

\begin{lemma}
 Let Alice have an $n$-bit string $\mathbf{d}$, and a $k$-bit string $\mathbf{r}$.
 Consider a $(n,k)$ linear error correcting code with a $(n-k) \times n$ parity check matrix $\mathbf{H}$ and code words $\{\mathbf{c_i}\}_{\mathbf{i}=00\ldots0}^{11\ldots1}$ indexed by the $k$-bit binary strings. 
 The following two announcements reveal the same information about $\mathbf{d}$ and $\mathbf{r}$:\\
 1. Announce $\mathbf{r}$ and $\mathbf{a}$ where
 \begin{equation}
 \label{eq:computedparities}
  \mathbf{a} = \mathbf{H} \mathbf{d}.
 \end{equation}
 2. Let $\mathbf{c_j}$ be the codeword that gives $\mathbf{c_j} \oplus \mathbf{d}$ the lowest numerical value when interpreted as a binary number.
    Announce $\mathbf{m} := \mathbf{c_r} \oplus \mathbf{c_j} \oplus \mathbf{d}$.
\end{lemma}
\begin{proof}
 We first show that given $\mathbf{m}$, we may compute $\mathbf{a}$ and $\mathbf{r}$.
 Since $\mathbf{Hc} = \mathbf{0}$ for any valid codeword $\mathbf{c}$, we have 
 $\mathbf{Hm} = \mathbf{Hc_r} \oplus \mathbf{Hc_j} \oplus \mathbf{Hd} 
 = \mathbf{a}$.
 We can now compute all the $2^k$ possible $\mathbf{d}_j$ compatible with $\mathbf{Hd}_j = \mathbf{a}$. 
 Let $\mathbf{d}_0$ be the one with lowest numerical value. 
 Since $\mathbf{d}_0 = \mathbf{c_j} \oplus \mathbf{d}$, we have $\mathbf{m} = \mathbf{c_r} \oplus \mathbf{d_0}$ and therefore $\mathbf{c_r} = \mathbf{m} \oplus \mathbf{d_0}$.
 Since we know the codeword $\mathbf{c_r}$, we also know its index string $\mathbf{r}$.
 
 Next, we show that given $\mathbf{r}$ and $\mathbf{a}$, we can compute $\mathbf{m}$.
 Again, we let $\mathbf{d}_0$ be the solution to $\mathbf{Hd} = \mathbf{a}$ which has lowest numerical value. 
 Since $\mathbf{c_j}$ is defined such that $\mathbf{c_j} \oplus \mathbf{d} = \mathbf{d_0}$, we get 
 $\mathbf{m} = \mathbf{c_r} \oplus \mathbf{d_0}$, where both $\mathbf{c_r}$ and $\mathbf{d_0}$ are known.
\end{proof}

For the purpose of advantage distillation, the $k$-bit string $\mathbf{r}$ is chosen uniformly at random. 
In the original AD proposal, the announcement that is made is $\mathbf{c_s} \oplus \mathbf{d}$, where $\mathbf{c_s}$ is chosen uniformly at random. 
If this choice is done by computing $\mathbf{c_s} := \mathbf{c_r} + \mathbf{c_j}$, the choice of $\mathbf{c_s}$ is still uniform as long as $\mathbf{c_j}$ does not depend or $\mathbf{r}$, and here it only depends on $\mathbf{d}$.
In our approach, the random string $\mathbf{r}$ is never actually generated and announced, but a public announcement of uncorrelated random data does not influence anyone's knowledge about the data that Alice and Bob will later use for generating a secret key.
So with the exception that the original approach broadcasts $k$ unrelated bits of randomness, the two descriptions of the advantage distillation process are equivalent.

\chapter{Proofs of lemmas from chapter~\protect{\ref{ch:k2analytical}}}

This appendix collects the proofs of the lemmas in chapter \ref{ch:k2analytical}.
The proofs are somewhat tedious, but not very deep, and the main line of arguments can be followed in the main text.

\section{Proof of lemma \protect{\ref{lem:twoencodedsimp}}}
\label{proof:twoencodedsimp}

 The parity check matrix $\mathbf{H}$ used in any linear advantage distillation scheme can be converted into an equivalent one on systematic form $\mathbf{H} = [\mathbf{P}|\calone]$. 
 When $k = 2$, the parity matrix is a $(n-2) \times 2$ matrix, so all the rows in $\mathbf{P}$ are one of the 3 rows $01$, $10$, and $11$ 
 (any $00$ rows would mean that the corresponding raw bits are publicly revealed, which means that it is reducible to a parity matrix with those rows removed, see section \ref{sec:reducibleannouncements}).
 The parity matrix is equivalent under permuting the rows so the generic form of the systematic parity check matrix is
 \begin{equation}
 \mathbf{H} = \left[
\begin{array}{cc|cccccccccc}
 1      &    1   & 1 &   &  \\
 1      &    1   &   & 1 &   \\
 \vdots & \vdots &   &   & \ddots &\\
 1      &    1   &   &   &        & 1 &\\
 1      &    0   &   &   &        &   & 1 &\\
 \vdots & \vdots &   &   &        &   &   & \ddots &\\
 1      &    0   &   &   &        &   &   &        & 1 &\\
 0      &    1   &   &   &        &   &   &        &   & 1 &\\
 \vdots & \vdots &   &   &        &   &   &        &   &   & \ddots &\\
 0      &    1   &   &   &        &   &   &        &   &   &        & 1
 
\end{array}
\right].
\end{equation}
The parity check matrix is fully characterized by the number of different 2-bit strings in the parity matrix, $n_{11}$, $n_{10}$, and $n_{01}$.
Recall the form of the repetition code parity check matrix from Eq.~\eqref{eq:repetitionparitycheck}.
In the special case that $n_{11} = 0$ this is already case 1 in the lemma.
The $n_{01}$ last rows are then the repetition code parity check matrix on the second bit together with the last $n_{01}$ bits. 
The $n_{10}$ first rows is the same on bits 1 and from 3 to $n_{10}+2$. 

In the generic case, we have $n_{11} > 0$. 
We can then add the first row of $\mathbf{H}$ to all the other rows that start with $11$, if any. 
We then get the equivalent parity check matrix which is no longer on systematic form due to the 1 entries in the third column
 \begin{equation}
 \mathbf{H}_\textrm{nonsyst} = \left[
\begin{array}{ccc|ccccccccc}
 1      &    1   & 1      &   &  \\
 0      &    0   & 1      & 1 &   \\
 \vdots & \vdots & \vdots & &\ddots &\\
 0      &    0   & 1      &   &        & 1 &\\
 1      &    0   & 0      &   &        &   & 1 &\\
 \vdots & \vdots & \vdots &   &        &   &   & \ddots &\\
 1      &    0   & 0      &   &        &   &   &        & 1 &\\
 0      &    1   & 0      &   &        &   &   &        &   & 1 &\\
 \vdots & \vdots & \vdots &   &        &   &   &        &   &   & \ddots &\\
 0      &    1   & 0 &   &        &   &   &        &   &   &        & 1
 
\end{array}
\right].
\end{equation}
If we start announcing parities from the bottom row, we first announce the repetition code advantage distillation information on the block consisting of bit 2 together with the last $n_{01}$ bits, then the same information on bit 1 together with the next $n_{10}$ bits, and then the same information on bits 3 to $n_{11} + 2$. 
After this announcement, all but the first 3 bits can be discarded, since their values can anyway be inferred from the announcement and the first three bits.
Finally, the first row specify that we announce the parity of those three bits.
In other words, this is case 2 in the lemma.

\section{Proof of lemma \ref{lem:RCADoutform}}
\label{proof:RCADoutform}

On a general Bell-diagonal state,
the state after repetition code advantage distillation is characterized by the new eigenvalues from Eqs.~\eqref{eq:RCADstatefirst}-\eqref{eq:RCADstatelast}.
\begin{subequations}
\begin{align}
\label{eq:firstrecursion}
p_I^{(n)} &= \frac{1}{2}\left( (p_I + p_z)^n + (p_I - p_z)^n\right),\\ 
p_z^{(n)} &= \frac{1}{2}\left( (p_I + p_z)^n - (p_I - p_z)^n\right),\\ 
\label{eq:xrecursion} 
p_x^{(n)} &= \frac{1}{2}\left( (p_x + p_y)^n + (p_x - p_y)^n\right),\\
p_y^{(n)} &= \frac{1}{2}\left( (p_x + p_y)^n - (p_x - p_y)^n\right),
\label{eq:lastrecursion}
\end{align}
\end{subequations}
where $n$ is the block size and the state is normalized such that its trace is the success probability of the advantage distillation step.

The starting state with $p_x = p_y = p_z = (5-\sqrt{5})/20,$ $p_I = 1 -p_x - p_y - p_z$ satisfies 
\begin{equation}
\label{eq:chau-distillability-zero}
 D_C = \log_2 \left(  \frac{(p_I - p_z)^2}{(p_I + p_z)(p_x + p_y)} \right) = 0.
\end{equation}
As shown in section \ref{sec:stateafterrcad}, the value of this parameter after RCAD is $D_C^\text{RCAD} = n D_C = 0$.
The set of states with $D_C = 0$ is those states that satisfy
\begin{equation} 
\label{eq:chau-border}
(p_I - p_z)^2 = (p_I + p_z)(p_x + p_y),
\end{equation}
With the parameters we defined in Eqs. \eqref{eq:alpha-transformation-first}-\eqref{eq:alpha-transformation-last},
\begin{subequations}
\begin{align}
 \label{eq:alpha-transformation-proof-first}
 \alpha_0 &= p_I + p_x + p_y + p_z = 1,\\
 \label{eq:alpha-transformation-proof-alpha1}
 \alpha_1 &= p_I - p_x - p_y + p_z,\\
 \label{eq:alpha-transformation-proof-alpha2}
 \alpha_2 &= \sqrt{2} ( p_I - p_z ),\\
 \label{eq:alpha-transformation-proof-last}
 \alpha_3 &= \sqrt{2} ( p_x - p_y ).
\end{align}
\end{subequations}
the $D_C = 0$ condition translates into
\begin{equation}
\label{eq:chauzero}
 \alpha_1^2 + 2 \alpha_2^2 = \alpha_0^2 = 1
\end{equation}
which is the equation for an an ellipse with center in $(\alpha_1, \alpha_2) = (0,0)$, $\alpha_1$-semiaxis $1$, and $\alpha_2$-semiaxis $1/4$.
This ellipse can be parametrized as
\begin{subequations}
 \begin{align}
 \label{eq:alpha1param}
  \alpha_1 &= \cos \theta\\
  \label{eq:alpha2param}
  \alpha_2 &= \tfrac{1}{\sqrt{2}} \sin \theta
 \end{align}
\end{subequations}
The initial state also has $p_x = p_y$ and from Eqs.~\eqref{eq:xrecursion}-\eqref{eq:lastrecursion} we see that this property is also conserved after RCAD.
Putting together $p_x = p_y$ and $p_I + p_x + p_y + p_z = 1$ with Eqs.~\eqref{eq:alpha-transformation-proof-alpha1}-\eqref{eq:alpha-transformation-proof-alpha2} and Eqs.~\eqref{eq:alpha1param}-\eqref{eq:alpha2param}, we get the parametrization \eqref{eq:RCADoutparametrizationI}-\eqref{eq:RCADoutparametrizationZ} from the lemma.

The next step is to verify the values for $\cos \theta$ given in the lemma.
It is straightforward to insert $\cos \theta = \frac{1}{\sqrt{5}}$ and $\sin \theta = \sqrt{1 - \cos^2 \theta} = \frac{2}{\sqrt{5}}$ and verify that this gives the initial state from Eq.~\eqref{eq:6stateisotropic} with $p = (5 - \sqrt{5})/20$.
For the state after RCAD, we rewrite Eqs. \eqref{eq:firstrecursion}-\eqref{eq:lastrecursion} in terms of $\alpha_i$ and get
\begin{subequations}
\begin{align}
\label{eq:firstrecursionalpha}
\alpha_0^{(n)} &= \frac{1}{2^n}\left( (\alpha_0 + \alpha_1)^n + (\alpha_0 - \alpha_1)^n\right),\\ 
\alpha_1^{(n)} &= \frac{1}{2^n}\left( (\alpha_0 + \alpha_1)^n - (\alpha_0 - \alpha_1)^n\right),\\ 
\frac{\alpha_2^{(n)}}{\sqrt{2}} &= \left( \frac{\alpha_2}{\sqrt{2}} \right)^n,\\ 
\frac{\alpha_3^{(n)}}{\sqrt{2}} &= \left( \frac{\alpha_3}{\sqrt{2}} \right)^n.
\label{eq:lastrecursionalpha}
\end{align}
\end{subequations}
For the starting state, we have $\alpha_0 = 1$ and $\alpha_1 = 1-4p = 1/\sqrt{5}$.
For blocksize $n$ we then have 
\begin{equation}
 \cos \theta = \frac{\alpha_1^{(n)}}{\alpha_0^{(n)}}  
 = \frac{(\alpha_0 + \alpha_1)^n + (\alpha_0 - \alpha_1)^n}{(\alpha_0 + \alpha_1)^n + (\alpha_0 - \alpha_1)^n} = \frac{1 - x^n}{1 + x^n}
\end{equation}
for $x = (3 - \sqrt{5})/2 \approx 0.38 < 1$.
Inserting $n=2$ and $n=3$ gives $\cos \theta = \frac{\sqrt{5}}{3}$ and $\cos \theta = \frac{2}{\sqrt{5}}$, respectively.
We also note that $(1 - x^n)/(1 + x^n)$ increases when $n$ increases for $0 < x < 1$, so for $n > 3$ we will have $\cos \theta >  \frac{2}{\sqrt{5}}$.

\section{Proof of lemma \ref{lem:threeparityoutput}}
\label{proof:threeparityoutput}

The indices of the eigenvalues indicate which error pattern would result in the corresponding eigenvector, e.g. $p_{Iyx}$ indicates no error on qubit $B_1$, a $\sigma_y$ on qubit $B_2$, and a $\sigma_x$ on qubit $B_3$.
We assume that all errors happen on Bob's side and regard Alice's outcomes as correct.
We can split the errors into bit errors and phase errors. 
A $\sigma_x$ is a bit error without a phase error, $\sigma_y$ is both bit and phase error, and $\sigma_z$ is a phase error without a bit error.

Following the implementation in section \ref{sec:quantumimplementation}, Alice and Bob could use the circuit 
\begin{center}
\rotatebox{-90}{
\Qcircuit @C=1em @R=1em {
& \targ    & \targ   & \measureD{\vphantom{M}} \\
& \qw      &\ctrl{-1}& \qw  \\
& \ctrl{-2}& \qw     & \qw 
}
}
\end{center}
to implement the measurement corresponding to the parity check matrix 
$\begin{bmatrix} 1 & 1 & 1 \end{bmatrix}$,
on their three qubits, so that the qubits $A_3$ and $B_3$ are measured while  $A_1$, $A_2$, $B_1$, and $B_2$ remain.
Bit errors propagate from the two left qubits $B_1$ and $B_2$ to the right qubit $B_3$ that is subsequently measured and phase errors propagate from $B_3$ to both $B_1$ and $B_2$.
There are four ways of ending up with any specific error pattern on $B_1$ and $B_2$---one for each possible Pauli error on $B_3$.
Two of those possibilities will leave a bit error on $B_3$ before measurement. 
In that case the remaining qubits will be discarded, so only two of the error possibilities on $B_3$ give rise to any specific error pattern on $B_1$ and $B_2$ after postselection.
For example, in order to have no errors on the two left qubits before the measurement, the four possibilities are 
$\calone^{B_1} \otimes \calone^{B_2} \otimes \calone^{B_3}$, 
$\calone^{B_1} \otimes \calone^{B_2} \otimes \sigma_x^{B_3}$, 
$\sigma_z^{B_1} \otimes \sigma_z^{B_2} \otimes \sigma_y^{B_3}$, and 
$\sigma_z^{B_1} \otimes \sigma_z^{B_2} \otimes \sigma_z^{B_3}$, which has the probability
$p_{III} + p_{IIx} + p_{zzy} + p_{zzz}$.
However, in the cases where there is a bit error ($\sigma_x$ or $\sigma_y$) on $B_3$, it will stay there since there is an even number of bit errors (i.e. 0) on $B_1$ and $B_2$, so the probability that there are no bit- or phase errors \emph{and} the postselection succeeds is $p_{II}^\textrm{out} = p_{III} + p_{zzz}$.
The pattern is the same for all output probabilities that have an even number of bit errors, 
$p_{ij}^\textrm{out} = p_{ijI} + p_{mnz}$, where $(m,n)$ has the same bit error pattern as $(i,j)$ are, but the opposite phase error pattern.
For output probabilities with an odd number of bit errors, the pattern is similar, only that $I$ on qubit $B_3$ is swapped with $x$ and $z$ is swapped with $y$, i.e. there must be a pre-existing bit error on $B_3$ for it to have no bit error after the CNOTs so that it survives postselection.
This gives the relations that are given in the lemma.

\section{Proof of lemma \ref{lem:LAD2outform}}
\label{proof:LAD2outform}

As already mentioned in section \ref{sec:quantumimplementation}, the states after linear advantage distillation are still multi-qubit Bell-diagonal.
This means that the state is invariant under $\sigma_x \otimes \sigma_x$ and $\sigma_z \otimes \sigma_z$ on each qubit pair. 
In this case we have only two qubit-pairs.
On each side we define the new basis
$\ket{0}_\text{local} := \ket{00}$, 
$\ket{1}_\text{local}:= \ket{01}$, 
$\ket{2}_\text{local} := \ket{10}$, 
and $\ket{3}_\text{local} := \ket{11}$ 
and the operators
$X_1 := \sigma_x \otimes \calone$,
$X_2 := \calone \otimes \sigma_x$,
$Z_1 := \sigma_z \otimes \calone$, and
$Z_2 := \calone \otimes \sigma_z$.
On the local 4-dimensional system, the two-qubit operators can be expressed as
\begin{subequations}
\begin{align}
X_1 = \ketbra{2}{0} + \ketbra{3}{1} + \ketbra{0}{2} + \ketbra{1}{3}, \\
X_2 = \ketbra{1}{0} + \ketbra{0}{1} + \ketbra{3}{2} + \ketbra{2}{3}, \\
Z_1 = \ketbra{0}{0} + \ketbra{1}{1} - \ketbra{2}{2} - \ketbra{3}{3}, \\
Z_2 = \ketbra{0}{0} - \ketbra{1}{1} + \ketbra{2}{2} - \ketbra{3}{3}.
\end{align}
\end{subequations}

The state is invariant under $Z_1 \otimes Z_1$ and $Z_2 \otimes Z_2$. 
Each of these two operators splits the 16-dimensional Hilbert space into an 8-dimensional positive eigenspace and an 8-dimensional negative eigenspace, so there are 4 joint eigenspaces: 
($++$) eigenspace spanned by $\ket{00}, \ket{11}, \ket{22}$, and $\ket{33}$, 
($+-$) spanned by $\ket{01}, \ket{10}, \ket{23}$, and $\ket{32}$, 
($-+$) spanned by $\ket{02}, \ket{13}, \ket{20}$, and $\ket{31}$, and 
($--$) spanned by $\ket{03}, \ket{12}, \ket{21}$, and $\ket{30}$.
Since any vector which is not on one of these joint eigenspaces will be changed by applying 
$Z_1 \otimes Z_1$ or $Z_2 \otimes Z_2$,
the invariant density matrix will be a direct sum of operators on these 4-dimensional subspaces.

The state is also invariant under $X_1 \otimes X_1$ and $X_2 \otimes X_2$.
These operations take a matrix element on one of the joint eigenspaces into another matrix element on the same joint eigenspace. 
The matrix elements therefore have to be equal.
This gives the following equality relations on each of the joint eigenspaces.
\begin{subequations}
\begin{align}
  \label{eq:equalmatrixelementsfirst}
  \rho_{00,00} = \rho_{11,11} = \rho_{22,22} = \rho_{33,33} = \tfrac{1}{4}( p_{II} + p_{zI} + p_{Iz} + p_{zz} )\\
  \rho_{00,11} = \rho_{11,00} = \rho_{22,33} = \rho_{33,22} = \tfrac{1}{4}( p_{II} + p_{zI} - p_{Iz} - p_{zz} )\\
  \rho_{00,22} = \rho_{11,33} = \rho_{22,00} = \rho_{33,11} = \tfrac{1}{4}( p_{II} - p_{zI} + p_{Iz} - p_{zz} )\\
  \rho_{00,33} = \rho_{11,22} = \rho_{22,11} = \rho_{33,00} = \tfrac{1}{4}( p_{II} - p_{zI} - p_{Iz} + p_{zz} )
\end{align}
\begin{align}
  \rho_{02,02} = \rho_{13,13} = \rho_{20,20} = \rho_{31,31} = \tfrac{1}{4}( p_{xI} + p_{yI} + p_{xz} + p_{yz} )\\
  \rho_{02,13} = \rho_{13,02} = \rho_{20,31} = \rho_{31,20} = \tfrac{1}{4}( p_{xI} + p_{yI} - p_{xz} - p_{yz} )\\
  \rho_{02,20} = \rho_{13,31} = \rho_{20,02} = \rho_{31,13} = \tfrac{1}{4}( p_{xI} - p_{yI} + p_{xz} - p_{yz} )\\  
  \rho_{02,31} = \rho_{13,20} = \rho_{20,13} = \rho_{31,02} = \tfrac{1}{4}( p_{xI} - p_{yI} - p_{xz} + p_{yz} )
\end{align}
\begin{align}
  \rho_{01,01} = \rho_{10,10} = \rho_{23,23} = \rho_{32,32} = \tfrac{1}{4}( p_{Ix} + p_{zx} + p_{Iy} + p_{zy} )\\
  \rho_{01,10} = \rho_{10,01} = \rho_{23,32} = \rho_{32,23} = \tfrac{1}{4}( p_{Ix} + p_{zx} - p_{Iy} - p_{zy} )\\
  \rho_{01,23} = \rho_{10,32} = \rho_{23,01} = \rho_{32,10} = \tfrac{1}{4}( p_{Ix} - p_{zx} + p_{Iy} - p_{zy} )\\
  \rho_{01,32} = \rho_{10,23} = \rho_{23,10} = \rho_{32,01} = \tfrac{1}{4}( p_{Ix} - p_{zx} - p_{Iy} + p_{zy} )
\end{align}
\begin{align}
  \rho_{03,03} = \rho_{12,12} = \rho_{21,21} = \rho_{30,30} = \tfrac{1}{4}( p_{xx} + p_{yx} + p_{xy} + p_{yy} )\\
  \rho_{03,12} = \rho_{12,03} = \rho_{21,30} = \rho_{30,21} = \tfrac{1}{4}( p_{xx} + p_{yx} - p_{xy} - p_{yy} )\\
  \rho_{03,21} = \rho_{12,30} = \rho_{21,03} = \rho_{30,12} = \tfrac{1}{4}( p_{xx} - p_{yx} + p_{xy} - p_{yy} )\\
  \label{eq:equalmatrixelementslast}
  \rho_{03,30} = \rho_{12,21} = \rho_{21,12} = \rho_{30,03} = \tfrac{1}{4}( p_{xx} - p_{yx} - p_{xy} + p_{yy} )
\end{align}
\end{subequations}
The joint eigenspaces are subspaces with no bit errors, bit error on first qubit, bit error on second qubit, and bit error on both qubits, respectively.
The relation to the eigenvalues are also given by Eqs.~\eqref{eq:equalmatrixelementsfirst}-\eqref{eq:equalmatrixelementslast}.
We can use the relations \eqref{eq:firstthreeparityoutput}-\eqref{eq:lastthreeparityoutput} from lemma \ref{lem:threeparityoutput} and the fact that $p_{ijk}$ is a product distribution, $p_{ijk} = q_i r_j s_k$ to write the eigenvalues in terms of the eigenvalues of the three original Bell-diagonal states.
That gives us
 \begin{subequations}
  \begin{align}
    \rho_{00,00} = \rho_{11,11} = \rho_{22,22} = \rho_{33,33}
    &= \tfrac{1}{4} (q_I + q_z)(r_I + r_z)(s_I + s_z)\\
    \rho_{00,11} = \rho_{11,00} = \rho_{22,33} = \rho_{33,22}
    &= \tfrac{1}{4} (q_I + q_z)(r_I - r_z)(s_I - s_z)\\
    \rho_{00,22} = \rho_{11,33} = \rho_{22,00} = \rho_{33,11}
    &= \tfrac{1}{4} (q_I - q_z)(r_I + r_z)(s_I - s_z)\\
    \rho_{00,33} = \rho_{11,22} = \rho_{22,11} = \rho_{33,00}
    &= \tfrac{1}{4} (q_I - q_z)(r_I - r_z)(s_I + s_z)
  \end{align}
  \begin{align}
    \rho_{02,02} = \rho_{13,13} = \rho_{20,20} = \rho_{31,31}
    &= \tfrac{1}{4} (q_x + q_y)(r_I + r_z)(s_x + s_y)\\
    \rho_{02,13} = \rho_{13,02} = \rho_{20,31} = \rho_{31,20}
    &= \tfrac{1}{4} (q_x + q_y)(r_I - r_z)(s_x - s_y)\\
    \rho_{02,20} = \rho_{13,31} = \rho_{20,02} = \rho_{31,13}
    &= \tfrac{1}{4} (q_x - q_y)(r_I + r_z)(s_x - s_y)\\
    \rho_{02,31} = \rho_{13,20} = \rho_{20,13} = \rho_{31,02}
    &= \tfrac{1}{4} (q_x - q_y)(r_I - r_z)(s_x + s_y)
  \end{align}
  \begin{align}
    \rho_{01,01} = \rho_{10,10} = \rho_{23,23} = \rho_{32,32}
    &= \tfrac{1}{4} (q_I + q_z)(r_x + r_y)(s_x + s_y)\\
    \rho_{01,10} = \rho_{10,01} = \rho_{23,32} = \rho_{32,23}
    &= \tfrac{1}{4} (q_I + q_z)(r_x - r_y)(s_x - s_y)\\
    \rho_{01,23} = \rho_{10,32} = \rho_{23,01} = \rho_{32,10}
    &= \tfrac{1}{4} (q_I - q_z)(r_x + r_y)(s_x - s_y)\\
    \rho_{01,32} = \rho_{10,23} = \rho_{23,10} = \rho_{32,01}
    &= \tfrac{1}{4} (q_I - q_z)(r_x - r_y)(s_x + s_y)
  \end{align}
  \begin{align}
    \rho_{03,03} = \rho_{12,12} = \rho_{21,21} = \rho_{30,30}
    &= \tfrac{1}{4} (q_x + q_y)(r_x + r_y)(s_I + s_z)\\
    \rho_{03,12} = \rho_{12,03} = \rho_{21,30} = \rho_{30,21}
    &= \tfrac{1}{4} (q_x + q_y)(r_x - r_y)(s_I - s_z)\\
    \rho_{03,21} = \rho_{12,30} = \rho_{21,03} = \rho_{30,12}
    &= \tfrac{1}{4} (q_x - q_y)(r_x + r_y)(s_I - s_z)\\
    \rho_{03,30} = \rho_{12,21} = \rho_{21,12} = \rho_{30,03}
    &= \tfrac{1}{4} (q_x - q_y)(r_x - r_y)(s_I + s_z)
  \end{align}
 \end{subequations}

Now we can insert the form of the states before the last announced bit (from lemma \ref{lem:RCADoutform} with different angles)
 \begin{align*}
  q_I + q_z &= \frac{1}{2}(1+\cos \theta), & r_I + r_z &= \frac{1}{2}(1+\cos \phi), & s_I + s_z &= \frac{1}{2}(1+\cos \alpha),\\
  q_I - q_z &= \frac{1}{2} \sin \theta, & r_I - r_z &= \frac{1}{2} \sin \phi, & s_I - s_z &= \frac{1}{2} \sin \alpha,\\
  q_x + q_y &= \frac{1}{2}(1-\cos \theta), & r_x + r_y &= \frac{1}{2}(1-\cos \phi), & s_x + s_y &= \frac{1}{2}(1-\cos \alpha),\\
  q_x - q_y &= 0, & r_x - r_y &= 0, & s_x - s_y &= 0.
 \end{align*}
 This gives zero for the matrix elements that have $(q_x - q_y)$, $(r_x - r_y)$, or $(s_x - s_y)$ in their expression. 
 The other matrix elements are given by the expressions in the lemma.

\chapter{Sketch of generalized calculation for two encoded bits}

In chapter \ref{ch:rcad} we assumed that our starting state was any \textindex{Bell-diagonal state} on the threshold where the a secret key could no longer be distilled using RCAD.
We showed that after RCAD, the states always had a symmetric extension.
In chapter \ref{ch:k2analytical}, when considering LAD with two encoded bits, we restricted the starting state to being an \textindex{isotropic state} on the threshold.
This is fine for the six-state protocol where the error parameter is the average QBER in the three bases (see section \ref{sec:parameterestimation}). 
However, when the QBER in each basis is recorded in the \textindex{six-state protocol} or in the \textindex{BB84 protocol} where one parameter of the Bell-diagonal state has to be optimized over, the appropriate starting state cannot be assumed to be of this form.

An analogue calculation to the one in in chapter \ref{ch:k2analytical} can be done without the assumption of the isotropic starting state.
Only at the last step where we give analytic parameters for the simplified problem does this more general approach result in a more difficult problem.
In this appendix, we sketch how the calculation analogue to the one in chapter \ref{ch:k2analytical} goes when we do not assume an isotropic starting state.

\section{Reduction to starting state on the symmetric subspace}

The main advantage of using an isotropic starting state was that it satisfies $p_x = p_y$ (i.e. it is invariant under $S^\dagger \otimes S$) and that this property also holds for the state after RCAD.
This state can therefore be parametrized by a single parameter, which depends only on the blocksize used in the RCAD step. 
Instead, we here show that it is sufficient to start with states that satisfy $p_y = 0$, which means that their support is on the \textindex{symmetric subspace}, $P_{AB} \rho^{AB} = \rho^{AB}$.

We now assume that our starting state is any Bell-diagonal state,
\begin{equation}
  \label{eq:belldiagstateapp}
  \rho = p_I \proj{\Phi^+} + p_x \proj{\Psi^+} + p_y \proj{\Psi^-} + p_z \proj{\Phi^-},
\end{equation}
which satisfies $D_C = 0$ (see section \ref{sec:stateafterrcad}), which means that
\begin{equation} 
\label{eq:chau-border-app}
(p_I - p_z)^2 = (p_I + p_z)(p_x + p_y).
\end{equation}

Recall from lemma \ref{lem:twoencodedsimp} that linear advantage distillation steps with two encoded bits can be implemented by first performing RCAD on three subblocks followed by an announcement of the parity of the resulting bits.
In section \ref{sec:stateafterrcad} we showed that the state after RCAD also satisfies $D_C = 0$ when the original state does.
Any state on the $D_C = 0$ border satisfy 
\begin{subequations}
\label{eq:RCADoutparametrization-app}
 \begin{align}
\label{eq:RCADoutparametrizationI-app}
  p_I &= \tfrac{1}{4}(1 + \cos \theta + \sin \theta),\\
\label{eq:RCADoutparametrizationXY-app}
  p_x + p_y &= \tfrac{1}{2}(1 - \cos \theta)\\
\label{eq:RCADoutparametrizationZ-app}
  p_z &= \tfrac{1}{4}(1 + \cos \theta - \sin \theta),
 \end{align}
\end{subequations}
which is similar to lemma \ref{lem:RCADoutform}, but 
$p_x - p_y$ can be anything between $-\tfrac{1}{2}(1 - \cos \theta)$ and $\tfrac{1}{2}(1 - \cos \theta)$ (corresponding to $p_x = 0$ and $p_y = 0$, respectively).
We can write any such state as a convex combination of one that has $p_x = 0$ and one that has $p_y = 0$ (which is why we only needed to verify that we always had a symmetric extension for those states in section \ref{sec:cross-section-symext}).
This means that the final state is a convex combination of the eight states we get by choosing $p_x = 0$ or $p_y = 0$ in each of the three qubit pairs before announcing the final parity.
These states are all connected by local unitary operators. 
Therefore, if one of the states has a symmetric extension, then they all have one and any convex combination also has one.
It is therefore sufficient to construct a symmetric extension for the state that we get by starting with three copies of the state with $p_y = 0$ and announcing the parity of the three bits.
Instead of Eqs. \eqref{eq:RCADoutparametrizationI}--\eqref{eq:RCADoutparametrizationZ} from lemma \ref{lem:RCADoutform} we then get that the form of the Bell-diagonal state after RCAD is 
\begin{subequations}
\label{eq:RCADoutparametrization-symsub}
 \begin{align}
\label{eq:RCADoutparametrizationI-symsub}
  p_I &= \tfrac{1}{4}(1 + \cos \theta + \sin \theta),\\
\label{eq:RCADoutparametrizationX-symsub}
  p_x &= \tfrac{1}{2}(1 - \cos \theta),\\
\label{eq:RCADoutparametrizationY-symsub}
  p_y &= 0,\\
\label{eq:RCADoutparametrizationZ-symsub}
  p_z &= \tfrac{1}{4}(1 + \cos \theta - \sin \theta).
 \end{align}
\end{subequations}

From Eqs~\eqref{eq:firstthreeparityoutput}--\eqref{eq:lastthreeparityoutput} in lemma \ref{lem:threeparityoutput} we can see that when there is zero probability of having a $\sigma_y$ error on any of the three pairs in the starting state, all $p_{ij}^{out}$ where one of $i$ and $j$ (but not both) are $y$ also become zero. 
The vanishing terms are exactly those that have an eigenvector on the \textindex{antisymmetric subspace}, so the resulting state is supported only on the \textindex{symmetric subspace} (the eigenvector $\ket{\Psi^- \Psi^-}$ corresponding to the eigenvalue $p_{yy}^{out}$ is on the symmetric subspace, since each $\ket{\Psi^-}$ is antisymmetric and both subsystems pick up a minus sign upon swapping $A$ and $B$).

Similar to Eqs~\eqref{eq:firstABstate}--\eqref{eq:lastABstate} in lemma \ref{lem:LAD2outform}, we can write down the nonzero elements of the \textindex{density operator} parametrized by three angles, 
\begin{subequations}
 \begin{align}
  \label{eq:firstABstate-symsub}
  \rho_{00,00} = \rho_{11,11} = \rho_{22,22} = \rho_{33,33} 
  &= \frac{1}{32} (1+\cos \theta)(1+\cos \phi)(1+\cos \alpha)\\
  \rho_{00,11} = \rho_{11,00} = \rho_{22,33} = \rho_{33,22} 
  &= \frac{1}{32} (1+\cos \theta) \sin \phi \sin \alpha\\
  \rho_{00,22} = \rho_{11,33} = \rho_{22,00} = \rho_{33,11} 
  &= \frac{1}{32} \sin \theta (1+\cos \phi) \sin \alpha\\
  \rho_{00,33} = \rho_{11,22} = \rho_{22,11} = \rho_{33,00} 
  &= \frac{1}{32} \sin \theta \sin \phi (1+\cos \alpha)
 \end{align}
 \begin{align}
  \rho_{01,01} = \rho_{10,10} = \rho_{23,23} = \rho_{32,32} 
  &= \frac{1}{32} (1+\cos \theta)(1-\cos \phi)(1-\cos \alpha)\\
  \rho_{01,10} = \rho_{10,01} = \rho_{23,32} = \rho_{32,23} 
  &= \frac{1}{32} (1+\cos \theta)(1-\cos \phi)(1-\cos \alpha)\\
  \rho_{01,23} = \rho_{10,32} = \rho_{23,01} = \rho_{32,10} 
  &= \frac{1}{32} \sin \theta (1-\cos \phi)(1-\cos \alpha)\\
  \rho_{01,32} = \rho_{10,23} = \rho_{23,10} = \rho_{32,01} 
  &= \frac{1}{32} \sin \theta (1-\cos \phi)(1-\cos \alpha)
 \end{align}
 \begin{align}
  \rho_{02,02} = \rho_{13,13} = \rho_{20,20} = \rho_{31,31} 
  &= \frac{1}{32} (1-\cos \theta)(1+\cos \phi)(1-\cos \alpha)\\
  \rho_{02,20} = \rho_{13,31} = \rho_{20,02} = \rho_{31,13} 
  &= \frac{1}{32} (1-\cos \theta)(1+\cos \phi)(1-\cos \alpha)\\
  \rho_{02,13} = \rho_{13,02} = \rho_{20,31} = \rho_{31,20} 
  &= \frac{1}{32} (1-\cos \theta)\sin \phi(1-\cos \alpha)\\
  \rho_{02,31} = \rho_{13,20} = \rho_{20,13} = \rho_{31,02} 
  &= \frac{1}{32} (1-\cos \theta)\sin \phi(1-\cos \alpha)
 \end{align}
 \begin{align}
  \rho_{03,03} = \rho_{12,12} = \rho_{21,21} = \rho_{30,30} 
  &= \frac{1}{32} (1-\cos \theta)(1-\cos \phi)(1+\cos \alpha)\\
  \rho_{03,30} = \rho_{12,21} = \rho_{21,12} = \rho_{30,03} 
  &= \frac{1}{32} (1-\cos \theta)(1-\cos \phi)(1+\cos \alpha)\\
  \rho_{03,12} = \rho_{12,03} = \rho_{21,30} = \rho_{30,21} 
  &= \frac{1}{32} (1-\cos \theta)(1-\cos \phi)\sin \alpha\\
  \rho_{03,21} = \rho_{12,30} = \rho_{21,03} = \rho_{30,12} 
  &= \frac{1}{32} (1-\cos \theta)(1-\cos \phi)\sin \alpha
  \label{eq:lastABstate-symsub}
 \end{align}
\end{subequations}
Unlike the situation for an isotropic starting state, the only vanishing matrix elements are those that have to vanish because of the Bell-diagonal form of the state.

\section{Simplifying the symmetric extension problem}

As argued in section \ref{sec:simplifying-k2analytical}, the Bell-diagonal form of the states implies that we only need to construct the symmetric extension on the intersection of the positive eigenspaces of $\sigma_z^{A_i} \otimes \sigma_z^{B_i} \otimes \sigma_z^{B'_i}$, spanned by the vectors listed in Eq.~\eqref{eq:basisorder}.
The resulting matrix with the constraints is illustrated in figure \ref{fig:16x16constraints} on page \pageref{fig:16x16constraints}, and is the same as for the case with isotropic starting state, except that in this case the circular symbols do not represent constraints that sum to zero.

Since we cannot in this case set most of the off-diagonal matrix elements to zero, it looks harder to simplify this matrix.
However, since the state $\rho^{AB}$ resides on the \textindex{symmetric subspace}, we are guaranteed that if a symmetric extension exist, also a symmetric extension on the \textindex{fully symmetric subspace} exist. 
This means that $P_{AB} \rho^{ABB'} = P_{AB'} \rho^{ABB'} = P_{BB'} \rho^{ABB'} = \rho^{ABB'}$.
The one-sided swap operator swaps only the rows in $\rho^{ABB'}$.
The last six rows of the matrix correspond to the vectors $\ket{123}$, $\ket{231}$, $\ket{312}$, $\ket{132}$, $\ket{213}$, $\ket{321}$ and the three swap operators will swap different pairs of rows such that all the six last rows have to be equal.
Since the matrix is Hermitian, the last six columns also have to be equal.
Similarly, the tuple of vectors $(\ket{110}, \ket{220}, \ket{330})$ can be changed into both $(\ket{101}, \ket{202}, \ket{303})$ and $(\ket{011}, \ket{022}, \ket{033})$, so all the $3 \times 3$ blocks in rows and columns 2--10 must be equal.
In total, this symmetry means that the $16 \times 16$ matrix must have the form
\begin{equation}
\left[
\begin{array}{c|ccc|ccc|ccc|ccc|ccc}
  B   &  a  &   b   &  c   & a  &   b   &  c & a  &   b   &  c   &  s  &  s  &  s  &  s  &  s  &  s\\
 \hline
  a   &  &              &  & &              & & &              & & k_1 & k_1 & k_1 & k_1 & k_1 & k_1\\
  b   &  & \mathbf{M}_3 &  & & \mathbf{M}_3 & & & \mathbf{M}_3 & & k_2 & k_2 & k_2 & k_2 & k_2 & k_2\\
  c   &  &              &  & &              & & &              & & k_3 & k_3 & k_3 & k_3 & k_3 & k_3\\
 \hline
  a   &  &              &  & &              & & &              & & k_1 & k_1 & k_1 & k_1 & k_1 & k_1\\
  b   &  & \mathbf{M}_3 &  & & \mathbf{M}_3 & & & \mathbf{M}_3 & & k_2 & k_2 & k_2 & k_2 & k_2 & k_2\\
  c   &  &              &  & &              & & &              & & k_3 & k_3 & k_3 & k_3 & k_3 & k_3\\
   \hline
  a   &  &              &  & &              & & &              & & k_1 & k_1 & k_1 & k_1 & k_1 & k_1\\
  b   &  & \mathbf{M}_3 &  & & \mathbf{M}_3 & & & \mathbf{M}_3 & & k_2 & k_2 & k_2 & k_2 & k_2 & k_2\\
  c   &  &              &  & &              & & &              & & k_3 & k_3 & k_3 & k_3 & k_3 & k_3\\
   \hline
  s   & k_1 &  k_2  & k_3  & k_1&  k_2  & k_3& k_1&  k_2  & k_3  &  t  &  t  &  t  &  t  &  t  &  t \\
  s   & k_1 &  k_2  & k_3  & k_1&  k_2  & k_3& k_1&  k_2  & k_3  &  t  &  t  &  t  &  t  &  t  &  t \\
  s   & k_1 &  k_2  & k_3  & k_1&  k_2  & k_3& k_1&  k_2  & k_3  &  t  &  t  &  t  &  t  &  t  &  t \\
   \hline
  s   & k_1 &  k_2  & k_3  & k_1&  k_2  & k_3& k_1&  k_2  & k_3  &  t  &  t  &  t  &  t  &  t  &  t \\
  s   & k_1 &  k_2  & k_3  & k_1&  k_2  & k_3& k_1&  k_2  & k_3  &  t  &  t  &  t  &  t  &  t  &  t \\
  s   & k_1 &  k_2  & k_3  & k_1&  k_2  & k_3& k_1&  k_2  & k_3  &  t  &  t  &  t  &  t  &  t  &  t 
\end{array}
\right],
\end{equation}
where $\mathbf{M}_3$ is a $3 \times 3$ matrix. 
This matrix is positive semidefinite if and only if the $5 \times 5$ matrix on the subspace spanned by 
$\ket{000}$, $\ket{110}$, $\ket{220}, \ket{330}$, and $\ket{123}$,
\begin{equation}
\mathbf{M}_5 = 
\left[
\begin{array}{c|ccc|ccc|ccc|ccc|ccc}
  B   &  a  &   b   &  c   &  s\\
 \hline
  a   &  &              &  & k_1\\
  b   &  & \mathbf{M}_3 &  & k_2\\
  c   &  &              &  & k_3\\
 \hline
  s   & k_1 &  k_2  & k_3  &  t \\
\end{array}
\right],
\end{equation}
is positive semidefinite.
When the constraints on the $16 \times 16$ matrix is translated to the condensed matrix $\mathbf{M}_5$, they can be represented as
\begin{equation}
\label{eq:5x5constraints}
\left[
\begin{array}{c|ccc|c}
 \diamondsuit & \heartsuit & \spadesuit & \clubsuit &  \\
 \hline
 \heartsuit & (\diamondsuit\vartriangle) & \clubsuit & \spadesuit  & k_1\\
 \spadesuit & \clubsuit & (\diamondsuit\triangledown) & \heartsuit  & k_2\\
 \clubsuit & \spadesuit & \heartsuit & (\diamondsuit\Box) & k_3\\
           &    k_1     &    k_2     & k_3 & (\vartriangle \triangledown \  \Box)
\end{array}
\right].
\end{equation}
The parameters $k_1, k_2$, and $k_3$ are fixed by the constraints,
\begin{subequations}
\begin{gather}
 k_1 = \frac{1}{128} \sin \theta (1- \cos \phi) (1 - \cos \alpha),\\
 k_2 = \frac{1}{128} (1- \cos \theta) \sin \phi (1 - \cos \alpha),\\
 k_3 = \frac{1}{128} (1- \cos \theta) (1 - \cos \phi) \sin \alpha. 
\end{gather}
\end{subequations}
The other constraints on this subspace are
\begin{gather*}
 \label{eq:firstABBconstraint5x5}
 \rho_{000,000} + \rho_{110,110} + \rho_{220,220} + \rho_{330,330} 
 = \frac{1}{32} (1+\cos \theta)(1+\cos \phi)(1+\cos \alpha)\tag{$\diamondsuit$},\\
 \label{eq:heartconstraint5x5}
 \rho_{000,110} + \rho_{110,000} + \rho_{220,330} + \rho_{330,220} 
 = \frac{1}{32} (1+\cos \theta)\sin \phi \sin \alpha\tag{$\heartsuit$},\\
 \label{eq:spadeconstraint5x5}
 \rho_{000,220} + \rho_{110,330} + \rho_{220,000} + \rho_{330,110} 
 = \frac{1}{32} \sin \theta(1+\cos \phi)\sin \alpha\tag{$\spadesuit$},\\
 \label{eq:clubconstraint5x5}
 \rho_{000,330} + \rho_{110,220} + \rho_{220,110} + \rho_{330,000} 
 = \frac{1}{32} \sin \theta \sin \phi (1+\cos \alpha)\tag{$\clubsuit$},\\
 \label{eq:firstABBchangedconstraint}
 \rho_{110,110} + \rho_{123,123}  
 = \frac{1}{64} (1+\cos \theta)(1-\cos \phi)(1-\cos \alpha)\tag{$\vartriangle'$},\\
 \rho_{220,220} + \rho_{123,123}
 = \frac{1}{64} (1-\cos \theta)(1+\cos \phi)(1-\cos \alpha)\tag{$\triangledown'$},\\
 \label{eq:lastABBconstraint5x5}
 \rho_{330,330} + \rho_{123,123}
 = \frac{1}{64} (1-\cos \theta)(1-\cos \phi)(1+\cos \alpha)\tag{$\Box'$}.
\end{gather*}
The difference from the constraints on page \pageref{eq:firstABBconstraint4x4} is the three last constraints, \eqref{eq:firstABBchangedconstraint}--\eqref{eq:lastABBconstraint5x5}.
In this case, the matrix element $t = \rho_{123,123}$ cannot be any nonnegative value including zero, since the the elements $k_1 = \rho_{011,123}, k_2 = \rho_{022,123}$, and $k_3 = \rho_{033,123}$ are strictly positive.
In practice, $\rho_{123,123}$ can be small enough that this does not change the difficulty of fulfilling the constraints. 
The main difference is that the sum in these constraints is only half of what it was in chapter \ref{ch:k2analytical}.
This comes about since the matrix elements $\rho_{101,101}$, $\rho_{202,202}$, and $\rho_{303,303}$ no longer be can set to zero, since on the fully symmetric subspace these elements are equal to 
$\rho_{110,110}$, $\rho_{220,220}$, and $\rho_{330,330}$, respectively.
Since these constraints are on the diagonal matrix elements, this limits our possibility to choose a positive semidefinite $5 \times 5$ matrix of a nice form.

\section{Finding an explicit symmetric extension}

Even though we have reduced the problem to a $5 \times 5$ matrix with very similar constraints to the ones we were able to fulfill on the $4 \times 4$ matrix in section \ref{sec:explicitextension}, we have found no analytical parametrization that fulfills the linear constraints and yet have a nice enough form that positivity can be shown.
Still, we have found a procedure that will produce a matrix that fulfills the linear constraints and has produced a positive semidefinite matrix for the values of the parameters $\theta, \phi, \alpha$ that we have tested.
Unfortunately, the procedure does not allow the matrix elements to be expressed as an analytical function of $\theta, \phi$, and $\alpha$. 
The construction and verification of positivity has therefore only been done numerically.
The procedure expresses the matrix as 
\begin{equation}
 \mathbf{M}_5 = \proj{\psi_1} + \proj{\psi_2} + \mathbf{D},
\end{equation}
where $\mathbf{D}$ is a diagonal matrix and from this form it is evident that $\mathbf{M}_5$ is positive semidefinite if the matrix elements of $\mathbf{D}$ are nonnegative.
The procedure is the following.
\begin{description}
 \item[Step 1] Set $t = \rho_{123,123} := (1-\cos\theta)(1-\cos\phi)(1-\cos\alpha)/64$.
 \item[Step 2] Construct the column vector 
  \begin{equation*}
    \ket{\psi_1} := (0,k_1/\sqrt{t},k_2/\sqrt{t},k_3/\sqrt{t},\sqrt{t}),   
  \end{equation*}
 where $k_1, k_2$, and $k_3$ are fixed by the constraints.
 The matrix $\proj{\psi_1}$ now has the final values for the last row and column, so $\proj{\psi_2}$ and $\mathbf{D}$ will have 0 in those elements.
 \item[Step 3] Construct a column vector with nonnegative elements 
 \begin{equation*}
   \ket{\psi_2} := (\sqrt{B},v_1,v_2,v_3,0),
 \end{equation*}
which fulfill 
\begin{gather}
\begin{aligned}
 B = \frac{1}{64}[
      2  (1+\cos\theta)(1+\cos\phi)(1+\cos\alpha&)\\
    {} -   (1-\cos\theta)(1+\cos\phi)(1+\cos\alpha&)\\
    {} -   (1+\cos\theta)(1-\cos\phi)(1+\cos\alpha&)\\
    {} -   (1+\cos\theta)(1+\cos\phi)(1-\cos\alpha&)\\
    {} + 3 (1-\cos\theta)(1-\cos\phi)(1-\cos\alpha&)],
\end{aligned}\\
    \label{eq:psi2constraint1}
    \sqrt{B} v_1 + v_2 v_3 = \frac{1}{64} (1+\cos \theta) \sin \phi \sin \alpha - \frac{k_2k_3}{t},\\
    \label{eq:psi2constraint2}
    \sqrt{B} v_2 + v_1 v_3 = \frac{1}{64} \sin \theta (1+\cos \phi) \sin \alpha - \frac{k_1k_3}{t},\\
    \label{eq:psi2constraint3}
    \sqrt{B} v_3 + v_2 v_3 = \frac{1}{64} \sin \theta \sin \phi (1+\cos \alpha) - \frac{k_2k_3}{t}.
\end{gather}
This value of $B$ ensures that when $\rho_{110,110}$, $\rho_{220,220}$, and $\rho_{330,330}$ take values such that the diagonal constraints 
\eqref{eq:firstABBchangedconstraint}--\eqref{eq:lastABBconstraint5x5}
are satisfied, then also the diagonal constraint \eqref{eq:firstABBconstraint5x5} is satisfied.
The other conditions are such that $\proj{\psi_1} + \proj{\psi_2}$ satisfies the off-diagonal constraints \eqref{eq:heartconstraint5x5}, \eqref{eq:spadeconstraint5x5}, and \eqref{eq:clubconstraint5x5}.
  \item[Step 4] Construct a diagonal matrix $\mathbf{D} := \diag(0,d_1,d_2,d_3,0)$ where the three parameters are chosen such that the constraints
  \eqref{eq:firstABBchangedconstraint}--\eqref{eq:lastABBconstraint5x5}
  are satisfied by $\mathbf{M}_5 = \proj{\psi_1} + \proj{\psi_2} + \mathbf{D}$.
  If the three matrix elements $d_1$, $d_2$, and $d_3$ are nonnegative, the matrix $\mathbf{M}_5$ is positive semidefinite.
\end{description}
The part of this construction that cannot be done analytically is step 3. 
Solving for any of the unknown $v_i$ here gives an 8th order equation.
However, it is easy to construct $\proj{\psi_2}$ numerically by iteration.
Start with the matrix 
\begin{equation}
\begin{bmatrix}
 B & r_0 & s_0 & t_0 & 0\\
 r_0 & 0 & 0 & 0 & 0\\
 s_0 & 0 & 0 & 0 & 0\\
 t_0 & 0 & 0 & 0 & 0\\
 0 & 0 & 0 & 0 & 0\\
\end{bmatrix},
\end{equation}
where $r_0$, $s_0$, and $t_0$ are the shorthand for the RHS of Eqs. \eqref{eq:psi2constraint1},\eqref{eq:psi2constraint2}, and \eqref{eq:psi2constraint3}.
This matrix satisfies the linear constraints given by Eqs~\eqref{eq:psi2constraint1},\eqref{eq:psi2constraint2}, and \eqref{eq:psi2constraint3}, but it is not of rank one.
Next step is to update rows and columns 2--4 so that the matrix becomes of rank one.
This gives
\begin{equation}
\begin{bmatrix}
 B   & r_0        & s_0       & t_0        & 0\\
 r_0 & r_0^2/B    & r_0 s_0/B & r_0 t_0/B  & 0\\
 s_0 & r_0 s_0/B  & s_0^2/B   & s_0 t_0/B  & 0\\
 t_0 & r_0 t_0/B  & s_0 t_0/B & t_0^2/B    & 0\\
 0   & 0          & 0         & 0          & 0\\
\end{bmatrix}.
\end{equation}
This matrix is rank one, but does not fulfill the constraints.
The constraints can be fulfilled again my updating the first row and column to get 
\begin{equation}
\begin{bmatrix}
 B & r_0 - s_0 t_0/B & s_0 - r_0 t_0/B & t_0 - r_0 s_0/B  & 0\\
 r_0 - s_0 t_0/B & r_0^2/B    & r_0 s_0/B & r_0 t_0/B  & 0\\
 s_0 - r_0 t_0/B & r_0 s_0/B  & s_0^2/B   & s_0 t_0/B  & 0\\
 t_0 - r_0 s_0/B & r_0 t_0/B  & s_0 t_0/B & t_0^2/B    & 0\\
 0               & 0          & 0         & 0          & 0\\
\end{bmatrix}.
\end{equation}
This is the first correction to the original matrix. 
From this, the rows and columns 2--4 can be replaced with the values that give a rank-one matrix and the first row and column can be updated to fulfill the constraints again. 
This gives the iteration
\begin{gather}
 r_{i+1} = r_0 - \frac{s_i t_i}{B}\\
 s_{i+1} = s_0 - \frac{r_i t_i}{B}\\
 t_{i+1} = t_0 - \frac{r_i s_i}{B}
\end{gather}
Typically, $B$ is significantly larger than the parameters $r$, $s$, and $t$, so the iteration converges quickly.
The resulting matrix,
\begin{equation}
\begin{bmatrix}
 B             & r_\infty              & s_\infty             & t_\infty              & 0\\
 r_\infty      & r_\infty^2/B          & r_\infty s_\infty/B  & r_\infty t_\infty /B  & 0\\
 s_\infty      & r_\infty s_\infty /B  & s_\infty^2/B         & s_\infty t_\infty /B  & 0\\
 t_\infty      & r_\infty t_\infty /B  & s_\infty t_\infty /B & t_\infty ^2/B         & 0\\
 0             & 0                     & 0                    & 0                     & 0\\
\end{bmatrix},
\end{equation}
is both rank one and fulfills the constraints.

If we had had analytical expressions for $r_\infty$, $s_\infty$, and $t_\infty$, we would also be able to find analytical expressions for $d_1$, $d_2$, and $d_3$. 
By asking when those expressions are nonnegative, we would have had three analytical inequalities which when fulfilled would guarantee that we had a symmetric extension. 
It is possible that some other construction may give a provably positive semidefinite $\mathbf{M}_5$, but it seems that the constraints do not leave us as much freedom as we might need to allow for a simple construction.

\cleardoublepage

\backmatter

\providecommand{\urlprefix}{}
\newcommand{\doi}[1]{\href{http://dx.doi.org/#1}{\discretionary{}{}{}\nolinkurl{doi:#1}}}
\newcommand{\eprint}[2][]{\href{http://arxiv.org/abs/#2}{\nolinkurl{arXiv:#2}}}
\bibliography{journal_abbr,gomphd,gomphd-extra}
\bibliographystyle{gomphd}
\cleardoublepage

\printindex

\end{document}